\documentclass{myjpp}

\usepackage{amsmath}
\usepackage{bm}
\usepackage{makeidx}

\newcommand{\Rau}{R_\mathrm{AU}}
\newcommand{\amu}{a_{\mu\mathrm{m}}}
\newcommand{\bhat}{\hat{\bm{b}}}
\newcommand{\Ecm}{\bm{E}_\mathrm{cm}}
\newcommand{\oA}{\omega_\mathrm{A}}
\newcommand{\oH}{\omega_\mathrm{H}}
\newcommand{\tA}{\tau_\mathrm{A}}
\newcommand{\lH}{\ell_\mathrm{H}}
\newcommand{\Va}{V_\mathrm{A}}
\newcommand{\Vaz}{V_{\mathrm{A}z}}
\newcommand{\Vay}{V_{\mathrm{A}y}}
\newcommand{\bmid}{\beta_\mathrm{mid}}
\newcommand{\bmean}{\beta_\mathrm{mean}}
\newcommand{\Eni}{\bm{\mathcal{E}}_\mathrm{NI}}

\newcommand{\imp}[1]{#1}

\newcommand{\add}[1]{#1}

\usepackage{natbib}

\usepackage{comment}
\usepackage{booktabs}
\usepackage{graphicx}
\usepackage[figuresright]{rotating}
\usepackage{colortbl}
\usepackage{color}
\usepackage{wrapfig}
\usepackage{hyperref}
\hypersetup{%
  pdftitle    = {Dynamics of protoplanetary discs},
  pdfsubject  = {Habilitation a diriger des recherches},
  pdfkeywords = {protoplanetary discs},
  pdfauthor   = {Geoffroy Lesur},
  pdfcreator  = {\LaTeX\ with package \flqq hyperref\frqq},
  pdfproducer = {pdfeTeX-0.\the\pdftexversion\pdftexrevision},
}

\shorttitle{MHD of protoplanetary discs}
\shortauthor{G. R. J. Lesur}

\title{Magnetohydrodynamics of protoplanetary discs}

\author{Geoffroy R. J. Lesur\aff{1}
  \corresp{\email{geoffroy.lesur@univ-grenoble-alpes.fr}}}

\affiliation{\aff{1}Univ. Grenoble Alpes, CNRS, IPAG, 38000 Grenoble, France}

\makeindex

\begin{document}

\maketitle


\begin{abstract}
Protoplanetary discs are made of gas and dust orbiting a young star. They are also the birth place of planetary systems, which motivates a large amount of observational and theoretical research.

In these lecture notes, I present a review of the magnetic mechanisms applied to the outer regions ($R\gtrsim 1~\mathrm{AU}$) of these discs, which are the planet-formation regions. In contrast to usual astrophysical plasmas, the gas in these regions is noticeably cold ($T<300\,\mathrm{K}$) and dense, which implies a very low ionisation fraction close to the disc midplane. \add{In these notes, I deliberately ignore the innermost $(R\sim 0.1~\mathrm{A.U.})$ region which is influenced by the star-disk interaction and various radiative effects.
}. 

I start by presenting a short overview of the observational evidence for the dynamics of these objects. I then introduce the methods and approximations used to model these plasmas, including non-ideal MHD, and the uncertainties associated with this approach. In this framework, I explain how the global dynamics of these discs is modelled, and I present a stability analysis of this plasma in the local approximation, introducing the non-ideal magneto-rotational instability. Following this mostly analytical part, I discuss numerical models which have been used to describe the saturation mechanisms of this instability, and the formation of large-scale structures by various saturation mechanisms. Finally, I show that local numerical models are insufficient since magnetised winds are also emitted from the surface of these objects. After a short introduction on winds physics, I present global models of protoplanetary discs, including both a large-scale wind and the non-ideal dynamics of the disc. 

\end{abstract}

\setcounter{tocdepth}{3}
\tableofcontents

\newpage

\part{Observations \& physical description\label{partii}}

\section{Observational context\label{parti}}

Recent years have seen a dramatic change in our understanding of protoplanetary discs, both from an observational and a theoretical point of view. Observations are now able to resolve the outer regions (radii larger than 1 astronomical unit [a.u.]) and show the existence of many unexpected features: spiral arms, rings and crescent-like structures. Although these observations mostly probe the distribution of dust grains, they indicate that \imp{the gaseous structure of protoplanetary discs is much more complex and rich than initially anticipated}. In this part, we review the most recent evidence for proto-planetary disc structure and evolution, which can be used to constrain the most recent theoretical models.

\subsection{Observational diagnostics}

\begin{figure}[h!]
\centering
\includegraphics[width=1.0\hsize]{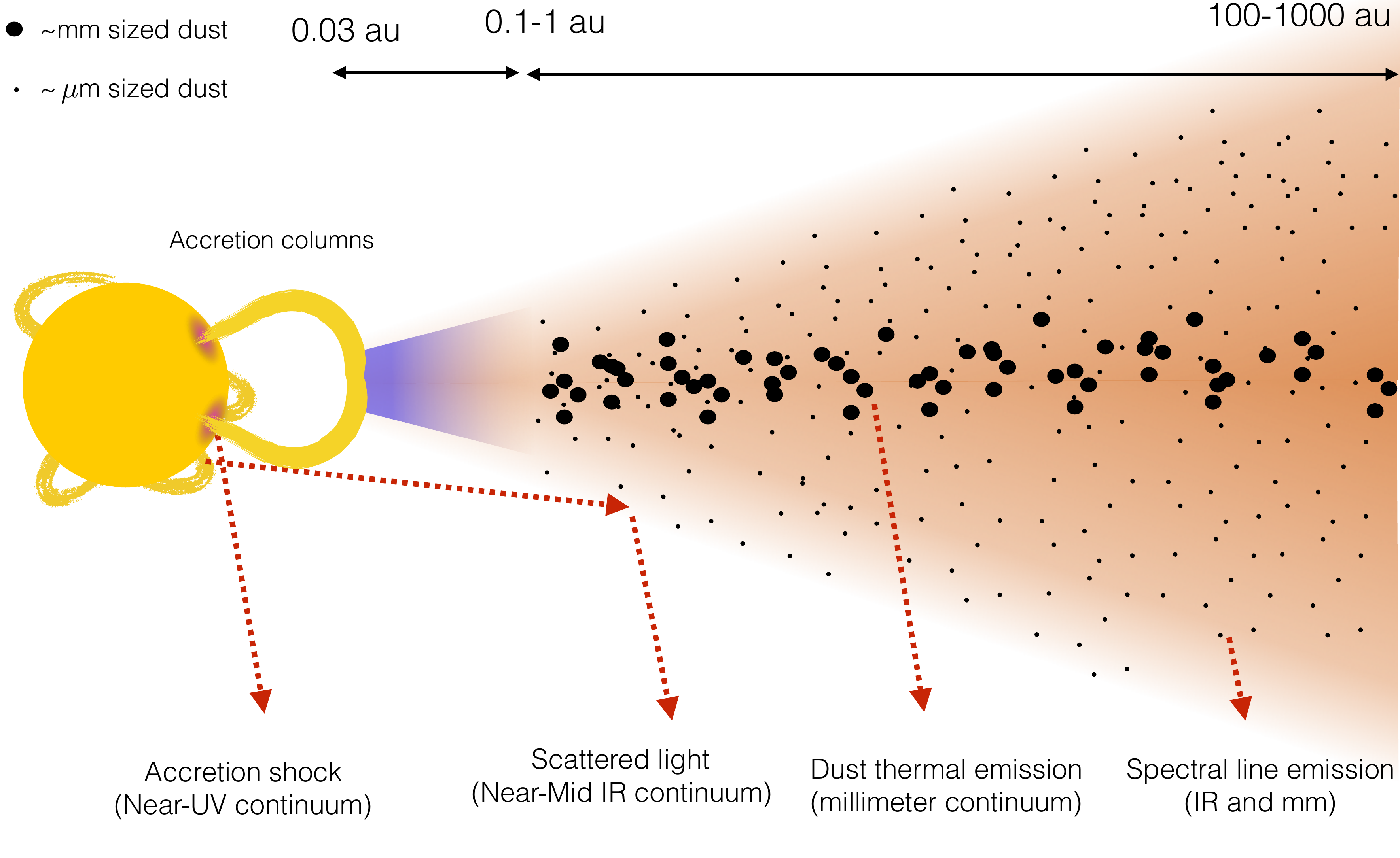}
\caption{Protoplanetary disc diagram showing the various observational diagnostics. Disc winds have been omitted for clarity.}
\label{fig:parti:bigpicture}
\end{figure}

Today observations probe different regions of the disc. In order to interpret these observations and constrain theoretical models, it is essential to clearly understand the quantities and limitations of each kind of observation. A typical protoplanetary disc can be separated into two parts: an inner dust-free disc (from a few stellar radii to the dust sublimation radius) made of hot gas (typically more than 1000 K) and an outer disc of gas and dust (Fig.~\ref{fig:parti:bigpicture}). The disc outer edge can range from 100 AU to more than 1000 AU depending on the object under consideration.

Observations typically probe the following regions:
\begin{itemize}
\item The UV excess is a signature of the accretion shock at the foot of accretion columns. It is very often the only way to deduce the accretion rate in a specific disc.
\item The near and mid-infrared continuum (also known as infrared excess) is due to stellar photons scattered by small dust grains (typically less than a $\mu \mathrm{m}$ in size). Scattered light probes the very surface of the dust layer as the dust disc is very optically thick at these wavelengths. For this reason, the intensity of scattered light is not related to the column density but to the amount of stellar light received by the layer. It therefore characterizes the disc \emph{geometry}.
\item The (sub)-millimetre continuum probes the thermal emission of bigger dust grains (typically with a size of the order of a mm). If the dust layer is optically thin at these wavelengths (as usually assumed), the emissivity is related to the column density of dust, but also to its temperature.
\item Spectral lines, both in the infrared and at radio wavelengths probe specific gas tracers such as gas or molecular transitions. These lines are usually optically thick which implies that they only probe the surface of the gas layer. For this reason, direct estimates of the gas mass in the disc is very difficult, and one has to rely on proxies.
\end{itemize}
These observational properties are then used to derive several useful dynamical quantities.
\subsection{Accretion\label{obs:acc}}
\index{Accretion!observations}
\begin{figure}
\centering
\includegraphics[width=0.45\hsize]{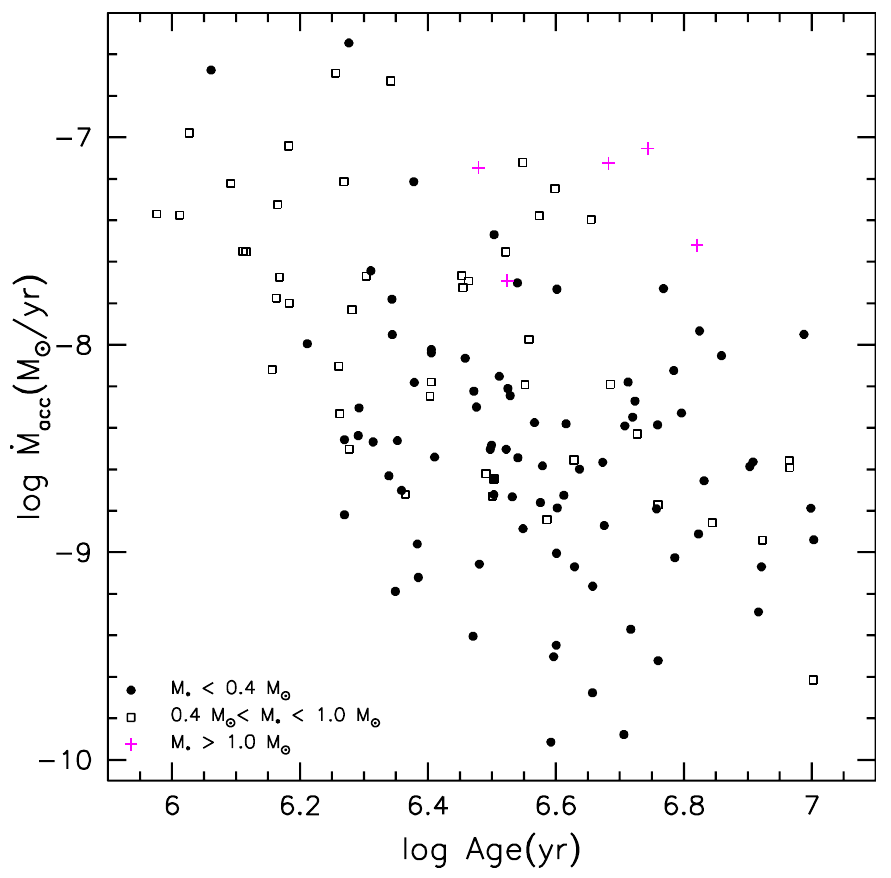}
\quad 
\includegraphics[width=0.45\hsize]{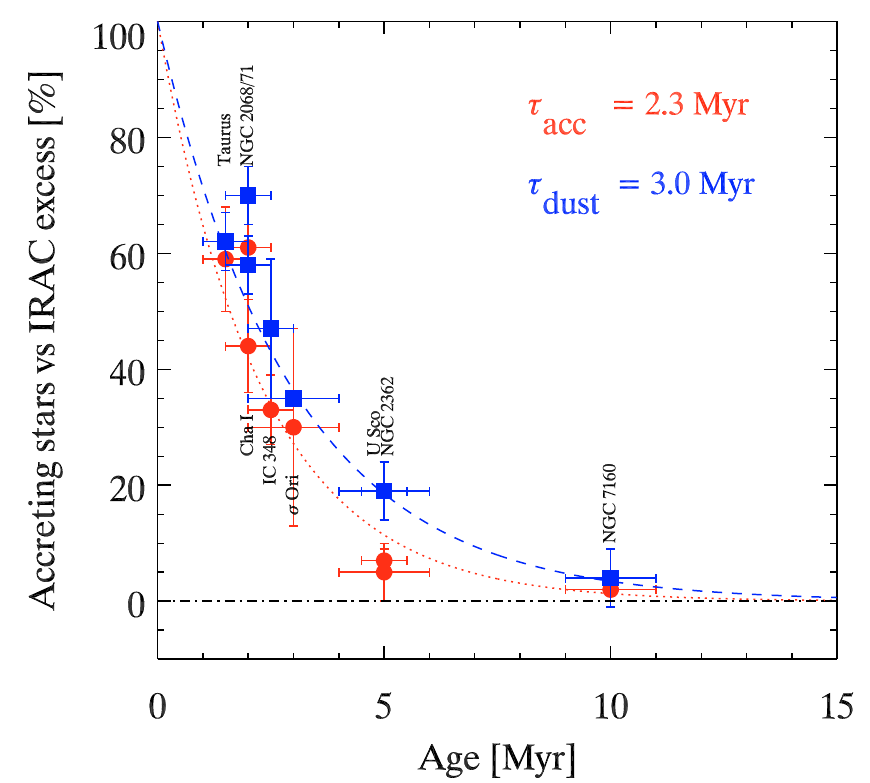}
\caption{(left) Measurement of the accretion rate as a function of stellar age in NGC 2264 using the excess UV due to accretion columns. From \cite{VB14}. \newline (right) Fraction of disc signature (accretion) and dust signature (infrared excess) as a function of the cluster age. Both show that discs have an average lifetime of a few million years. From \cite{FA10}.}
\label{fig:parti:mdot-age}
\end{figure}

Because the thermal equilibrium of protoplanetary discs for $R\gtrsim1\,\mathrm{A.U.}$ is dominated by the illumination of the central star \citep{A98},  
a direct measurement of the accretion rate through viscous heating is not possible. For this reason, observational evidence of accretion in these regions
are scarce and plagued by uncertainties. There are mainly two classes of accretion signature, which are all indirect. 

The first is the observational signature of accretion columns at the stellar surface. These accretion columns are formed when the disc material gets lifted up and accreted by the stellar magnetic field. The gas then ends up in a nearly free fall speed and hits the stellar surface, forming an accretion shock. The luminosity of this accretion shock observed in UV bands is directly related to the accretion rate in the accretion columns and therefore in the innermost disc. It should be kept in mind that accretion rates deduced by this method are not necessarily accretion rates in the entire disc, which can in principle vary with radius if the disc is not in steady state, or if the disc is losing mass from a wind. Typical results show accretion rates of the order of $10^{-8}\, M_\odot/\mathrm{yr}$ with uncertainties of the order of an order of magnitude depending on the object under consideration\footnote{Additional sources of uncertainties (not shown here) arise from the method used to reconstruct the mass accretion rate from the UV excess, and from the intrinsic accretion variability of the object \citep[e.g.][]{VB14}.} (e.g. Fig.~\ref{fig:parti:mdot-age}-left). These accretion rates tend to decrease over timescales of a few million years.

The second observational evidence lies in the proportion of stars showing disc features (accretion on the stellar surface, or infrared excess signifying the presence of dust around the star) as a function of the stellar age. The disappearance of these signatures in older stars allows one to evaluate the typical gas and dusty disc lifetimes. These two time scales do not necessarily match as the gas disc could, for instance, disappear before the dusty disc. However, they both show the same trend: disc tends to disappear on a timescale of a few million years (Fig.~\ref{fig:parti:mdot-age}-right).

By combining this information, and assuming that accretion is approximately constant during the lifetime of these objects, one deduces that typical protoplanetary disc masses range from $10^{-3}\,M_\odot$ to $10^{-1}\,M_\odot$, \add{which is consistent with mass inferred from the total dust content of the disc \citep{ARK13}}. 

\subsection{Ejection: winds and jets\label{obs:ej}}
\index{Jets}
\index{Winds}
\begin{figure}
\centering

\includegraphics[width=0.6\linewidth]{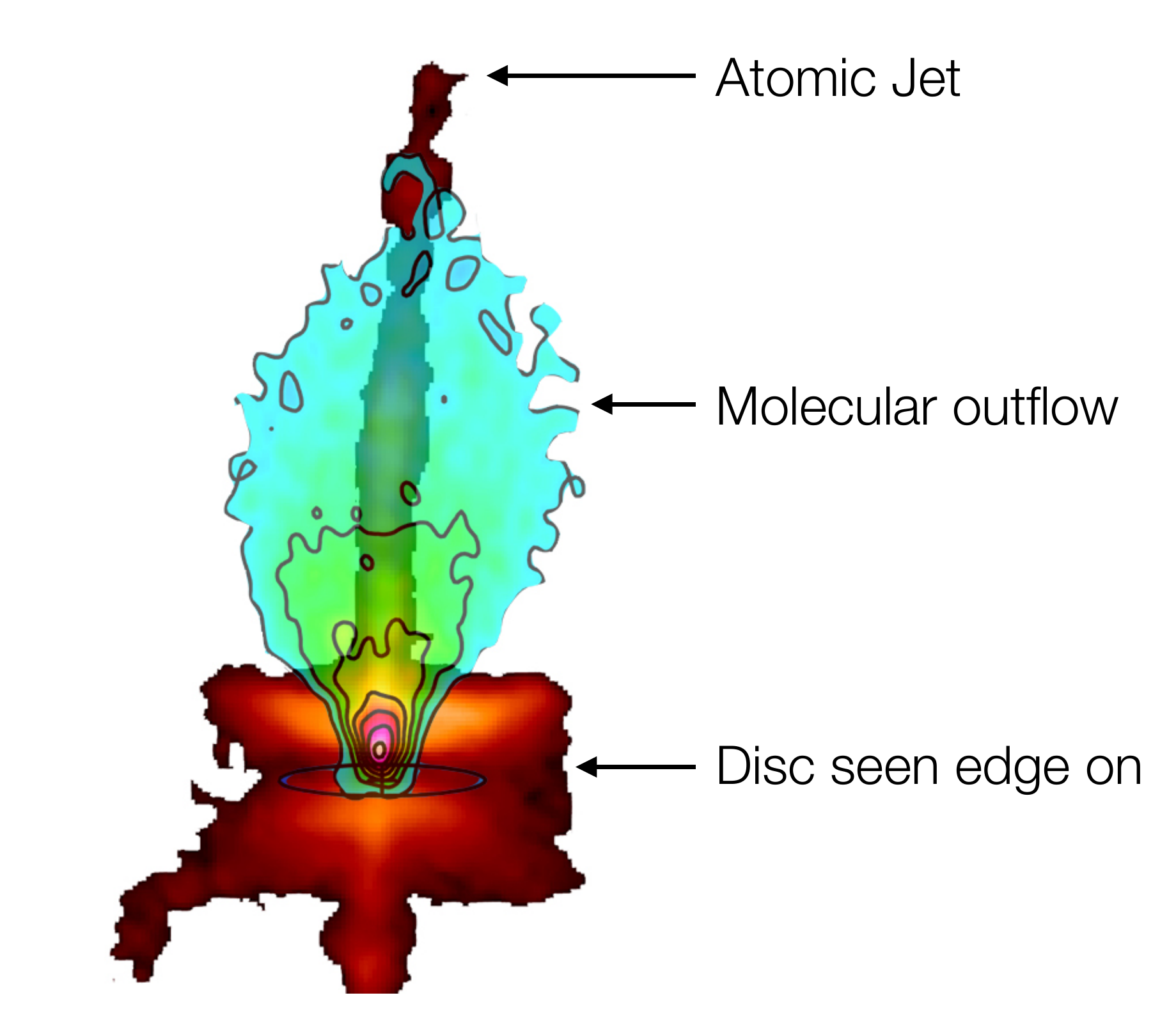}
\caption{Observation of a disc and an atomic jet seen by the Hubble Space Telescope \citep{BSW96} and a molecular wind observed in CO(2-1) by ALMA \citep{LD18} in HH30, a protoplanetary disc seen edge-on. Courtesy of F. Louvet.}
\label{fig:parti:winds}
\end{figure}

Protoplanetary discs are often observed in association with large-scale winds and jets. Jets are often seen in forbidden emission lines and correspond to fast collimated flow ($v>100\,\mathrm{km}.\mathrm{s}^{-1}$). Their high velocity suggests they are launched from the inner few AU of the disc \citep{FR14}. The typical outflow rate is estimated to be of the order of 10\% of the accretion rate in classical T-tauri stars \citep{FR14}.

In addition to these jets, a slower component is also observed in molecular lines. This ``molecular outflow'' is denser and reach velocities $v\sim 1-10 \,\mathrm{km}.\mathrm{s}^{-1}$ (Fig.~\ref{fig:parti:winds}). They could be due to the interaction of the jet with its environment, or they could be a genuine outflow component, emitted from the disc at $R\gtrsim 1\,\mathrm{A.U.}$.

\subsection{Structures\label{obs:struct}}

\begin{figure}
\centering
\includegraphics[width=0.45\hsize]{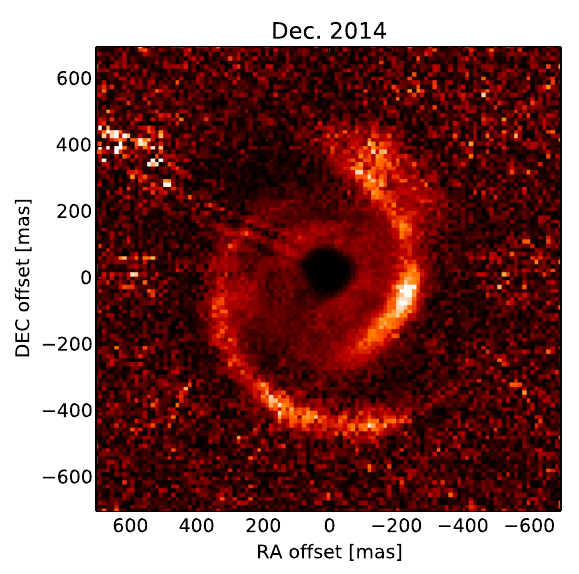}
\quad 
\includegraphics[width=0.45\hsize]{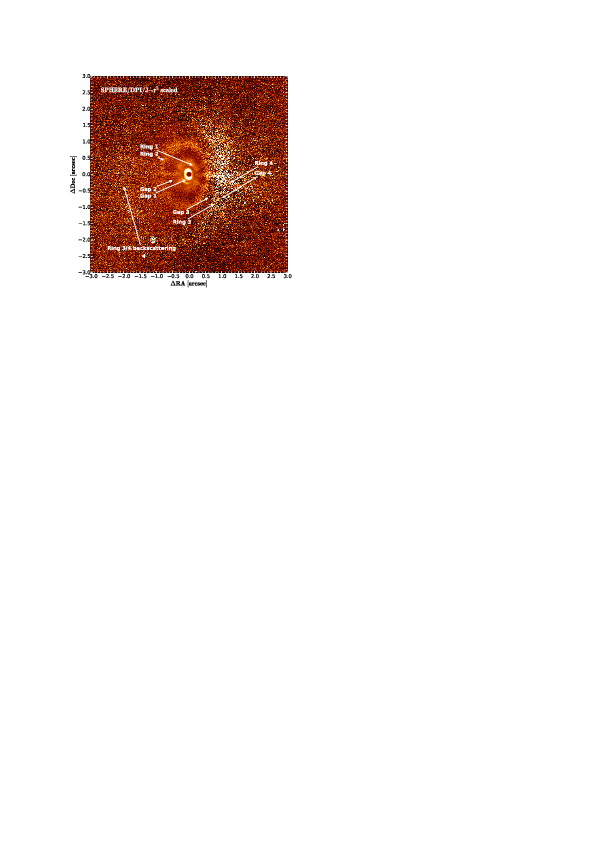}
\caption{Scattered light images in the near infrared using polarimetric differential imaging (PDI). Left: spiral structures observed in MWC758, from \cite{BJ15}. Right: multiple ring structures observed in HD97048, from \cite{GS16}.}
\label{fig:parti:scat-struct}
\end{figure}

The progress in observational techniques (adaptative optics, interferometry) now allows astronomers to resolve the disc and look for signatures of planet formation, accretion or other unexpected processes. The first class of observations relies on polarimetric differential imaging (PDI) of scattered light emission in the near infrared. This technique allows one to obtain only the light scattered by dust grains (which is naturally polarised) and not the light of the central object. They have been used to probe the disc surface of various disc (mainly transitional discs). Stunning structures such as spiral and rings were found\footnote{These structures trace the disc surface and not the column density} in several objects (Fig.~\ref{fig:parti:scat-struct}). 

\begin{figure}
\centering
\includegraphics[width=0.45\hsize]{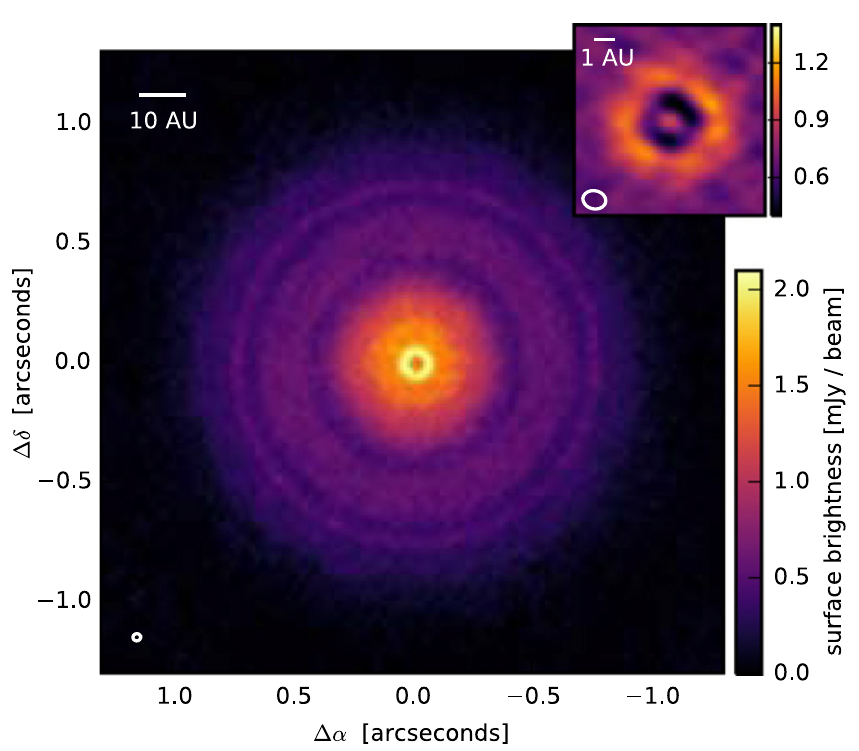}
\quad 
\includegraphics[width=0.45\hsize]{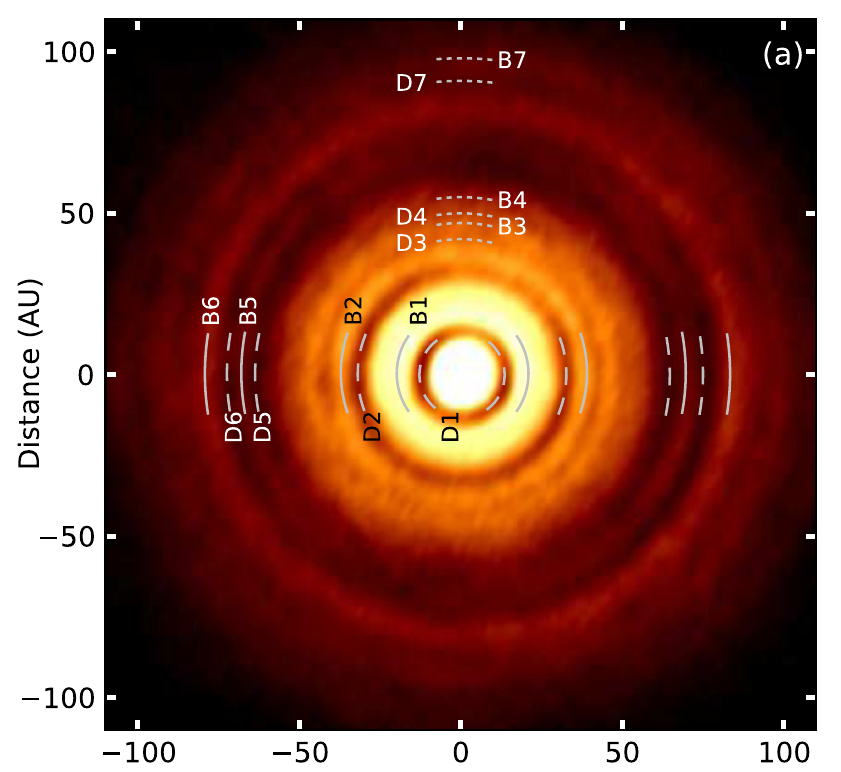}
\newline
\quad\newline
\includegraphics[width=0.45\hsize]{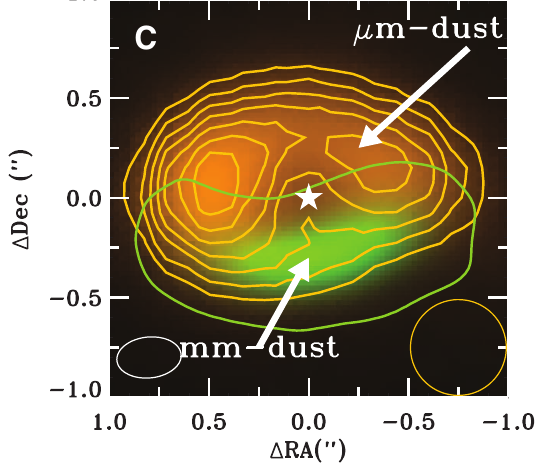}
\quad
\includegraphics[width=0.45\hsize]{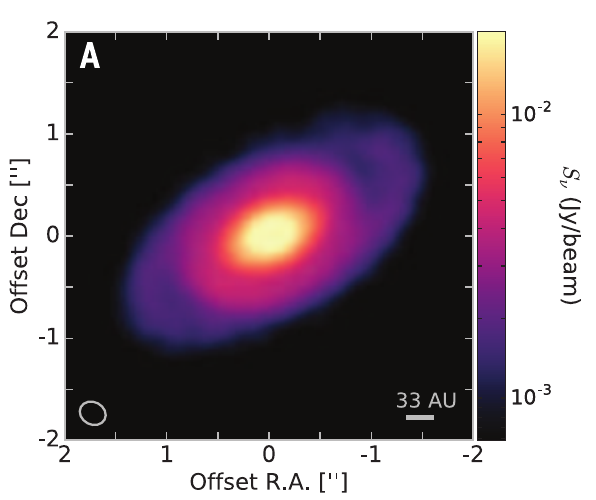}
\caption{Top left: Ring-like structures observed in TW Hydra. From \cite{AW16}. Top right: multiple ring structure in a deprojected image of HL-Tau from \cite{ALMA15}. Bottom left: horsehoe-like structure observed in Oph IRS 48 at sub-mm wavelengths (green, tracing mm-sized dust) and corresponding scattered light infrared emission (yellow, tracing $\mu\mathrm{m}$ size dust) from \cite{MD13}. Bottom right: spiral structures seen at sub-mm wavelengths in the young and massive disc of Elias 2-27, from \cite{PC16}}
\label{fig:parti:mm-struct}
\end{figure}

The second class of observation is based on interferometry at millimetric and submillimetric wavelengths. The ALMA observatory has been very successful at probing the very structure of protoplanetary discs with incredible resolution and unexpected results (Fig.~\ref{fig:parti:mm-struct}, \citealt{AH18}).

\index{Rings}
\index{Spiral}

\add{Although these observations probe the dust distribution in the disc, they also tell us about the gas distribution and dynamics, because the grains which are observed are tightly coupled to the gas through a drag force. Such a direct connection has been recently confirmed observationally by simultaneously looking at the continuum (dust) and line emissions (probing the gas kinematics) \citep{TBB19}. }

All these observations indicate that discs are not smooth and symmetrical. They are instead structured on length scales comparable to our solar system. Structures are categorised in spirals, rings and horseshoes, which can be associated with specific physical processes in the disc. It should be noted some of these structures are found in transitional discs, i.e. truncated discs which are presumably in the final evolution stage of protoplanetary discs. All of these structures could be the signature of embedded planets perturbing the disc structure by gravitational interaction. However, other processes have been proposed which do not assume planets. One of the key questions is therefore whether or not these structures are necessarily a signature of embedded planets. 
\subsection{Turbulence\label{obs:turb}}

\index{Turbulence!measure}
Turbulence is likely one of the key elements of any dynamical theory for the evolution of discs. Theoretical arguments (see \S \ref{sec:alpha-disc}) show that turbulence should be subsonic in these systems i.e. that chaotic motions of the gas are slower than the sound speed. This implies that turbulence is difficult to detect since the turbulent broadening of spectral lines is comparable to the thermal spreading of the molecules constituting the gas. For this reason, heavy molecules such as CN and CO tend to be preferred to detect turbulence, since their thermal velocity is lower compared to lighter molecules at a given equilibrium temperature. High-resolution spectra obtained from ALMA for CO lines indicates that turbulence is very weak, or inexistent \citep{FM15,FHR17}. Spectral broadening smaller than 3\% of the local sound speed are found as best fits to observational data at large distances (typically more than 30 au).  This turbulent broadening is way smaller than the typical values expected from ideal MHD turbulence which typically predicts $\delta v\gtrsim 0.1 c_s$.

Another signature of turbulence (or more precisely the lack of turbulence) lies in the dust vertical distribution. Indeed, dust grains naturally tend to settle towards the midplane, unless turbulence stirs them up into the disc atmosphere. Direct measurements of the thickness of the dust layer allow one to deduce the level of hydrodynamical turbulence in the disc. Such a measurement has been done in the case of HL-tau, where the thickness of the rings is used as a tracer for the disc thickness \citep{P16}. The result is that $100\,\mu \mathrm{m}$ grains have settled towards the midplane, with a vertical dust scale height about 10 times smaller than the gas scale height. This implies a very low level of turbulence in the disc, with typically $\delta v\sim 10^{-2}\,c_s$ ($\alpha \sim 10^{-4}$, see \S \ref{sec:alpha-disc}). 

\subsection{Magnetic fields\label{obs:mag}}
\index{Magnetic field!measure}
Evidence for magnetic fields in protoplanetary discs is scarce. Typical values are expected to be of the order of a Gauss at 1 au down to a few mG at a few tens of astronomical units \citep{W07}, although these theoretical values could vary by several orders of magnitude. For this reason, measurement through Zeeman effect is unfeasible except in the very inner disc. In this region, toroidal magnetic fields of a few kG have been measured, although it is not clear whether this field belongs to the host star or to the disc itself \citep{DP05}. \add{At larger distances (10s of AU), attempts at measuring the field strength through Zeeman splitting in molecular lines have only led to upper limits, with $B_z<0.8\mathrm{mG}$ and $B<30\mathrm{mG}$ \citep{VLC19}}.

Topological information on the field is also accessible through polarisation in the continuum (i.e. dust thermal emission). It is assumed that dust grains tend to align perpendicularly to magnetic field lines, thereby emitting thermal radiation with a preferred polarisation, perpendicular to the local field orientation \citep{CL07,SLK14}. However, polarisation in submillimetric radiation can also be due to self-scattering by dust grains \citep{KM15,YLL16}. Campaigns using multiple wavelengths observations have attempted to disentangle these two effects \citep{SYL17}, but the interpretation of the results in terms of magnetic topology remains very uncertain.

\add{Finally, magnetic field intensities can be deduced from meteoritic and cometary evidences in our own solar system, assuming that the field gets frozen in the body during its formation in the parent disc. Field strength of the order of $0.1\mathrm{G}$ around 1 AU are inferred from remnant magnetisation in meteorites following this idea \citep{FWL14}, while upper limits with $B<30\mathrm{mG}$ in the region around 15-45 AU is deduced from the magnetisation of Comet 67P/Churyumov-Gerasimenko \citep{BWH19}}.

\section{Disc prototype\label{sec:discprototype}}

\subsection{\label{sec:profile}Fluid properties}

Protoplanetary discs (PPDs) are rather cold objects, with temperatures ranging from 1000\,K in the inner ($0.1$ AU) disc down to 10\,K in the outer (100 AU) disc. In order to characterise these discs, It is important to quantify the typical length scales and time scales relevant to the problem. Let us start with a typical disc model which matches disc observations \citep{AWH09}:

\begin{align*}
\Sigma&=300\,\Rau^{-1}\,\mathrm{g.cm}^{-2}\\
T&=280\,\Rau^{-1/2}\,\mathrm{K}\\
\Omega&=2\times 10^{-7}\Rau^{-3/2}\,\mathrm{s}^{-1}
\end{align*}

\add{Here, we have defined the main physical properties of a disc: its surface density $\Sigma$, which correspond to the usual mass density integrated in the direction perpendicular to the disc plane, its temperature $T$, and its angular velocity $\Omega$ around the central object. We also define for convenience a dimensionless distance from the central object, in astronomical unit: $\Rau\equiv R/1\,\mathrm{AU}$.}
 
This simple model leads to a $0.04\,M_\odot$ mass disc, extending from 0.07 to 200 AU, rotating around a solar mass star, typical of discs which have been observed. We can deduce some useful dynamical parameters associated from this simplified models. Defining the isothermal sound speed as $c_s\equiv \sqrt{P/\rho}$ and using the vertical hydrostatic equilibrium to define the disc vertical scale height (\S\ref{sec:equilibrium}) $H=c_s/\Omega$, one gets:
\begin{align*}
c_s&=10^5\,\Rau^{-1/4}\,\mathrm{cm.s}^{-1}\\
H&=5\times 10^{11}\,\Rau^{5/4}\mathrm{cm}\\
\frac{H}{R}&=0.03\,\Rau^{1/4}\\
\rho_\mathrm{mid}&=6\times 10^{-10}\,\Rau^{-9/4}\,\mathrm{g.cm}^{-3}\\
n_\mathrm{mid}&=1.5\times 10^{14}\,\Rau^{-9/4}\,\mathrm{cm}^{-3}\\
P_\mathrm{mid}&=6\,\Rau^{-11/4}\,\mathrm{dyn.cm}^{-2}
\end{align*}

\subsection{Magnetic fields}
\index{Magnetic field!model}
Magnetic fields in PPDs are poorly constrained (\S\ref{obs:mag}). It is widely believed that fields are largely subthermal: the thermal pressure of the fluid dominates over the magnetic pressure (this requirement follows from the fact that the discs are approximately in Keplerian rotation). This translates into a plasma $\beta$ parameter
\index{$\beta$, plasma}
\begin{align*}
\beta&\equiv\frac{P_\mathrm{th}}{P_\mathrm{mag}}\\
&=\frac{8\pi P}{B^2}\gg 1.
\end{align*}
In practice, $\beta\simeq 1$ constitutes a lower limit for the MRI to operate in geometrically thin discs (see \ref{sec:linstrat}). Note also that if dynamo action is generating a field (both ordered or disordered), then $\beta$ cannot reach a value lower than $\beta\sim 1$, hence this value is actually a lower limit for the typical plasma $\beta$ expected in these discs. It is possible to connect the field strength to the plasma $\beta$ using the properties defined above and obtain:
\begin{align*}
B=12\,\Rau^{-11/8}\beta^{-1/2} \,\mathrm{G}    
\end{align*}
\add{The upper bound $B\lesssim 10\,\mathrm{mG}$ for $R\sim 10\,\mathrm{AU}$ mentioned in \S\ref{obs:mag} tend to suggest $\beta\gtrsim 10^4$ in these regions, which confirms that the field strength is expected to be strongly subthermal}.
\subsection{Fluid approximation}

Protoplanetary discs are mostly constituted of neutral gas. In order to describe this gas, it is tempting to use the fluid approximation. For this approximation to be valid, the gas under consideration needs to be collisional, i.e., gas particles need to be subject to many collisions during one dynamical timescale. This ensures that at the microphysical level, the velocity distribution of the gas phase can be approximated by a \add{Maxwellian} distribution, allowing us to use a scalar pressure field.

Assuming the gas is mainly made of $H_2$ molecules of radius $10^{-8}\,\mathrm{cm}$, we can estimate the cross section of neutral molecules as $\sigma_{nn}=3\times 10^{-16}\,\mathrm{cm}^2$. This gives us an approximate mean free path $\ell_\mathrm{mfp}$ and collision frequency $\omega_\mathrm{coll}$

\index{Mean free path}
\begin{align*}
    \ell_\mathrm{mfp}&=\frac{1}{n\sigma_{nn}}=22\,\Rau^{9/4}\,\mathrm{cm}\\
    \omega_\mathrm{coll}&=\frac{v_\mathrm{th}}{\ell_\mathrm{mfp}}=5\times 10^3\,\Rau^{-5/2}\,\mathrm{s}^{-1}.
\end{align*}
We therefore have $\ell_\mathrm{mfp}\ll R$ and $\omega_\mathrm{coll}\gg \Omega$, which validate the fluid approximation to describe the dynamics of protoplanetary discs to a very good approximation. \add{It should be noted that these quantities are evaluated at the disc midplane. If one looks at regions well above the disc, as in the case of outflows, $\ell_\mathrm{mfp}$ increases significantly. One finds that $\ell_\mathrm{mfp}\gtrsim H$ when $n\lesssim 10^4\,\mathrm{cm}^{-3}$, i.e. when the atmosphere is $10^{10}$ times less dense than the midplane at 1 AU. Such a strong density contrast is almost never reached in outflow models, where one finds density contrasts between $10^4$ and $10^7$ (e.g. Fig.~\ref{fig:selforg_glob}). Nevertheless, it should be kept in mind that very weak outflows in the outermost parts of the disc can be close to the collisionless regime}.

\subsection{\label{sec:grain_pop}Grain population}

\index{Grains!abundance}
The question of grains is of importance in PPDs. As it is usually assumed, we consider a constant dust to gas mass fraction, equal to that of the interstellar medium (1/100). We further assume that grains are spherical with a radius $a$ and made of olivine with a density $\rho_o=3\,\mathrm{g.cm}^{-3}$. The density of grains is therefore
\begin{align*}
\rho_\mathrm{grain}&=6\times 10^{-12}\,\Rau^{-9/4}\,\mathrm{g.cm}^{-3}\\
n_\mathrm{grain}&=1.4\,\Rau^{-9/4}\amu^{-3}\,\mathrm{cm}^{-3}
\end{align*}
In this last estimate, we have assumed that all the grains had the same size. This is an over-estimation since the sizes are actually distributed over a wide range of scales. \add{In addition, the grain size distribution is expected to evolve with time as grains are known to be growing in protoplanetary discs}. However, this order of magnitude estimate points to an important fact: the abundance of grains $n_\mathrm{grain}/n\sim 10^{-14}\amu^{-3}$. Hence, if grains are smaller than $1\,\mu\mathrm{m}$, the typical ionisation fraction of protoplanetary discs ($~10^{-14}$) suggest that \emph{grains are more abundant than free charge carriers}. As we will see in \S \ref{sec:roleofgrains}, this has a huge impact on the plasma conductivity tensor as grains can become the main charge carriers.

\subsection{Ionisation fraction}
\index{Ionisation}
\index{$\xi$ ionisation fraction}
The ionisation fraction  $\xi\equiv n_{-}/n_n$, where $n_{-}$ is the number of free negative charge carriers, is a highly uncertain quantity, with very little constraints coming from observations. The ionisation fraction typically range from $10^{-16}$---$10^{-13}$ at 1 au to $10^{-13}$---$10^{-10}$ at 100 au. However, the resulting plasma is not necessarily a plasma made of electrons and molecular ions. Indeed, if dust grains are present and sufficiently abundant, they tend to suck electrons and ions in the gas phase, leading to a plasma made of positively and negatively charged grains \citep{SM00}.

Here, we illustrate how each physical process affects the ionisation fraction by considering a simple chemical network which includes singly charged grains. We combine this network with ionisation rate prescriptions for the various ionisation sources (X-Ray, UVs, cosmic rays and radioactive decay). 

\subsubsection{\label{sec:ion_source}Sources of ionisation}
\index{Ionisation!rates}
\index{$\zeta$ ionisation rate}
Since we focus on the outer part of protoplanetary discs ($R>1~\mathrm{AU}$), the gas is mostly cold with $T<300~\mathrm{K}$. This implies that thermal ionisation (due to collision between molecules) is inefficient, and one has to rely on non-thermal ionisation processes. Here, we consider the following effects with their associated ionisation rate $\zeta$:
\begin{itemize}
\item X-ray ionisation due to \add{bremsstrahlung emission from an isothermal $T=5\,\mathrm{keV}$ corona localised around the central protostar} \citep[][see their Eq.~21]{IG99,BG09};
\item cosmic-ray ionisation with $\zeta_\mathrm{CR}=\zeta_{\mathrm{CR},0} \exp(-\Sigma/96~\mathrm{g~cm}^{-2})~\mathrm{s}^{-1}$ \citep[e.g.][]{UN81} and $\zeta_{\mathrm{CR},0}=10^{-17}~\mathrm{s}^{-1}$, corresponding to the interstellar value ;
\item radioactive decay with $\zeta_\mathrm{rad}=10^{-19}~\mathrm{s}^{-1}$ \citep{UN09}.
\end{itemize}
The amount of ionisation due to CRs is highly disputed. Some authors have proposed that due to the wind coming from the young star, CRs are magnetically mirrored from the protoplanetary disc, resulting in a significantly reduced ionisation rate due to CRs ($\zeta_{\mathrm{CR},0}\sim 10^{-20}~\mathrm{s}^{-1}$, \citealt{CA13}). On the opposite, it has been proposed that CRs could be accelerated in shocks produced in the protostellar jet by a Fermi process. This could result in ionisation rates as high as $\zeta_{\mathrm{CR},0}\sim 10^{-13}~\mathrm{s}^{-1}$ \citep{PI18}. Observations of TW Hya tend to suggest a low ionisation rate due to CR ($\zeta_{\mathrm{CR},0}\lesssim 10^{-19}~\mathrm{s}^{-1}$, \citealt{CB15}), though this is still highly model dependent. Because of these uncertainties, some authors (e.g. \citealt{IN06}) have simply omitted CR ionisation and consider only X-rays as the main source of ionisation. These difference and uncertainties in the treatment of the ionisation rate have to be kept in mind when comparing the results of different research groups.

\begin{figure}
\centering
\includegraphics[width=0.80\hsize]{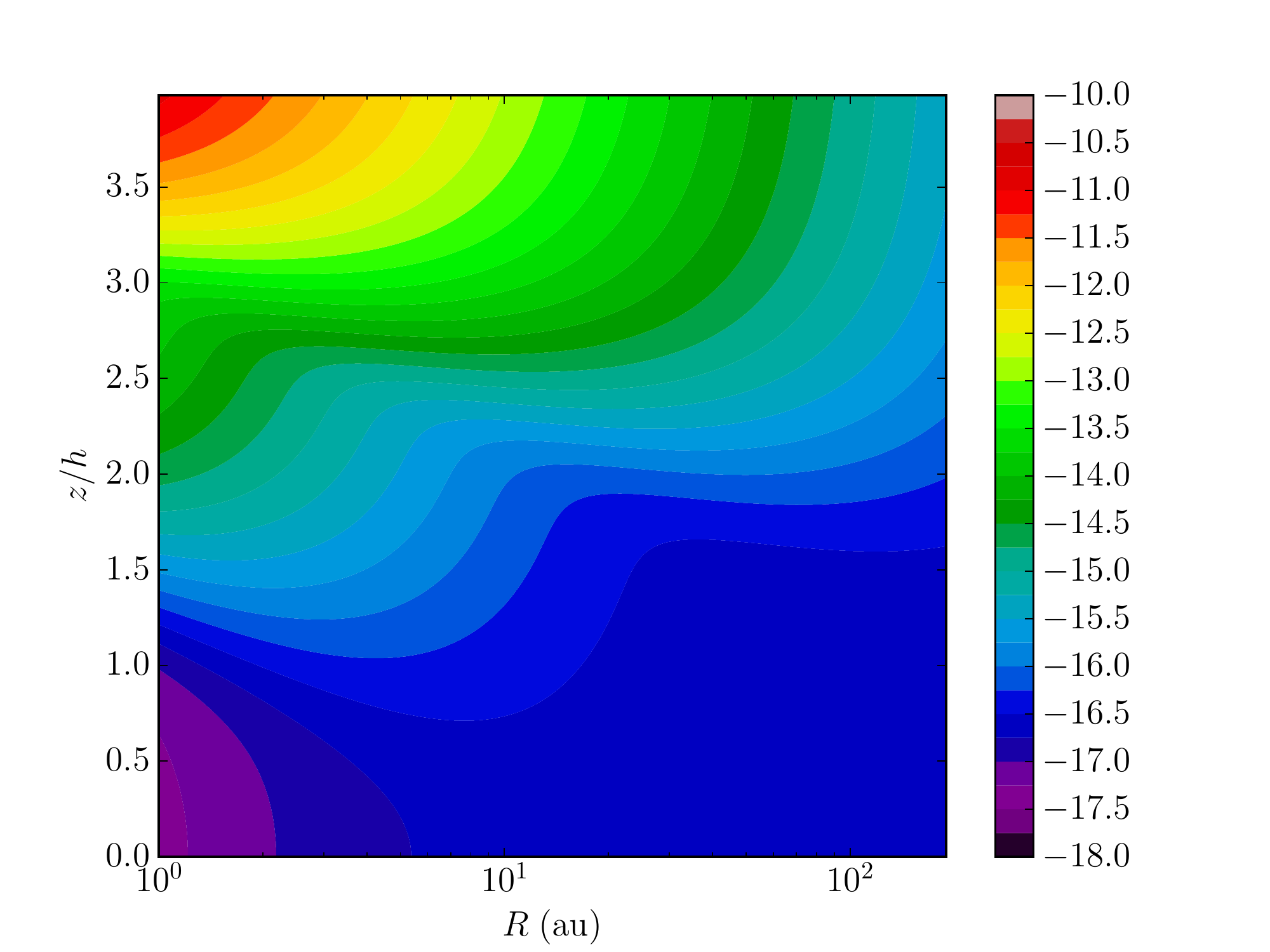}
\caption{Ionisation rate $\log(\zeta)$ ($\mathrm{s}^{-1}$) as a function of radius and altitude (in disc scale height) resulting from X-ray, cosmic rays and radioactive decay. }
\label{fig:partii:ionisation_rate}
\end{figure}

We show in Fig.~\ref{fig:partii:ionisation_rate} the resulting ionisation rate following the disc structure presented in \S\ref{sec:profile}. We find that cosmic rays are shielded only in the innermost parts of the disc, where the column density goes above $100~\mathrm{g~cm}^{-2}$. Most of the disc midplane up to $z\sim h$ has $\zeta\simeq \zeta_{\mathrm{CR},0}$, indicating that CRs are indeed the dominant source of ionisation in this region. Above $z\sim h$, X-rays start to penetrate the disc and the ionisation rate rises. 

\subsubsection{A simple chemical model}
To illustrate the typical ionisation fractions expected in protoplanetary discs, we follow \cite{OD74}, \cite{FTB02} and \cite{IN06} defining the following reaction network  and rates with free electrons, neutral molecules $m$, molecular ions ${m^+}$ and metal atoms ${M}$:
\begin{align*}
\mathrm{m} + \textrm{ionising radiation} &\rightarrow  \mathrm{m}^{+} + \mathrm{e}^{-}&\zeta\\
\mathrm{m}^{+} + e^{-} &\rightarrow \mathrm{m}&\delta\\
\mathrm{M}^{+} + e^{-} &\rightarrow  \mathrm{M}&\delta_r \\
\mathrm{m}^{+} + \mathrm{M} &\rightarrow  \mathrm{m} + \mathrm{M}^{+}&\delta_t
\end{align*}
where $\zeta$ is the ionisation rate, $\delta$ is the dissociative recombination rate for molecular ions, $\delta_r$ the radiative recombination rate for metal atoms, and $\delta_t$ the rate of charge transfer from molecular ions to metal atoms.  Following \cite{FTB02}, we take
\begin{align*}
\delta_r&=3\times 10^{-11} T^{-1/2}\,\mathrm{cm}^3\mathrm{s}^{-1}\\
\delta&=3\times 10^{-6} T^{-1/2}\,\mathrm{cm}^3\mathrm{s}^{-1}\\
\delta_t&=3\times 10^{-9}\,\mathrm{cm}^3\mathrm{s}^{-1}    
\end{align*}

In the absence of metals and grains, the rate equations admit a simple solution in steady state
\begin{align}
\label{eq:zeta_chem}
    \xi=\sqrt{\frac{\zeta}{\delta n_n}}.
\end{align}
In the opposite metal-dominated limit, still without grains, one gets
\begin{align*}
    \xi=\sqrt{\frac{\zeta}{\delta_r n_n}}.
\end{align*}
Since $\delta_r\ll \delta$, one clearly sees that the absence of metals leads to a dramatic decrease in the ionisation fraction \citep{FTB02}.

\subsubsection{\label{sec:ion_profile}Typical ionisation fraction profile}
\index{Ionisation!fraction}
\begin{description}
\item[Grain-free case, Metal-free case: ]
combining (\ref{eq:zeta_chem}) to the ionisation rate in \S\ref{sec:ion_source}, one can obtain the ionisation fraction in the disc. However, this ionisation fraction depends not only on the disc chemistry one assumes but also on the disc \emph{structure}. A lot of theoretical work has focused on the minimum mass solar nebula (MMSN) model which assumes $\Sigma=1700~ \Rau^{-3/2}\,\mathrm{g~cm}^{-2}$ \citep{W07,BS13,LKF14}. This makes the disc much denser in the inner part, resulting in a stronger shielding of cosmic rays and a lower ionisation fraction than less dense discs. As an illustration, we show in Fig.~\ref{fig:partii:ionisation_fraction} the resulting ionisation fraction with the disc structure presented in \S\ref{sec:profile} and with a MMSN disc model.

\begin{figure}
\centering
\includegraphics[width=0.49\hsize]{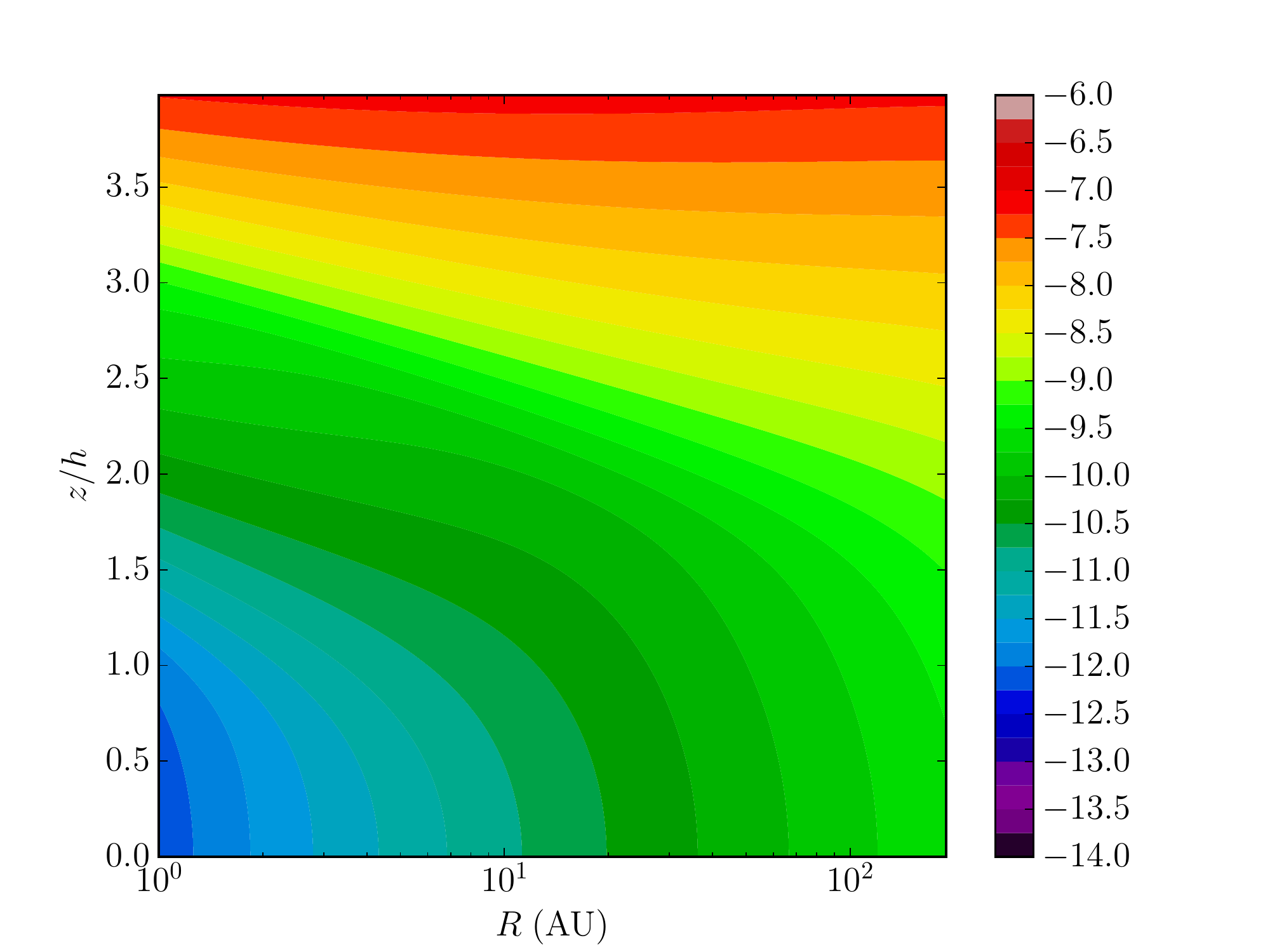}
\includegraphics[width=0.49\hsize]{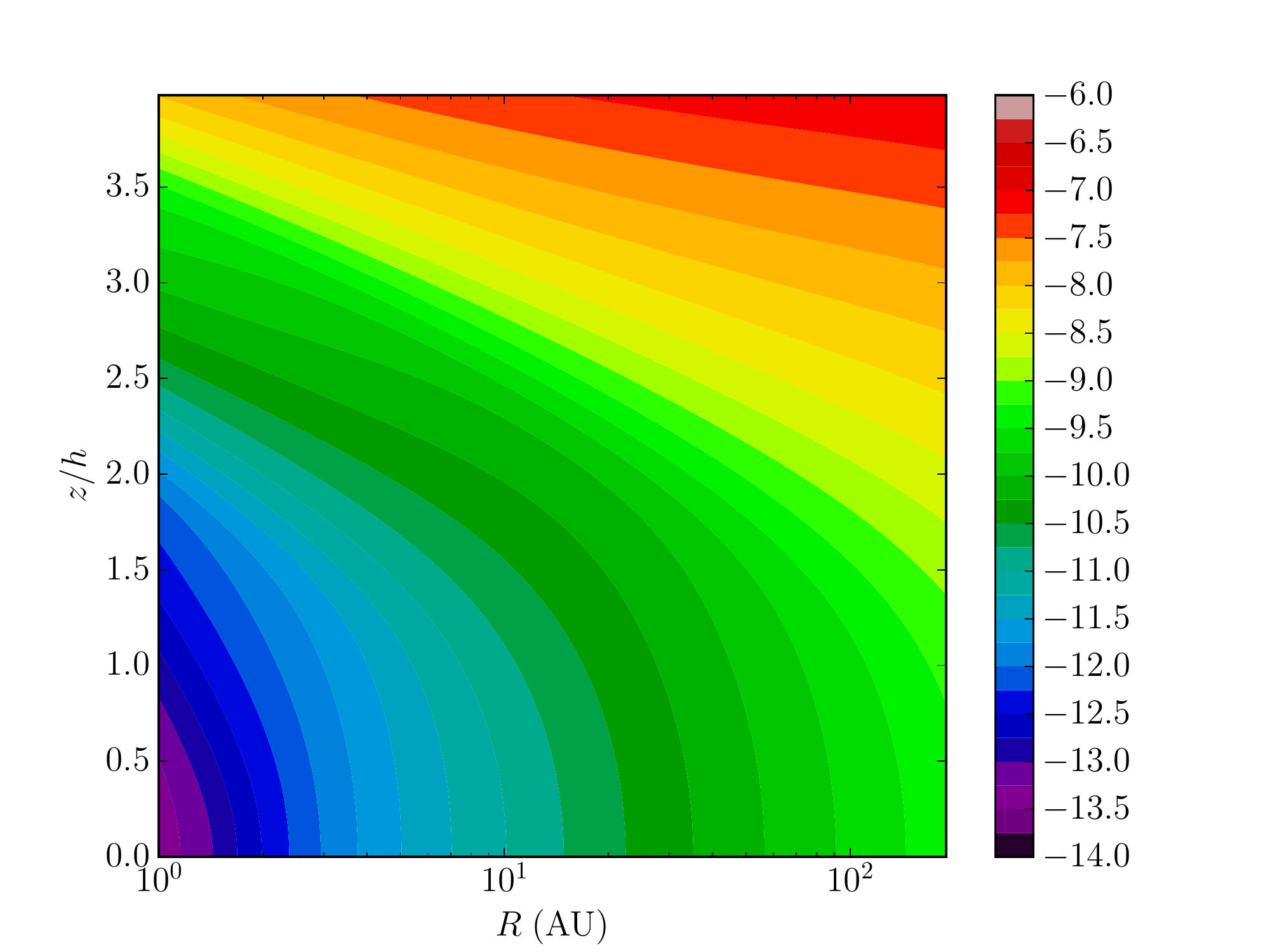}
\caption{Ionisation fraction $\log(\xi)$ as a function of position in our disc model (\S\ref{sec:profile}) (left) and in a minimum mass solar nebula (right). Note the difference in ionisation fraction close to the disc midplane for $R<10~\mathrm{AU}$ }
\label{fig:partii:ionisation_fraction}
\end{figure}

We observe that the lowest ionisation fraction reach $10^{-14}$ in the MMSN case, or $10^{-13}$ in our disc model. The lowest ionisation fractions are reached in the innermost parts of the disc, where the recombination is the fastest and CRs+X-rays are efficiently shielded. The ionisation fraction progressively increases when X-rays start to penetrate, until one reach ionisation fractions as high as $10^{-6}$ at a few scale heights. Note that the differences between these models are only significant for $R<10\,\mathrm{AU}$ since the column densities between the MMSN and our disc model are similar above this radius.

\item[Inclusion of grains and metals: ]

\index{Grains!ionisation}
as demonstrated \add{by \cite{E79} and \cite{UN80} in the context of molecular clouds , and later applied to protoplanetary discs} \citep{SM00,IN06,W07}, grains tend to accelerate the recombination of electrons \add{by removing them from the gas phase}, resulting in a lower global ionisation fraction, which we have ignored here. To illustrate the impact of grains, let us add the following reactions to our simplified reaction network:

\begin{align*}
\mathrm{grain} + \mathrm{m}^{+} &\rightarrow  \mathrm{grain}^{+} + \mathrm{m}\\
\mathrm{grain}^{-} + \mathrm{m}^{+} &\rightarrow  \mathrm{grain} + \mathrm{m}\\
\mathrm{grain} + \mathrm{e}^{-} &\rightarrow  \mathrm{grain}^{-}\\
\mathrm{grain}^{+} + \mathrm{e}^{-} &\rightarrow  \mathrm{grain}\\
\mathrm{grain} + \mathrm{M}^{+} &\rightarrow  \mathrm{grain}^{+} + M\\
\mathrm{grain}^{-} + \mathrm{M}^{+} &\rightarrow  \mathrm{grain} + M\\
\mathrm{grain}^{+} + \mathrm{grain}^{-} &\rightarrow  \mathrm{grain} + \mathrm{grain}\\
\end{align*}

The above reaction network only considers singly charged grains, while it is well known that grains can have many charges \citep{I12}. We chose this approach to illustrate in the simplest model the impact of grains on the ionisation fraction, and later on the diffusivities since the abundance of multiply charged grains is usually lower than that of singly charged grains for $z<h$ \citep{W07}.
 
The rates for these reactions are computed by assuming each species collides at its thermal velocity with a spherical grain of radius $a$ (see \S\ref{sec:grain_pop} for more details). We assume a fixed sticking probability of electrons on grains, which corresponds to the probability of bouncing back from a grain\footnote{This probability varies greatly in the literature, from fixed values in \citep{W07} to various temperature and charge-dependent fits in \cite{IN06} and \cite{B11b}. Choosing a fixed sticking probability, as we do, tends to increase the effect of grains at high temperature, so the results presented here are a limit case of extreme grain sticking efficiency.}.

The resulting ionisation fraction due to electrons and charged grains are presented in Fig.~\ref{fig:partii:ionisation_fraction_grains}. We observe essentially two trends. First, the smallest ionisation fraction is found when grains are present, while the highest ionisation fractions correspond to grain-free metal-rich cases, with variations due to this composition effect of the order of 3 orders of magnitude. Second, the ionisation fraction increases with increasing radius. This effect is not only because the ionisation rate increases, but also because the recombination rate decreases due to lower densities. Let us finally point out that when grains are present, they can become the main charge carrier, as is the case at $R=5\,AU$. 

In the following, we will use the value $\xi=10^{-13}$ to evaluate several plasma parameters, keeping in mind this corresponds to a lower bound in our disc model.

\begin{figure}
\centering
\includegraphics[width=0.48\hsize]{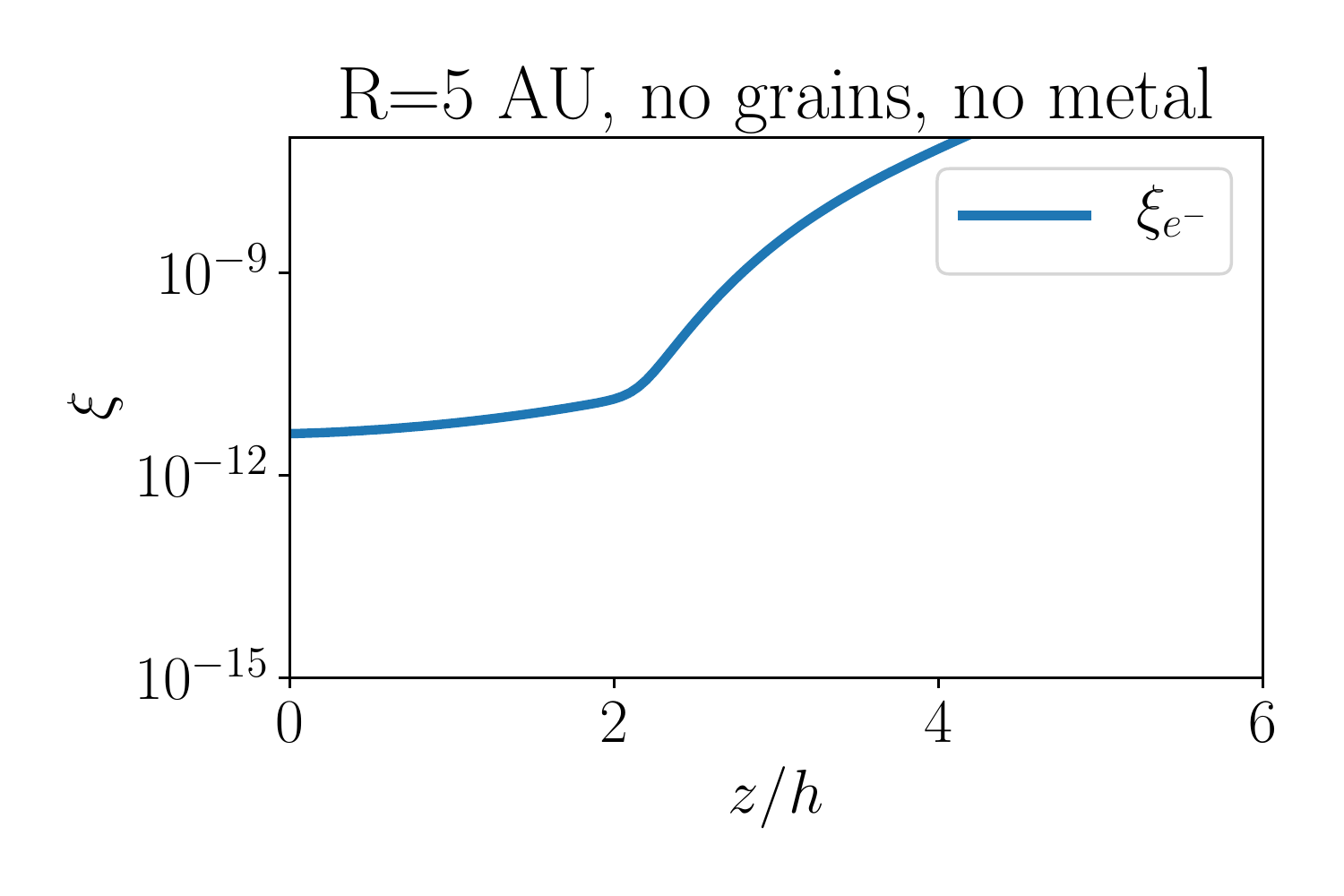}
\includegraphics[width=0.48\hsize]{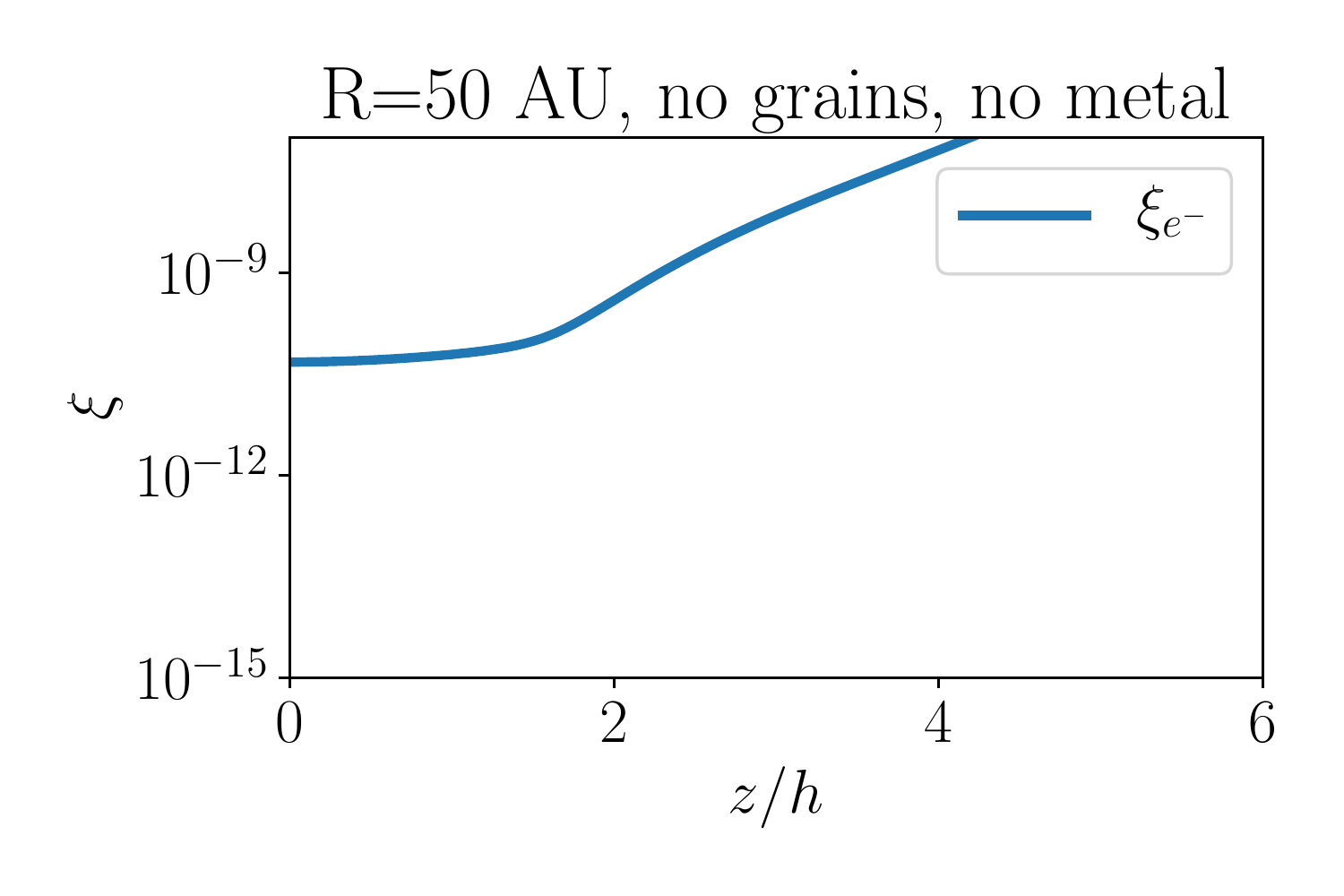}
\includegraphics[width=0.48\hsize]{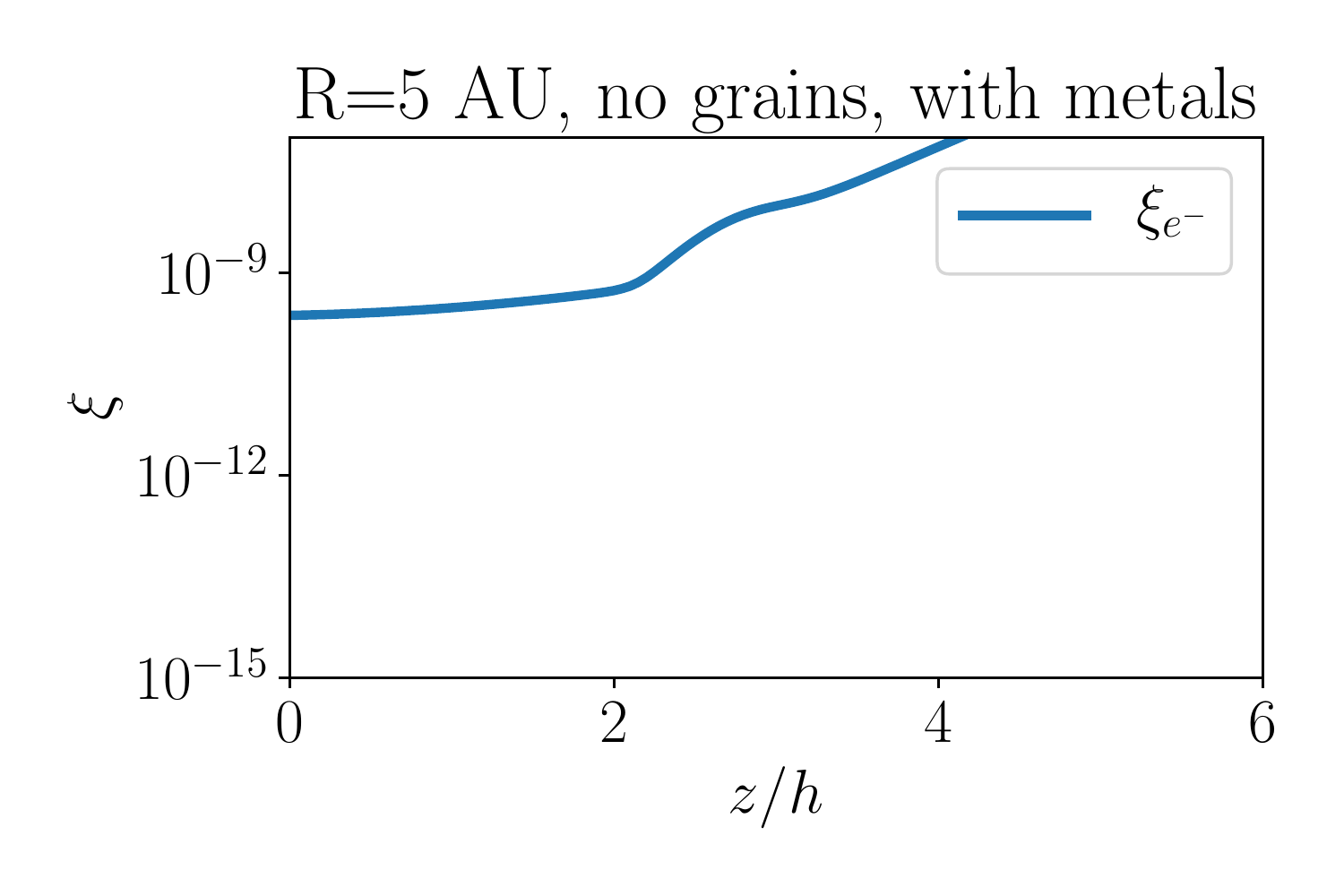}
\includegraphics[width=0.48\hsize]{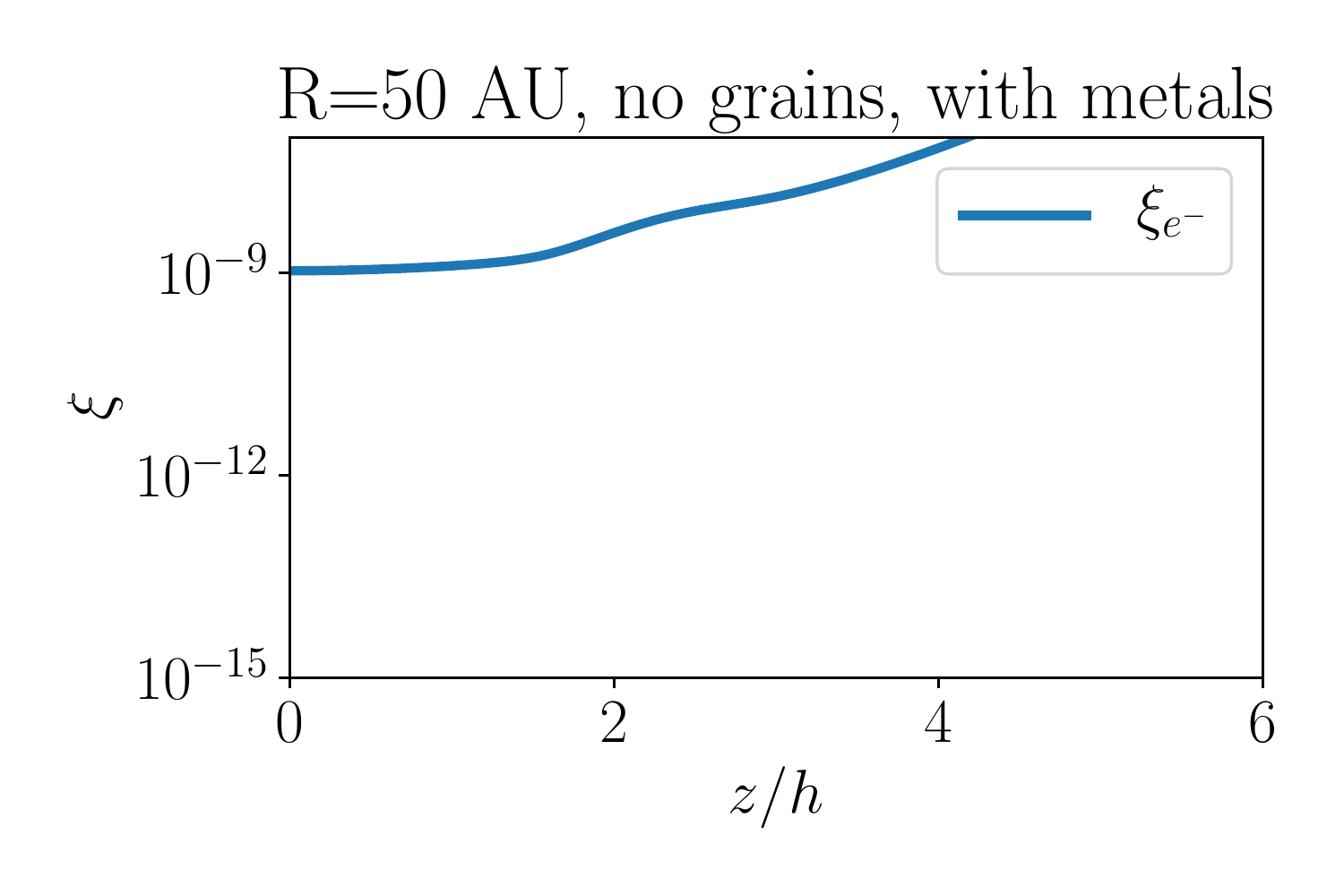}
\includegraphics[width=0.48\hsize]{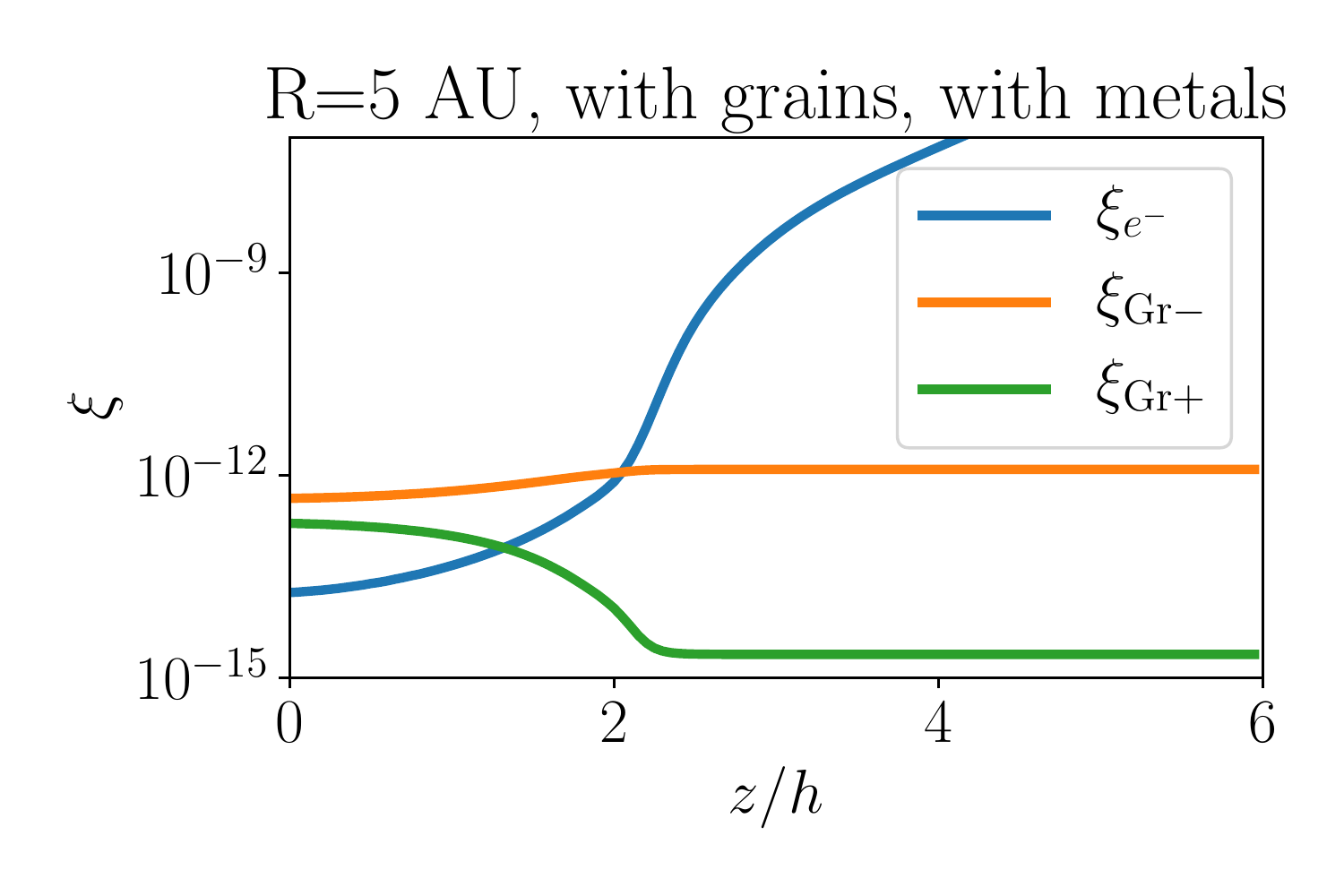}
\includegraphics[width=0.48\hsize]{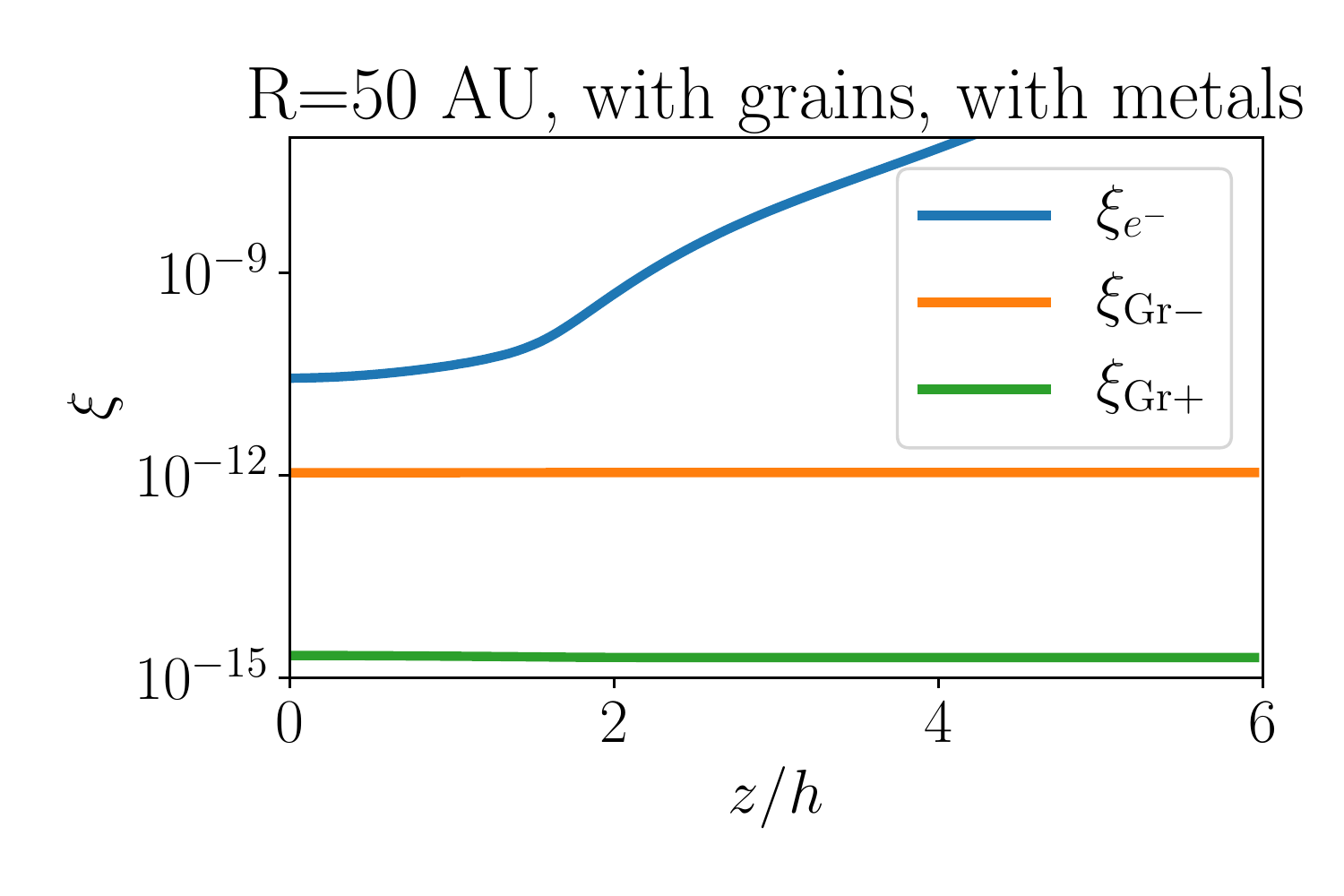}
\caption{Ionisation fraction $\xi$ for 3 different compositions: row 1: no grains, no metals ; row 2: no grains with [M]=$10^{-8}$ ; row 3: with $a=0.1\mu\mathrm{m}$ grains and metal atoms. The first column corresponds to $R=5~\mathrm{AU}$ and the second column $R=50~\mathrm{AU}$}
\label{fig:partii:ionisation_fraction_grains}
\end{figure}

\end{description}

\section{Plasma description in protoplanetary discs\label{sec:plasma}}
In this section we explore the properties of the plasma constituting protoplanetary discs and ask whether they can be described using non-ideal MHD. For this limit to be valid, we have to satisfy 3 criteria which we discuss below
\begin{enumerate}
\item \add{Binary Coulomb} interactions should be negligible. This implies that the plasma parameter (defined below) is much larger than 1.
\item Electro-neutrality is satisfied on timescales of interest, i.e. any charge separation is quickly eliminated by electrostatic interactions.
\item The behaviour of each fluid component (electrons, ions, neutrals, charged grains) can be described using a single fluid approximation.
\end{enumerate}

\subsection{Plasma parameter}

Several quantities allow one to characterise a plasma, the first one being the nature of the electromagnetic interaction. The most fundamental quantity characterising a plasma is the Debye length which may be written in an electron-ion plasma
\index{Debye length}
\begin{align*}
    \lambda_\mathrm{D}&\equiv\sqrt{\frac{k_BT_e}{4\pi (1+Z) n_e e^2}}\\
    &=30\,\bigg(\frac{\xi}{10^{-13}}\bigg)^{-1/2} \Rau^{7/8}(1+Z)^{-1/2}\,\mathrm{cm}
\end{align*}
where $Z$ is the averaged number of charges on the ions and we have assumed electro-neutrality so that $n_i=n_e/Z$. The Debye length is clearly below the scales of interest in protoplanetary discs. Even if one considers charged grains, the same Debye length can be derived since it does not depend on the particle mass. In addition to this characteristic length, a ``good'' plasma should have many particles in a Debye sphere, ensuring that screening of short-range Coulomb interaction. This is quantified by the plasma parameter $\Upsilon$, equals to the number of charge carriers in a Debye sphere
\index{Plasma!parameter}
 \begin{align*}
 \Upsilon &= 4\pi n_e\lambda_\mathrm{D}^3    \\
 &=4\times 10^5\, (1+Z)^{-3/2}\Rau^{3/8}\bigg(\frac{\xi}{10^{-13}}\bigg)^{-1/2}.
 \end{align*}
Hence, despite the low ionisation fraction and low temperature of these objects, they are still very much in the plasma regime where short-range Coulomb interactions can be neglected. Note however that reducing the ionisation fraction and at the same time, increasing the number of charges could change this picture, breaking the plasma approximation altogether. However, this would require $Z\gtrsim 10^3$ in protoplanetary discs, a value which is never encountered, even in chemical models including grains (e.g. \citealt{W07}). 


\subsection{Electro-neutrality and drag}
\index{Electro-neutrality}
Protoplanetary discs are weakly ionised objects. This implies that the dynamical equations describing the flow and the approximations underlying their derivation should be clearly stated. In this section, we derive these equations, starting from the multi-fluid plasma description. We assume the gas is made of neutral and charged ``particles'' (particle could mean electron, ion, or charged grain, indifferently). The multi-fluid approximation is valid since the collision timescales are short, as demonstrated above. We therefore start from the following dynamical equations:
\begin{align}
\label{eq:fond_mass}\frac{\partial n_j}{\partial t}+\bm{\nabla\cdot }n_j\bm{v}_j&=0,\\
\label{eq:fond_mom}\frac{\partial n_jm_j \bm{v}_j}{\partial t}+\bm{\nabla\cdot }(n_jm_j\bm{v}_j\otimes\bm{v}_j)&=-\bm{\nabla}P_j+\bm{f}_j+n_jq_j\Bigg(\frac{\bm{v}_j\bm{\times B}}{c}+\bm{E}\Bigg)+\bm{R}_j,
\end{align}
where $n_j$, $m_j$, $\bm{v}_j$, $P_j$, $q_j$, $\bm{f}_j$ are the number density, mass, velocity, pressure, charge, and additional forces (gravity, etc) on species $j$. We have also included a drag force $\bm{R}_j$ between this species and all of the other species. This force is due to inter-species collisions and can be written as
\begin{align*}
\bm{R}_j=\sum_k \gamma_{jk}\rho_j\rho_k(\bm{v}_k-\bm{v}_j),
\end{align*}
since each fluid component is collisional and therefore has a Maxwellian velocity distribution. Here, $\gamma=\langle \sigma v\rangle_{jk}/(m_j+m_k)$ and $\langle \sigma v\rangle_{jk}$ is the momentum exchange rate between species $j$ and $k$. As expected from momentum conservation, we have $\sum_j \bm{R}_j=0$. 

It is usually assumed that electro-neutrality follows from the fact that the plasma frequency $\omega_p$ is much larger than any frequency of interest. While this is indeed a good criterion for a fully ionised plasma, it is not necessarily true for a weakly ionised plasma. Let us therefore revisit this criterion, starting from the linearised multi-fluid equations. We perturb only one species along the $x$ axis, leaving the other ones unperturbed. We moreover assume that the fluid pressure and other external forces are negligible compared to electromagnetic forces. The linearised equation of motion reads
\begin{align*}
\frac{\partial \delta n}{\partial t}+n_0\partial_x v_x &=0,\\
n_0m\frac{\partial v_x}{\partial t}&=n_0qE_x-\gamma m n_0\rho v_x.
\end{align*}
Solving these equations requires an equation for the electric field, which is obtained from one of Maxwell's equation
\begin{align*}
\partial_x E&=4\pi q \delta n.
\end{align*}
We can combine these equations to obtain a second order relation on the density fluctuation
\begin{align*}
-\frac{\partial^2\delta n}{\partial t^2}=\omega_p^2\delta n+\frac{1}{\tau_s}\frac{\partial \delta n}{\partial t},    
\end{align*}
\index{Plasma!frequency}
where we have introduced the plasma frequency $\omega_p$ and the stopping time $\tau_s$ as
\begin{align*}
\omega_p&\equiv \Bigg(\frac{4\pi n q^2}{m}\Bigg)^{1/2}\\
\tau_s&\equiv \frac{1}{\gamma \rho}.
\end{align*}
Dynamically, this equation describes damped plasma oscillations with frequencies
\begin{align*}
\omega_{\pm}=\frac{i\tau_s^{-1}\pm\sqrt{4\omega_p^2-\tau_s^{-2}}}{2},
\end{align*}
for which we can distinguish two physical limits. 
\begin{enumerate}
\item $\omega_p\gg \tau_s^{-1}$ in which case the plasma is subject to plasma oscillations at frequency $\omega_p$ with a damping timescale equal to $\tau_s$. If we consider phenomena on frequencies much lower than $\omega_p$, we can average out the highest order time derivative and get a simple closure relation between $v_x$ and $E_x$: $v_x=qE_x/\gamma m \rho$ which constitutes the base of Ohm's law. \add{Once these oscillations are time-averaged, the plasma can be assumed to be electrically neutral.}
\item $\omega_p\ll \tau_s^{-1}$ in which case the plasma is subject to over-damped oscillations with two imaginary frequencies $\omega_+=i\tau_s^{-1}$ and $\omega_-=i\omega_p^2\tau_s\ll\omega_+$ associated to two damping timescales $\tau_\pm=(\omega_\pm)^{-1}$. \add{To interpret physically these timescales, let us consider a plasma at rest in which we introduce a localised charge deficit. First, the plasma is going to start moving to ``fill'' the charge deficit. Because of the drag, however, it very rapidly reaches an asymptotic velocity, given by $v_x=qE_x/\gamma m \rho$. $\tau_+$ is the time needed by the system to be put in motion and reach this quasi-stationary velocity. This velocity fluctuation, however, is smaller than that which would be obtained in a pure plasma oscillation, because the drag prevents the plasma from reaching high velocities. Hence, it takes a time $\tau_-$ to actually fill the charge deficit. This implies that Ohm's law, given by the asymptotic velocity, is valid on timescales longer than $\tau_+$, and that charge inertia can be neglected in that limit. However, charge neutrality is restored on the much longer timescale $\tau_-$. }
\end{enumerate}

To summarise, it is possible to neglect \add{inertia} for the charged species in the momentum equation provided that the timescales under consideration are larger than $\mathrm{max}(\tau_s,\omega_p^{-1})$, and recover Ohm's law without time derivative. \add{Note that this condition is different from electroneutrality, which requires timescales longer than $\mathrm{max}\big(\omega_p^{-1},(\omega_p^2\tau_s)^{-1}\big)$, which are significantly longer than $\tau_s$ when $\omega_p\tau_s\ll 1$. It should be pointed out that this analysis was done for a single species, while plasmas in protoplanetary discs can be made of many different species. Hence, the condition for electroneutrality need to be satisfied only by the most mobile specie of the plasma, which can then compensate for charge fluctuations, and not necessarily by all of the species present. }

\index{Grains!plasma frequency}
In protoplanetary discs, we get the following values for the plasma frequency, depending on the type of charge carrier
\begin{align*}
\omega_{p,e}&=2.2\times 10^{5} \bigg(\frac{\xi}{10^{-13}}\bigg)^{1/2}\Rau^{-9/8}    \,\mathrm{s}^{-1}\\
\omega_{p,i}&=9.3\times 10^{2} \bigg(\frac{\xi}{10^{-13}}\bigg)^{1/2}\Rau^{-9/8}    \,\mathrm{s}^{-1}\\
\omega_{p,g}&=1.8\times 10^{-3} \bigg(\frac{\xi}{10^{-13}}\bigg)^{1/2}\Rau^{-9/8}\amu^{-3/2}\,\mathrm{s}^{-1}
\end{align*}
where $e,i,g$ stands for electrons, ions and grains. As it can be seen this frequency is always short compared to the timescales of interest, but grains tend to have significantly lower frequencies due to their higher inertia.

The stopping times can be estimated starting from the momentum exchange rates $\langle \sigma v\rangle_{ij}$. Since we are interested only in weakly ionised plasmas, collisions between charged species will be extremely rare. We will therefore only consider neutral-charge collisions.

The ``collision'' between electron/ions and neutrals are mainly a result of the electrostatic interaction between the approaching charge and the dipole induced on the neutral by the charge. This is estimated by 
\begin{align*}
\langle \sigma v\rangle_e&=8.3\times 10^{-9}\times \mathrm{max}\Big[1,\Big(\frac{T}{100\,K}\Big)^{1/2}\Big]\mathrm{cm}^3s^{-1}\\
\langle \sigma v\rangle_i&=2.4\times 10^{-9}\Big(m_H/m_n\Big)^{1/2}\,\mathrm{cm}^3\mathrm{s}^{-1}
\end{align*}
where \add{$\langle \sigma v\rangle_e$ is deduced from \cite{DRD83} and $\langle \sigma v\rangle_i$ is obtained from \cite{DR11}, following \cite{B11b}}\footnote{The momentum exchange rate $\langle \sigma v\rangle_i$ estimated by \cite{B11b} is actually the collision rate. The momentum exchange rate quoted here is larger by a factor approximately 1.21 than the collision rate quoted by \cite{B11b} (see \cite{DR11}, eq.~2.39)}. For grains above a size of a few $10^{-2}\,\mu\mathrm{m}$, collisions mainly behave as billiard balls. In other words, $\sigma v$ is \add{roughly} equal to the velocity of the incident neutral times the cross-section of the grain. For spherical grains, this leads to\footnote{Note that the expression provided by \cite{B11b} for this rate is incorrect by more than 3 orders of magnitude}
\begin{align*}
\langle \sigma v\rangle_g&=\pi a^2 \sqrt{\frac{2k_B T}{m_n}}\\
&=2.6\times 10^{-3} \amu \Big(\frac{T}{100\,K}\Big)^{1/2}\,\mathrm{cm}^3\mathrm{s}^{-1}.
\end{align*}
\index{Stopping time}
\index{Grains!stopping time}
These rates allow us to compute stopping times for each species following the definition above.
\begin{align*}
\tau_{s,e}&=6.7\times 10^{-7}\,\Rau^{9/4}\,\mathrm{s},\\
\tau_{s,i}&=4.9\times 10^{-5}\,\Rau^{9/4}\,\mathrm{s},\\
\tau_{s,g}&=8.1\times 10^4\,\Rau^{9/4}\amu^2\Bigg(\frac{100\,K}{T}\Bigg)\,\mathrm{s}    ,
\end{align*}
which shows that because of the low ionisation fraction and the neutral drag, $\omega_{p,j}\tau_{s,j}<1$ \add{for ions and electrons, while it is $> 1$ for grains. This means that plasma oscillations are over-damped for ions and electrons (case ii above) and are not directly relevant for quasi-neutrality. Nevertheless, $\omega_p\tau_s>10^{-2}$, so even in this case, electroneutrality is recovered on timescales shorter than a second. Grains, on the other hand, are usually in regime (i), with a relatively low plasma frequency (period of a few days for 1$\mu$m size grains), decreasing rapidly with increasing grain size. Grains are usually not the only charge carrier in discs, so electro-neutrality is guaranteed by ions and electrons, but it should be kept in mind that, in a hypothetical situation where grains would be the only charge carrier, electro-neutrality could be violated, leading to phenomena similar to lightning. This however will not be explored here, and we will only consider situations where ions and electrons are still present in the system.}

\subsection{Single fluid approximation}
\subsubsection{Dynamical equation for the center of mass}
\index{Plasma!single fluid approximation}
The set of equations (\ref{eq:fond_mass}-\ref{eq:fond_mom}) can in principle be solved simultaneously \citep{KD14}. However, it is numerically expensive since the numerical time steps are usually limited by $\tau_s$ which is much smaller than the timescales of interest (see above). \add{Note however that there are situations where the multifluids approach cannot be avoided, such as when the timescale to reach the ionisation/recombination equilibrium becomes of the order of the timescales of interest \citep[e.g.][]{IN08}, or when the neutral density is so low that the collision timescale $\tau_s$ becomes of the order of the timescales of interest, which can occur well above the disc in the early phases of star formation, when X-rays and UVs are not yet produced by the central body.}

\add{However, if one focuses on disc dynamics and its immediate environment once the central star is formed, the single fluid approximation is a perfectly reasonable approximation, as multi-fluid approaches tend to confirm \citep{RRD16}.} For this reason, I will focus here on the single fluid approximation. To derive this single fluid approximation, let us consider the dynamical equations for the center of mass of the fluid, defining the total mass density $\rho=\sum_jn_j m_j$, the flow velocity $\bm{v}=\sum_j n_j m_j\bm{v}_j/\rho$ and the drift speed for each species $\bm{w}_j=\bm{v}_j-\bm{v}$ we sum equations (\ref{eq:fond_mass}) and (\ref{eq:fond_mom}) to obtain
\begin{align*}
\frac{\partial \rho}{\partial t}    +\bm{\nabla\cdot }\rho \bm{v}&=0\\
\frac{\partial \rho \bm{v}}{\partial t}+\bm{\nabla\cdot }(\rho\bm{v}\otimes\bm{v})&=\bm{\nabla\cdot }\Big(\sum_j n_jm_j\bm{w}_j\otimes\bm{w}_j\Big)-\bm{\nabla}P+\bm{f}+\frac{\bm{J\times B}}{c}+\sum_j{n_j q_j}\bm{E},
\end{align*}
where we have introduced the total pressure and force $P$ and $\bm{f}$ as well as the total current $\bm{J}=\sum_j n_j q_j v_j$. These equations are exact. However, they don't correspond to the usual dynamical equations one is used to, and it's important to understand why each extra term can be neglected.

The first term on the righthand side corresponds to transport of momentum by the drift velocity. Physically, it can be interpreted as a diffusion of momentum due to drifting particles.  It can be neglected, provided that drift velocities are small, i.e. that $w_j<L \Omega\sqrt{\rho/\rho_j}$ where $L$ is the typical length scale of interest and $\Omega$ is the typical frequency\footnote{Comparing the drift velocity to the mean velocity $\bm{v}$, as it is sometimes done, is meaningless since by a Galilean boost, any drift velocity can be made negligible compared to the mean.}. The presence of the density ratio ensures that even for drift velocities comparable to $L\Omega$, this term is negligible.

We also have a term involving the total charge of the flow $\sum_j n_j q_j$. As shown above, this term is negligible provided that the timescale of interest is sufficiently long to recover charge neutrality, which is usually the case. We can therefore drop this term altogether to obtain the usual single fluid equations
\begin{align*}
\frac{\partial \rho}{\partial t}    +\bm{\nabla\cdot }\rho \bm{v}&=0\\
\frac{\partial \rho \bm{v}}{\partial t}+\bm{\nabla\cdot }(\rho\bm{v}\otimes\bm{v})&=-\bm{\nabla}P+\bm{f}+\frac{\bm{J\times B}}{c}.
\end{align*}

\subsubsection{Ohm's law}
\index{Plasma!Ohm's law}
In the equation of motion for the center of mass, we have left aside the fact that additional equations were required to obtain $\bm{B}$ and $\bm{J}$. Indeed, Maxwell's equations give us
\begin{align*}
\frac{\partial \bm{B}}{\partial t}&=-c\bm{\nabla\times E},\\
\bm{J}&=\frac{c}{4\pi}\bm{\nabla \times B}.
\end{align*}
The remaining unknown is therefore the electric field. Because of our assumption of electro-neutrality, we cannot use Gauss's law to compute the electric field (since under our scheme of approximation, the total charge density is zero). However, we can use the dynamical equation for charged species to deduce the electric field \emph{that is consistent with quasi-neutrality}.

Let us start with (\ref{eq:fond_mass}), and let us separate the velocity into a velocity for the center of mass, and the drift velocity for species $j$. 
\begin{align*}
\rho_j \frac{\mathrm{d} \bm{w}_j}{\mathrm{d}\,t}&=-\bm{\nabla}P_j+\bm{f}_j+n_jq_j\Bigg(\frac{\bm{w}_j\bm{\times B}}{c}+\bm{E}_\mathrm{cm}\Bigg)+\bm{R}_j\\
& \quad -\rho_j\Bigg[\bm{w}_j\bm{\cdot \nabla v}+\bm{v}\bm{\cdot \nabla w}_j+\frac{\bm{F}_\mathrm{cm}}{\rho}\Bigg]
\end{align*}
where we have defined the electric field in the center of mass frame $\bm{E}_\mathrm{cm}\equiv \bm{E}+\bm{v\times B}/c$ and the forces on the center of mass $\bm{F}_\mathrm{cm}=-\bm{\nabla}P+\bm{f}+\bm{J\times B}/c$. Several terms can be neglected here assuming that the stopping time for the species is short compared to the other timescales of the problem.
\begin{itemize}
\item     $\mathrm{d}_t \bm{w}_j$ can be neglected provided that $\Omega\ll \tau_s^{-1}$ (i.e. this assumption is identical to the quasi-neutrality assumption discussed above). In other words, the inertia of charged particles is negligible and they instantaneously reach their asymptotic velocity.
\item  Similarly, the inertial term (second line) and external forces $\bm{f}_j$ can be neglected since they modify the impulsion on timescales long compared to $\tau_s$.
\item $\bm{\nabla} P_j\sim \rho_j c_{s,j}^2/\Lambda$ is negligible provided that $c_{s,j}\lesssim \Omega\Lambda$.
\end{itemize}
The equations of motion for charged particles in the frame of the center of mass therefore read
\begin{align*}
    q_j\Bigg(\frac{\bm{w}_j\bm{\times B}}{c}+\bm{E}_\mathrm{cm}\Bigg)-\gamma_{jn}m_j\rho \bm{w}_j=0
\end{align*}
where we have assumed that dominant collisions were due to neutrals. This is usually recast as
\begin{align}
    \label{eq:ele_ohm}\bm{w}_j-\mu_j \bm{w}_j\bm{\times}\bhat=\frac{c\mu_j}{B}\Ecm,
\end{align}
where $\bhat$ is a unit vector parallel to $\bm{B}$ and
\begin{align*}
\mu_j\equiv \frac{q_jB}{\gamma_{jn}\rho m_j c},
\end{align*}
\index{Hall!parameter}
is the Hall parameter \citep{WN99}. Equation (\ref{eq:ele_ohm}) can be solved for $\bm{w}_j$, which gives the asymptotic velocity
\begin{align*}
\bm{w}_{j,\parallel}&=\frac{c\mu_j}{B}\bm{E}_{\mathrm{cm},\parallel},\\
\bm{w}_{j,\perp}&=\frac{c\mu_j}{B(1+\mu_j^2)}\Bigg[\bm{E}_{\mathrm{cm},\perp}+\mu_j\bm{E}_{\mathrm{cm},\perp}\bm{\times}\bhat\Bigg].
\end{align*}
We eventually obtain an expression closing our set of equations by relating the drift velocities to the current in the flow $\bm{J}=\sum_j n_j q_j \bm{w}_j$ and assuming quasi-neutrality $\sum_j n_j q_j=0$:
\begin{align*}
\bm{J}_\parallel&=\frac{c}{B} \Bigg(\sum_j q_j n_j \mu_j\Bigg) \bm{E}_\parallel, \\
\nonumber \bm{J}_\perp&=\frac{c}{B}\Bigg(\sum_j\frac{q_jn_j\mu_j}{1+\mu_j^2}\Bigg)\bm{E}_{\mathrm{cm},\perp}+\frac{c}{B}\Bigg(\sum_j \frac{q_j n_j}{1+\mu_j^2}\Bigg)\bhat\bm{\times E}_{\mathrm{cm},\perp}.
\end{align*}
\index{Hall!conductivity}
\index{Ohmic!conductivity}
\index{Petersen!conductivity}
These expression constitute the base of Ohm's law. We can identify 3 conductivity tensors, the Ohmic, Hall and Petersen conductivity tensors:
\begin{align*}
\sigma_O&=    \frac{c}{B}\sum_j q_j n_j \mu_j,\\
\sigma_H&=\frac{c}{B}\sum_j \frac{q_j n_j}{1+\mu_j^2},\\
\sigma_P&=\frac{c}{B}\sum_j\frac{q_jn_j\mu_j}{1+\mu_j^2},
\end{align*}
defined so that Ohm's law can be written in the more familiar form:
\begin{align*}
\bm{J}=\sigma_\parallel \bm{E}_{\mathrm{cm},\parallel}+\sigma_H\bhat\bm{\times}\bm{E}_{\mathrm{cm},\perp}+\sigma_P\bm{E}_{\mathrm{cm},\perp}.
\end{align*}
This relation can be inverted one final time to obtain the electric field in the observer frame and write the induction equation as
\begin{align*}
\frac{\partial \bm{B}}{\partial t}&=\bm{\nabla\times} \Big(\bm{v\times B}\Big)-\bm{\nabla\times} \Big(\eta_O \bm{\nabla \times B}+\eta_H(\bm{\nabla\times B})\bm{\times}\bhat+\eta_A(\bm{\nabla\times B})_\perp\Big)
\end{align*}
\index{Hall!diffusivity}
\index{Ohmic!diffusivity}
\index{Ambipolar!diffusivity}
where the magnetic diffusivities are defined as
\begin{align}
\label{eq:eta_O}
\eta_O&=\frac{c^2}{4\pi}\frac{1}{\sigma_O}\\
\label{eq:eta_H}
\eta_H&=\frac{c^2}{4\pi}\frac{\sigma_H}{\sigma_H^2+\sigma_P^2}\\
\label{eq:eta_A}
\eta_A&=    \frac{c^2}{4\pi}\Bigg(\frac{\sigma_P}{\sigma_H^2+\sigma_P^2}-\frac{1}{\sigma_O}\Bigg)
\end{align}
where the subscript O,H,A stands for Ohmic, Hall and ambipolar. 

\subsection{Non-ideal diffusivities\label{sec:nonideal}}
\subsubsection{\label{sec:twospecies}Simplified case of two charged species}
\index{Plasma!electron-ion}
In the simplest case of an plasma made of 2 singly charged species (+) and (-), we get the following simplified expressions from (\ref{eq:eta_O}), (\ref{eq:eta_H}) and (\ref{eq:eta_A})
\begin{align*}
    \eta_O&=\frac{cB}{4\pi e n_+}\Bigg(\frac{1}{\mu_+-\mu_-}\Bigg)\\
    \eta_H&=\frac{cB}{4\pi e n_+}\Bigg(\frac{\mu_++\mu_-}{\mu_--\mu_+}\Bigg)\\
    \eta_A&=\frac{cB}{4\pi e n_+}\Bigg(\frac{\mu_+\mu_-}{\mu_--\mu_+}\Bigg)
\end{align*}
First, all of these coefficients are proportional to $n_+^{-1}$, i.e. are inversely proportional to the ionisation fraction. Secondly, since $\mu_j\propto B$, we find that $\eta_O$ does not depend on $B$, while $\eta_H\propto B$ and $\eta_A\propto B^2$. Finally, we find that $\eta_H$ may have either sign. If $|\mu_-|>|\mu_+|$, we find $\eta_H>0$, and $\eta_H<0$ otherwise. Since $\mu$ is essentially a measure of the collisionality and mass of the charge carrier, it indicates that the sign of the Hall effect depends on the nature of the charge carriers. In the case where the positive and negative charge carriers have identical masses and $\gamma$ so that $\mu_-=-\mu_+$, the Hall effect vanishes. 

\begin{figure}
\centering
\includegraphics[width=0.85\hsize]{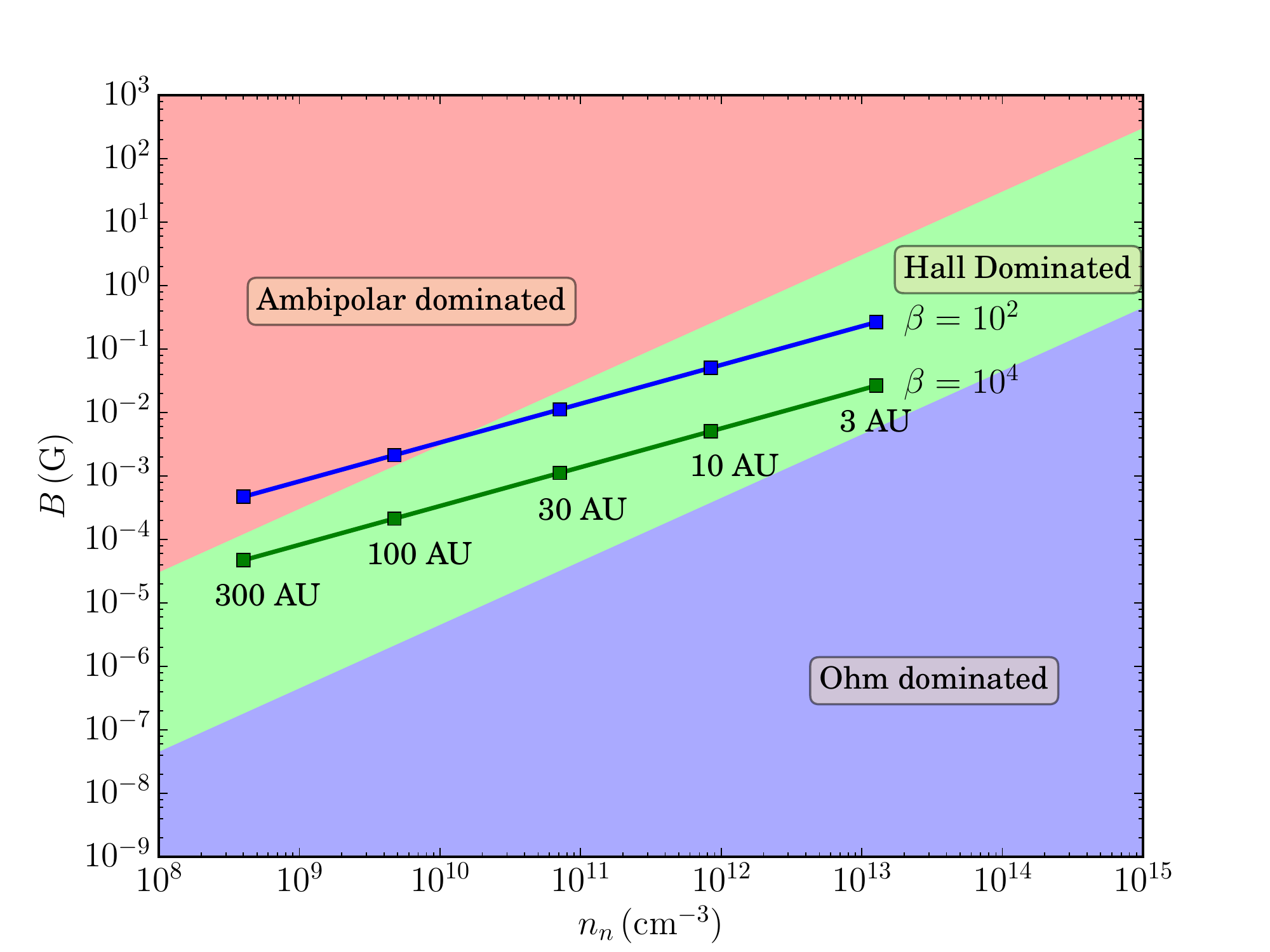}
\caption{Non-ideal regimes as a function of the neutral density and magnetic field intensity, computed for an electron-ion plasma and assuming $T<100\,\mathrm{K}$. The blue and green lines correspond to the typical values of a protoplanetary disc midplane, for various plasma $\beta$ parameters.}
\label{fig:partii:regimes}
\end{figure}

In the case of an electron-ion plasma, we have $|\mu_e|\simeq m_i/m_e|\mu_i|\gg|\mu_i| $. Hence $\eta_O\propto |\mu_e|^{-1}$ and $\eta_A\propto |\mu_i|$, which justifies the usual statement that Ohmic diffusion is due to electron-neutral collisions and Ambipolar diffusion to ion-neutral collisions. We also have $\eta_H=|\mu_e|\eta_O$ and $\eta_A=|\mu_e\mu_i|\eta_O$. Hence, we can distinguish three regimes depending on the Hall parameter of the ion and electrons:
\begin{itemize}
\item $1<\mu_i<|\mu_e|$ in which case $\eta_A>\eta_H>\eta_O$ and the regime is predominantly Ambipolar
\item $\mu_i<1<|\mu_e|$ in which case $\eta_H>(\eta_A,\eta_O)$ known as the Hall regime
\item $\mu_i<|\mu_e|<1$ where $\eta_O>\eta_H>\eta_A$ and which is dominated by Ohmic diffusion.
\end{itemize}
This allows us to delimit the Ohmic, Hall and Ambipolar regime as a function of the neutral density and the field intensity (Fig.~\ref{fig:partii:regimes}). As it can be seen, the midplane of protoplanetary discs is expected to lie mostly in the Hall regime and possibly in the ambipolar regime in the outer most parts of the disc.

A word of caution though: the physical nature of the Hall effect is different from the Ohmic and ambipolar counterparts (The Hall effect is dispersive but not diffusive since $\bm{J\times B\cdot B=0}$). Being in the Ohmic or ambipolar dominated regime does not automatically imply that the Hall effect is dynamically unimportant.

Finally, we obtain the usual expressions for the diffusivities in the electron-ion case:
\begin{align*}
    \eta_O&=\frac{c^2\gamma_{en}m_n m_e}{4\pi e^2}\frac{1}{\xi}\\
    &=2.3\times 10^{16} \bigg(\frac{\xi}{10^{-13}}\bigg)^{-1}\,\mathrm{cm}^2\mathrm{s}^{-1}\\
    \eta_H&=\frac{cB}{4\pi e n_e}\\
          &=5.0\times 10^{17}\,\bigg(\frac{\xi}{10^{-13}}\bigg)^{-1}\Bigg(\frac{B}{1\mathrm{G}}\Bigg)\Bigg(\frac{n_n}{10^{14}\,\mathrm{cm}^{-3}}\Bigg)^{-1}\,\mathrm{cm}^2\mathrm{s}^{-1}\\
    \eta_A&=\frac{B^2}{4\pi\gamma_{in}\rho \rho_i}\\
          &=1.6\times 10^{16}\, \bigg(\frac{\xi}{10^{-13}}\bigg)^{-1}\Bigg(\frac{B}{1\mathrm{G}}\Bigg)^2\Bigg(\frac{n_n}{10^{14}\,\mathrm{cm}^{-3}}\Bigg)^{-2}\,\mathrm{cm}^2\mathrm{s}^{-1}
\end{align*}

These values can be compared to diffusivities of everyday material such as iron ($\eta=8\times 10^2\,\mathrm{cm}^2/\mathrm{s}$), demineralised water ($\eta=1.4\times 10^{15}\,\mathrm{cm}^2/\mathrm{s}$) and dry air ($\eta=1.6\times 10^{24}\,\mathrm{cm}^2/\mathrm{s}$). \add{Even though one might wrongfully conclude from this that MHD effects are irrelevant, the time scales ($\sim\mathrm{year}$) and length scales ($\sim \mathrm{AU}$) are also much larger than conventional everyday experiments. This illustrates the fact that dimensionless numbers should be compared and not dimensional quantities. As we will see, one obtains magnetic Reynolds numbers of $O(1)$, which put these flows in a regime comparable to liquid sodium experiments on Earth.} 

\subsubsection{\label{sec:diffu_application}Dimensionless numbers and application to disc models}
It is customary to define dimensionless numbers in association with non-ideal effects in order to quantify their relative importance in the induction equation. First one can define Elsasser numbers
\index{Hall!Elsasser number}
\index{Ohmic!Elsasser number}
\index{Ambipolar!Elsasser number}
\index{Elsasser number}
\index{$\Lambda$ Elsasser number}
\begin{align*}
\Lambda_{O,H,A}\equiv \frac{V_A^2}{\Omega \eta_{O,H,A}}
\end{align*}

where $V_A$ is the Alfv\'en speed and $\Omega$ is the rotation rate of the system. Note however that $\Lambda_O\propto B^2$ and $\Lambda_H\propto B$, which makes these numbers less useful when it comes to predicting the saturation of MHD instabilities since $B$ is \emph{a priori} unknown. It is therefore useful to define two additional dimensionless numbers, the magnetic Reynolds number and the Hall Lundquist number
\index{Hall!Lundquist number}
\index{Ohmic!Reynolds number}
\index{Magnetic Reynolds number}
\index{$\mathrm{Rm}$ magnetic Reynolds number}
\index{$\mathcal{L}_\mathrm{H}$ Hall Lundquist number}
\begin{align*}
\mathrm{Rm}&\equiv\frac{\Omega H^2}{\eta_O},\\
\mathcal{L}_\mathrm{H}&\equiv \frac{V_A H}{\eta_H}.
\end{align*}
\begin{figure}
\centering
\includegraphics[width=0.95\hsize]{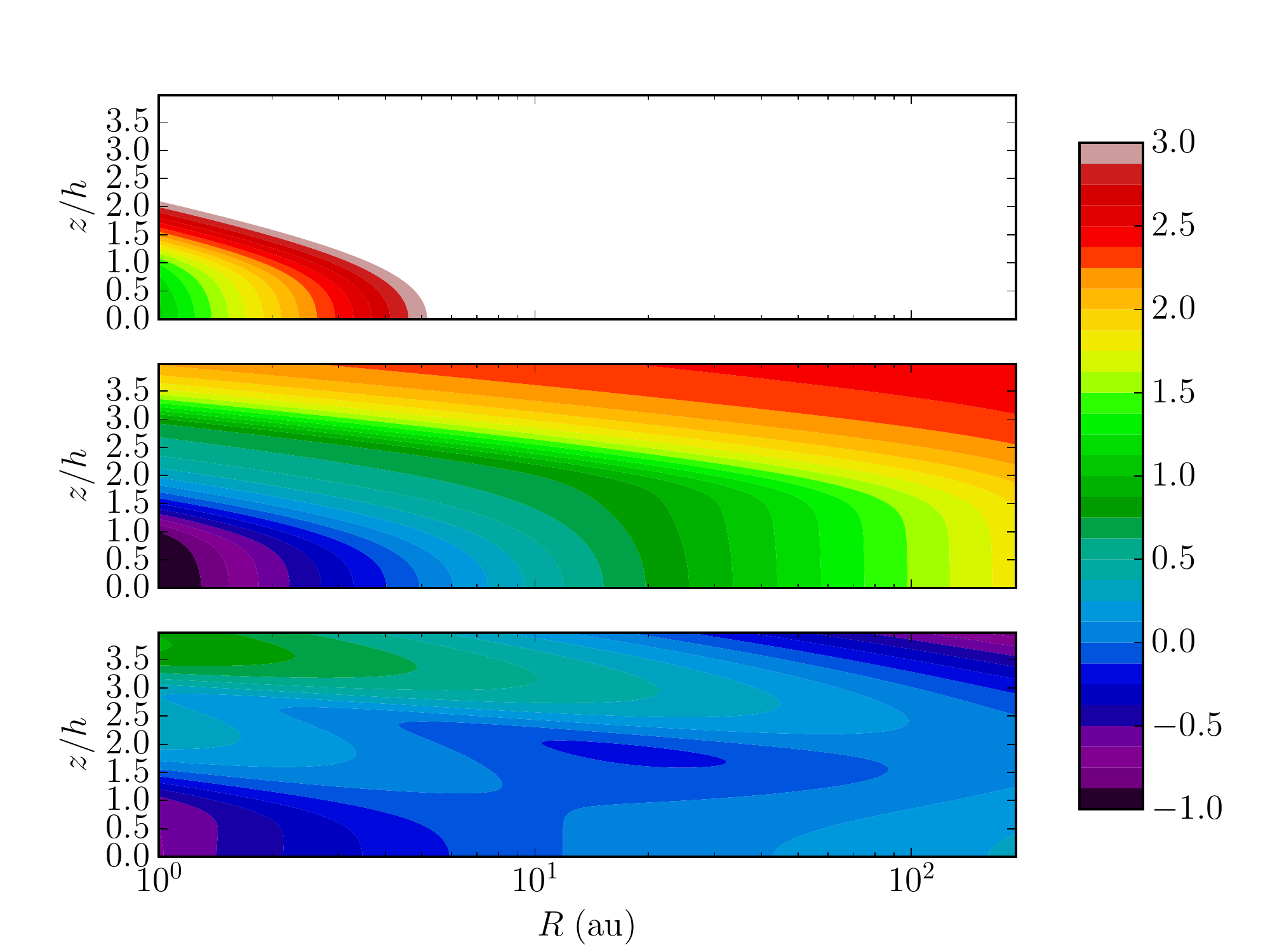}
\caption{Magnetic reynolds number $\mathrm{Rm}$ (top), Hall Lundquist number $\mathcal{L}_H$ (middle) and Ambipolar Elsasser number $\Lambda_A$ (bottom) in our disc model (\S\ref{sec:profile}), using a simple ion-electron approximation with a metal-free chemistry. White values are $>10^3$. }
\label{fig:partii:ni_effects}
\end{figure}

\index{Dead zone}
These two numbers do not depend on the field strength (at least in the two species plasma case), and they turn out to be excellent saturation predictors in the non-linear regime of the MRI. We show in Fig.~\ref{fig:partii:ni_effects} the dimensionless numbers resulting from our grain-free metal-free ionisation model. As it can be seen, $\mathrm{Rm}<10^3$ only in the innermost regions of the disc. This is the region which was historically defined as the ``dead zone'' \citep{G96}. In addition, we find $10^{-1}<\mathcal{L}_\mathrm{H}<10$ in most of the disc midplane with a sharp increase at the disc surface while $\Lambda_A\simeq 1$ in most of the disc.

\subsubsection{\label{sec:roleofgrains}The role of grains}
\index{Grains!diffusivity}
When it comes to the conductivity of protoplanetary discs, grains play essentially two roles.
\begin{enumerate}
\item By capturing free electrons, they become predominantly negatively charged, and they increase the recombination rate with ions thanks to their large cross-section and reaction rates at the grain surface. The end product is generally a reduced ionisation fraction, possibly by several orders of magnitude (see \S\ref{sec:ion_profile}).
\item Due to their high inertia, charged grains enter into the conductivity tensor as a very low Hall parameter species. In this case, the scaling laws obtained for the two species case do not hold anymore. The abundance of charged grains, therefore, changes the diffusion regime in which the system lies.
\end{enumerate}

\begin{figure}
\centering
\includegraphics[width=0.48\hsize]{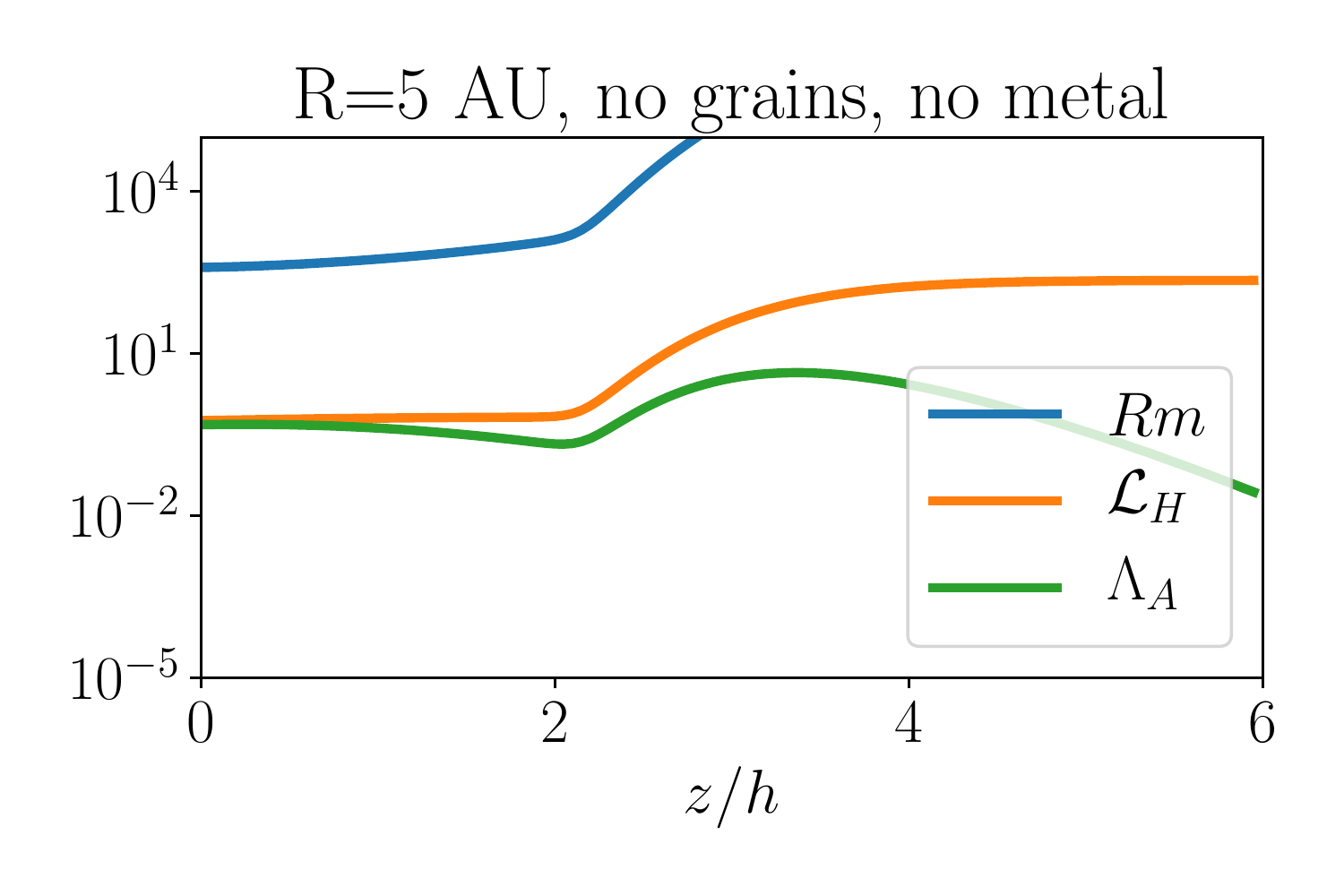}
\includegraphics[width=0.48\hsize]{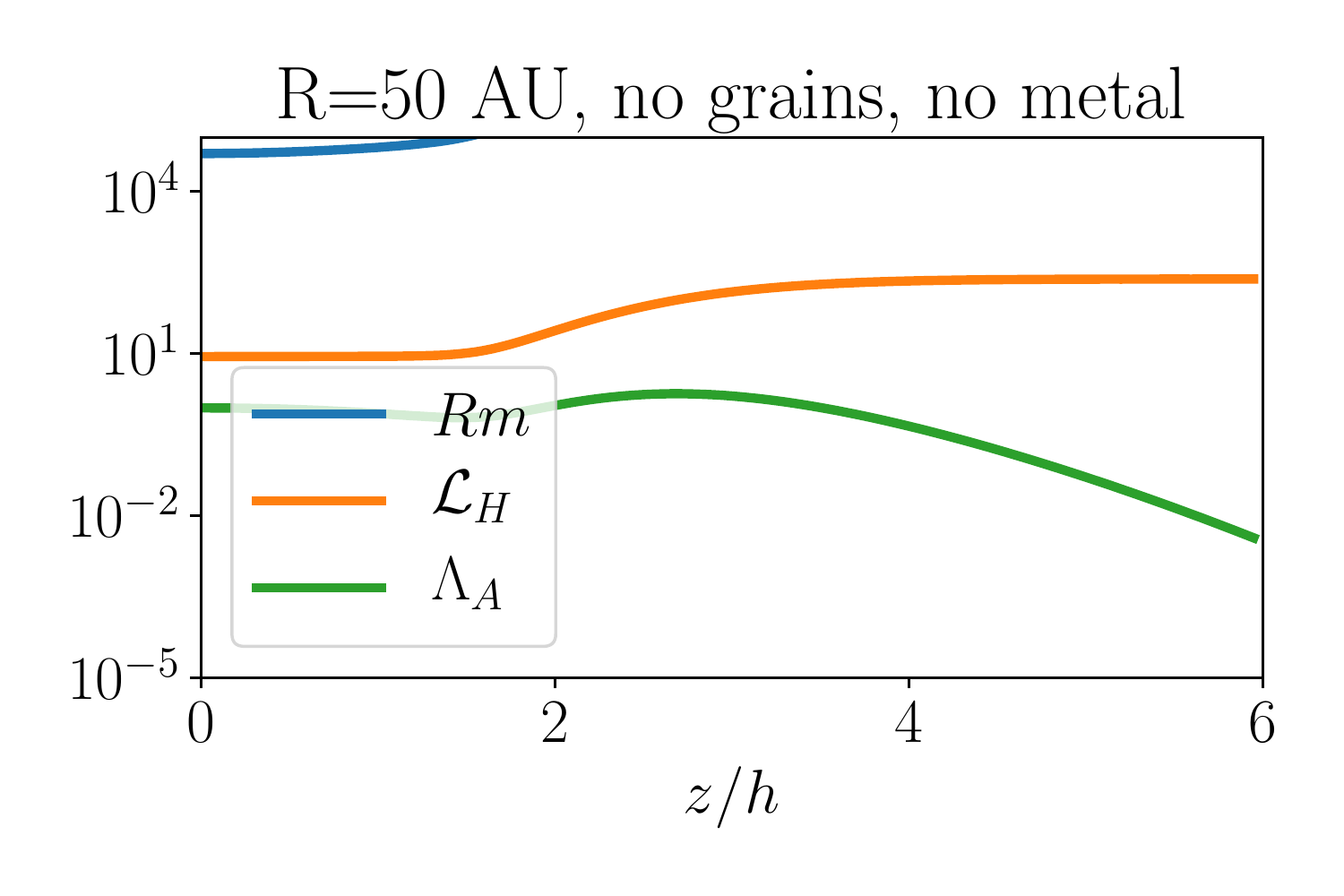}
\includegraphics[width=0.48\hsize]{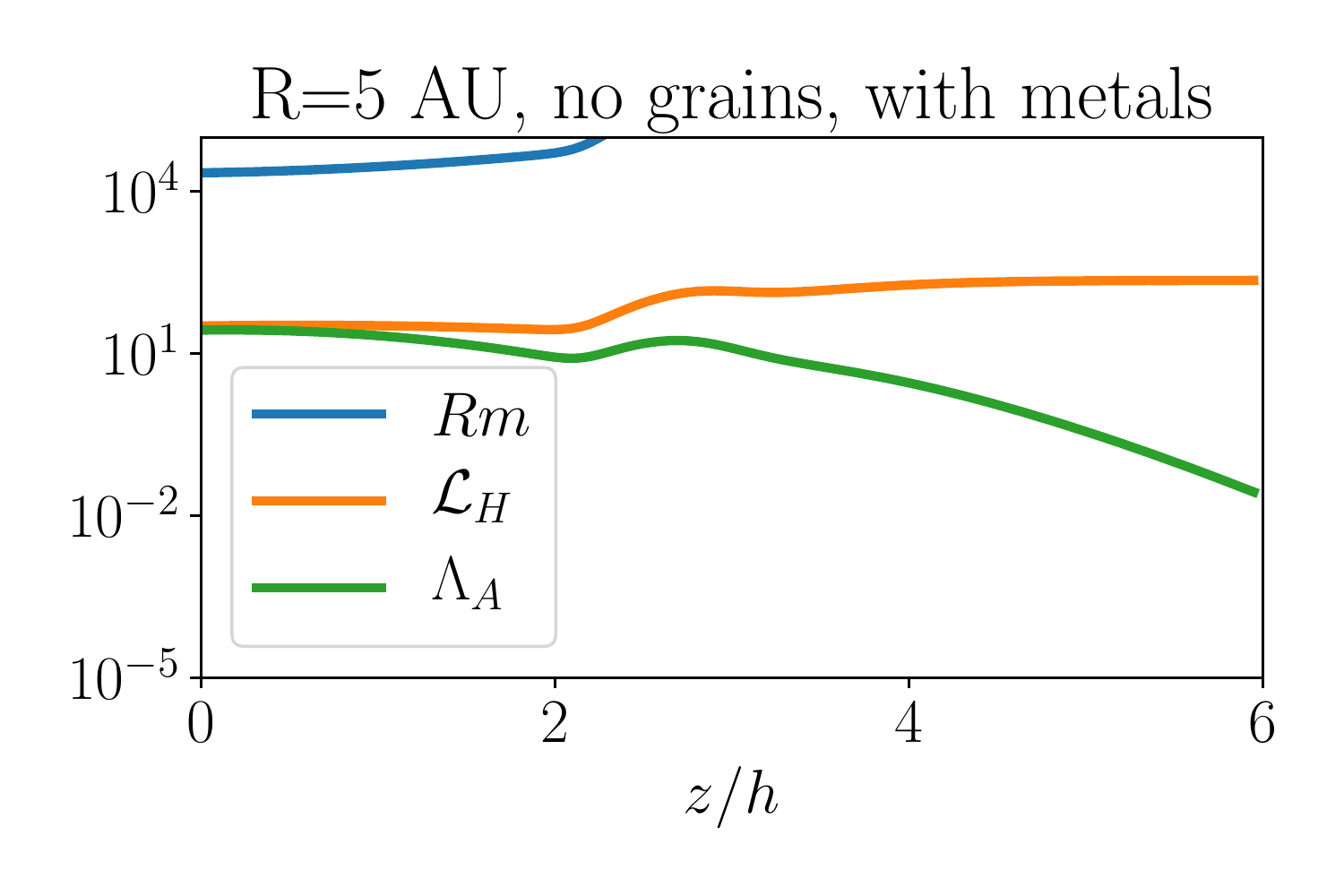}
\includegraphics[width=0.48\hsize]{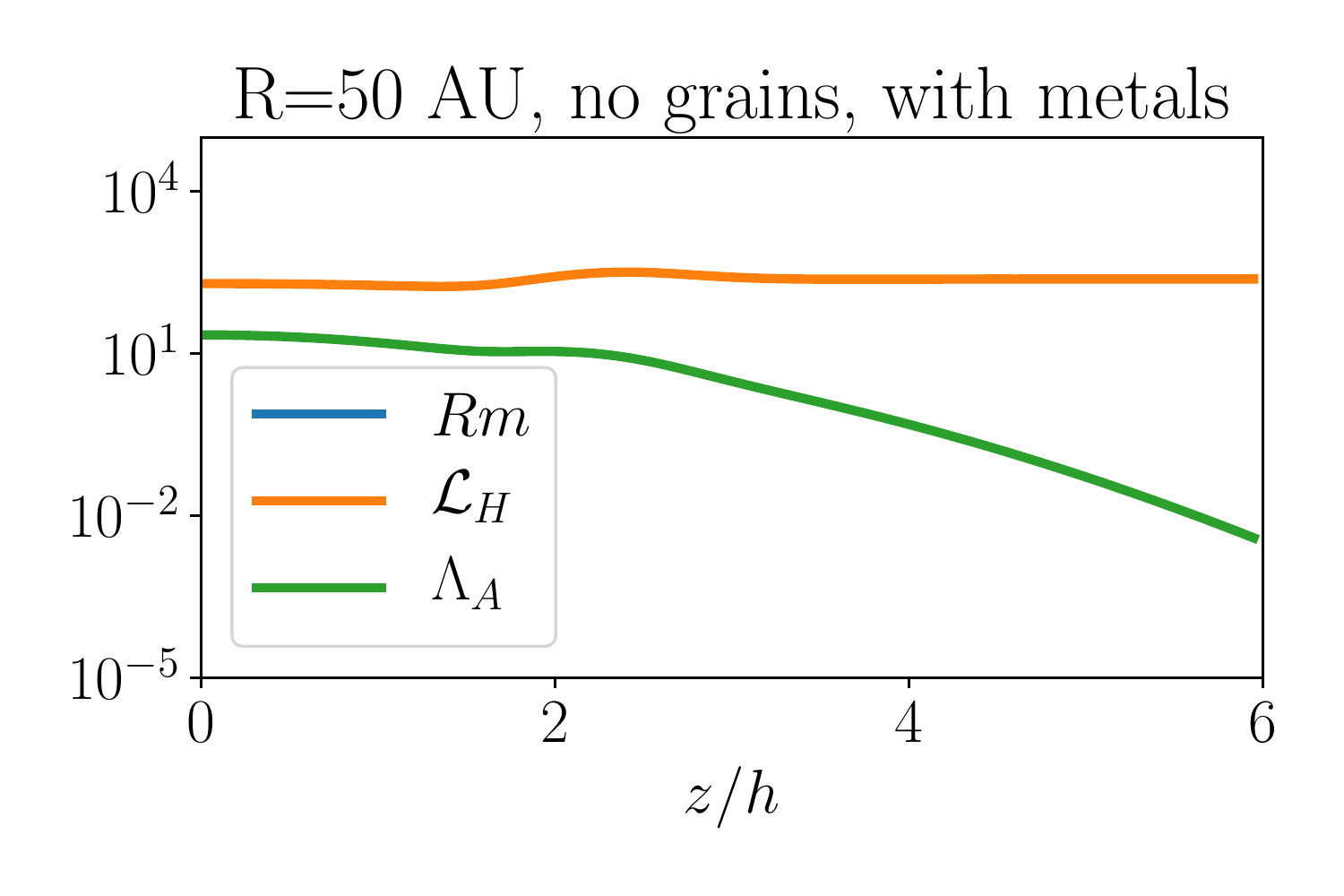}
\includegraphics[width=0.48\hsize]{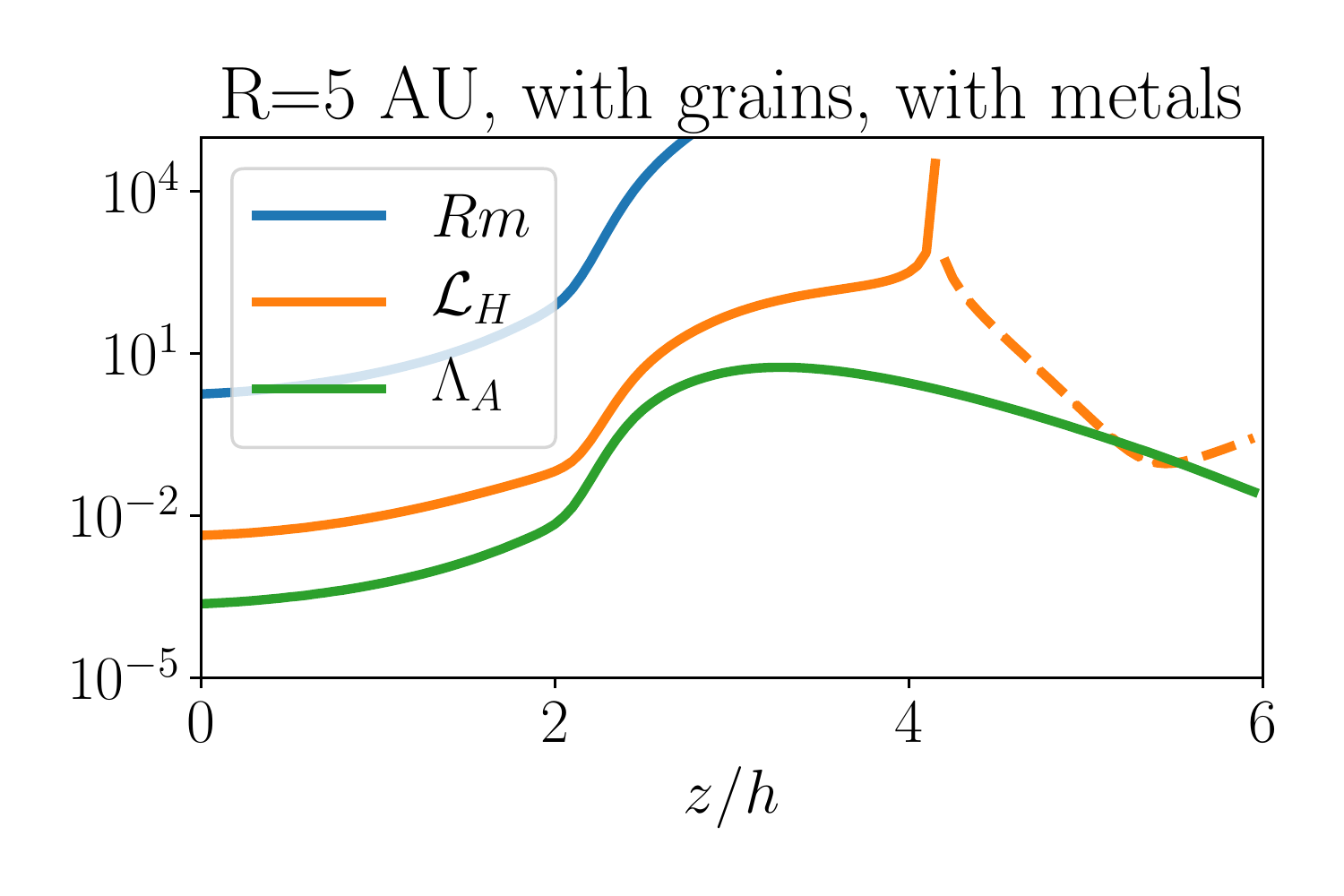}
\includegraphics[width=0.48\hsize]{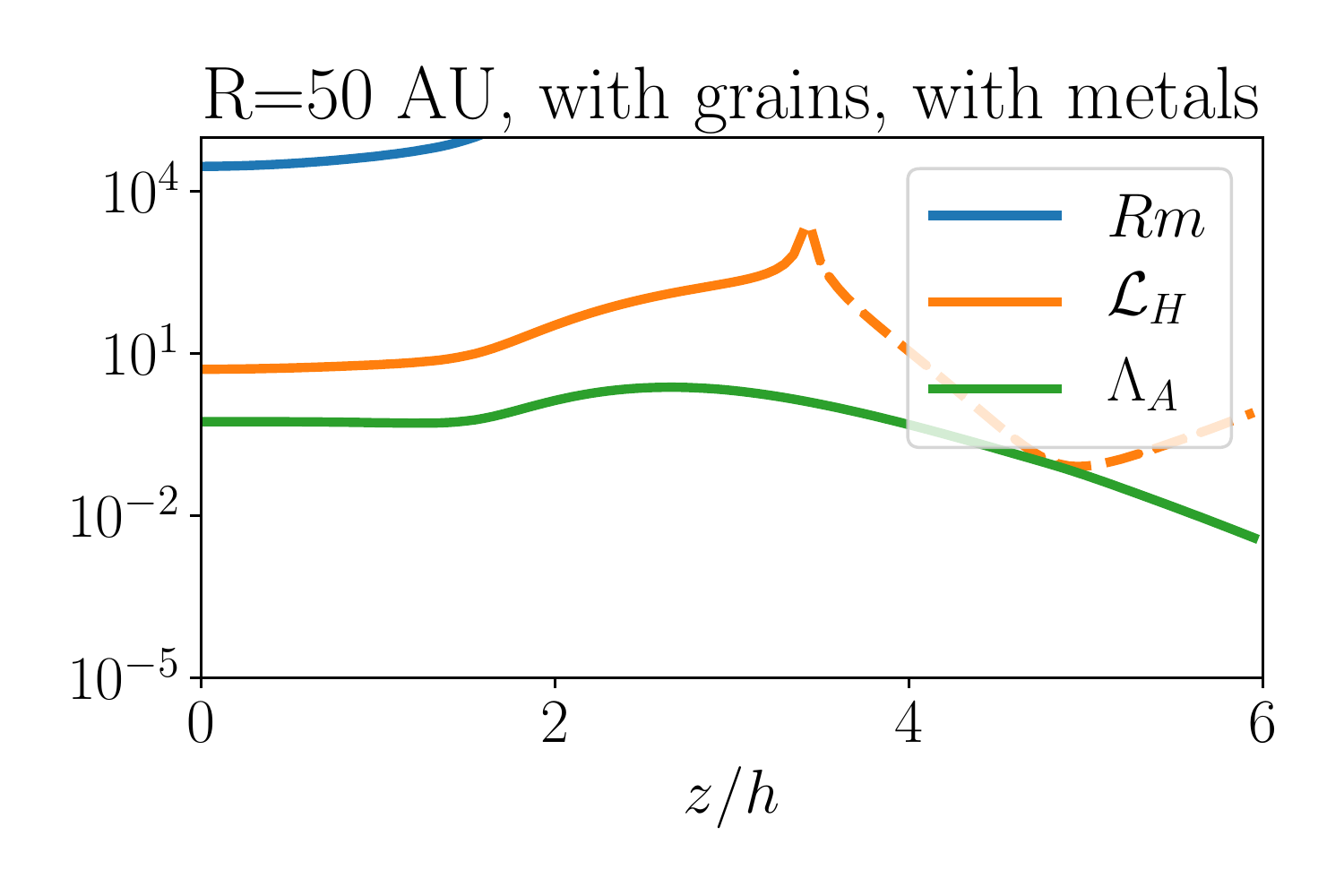}
\caption{Dimensionless diffusivity for 3 different compositions: row 1: no grains, no metals (identical to Fig.~\ref{fig:partii:ni_effects}); row 2: no grains with [M]=$10^{-8}$ ; row 3: with $a=0.1\mu\mathrm{m}$ grains and metal atoms. The first column corresponds to $R=5~\mathrm{AU}$ and the second column $R=50~\mathrm{AU}$. Dashed lines correspond to negative diffusivities for the Hall effect.}
\label{fig:partii:diffusivity_grains}
\end{figure}

To illustrate how the diffusivity depends on the presence of grains, we present in Fig.~\ref{fig:partii:diffusivity_grains} an example of diffusivity computation in plasmas of different compositions (these compositions are identical to the ones discussed in \S\ref{sec:ion_profile}). As it can be seen, the addition of $0.1~\mu\mathrm{m}$ size grains to the system has a dramatic impact on the diffusivities: all of the dimensionless numbers decrease by several orders of magnitude close to the midplane, while ambipolar diffusion becomes stronger than Hall in the case with grains. Overall, ambipolar diffusion increases by $10^4$ while Hall and Ohmic increase by $10^2$ compared to the fiducial metal-free case. The Hall effect also changes sign at the disc surface. \add{This arises because of the presence of negatively charged grains which contribute to the disc conductivity tensor by reducing the "effective" Hall parameter of negative charge carriers (electrons+grains-). In the end, the Hall conductivity becomes dominated by ions when they become sufficiently abundant: at the disc surface.}

In the grain-free case, the presence of metal atoms tends to decrease the diffusivities by typically 1-2 orders of magnitude. However, let us point out that the effect of metal atoms disappears once grains are sufficiently abundant \citep{IN06}. Our grain-free metal-rich model is, therefore, a best-case scenario for the ionisation fraction and the diffusivities.

The simplified grain model we have used is by no mean the final answer to this question. However, it demonstrates the strong impact of grains on the dynamics of the plasma. The intensity of this effect also depends on the grain size and grain abundance, a lower abundance or larger grain size leading to a smaller effect \citep{IN06,SW08}.  Here, we have purposely chosen a very low grain size with an interstellar abundance to illustrate a worst case scenario for the ionisation fraction. Finally, if one assumes polycyclic aromatic hydrocarbons (PAHs) are present in the gas phase, they then behave as very small grains, capturing all of the floating electrons and also affecting significantly the amplitude of non-ideal effects \citep{B11}.

\subsubsection{Conclusion on non-ideal MHD effects}
Overall, there is no general consensus on the quantitative strength of non-ideal MHD effects in the outer part ($R>1~\mathrm{AU}$) of protoplanetary discs. It is clear that these effects are qualitatively very important though, and that these objects are \emph{far} from the ideal MHD regime. Let us summarise here the source of uncertainty and their implication for the strength of non-ideal effects.
\begin{description}
\item[Ionisation rate:]\ because cosmic rays can be both shielded by the stellar wind, or``locally'' produced in shocks surrounding the forming star, there is an tremendous uncertainty of 6 orders of magnitude on the ionisation rate due to cosmic rays (see the discussion in \S\ref{sec:ion_source}). Since this mechanism is the main source of ionisation in the disc below two scale heights, this implies a 3 orders of magnitude uncertainty in the ionisation fraction $\xi$ and similarly on the diffusivity coefficients. Ionisation due to X-rays is also subject to caution since the X-ray flux coming from the star is largely variable, leading to order of magnitude fluctuations of the ionisation fraction close to the disc surface.

\item[Disc structure:]\ the disc structure is a fundamental parameter which determines the penetration depth of ionising radiations but also the recombination rate. Denser disc models, such as the minimum mass solar nebula, tend to have lower ionisation fractions and larger diffusion. We have shown that by comparing a theoretical MMSN disc model to a model favoured by observations (\S\ref{sec:profile}), one can change the ionisation fraction by 2 orders of magnitude (see \S\ref{sec:ion_profile}). Since the gas column density profile is largely unknown for $R\sim 10\,\mathrm{AU}$, one is forced to use the gas column density as a free parameter.

\item[Grains:]\ as shown above, grains affect both the ionisation fraction and the dependence of the diffusivities on $\xi$. Overall, grains tend to reduce the ionisation fraction by several orders of magnitude (typically 2-3). Because of the change in composition (grains become the dominant charge carrier close to the midplane), diffusivities increase by 2-4 orders of magnitude in our worst-case scenario, compared to the metal-free case. In addition, the Hall diffusivity $\eta_H$ can be reversed. The effect of grains naturally depends on the assumed grain size and abundance. It is usually found that grain size $a>1~\mu\mathrm{m}$ do not impact the conductivity tensor too much \citep{SW08} and that a significant depletion of small grains also reduces their effect \citep{IN06}. All of these calculations assume all of the grains have the same size, which most presumably overestimates the abundance of grains and their impact on the conductivity tensor. More realistic grain size distribution, including more complex chemical reaction networks \citep[e.g.][]{TL19} tend to obtain diffusivities which are within an order of magnitude of the diffusivities discussed in our grain-free metal-free scenarios. 

Last, the physics at grain surface is poorly understood which gives a lot of freedom to chemical models. For instance, some authors assume a fixed sticking coefficients of electrons and ions on dust grains as we do \citep{SM00,W07}, while others include dependencies as a function of the grain size, charge and temperature \citep[note however that the dependency of the sticking coefficient on the grain charge differs significantly between these authors]{IN06,B11b}. This can have an additional order of magnitude impact on the resulting diffusivities.
\end{description}

Overall, one is forced to conclude that the conductivity tensor of protoplanetary discs is plagued by uncertainties and that no chemical/ionisation/grain model is better than the other. Given the discussion above, the uncertainty on the diffusion coefficient is at least $\pm 3$ orders of magnitudes, which has dramatic effects on the dynamical behaviour of these objects. Until more constraints are obtained for these coefficients, theoreticians are forced to explore in a more or less systematic manner the parameter space of the conductivity tensor. 

Because of these uncertainties, we will focus in the following on the "intermediate" case of a metal-free grain-free case which we discussed in \S\ref{sec:diffu_application} and for which diffusivities are given in Fig.~\ref{fig:partii:ni_effects}.

\newpage
\part{Disc dynamics: global \& local views}
\section{Introduction}
\subsection{Motivations}
Explaining accretion in discs is a long-standing problem of modern astrophysics. Even though angular momentum transport equations have been known for a long time, the road to quantifying the level of stress in various discs has been paved by unforeseen difficulties. The main idea which has been followed since the pioneering work of \cite{SS73} is that accretion discs are somehow turbulent, and this turbulence generates a \emph{radial} stress. The key is then to relate this radial stress to the other large-scale quantities such as the disc \add{surface density $\Sigma$ and thickness $H$, the rotation rate $\Omega$, the diffusivities, the magnetic field strength, etc}. This approach is, in essence very similar to the mixing length theory of convection, except that in the disc case, one doesn't transport heat but angular momentum. In discs, it is called the $\alpha$-disc theory.

Here, we present the basic concepts behind accretion and the $\alpha$-disc theory. Then, we introduce the magnetorotational instability (MRI), which is probably the most promising instability to explain the origin of accretion in astrophysical discs. Finally, we apply the MRI in the context of protoplanetary discs, taking into account non-ideal MHD effects.

\subsection{Disc equilibrium\label{sec:equilibrium}}
A protoplanetary disc is typically made of gas (and possibly dust) orbiting a young stellar object of mass $M$. Here, we assume that the gravity of the orbiting gas onto itself (self-gravity) is negligible. This is not necessarily true in very massive discs or in the outer parts of young class 0 objects. Under these assumptions, the gravitational potential is simply that of the central object and the equilibrium may simply be written

\begin{align*}
0&=-\frac{1}{\rho}\frac{\partial P}{\partial R}-\partial_R\psi+\Omega^2R\\
0&=    -\frac{1}{\rho}\frac{\partial P}{\partial z}-\partial_z\psi,
\end{align*}
where $(R,z)$ are cylindrical coordinates and $\Omega$ is the angular velocity of the flow, which we assume only depends on $R$ and $\psi=-GM/(R^2+z^2)^{1/2}$ is the cylindrical potential. A useful quantity will be the Keplerian frequency which corresponds to the orbital frequency of a test particle on a circular orbit at radius $R$:
\index{Keplerian frequency}
\begin{align*}
\Omega_K(R)=\sqrt{\frac{GM}{R^3}}    
\end{align*}

In order to simplify the computation, let us assume that the disc is locally isothermal\footnote{This assumption is approximately valid since protoplanetary discs are passively irradiated. In turbulent discs, recent 3D RMHD simulations show a vertical temperature profile very close to isothermal, e.g. \citealt{FF13}}: $T(R)$. Under these assumptions, the sound speed may be written
\index{Sound speed}
\begin{align*}
c_s&\equiv\sqrt{\frac{P}{\rho}}=\sqrt{\frac{kT}{\mu}}
\end{align*}
where $k$ is Boltzmann's constant and $\mu$ is the mean molecular mass. Since the disc is locally isothermal, $c_s$ only depends on $R$, as the temperature does.

We start with the vertical equilibrium which we consider close to the disc midplane ($z\ll R$) since we assume the disc is thin:
\begin{align*}
c_s^2\partial_z\log \rho&=-\frac{GMz}{(R^2+z^2)^{3/2}}\\
&\simeq z\Omega_K ^2+O(z^3),
\end{align*}
where we have assumed $z\ll R$. We deduce from this the vertical density profile
\begin{align*}
\rho=\rho_0(r)\exp\Big(-\frac{z^2}{2H^2}\Big)    
\end{align*}
where we have defined the disc scale height
\index{Disc thickness}
\begin{align}
\label{eq:thickness}
H\equiv c_s/\Omega_K.
\end{align}
The thin disc approximation $H\ll R$ implies that the disc is cold, or in other words that $c_s\ll R\Omega_K$. 

In the radial direction, we first have to compare the radial pressure gradient to the gravitational potential
\begin{align*}
0&=\underbrace{-\frac{1}{\rho}\frac{\partial P}{\partial R}}_{\sim c_s^2/R}-\underbrace{\partial_R\psi}_{\sim \Omega_K^2R}+\Omega^2R
\end{align*}
The pressure gradient is $(H/R)^2$ smaller than the gravitational potential and can be neglected in the thin disc approximation. This means that the disc is \emph{to a very good approximation} a Keplerian disc $\Omega=\Omega_K$. Note however that local (i.e. on radial length scales of the order of $H$) pressure variations may exist leading to measurable deviation from the Keplerian rotation. These variations are typically responsible for zonal flows and local pressure maxima.

\subsection{Accretion theory}
\index{Accretion!theory}
\add{The energetics of MHD-driven discs has been extensively discussed by \cite{BGH94}, \cite{BH98} and \cite{BP99}. Here, we revisit this question, and include the possibility of wind-driven accretion in the system.} The accretion of mass in astrophysical discs is described by the equation of mass, angular momentum, and mechanical energy conservation equations:
\begin{align}
\label{eq:cont_fll}
\frac{\partial \rho}{\partial t}&+\bm{\nabla \cdot}\rho\bm{u}=0\\
 \frac{\partial R\rho u_\phi}{\partial t}&+\bm{\nabla\cdot}\Big[R\rho u_\phi\bm{u}-R\frac{B_\phi\bm{B}}{4\pi} \Big]=0\\
\nonumber\frac{\partial \frac{1}{2}\rho u^2+\rho \psi +\frac{B^2}{8\pi}}{\partial t}&
+\bm{\nabla\cdot}\Bigg[\Big(\frac{1}{2}\rho u^2+\rho \psi+P +\frac{B^2}{4\pi}\Big)\bm{u}\\
&\, \quad\quad\quad\quad-\frac{\bm{u\cdot B}}{4\pi}\bm{B}-\frac{\Eni\times \bm{B}}{4\pi}\Bigg]= P\bm{\nabla \cdot u}+\frac{\Eni\cdot \bm{J}}{c}
\end{align}
where $\Eni$ are electromotive forces due to non-ideal effects. \add{Note that molecular viscosity is usually negligible in these equations as it is several orders of magnitude smaller than non-ideal MHD effects. One notable exception is naturally when non-ideal MHD effects are absent, such as in ideal-MHD flows or in purely hydrodynamic flows subject to turbulence and/or spiral density waves. In these cases, viscosity becomes non-negligible in the energy equation because of the formation of small-scale structures, either through a direct turbulent cascade, or thanks to shocks. In any case, this viscosity then leads to an additional definite negative source term in the energy equation, which transforms mechanical energy into heat. The energy flux and angular momentum flux terms associated to viscosity are always negligible for practical applications.}

In order to capture the dynamics of the disc, we separate the gravitational potential $\psi$ as a midplane potential $\Psi$ and a deviation as one moves away from the disc midplane $\Phi$:
\begin{align}
\psi=\Psi(R)+\Phi(R,z)	
\end{align}
\add{We also separate the mean rotational motion of the disc from its deviations} (not necessarily small):
\begin{align*}
u_r=v_r\quad;\quad
u_\phi=\Omega R+v_\phi\quad;\quad
u_z=v_z,
\end{align*}
where we only assume that $\Omega$ satisfies the radial equilibrium in the disc midplane
\begin{align*}
\Omega^2R=\partial_R \Psi
\end{align*}
Under these assumptions, it is possible to rewrite the angular momentum conservation as
\begin{align}
\label{eq:ang_mom_fll}
\frac{\partial R\rho v_\phi}{\partial t}+\rho \bm{u}\bm{\cdot\nabla }(\Omega R^2)+\bm{\nabla\cdot}\Big[R\rho v_\phi\bm{u}-R\frac{B_\phi\bm{B}}{4\pi} \Big]=0.
\end{align}
where we have used the continuity equation to eliminate the terms proportional to $\Omega R^2$. A similar procedure can be followed for the energy equation, which can be written as
\begin{align*}
\big(\frac{1}{2}\Omega^2R^2+\Psi\big)\Big[	\frac{\partial \rho }{\partial t}&+\bm{\nabla \cdot}\rho\bm{u}\Big]+\rho \bm{u \cdot \nabla}\big(\frac{1}{2}\Omega^2R^2+\Psi\big)\\
+\Omega\Bigg(\frac{\partial R\rho v_\phi}{\partial t}+\bm{\nabla\cdot}\Big[R\rho v_\phi\bm{u}&-R\frac{B_\phi\bm{B}}{4\pi} \Big]\Bigg)+\Big[R\rho v_\phi\bm{u}-R\frac{B_\phi\bm{B}}{4\pi} \Big]\bm{\cdot\nabla} \Omega\\
+\frac{\partial \frac{1}{2}\rho v^2+\rho \Phi +\frac{B^2}{8\pi}}{\partial t}&+\bm{\nabla\cdot}\Bigg[\Big(\frac{1}{2}\rho v^2+\rho \Phi+P +\frac{B^2}{4\pi}\Big)\bm{v}\\
 &\,\qquad\qquad-\frac{\bm{v\cdot B}}{4\pi}\bm{B}-\frac{\Eni\times \bm{B}}{4\pi}\Bigg]=P\bm{\nabla \cdot u}+\frac{\Eni\cdot \bm{J}}{c}.
\end{align*}
We recognise the mass conservation equation in the first line, and the angular momentum conservation equation in the second line. Substituting (\ref{eq:cont_fll}) and (\ref{eq:ang_mom_fll}) in the equation above allows us to recast energy conservation as
\begin{align*}
\rho \bm{u \cdot}[\nabla \Psi-\Omega^2R\bm{\nabla}R]&+\Big[R\rho v_\phi\bm{u}-R\frac{B_\phi\bm{B}}{4\pi} \Big]\bm{\cdot\nabla} \Omega\\
+\frac{\partial \frac{1}{2}\rho v^2+\rho \Phi +\frac{B^2}{8\pi}}{\partial t}&+\bm{\nabla\cdot}\Bigg[\Big(\frac{1}{2}\rho v^2+\rho \Phi+P +\frac{B^2}{4\pi}\Big)\bm{v}\\
 &\,\qquad\qquad-\frac{\bm{v\cdot B}}{4\pi}\bm{B}-\frac{\Eni\times \bm{B}}{4\pi}\Bigg]=P\bm{\nabla \cdot u}+\frac{\Eni\cdot \bm{J}}{c}.
\end{align*}
where we recognise the radial equilibrium in the first term, which can be cancelled out. Hence, we get an energy equation for the velocity \emph{fluctuations} which reads
\begin{align}
\label{eq:energ_fll}
\nonumber 	\frac{\partial \frac{1}{2}\rho v^2+\rho \Phi +\frac{B^2}{8\pi}}{\partial t}+\bm{\nabla\cdot}\Bigg[\Big(\frac{1}{2}&\rho v^2+\rho \Phi+P +\frac{B^2}{4\pi}\Big)\bm{v}-\frac{\bm{v\cdot B}}{4\pi}\bm{B}-\frac{\Eni\times \bm{B}}{4\pi}\Bigg]\\&=P\bm{\nabla \cdot u}-\Big[R\rho v_\phi\bm{u}-R\frac{B_\phi\bm{B}}{4\pi} \Big]\bm{\cdot\nabla} \Omega+\frac{\Eni\cdot \bm{J}}{c}
\end{align}

\subsubsection{Averaged equations}
In order to compute averaged conservation equations, we define an azimuthal average as
\begin{align*}
\langle Q\rangle =\frac{1}{2\pi}\int d\phi \,Q	
\end{align*}
and a vertical integration of the azimuthal average
\begin{align}
\label{eq:avg}
\overline{Q}=\int_{z=-h}^{z=+h}dz\,\langle Q\rangle,
\end{align}
so that the continuity equation (\ref{eq:cont_fll}) reads
\begin{align}
\label{eq:mass}
\frac{\partial \Sigma}{\partial t}+\frac{1}{R}\frac{\partial}{\partial R}R \overline{\rho u_r}+\Big[\langle \rho v_z\rangle\Big]_{z=-h}^{+h}=0
\end{align}
where $\Sigma\equiv\overline{\rho}$ is the gas surface density.

The equation of angular momentum conservation (\ref{eq:ang_mom_fll}) can be recast using the same averaging procedure (\ref{eq:avg}) defined above to get an equation relating the mass accretion rate $\overline{\rho v_r}$ as to the radial and surface stresses
\index{Angular momentum conservation}
\index{Stress!accretion theory}
\begin{align}
\label{eq:angular_final}
\overline{\rho v_r}\frac{\partial}{\partial R}\Omega R^2+\frac{1}{R}\frac{\partial}{\partial R} R^2\Bigg[\underbrace{\overline{\rho v_\phi v_r}-\frac{\overline{B_\phi B_r}}{4\pi}}_{\textrm{Radial stress}}\Bigg]+\underbrace{\Bigg[R\langle \rho v_\phi v_z\rangle-R\frac{\langle B_\phi B_z\rangle}{4\pi}\Bigg]_{z=-h}^{+h}}_{\textrm{Surface stress}}=0.
\end{align}
where we have assumed that $v\ll\Omega R$ which allows us to neglect the remaining time derivative.

This demonstrates the close relationship between the accretion rate and the transport of angular momentum by the stresses. Angular momentum can be transported outward in the disc by the radial stress, or evacuated from the disc by a torque applied at the disc surface, as for example when a magnetised wind is present.

This link between accretion and stress can also be seen by averaging of the mechanical energy equation (\ref{eq:energ_fll}):
\begin{align}
\partial_t\overline{\mathcal{E}_m}+\frac{1}{R}\frac{\partial}{\partial R} R \overline{\mathcal{F}_{m,R}}	+\Big[\langle \mathcal{F}_{m,z}\rangle \Big]_{z=-h}^{+h}=\overline{P\bm{\nabla \cdot v}}-\underbrace{\Big[\overline{\rho v_\phi v_R}-\frac{\overline{B_\phi B_R}}{4\pi} \Big]\frac{d\Omega}{d\log R}}_{\textrm{Radial stress source term}}+\overline{\frac{\Eni\cdot \bm{J}}{c}}
\end{align}
where we have the mechanical energy of the \emph{fluctuations}
\begin{align*}
	\mathcal{E}_m=\frac{1}{2}\rho v^2+\rho \Phi +\frac{B^2}{8\pi}
\end{align*}
and its associated energy flux
\begin{align}
\Big(\mathcal{E}_m+P+\frac{B^2}{8\pi}\Big)\bm{v}-\frac{\bm{v\cdot B}}{4\pi}\bm{B}-\frac{\Eni\times \bm{B}}{4\pi}.
\end{align}
This energy equation demonstrates a very important fact: unless one assumes that the energy flux locally deposits energy (which implies that a source of energy is externally provided to the disc), then the only term which can balance diffusive (and viscous, when applicable) losses is the radial stress source term, which appears as a source term in the conservation of mechanical energy. Diffusive (and viscous) source terms being necessarily negative definite, we have
\begin{align}
	\Big[\overline{\rho v_\phi v_R}-\frac{\overline{B_\phi B_R}}{4\pi} \Big]\frac{d\Omega}{d\log R}<0.
\end{align}
Since this term balances losses (which convert mechanical energy into heat), it is also equal to the local heating rate of the disc is we assume the fluctuations are statistically steady (as in a saturated turbulent state) and no energy escapes via the vertical energy flux\footnote{When an outflow is present, one finds that the vertical energy flux $\mathcal{F}_{m,z}$ extracts energy from the disc, which implies that heating is actually smaller than the radial stress source term}.  Note that the surface stress does not appear as a source term, as it does not lead to any local heating, despite driving accretion. That's one of the key difference between radially-driven and vertically-driven accretion.

\subsection{$\alpha$ disc theory\label{sec:alpha-disc}}
\index{$\alpha$ disc}
This theory assumes no wind is present at the disc surface. In order to solve the long-term evolution of the disc, one needs to express the radial stress
\index{Stress!$\alpha$-disc theory}
\begin{align*}
\overline{W_{r\phi}}=\overline{\rho v_\phi v_r}-\frac{\overline{B_\phi B_r}}{4\pi},
\end{align*}
as a function of vertically averaged quantities such as $\Sigma$, $\overline{P}$, etc. Historically, and based on a purely dimensional argument \citep{SS73}, it is usually assumed that
\begin{align*}
    \overline{W_{r\phi}}=\alpha \overline{P},
\end{align*}
where $\alpha$ is a dimensionless constant. Physically, it can however be justified as a mixing length theory: let us consider turbulent velocity fluctuations $v$ in a thin disc. The fluctuations are confined in the disc thickness $H$ with a forcing frequency $\Omega_K$ (these two quantities are the only length and frequency accessible to an ideal system). Hence, we expect $v=\theta H\Omega_K$ where $\theta$ is a dimensionless constant, of order unity. Therefore $\overline{W_{r\phi}}=\theta^2 \overline{\rho v^2}=\theta^2 \overline{\rho H^2\Omega_K^2}$. Using (\ref{eq:thickness}), one gets $W_{r\phi}=\theta^2 \overline{\rho c_s^2}=\theta^2 \overline{P}$. Hence, thanks to the vertical equilibrium of a thin disc, \cite{SS73} prescription shows up as a mixing length theory with a length $H$, a frequency $\Omega_K$ and $\alpha=\theta^2$. 

\index{Mach number}
Interestingly, since $H=c_s/\Omega_K$, $\theta$ is actually a measure of the Mach number of the flow $\theta=v/c_s$. If the turbulence was strongly supersonic, then strong shocks would appear, dissipating rapidly turbulent fluctuations until they become subsonic. For this reason, and in the absence of any supersonic excitation, turbulence is expected to be essentially subsonic with $\theta\lesssim 1$ and therefore $\alpha<1$.

\index{Viscous disc}
This prescription may be seen as a viscous theory. Indeed, the $\alpha$-disc prescription leads to $\overline{W_{r\phi}}=\alpha\overline{P}=\alpha \Sigma c_s H \Omega_K$. Since $R\mathrm{d}\Omega_K/\mathrm{d}R=-3/2\Omega_K$, the stress can be recast as
\begin{align*}
    \overline{W_{r\phi}}=-\frac{2}{3}\nu_t\Sigma \frac{\mathrm{d}\Omega}{\mathrm{d}\log R}
\end{align*}
where we have defined an effective viscosity $\nu_t=\alpha c_s H$. Here, we clearly recognise the usual $R-\phi$ component of the viscous stress in the Navier-Stokes equations.

Plugging the $\alpha$ prescription in (\ref{eq:angular_final}) and neglecting surface (wind) contribution leads to
\begin{align*}
    \overline{\rho v_r}&=-\frac{1}{R\partial_R(\Omega_KR^2)}\frac{\partial}{\partial R} R^2\alpha c_s^2\Sigma.
\end{align*}
This allows us to express the mass accretion rate $\dot{M}\equiv -2\pi R \overline{\rho v_r}$ as
\begin{align*}
\dot{M}=\frac{4\pi}{R\Omega_K }	\frac{\partial}{\partial R} R^2\alpha c_s^2\Sigma.
\end{align*}

We can then use mass conservation (\ref{eq:mass}) to get an equation for $\Sigma$
\begin{align*}
\frac{\partial \Sigma}{\partial t}=\frac{1}{R}\frac{\partial}{\partial R}\Bigg[ \frac{1}{\partial_R(\Omega_KR^2)}\frac{\partial}{\partial R} R^2\alpha c_s^2\Sigma \Bigg],
\end{align*}
which essentially constitutes a diffusion equation for the surface density. The diffusion timescale associated to accretion can be estimated using $c_s=\Omega_K H$. One finds
\begin{align*}
\tau_{\mathrm{visc}}^{-1}\sim \alpha\Omega_K\Bigg(\frac{H}{R}\Bigg)^2\ll\Omega_K
\end{align*}
Accretion therefore occurs on timescales much longer than the orbital timescale in thin discs. This is usually a problem for simulations trying to capture the phenomenon of accretion. However, it allows us to separate accretion from dynamics occurring at the local orbital frequency, by stating that accretion is essentially inexistent on this timescale.

\subsection{$\alpha$-$\upsilon$ disc theory\label{sec:alpha-upsilon-disc}}
\index{$\alpha$-$\upsilon$ disc}
This theory is identical to the alpha disc theory for the radial stress part, but it also includes a contribution from the surface term, due to a hypothetical wind. To do so, let us define
\begin{align*}
W_{z\phi}=	\rho v_\phi v_z-\frac{B_\phi B_z}{4\pi}, 
\end{align*}
and in a way similar to the $\alpha$ prescription, we assume
\begin{align*}
\Big[\langle W_{z\phi} \rangle\Big]_{z=-h}^{+h}=\upsilon P_\mathrm{mid}
\end{align*}
\index{$\upsilon$ wind torque parameter}
where $P_\mathrm{mid}$ is the midplane pressure of the disc. Using the same procedure as for the $\alpha$ disc, we can express the mass accretion rate as a function of $\alpha$ and $\upsilon$
\begin{align}
\label{eq:mdot_wind}
\dot{M}=\frac{4\pi}{R\Omega_K }	\Bigg[\underbrace{\frac{\partial}{\partial R} R^2\alpha \overline{P}}_\mathrm{radial}+\underbrace{R^2\upsilon P_\mathrm{mid}}_\mathrm{vertical}\Bigg]
\end{align}
The comparison between the $\alpha$ term and the $\upsilon$ term is revealing as it compares the role played by the radial and vertical stresses. One can assume that in first approximation $\overline{P}\simeq P_\mathrm{mid} H$ so that the vertical contribution is $R/H (\upsilon/\alpha)$ times larger than the radial one. This implies, in particular in thin discs where $R/H\gg 1$, that magnetised winds can easily be the dominant source of accretion.

In addition, using (\ref{eq:mdot_wind}) in the continuity equation, the vertical stress term shows up as a first order radial derivative of $\Sigma$ (=advection) while the radial term appears as a second order derivative as in the usual alpha disc theory. For this reason, wind-driven discs cannot be treated as viscous discs, since the wind component appears as an advective term in the surface density evolution. 

\subsection{Beyond the $\alpha$ prescription}
The $\alpha$ disc model is useful as a starting point to characterise the evolution of discs. However, it is not based on first principles, and it would be desirable to compute directly the turbulent stress $W_{r\phi}$ from the equations of motion for the gas.

This is however a rather complicated task which often implies using numerical tools, as the equations of motion cannot, in general, be solved analytically. Since the disc is thin, and turbulence, in the $\alpha$ disc theory, is supposed to be confined by the scale height $H\ll R$, one can start by using this scale separation to look only at what is happening at the scale $H$, leaving the global scale ($R$) apart. This is the idea of local models, often called "shearing box" models, following \cite{HGB95}.

\section{Local models}
\subsection{The Hill's approximation\label{sec:Hills}}
\begin{figure}
\begin{center}
    \includegraphics[width=0.39\linewidth]{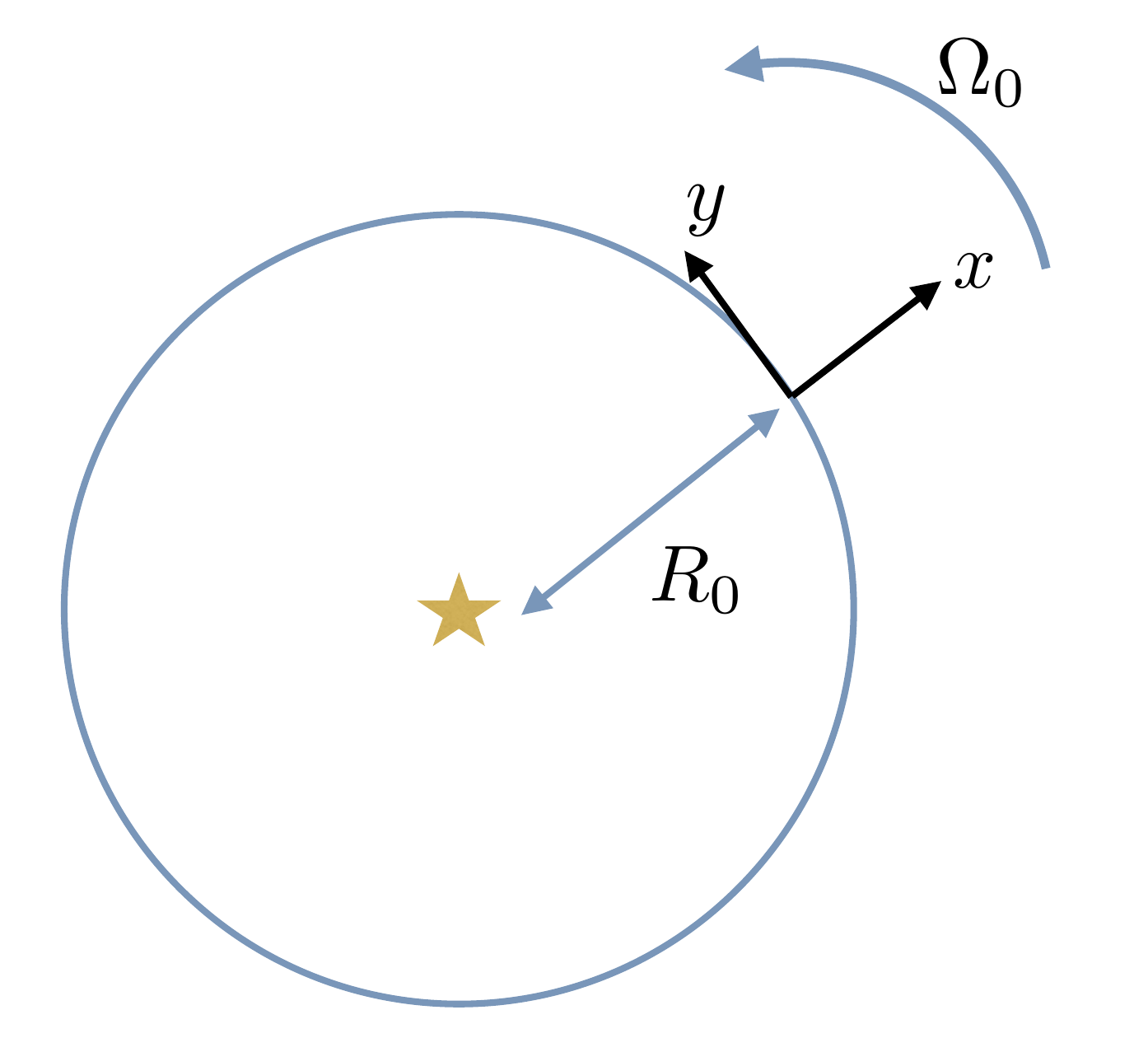}
    \caption{Rotating frame on a circular orbit at $R_0$ }
    \label{fig:rot_frame}
\end{center}
\end{figure}

\index{Hill's!approximation}
The Hill's approximation is a local view of the dynamics of an orbiting system, which was initially used by \cite{H78} to model the libration motions of the moon along its orbit. It's been used more recently as an efficient tool to model dynamics of gas or stars in gravitating systems (e.g. \citealt{GL65}) and it was later implemented numerically in the so-called "shearing-box" by \cite{HGB95}. In this model, one considers the dynamics of the flow around an equilibrium point $R_0$ which is rotating with the disc at the angular velocity $\Omega_O\equiv \Omega_K(R_0)$. We define a cartesian frame $(x,y,z$), attached to this point so that $x$ is aligned with the radius, $y$ with the azimuth and $z$ is aligned with the vertical direction (Fig.~\ref{fig:rot_frame}).

In this frame, the system follows the usual single-fluid equations of motions (MHD). Since it is rotating, we have in addition a Coriolis force and a centrifugal force, so that the equations of motion read
\begin{align*}
\partial_t\rho+\bm{\nabla \cdot }\rho \bm{v}&=0,\\
\partial_t \bm{v}+\bm{v\cdot \nabla v}&=-\frac{1}{\rho}\bm{\nabla}P+\frac{\bm{J\times B}}{\rho c}-2\Omega_0 \bm{e}_z\bm{\times v}+\Omega_0 R^2\bm{e}_R-\bm{\nabla}\psi,\\
\partial_t P+\bm{v\cdot \nabla}P&=-\gamma \bm{\nabla \cdot v},\\
\partial_t\bm{B}&=\bm{\nabla\times}\Bigg(\bm{v\times B}+c\bm{E}_\mathrm{NI}\Bigg),
\end{align*}
where $\psi$ is the gravitational potential, $\bm{E}_\mathrm{NI}$ is the non-ideal electromotive force and $R\equiv\sqrt{(R_0+x)^2+y^2}$ is the cylindrical radius. We also assume the gas follows an ideal equation of state with first adiabatic exponent $\gamma$. As is well known, the centrifugal force derives from a potential of the form $\psi_c=-\Omega_0^2 R^2    /2$. The effective potential (gravitational plus centrifugal) in the corotating frame therefore reads
\begin{align*}
\psi_\mathrm{eff}&=-\frac{GM}{\Big((R_0+x)^2+y^2+z^2\Big)^{1/2}}-\frac{1}{2}\Omega_0^2 \Big((R_0+x)^2+y^2\Big).
\end{align*}
The Hill's model focuses on a "small" (i.e. of the order of the disc scale height in the case of a gaseous disc) region around the fiducial point $R_0$. We therefore expand the effective potential around this point, assuming $x\sim y\sim z\lesssim H$ to get the Hill's potential
\index{Hill's!potential}
\begin{align}
\psi_\mathrm{eff,Hill}&=\Omega_0^2\Bigg[-\frac{3R_0^2}{2}-\frac{3}{2}x^2+\frac{1}{2}z^2+\mathcal{O}\Big(\frac{H^3}{R_0}\Big)\Bigg]
\end{align}
This effective potential has been truncated at the first non-trivial order. It is however interesting to note that it does not depends on $y$ and that is does not contain any cross term such as $xy$ or $xz$. It is also independent from $R_0$ (apart from the constant term). This simplicity in the effective potential is what makes this model so useful for analytical and numerical computation. Any higher order expansion will include curvature terms such as $x/R_0$ dependences and cross dependencies, making calculations much more tedious.

\begin{figure}
\begin{center}
    \includegraphics[width=0.45\linewidth]{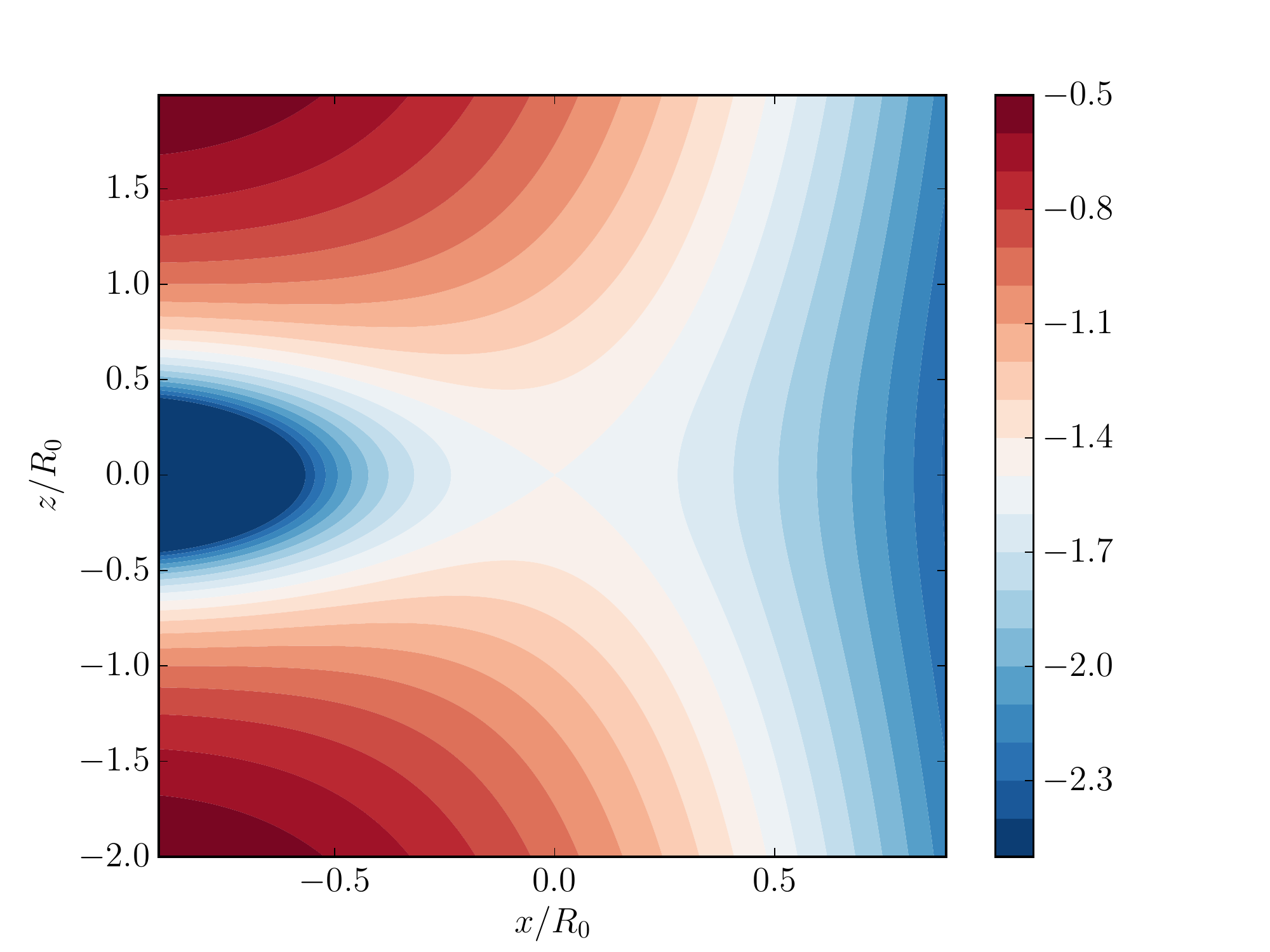}
    \includegraphics[width=0.45\linewidth]{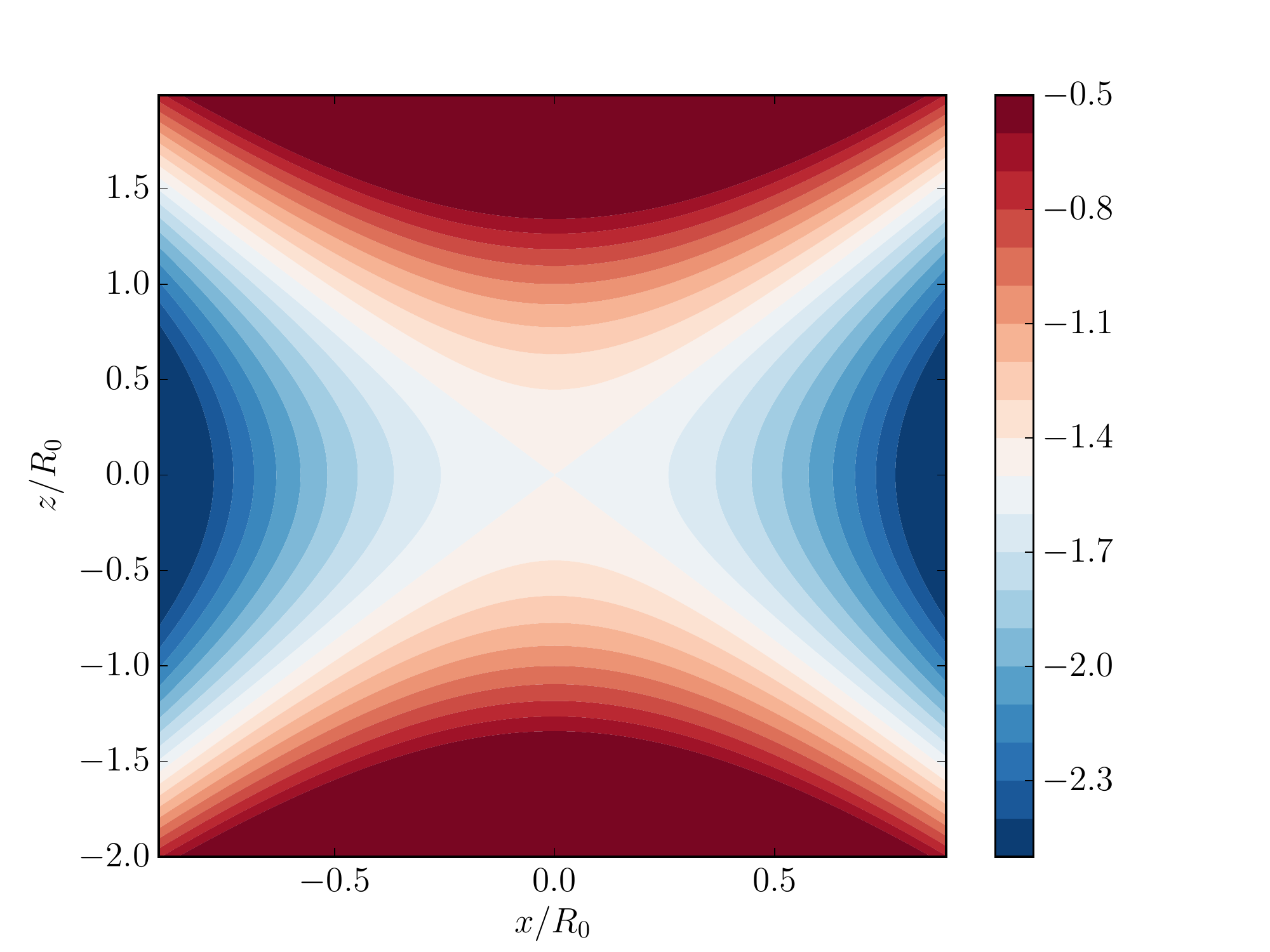}
    \caption{Effective potential in the rotating frame $\psi_\mathrm{eff}$ (left) and its Hill's approximation $\psi_\mathrm{eff,Hill}$ (right). Notice the $x\rightarrow -x$ symmetry of the Hill's potential, as well as the different asymptotic behaviour.}
    \label{fig:psi_comp}
\end{center}
\end{figure}

Let us emphasise already at this stage that the Hill's potential is \emph{not} adapted to global phenomenon. This can be seen by comparing the iso-potentials of $\psi_\mathrm{eff}$ and $\psi_\mathrm{eff,Hill}$ (Fig.~\ref{fig:psi_comp}). The Hill's approximation is found to be symmetrical in $x\rightarrow -x$, implying that one doesn't know where the center of attraction is located (both $x\rightarrow -\infty$ and $x\rightarrow +\infty$ are technically valid). Moreover, the neutral isopotential $\psi(x,z)=-3\Omega_0^2R_0^2/2$ has an asymptote for $z\rightarrow +\infty$ at $x=R_0(\sqrt{3}-1)$ which is absent in the Hill's approximation. This asymptote is key for outflows to be ejected to $z\rightarrow \infty$, and is the main reason why local models always produce outflows which depend on the location of the $z$ boundary conditions (see \S\ref{sec:meanfield_outflows}).

Finally, it can be shown that in the more general case of a central gravity of the form $\psi_\mathrm{grav}=r^{-2(q-1)}$, the equilibrium rotation profile is $\Omega\propto R^{-q}$ and the Hill's potential reads $\psi_\mathrm{eff,Hill}=\Omega_0^2R_0^2(qx^2+z^2)/2+\mathrm{constant}$. In the case of a gravitational potential from a central point mass, we simply have $q=3/2$.

The equations of motion in the Hill's approximation finally reads
\begin{align*}
\partial_t\rho+\bm{\nabla \cdot }\rho \bm{v}&=0,\\
\partial_t \bm{v}+\bm{v\cdot \nabla v}&=-\frac{1}{\rho}\bm{\nabla}P+\frac{\bm{J\times B}}{\rho c}-2\Omega_0 \bm{e}_z\bm{\times v}+\Omega_0^2(2q x\bm{e}_x-z\bm{e}_z)\\
\partial_t P+\bm{v\cdot \nabla}P&=-\gamma \bm{\nabla \cdot v}\\
\partial_t\bm{B}&=\bm{\nabla\times}\Bigg(\bm{v\times B}+c\bm{E}_\mathrm{NI}\Bigg),
\end{align*}
where we have differentiated the Hill’s potential to obtain the tidal acceleration $\Omega_0^2(q x\bm{e}_x-z\bm{e}_z)$. This set of equation admits a simple steady solution $\bm{V}_0$ for the velocity field by simply balancing the Coriolis and tidal forces in the $x$ direction:
\begin{align*}
\bm{V}_0=    -q\Omega_0 x \bm{e}_y.
\end{align*}
This velocity field represents a constant radial shear. It corresponds to the local representation of the Keplerian flow which is not in solid body rotation. Since the Hill's approximation only retains the first terms of the effective potential, the shear in the Hill's model is linear with $x$ and \emph{does not depend on $z$}. It is sometimes useful to work with the deviations from this velocity field. Let us define $\bm{w}\equiv \bm{v}-\bm{V}_0$ (where $w$ is not necessarily small compared to $v$), for which the equations of motion read
\begin{align}
\label{eq:Hill_density}
    D_t\rho+\bm{\nabla \cdot }\rho \bm{w}&=0,\\
    \label{eq:Hill_momentum}
D_t \bm{w}+\bm{w\cdot \nabla w}&=-\frac{1}{\rho}\bm{\nabla}P+\frac{\bm{J\times B}}{\rho c}-2\Omega_0 \bm{e}_z\bm{\times w}+q\Omega_0w_x\bm{e}_y- \Omega_0^2 z\bm{e}_z\\
D_t P+\bm{w\cdot \nabla}P&=-\gamma \bm{\nabla \cdot w}\\
\label{eq:Hill_induction}
D_t\bm{B}&=-q\Omega_0B_x\bm{e}_y+\bm{\nabla\times}\Bigg(\bm{w\times B}+c\bm{E}_\mathrm{NI}\Bigg),
\end{align}
\index{$\Omega$ effect}
\index{Lift up effect}
where we have defined the comoving derivative $D_t\equiv\partial_t-q\Omega x\partial_y$. It is worth noting that this last set of equation does not present any $x$ dependency, except in the comoving derivative. This property is the key allowing the definition of a numerical "shearing box" which we will present later. In exchange, new source terms have appeared when going from $v$ to $w$: the ``lift up" effect $q\Omega_0w_x\bm{e}_y$, which is essentially the advection of the mean keplerian flow by radial motions, and the ``$\Omega$ effect" of dynamo theory (actually due to shear, not rotation) $-q\Omega_0B_x\bm{e}_y$. As we will show later, these terms are key elements to the local physics of accretion discs. A local equivalent of the angular momentum equation can also be derived from (\ref{eq:Hill_momentum}). By defining $\mathcal{L}=w_y+(2-q)\Omega_0 x$, we get the conservation equation
\begin{align}
\label{eq:ang_mom_loc}
    D_t \mathcal{L}+\bm{w\cdot \nabla}\mathcal{L}&=-\frac{1}{\rho}\partial_y P+\frac{[\bm{J\times B}]_y}{\rho c},
\end{align}
where the Coriolis and tidal terms have vanished in the definition of the local angular momentum.

\subsection{The shearing box model\label{sec:shearingbox}}
\index{Shearing box}
\subsubsection{Introduction}
The shearing box model formally comes in two forms, the "large" shearing box (LSB, \add{also known as the stratified shearing box}) model which is essentially a numerical equivalent of the Hill's approximation in a finite size numerical domain, and the "small" shearing box (SSB, \add{also known as the unstratified shearing box}), which is a local approximation in the Hill's approximation. In essence, the SSB model assumes that one zooms on a region close to the disc midplane so that the box size $L\ll H$. In this case, vertical gravity can be neglected in (\ref{eq:Hill_momentum}) and since one expects $w\sim \Omega L$, the flow is strongly subsonic so that an incompressible approximation\footnote{Note however that in most of the unstratified shearing box models published to date, compressibility is kept in the equation of motion.} can be made in place of (\ref{eq:Hill_density}). The asymptotic of these two models are discussed in details by \cite{UR04}. Here we will mostly use the LSB, keeping in mind that an incompressible approximation is possible to study the basic effects of the shear.

\subsubsection{Boundary conditions}
\begin{figure}
\begin{center}
    \includegraphics[width=0.5\linewidth]{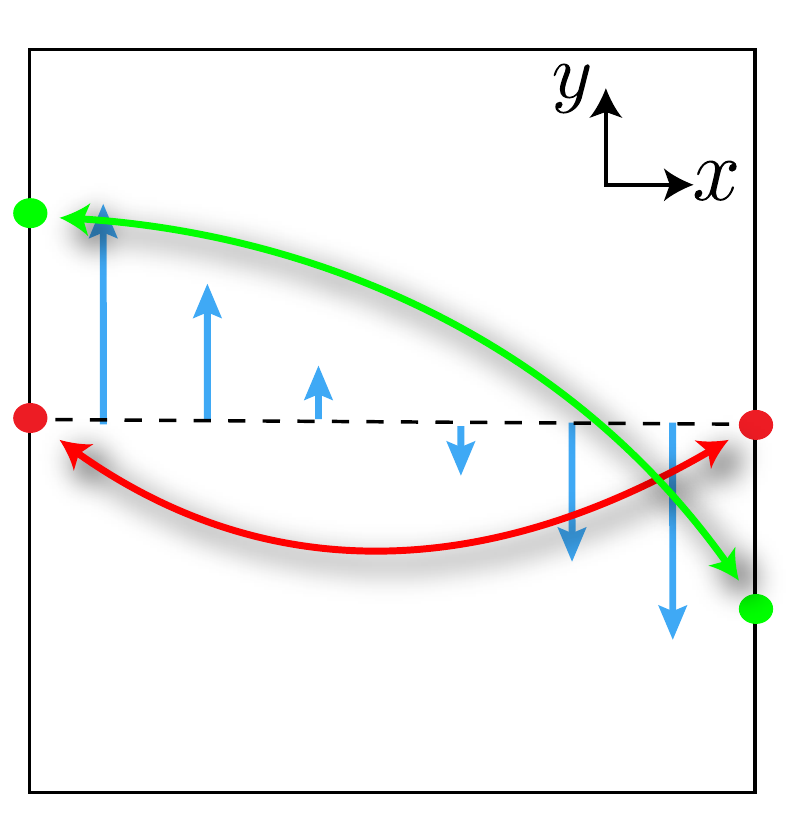}
    \caption{Radial boundary conditions in a shearing box. The background shear is represented in blue. At $t=0$ the boundary conditions are strictly periodic (red). At $t>0$, the periodic boundary conditions are shifted in time (green), according to the advection by the mean flow.}
    \label{fig:shearing_box}
\end{center}
\end{figure}

\index{Shearing sheet}
The shearing box model makes use of the Hill's approximation in the simplest possible numerical setup: a periodic box. Because the Hill's approximation is local, the shearing box model is also local and should satisfy the asymptotic rules shown above. In particular, for a box of size $L$, we should have $L\sim H\ll R_0$. However, the presence of the comoving derivative makes things a bit more difficult. The fields $\rho$, $w$, etc. are advected everywhere at the azimuthal velocity $-q\Omega x\bm{e}_y$. It is therefore physically inconsistent to assume $Q(-L_x/2,y,z)=Q(L_x/2,y,z)$ for any quantity $Q$ as one would do with $x$ periodic boundary conditions in a box of size $L_x$. In order to take into account the constant shear in the boundary conditions, one therefore enforces periodic boundary condition in a Lagrangian view known as "shearing sheet": $Q(-L_x/2,y,z)=Q(L_x/2,y+q\Omega L_x t,z)$, which can be represented graphically as in Fig.~\ref{fig:shearing_box}. 

In the azimuthal direction, periodic boundary conditions are the most natural choice. In the vertical direction, however, no obvious choice comes to mind. Depending on the problem at hand, one can use outflow, free-slip or even periodic boundary conditions. 

\subsubsection{Stress and accretion measurement}
One of the goals of the shearing box approach is to measure directly the turbulent stress in the global dynamical equations (\ref{eq:angular_final}). One therefore needs to quantify
\index{Stress!shearing box}
\begin{align*}
W_{xy}=\overline{\rho w_x w_y}-\frac{\overline{B_x B_y}}{4\pi}
\end{align*}
which is the Hill's equivalent of the radial stress term in (\ref{eq:angular_final}). Despite a non-zero radial stress, the shearing box model does not exhibit any accretion due to this component. This is because the shearing box model is a local asymptotic expansion. The terms leading to accretion are higher order terms in $H/R_0$, which have been neglected, and which would break the in-out symmetry of the Hill's potential. 

\subsubsection{Conserved quantities\label{sec:conservation}}
The shearing box model is very useful in the sense that it allows one to have a tight control on the conserved quantities of the flow: vertical and radial magnetic flux, momentum and energy are all conserved thanks to the relatively simple boundary conditions used in the horizontal direction. We thus define
\begin{align}
\langle \cdot \rangle\equiv \iiint\mathrm{d}x\,\mathrm{d}y\,\mathrm{d}z
\end{align}
Mass, momentum and flux conservation equation eventually read
\begin{align*}
\partial_t\langle \rho\rangle +\Big[\rho v_z\Big]_{z=z_b}&=0\\
\partial_t\langle \rho \bm{w}\rangle +\Big[\rho \bm{w}w_z+\Big(P+\frac{B^2}{8\pi}\Big)\bm{e}_z-\frac{1}{4\pi}\bm{B}B_z\Big]_{z=z_b}&=-2\Omega_0 \bm{e}_z\bm{\times}\langle \rho\bm{ w}\rangle\\ &\qquad\qquad  +q\Omega_0\langle \rho w_x\rangle \bm{e}_y- \Omega_0^2 \langle \rho z\rangle \bm{e}_z\\
\partial_t\langle \bm{B}\rangle +\Big[w_z\bm{B}-B_z\bm{w}\Big]_{z=z_b}&=-q\Omega_0\langle B_x\rangle\bm{e}_y \textrm{ + non-ideal terms}
\end{align*}
Interestingly, the momentum equation exhibits source terms connected to the effective potential and the Coriolis force. The vertical magnetic flux is found to be exactly conserved while the horizontal flux is not necessarily conserved. Horizontal flux can escape the box through the vertical boundary, or alternatively can be amplified thanks to the $\Omega$-effect which shows up as a source term of toroidal field. These simple conservation laws allows one to control very carefully the box physics. Moreover, since the vertical flux is conserved, one can classify shearing-box setup as a function of the average vertical flux. It is also possible to do this for the toroidal component in the $y$ direction, though conservation can be violated by flux escape at the boundary.

Energetics of the shearing box may be found by dotting the momentum equation with $\bm{w}$ and the induction equation with $\bm{B}$. One eventually gets an equation for the mechanical energy (kinetic plus magnetic) in the box
\begin{align}
\label{eq:sb_mech_conservation}
\partial_t \mathcal{E}_\mathrm{Mech}+\bm{\nabla\cdot} \bm{\mathcal{F}}_\mathrm{Mech}&=P\bm{\nabla \cdot w}+\bm{E}_\mathrm{NI}\bm{\cdot J}+q\Omega_0 \Bigg(\rho w_xw_y-\frac{B_xB_y}{4\pi}\Bigg)-\Omega_0^2\rho w_z z
\end{align}
where
\begin{align*}
    \mathcal{E}_\mathrm{Mech}&=\frac{1}{2}\rho w^2+\frac{1}{8\pi}B^2\\
    \bm{\mathcal{F}}_\mathrm{Mech}&=\frac{1}{2}\rho w^2\bm{w}+\frac{1}{4\pi}\Big(B^2\bm{w}-(\bm{w\cdot B})\bm{B}-c\bm{E}_\mathrm{NI}\bm{\times B}\Big)+P\bm{w}
\end{align*}
This set of equations if the local equivalent of (\ref{eq:energ_fll}). Several comments can be made on the energetics. First, we find an energy flux made of three contributions, kinetic energy, Poynting flux (split into advective, magnetic and non-ideal contributions) and a pressure term. Second, we find several source/sink terms:
\begin{itemize}
\item $PdV$    work. This term is usually small when thermal effects are negligible, but can become important in thermally-driven winds for example.
\item Nonideal effects. For Ohmic and ambipolar diffusion, $\bm{E}_{\mathrm{NI}}\sim -\bm{J}$, hence for these two terms, $\bm{E}_\mathrm{Ohm/Ambipolar}\bm{\cdot J}<0$. Meanwhile, the Hall effect has no contribution to this term since $\bm{J \times B\cdot J}=0$. Overall, the non-ideal term therefore appears as a sink of mechanical energy, as expected. Energy is dissipated into heat.
\item Radial stress. We recover the radial stress term of the global angular momentum equation (\ref{eq:angular_final}) which appears as a source term here. 
\item Potential work. The last term denotes the work done by the vertical gravity on the gas, which can become important in ejecting disc models. In this case, this term is negative.  
\item Viscous terms (not shown). When the flow is turbulent, viscous terms can become important at small scale. In this case, they show up as an additional sink term in the mechanical energy equation.
\end{itemize}
Overall, unless strong thermal effects are present, the only source of mechanical energy in this system is the radial stress term which is therefore positive definite, as already pointed out in the global version of this equation. The dynamics of the disc will therefore be dictated by how this source term is balanced by the various loss/flux terms in the energetics.

Let us finally point out that in a shearing box, the energy flux $\bm{\mathcal{F}}_\mathrm{Mech}$ is periodic/shear periodic. If we average the energy equation, we find that the only relevant flux component is the one escaping through the vertical boundaries, as for the mass and momentum fluxes.

\section{The linear MRI in local models}
\subsection{Lagrangian analysis}
\subsubsection{Linear Hydrodynamic stability}
Let us now consider a particle evolving in the Hill's effective potential under the influence of the effective gravity and the Coriolis force. The particle is initially at rest at $(x,y)=0$. Magnetic fields are neglected in this first approach. The equation of motion for the fluid particle may be written
\begin{align}
\frac{\mathrm{d}^2x}{dt^2}&=2q\Omega_0^2 x+2\Omega_0\frac{dy}{\mathrm{d}t}\\
\label{eq:h_y}
\frac{\mathrm{d}^2y}{\mathrm{d}t^2}&=-2\Omega_0\frac{\mathrm{d}x}{\mathrm{d}t}\\
\frac{\mathrm{d}^2z}{\mathrm{d}t^2}&=-\Omega_0^2 z
\end{align}
We first note that the vertical and horizontal equations of motion are separable. In the vertical direction, it describes oscillations of the fluid particle around the midplane at frequency $\Omega_0$.

\begin{figure}
\begin{center}
    \includegraphics[width=0.49\linewidth]{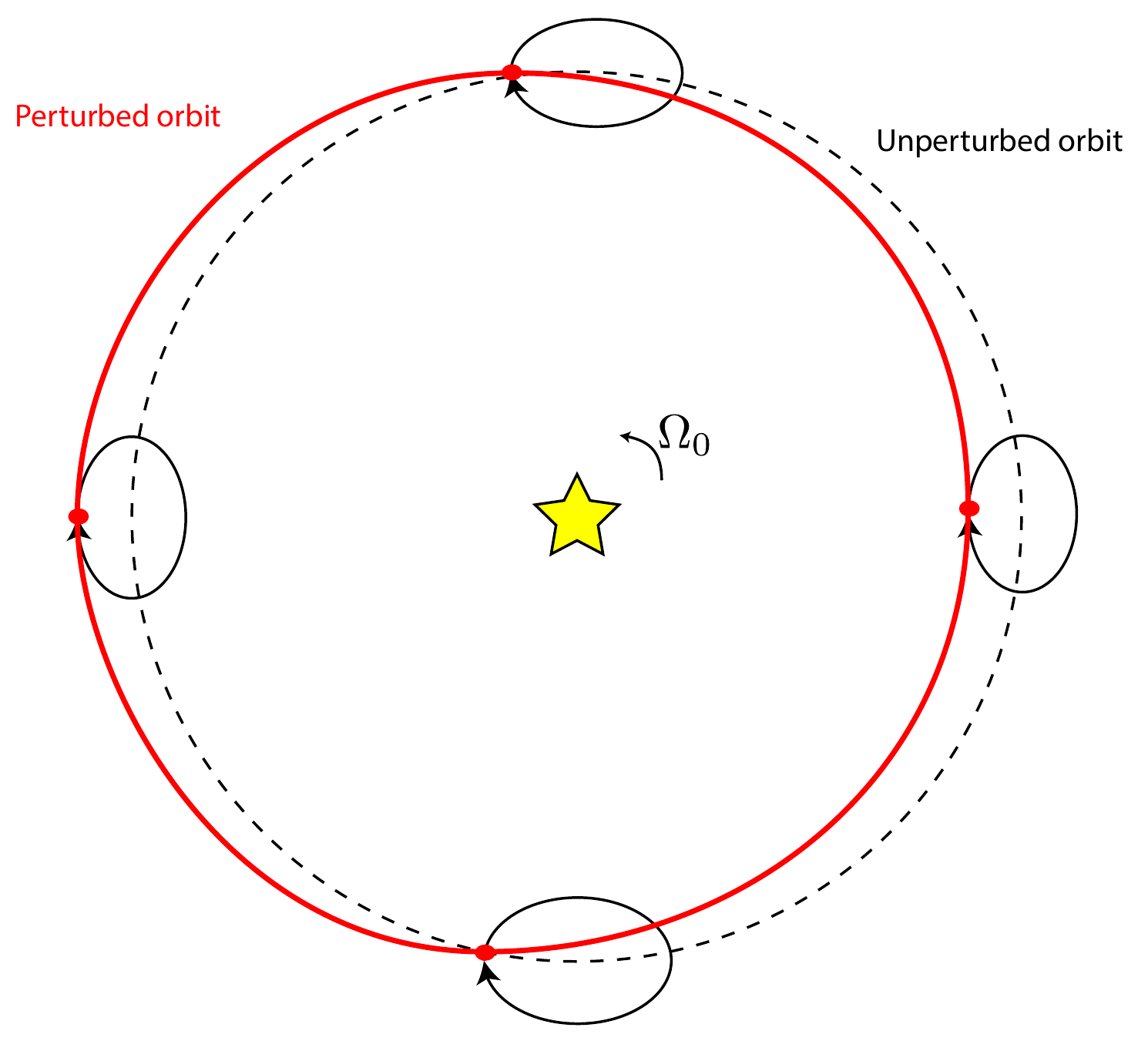}
    \caption{Epicyclic oscillations of a fluid particle orbiting a point mass resulting in an closed elliptic orbit. }
    \label{fig:epicyclic}
\end{center}
\end{figure}

In the horizontal direction, the equations describes epicycles. To show it, let us first integrate (\ref{eq:h_y}): 
\begin{align*}
\mathcal{L}=\frac{\mathrm{d}y}{\mathrm{d}t}+2\Omega_0 x
\end{align*}
where $\mathcal{L}$ is the local angular momentum of the particle, as defined in (\ref{eq:ang_mom_loc}). Our particle being initially in equilibrium at $(x,y)=0$, it has $\mathcal{L}=0$ and we can write the radial equation of motion as
\begin{align}
\frac{d^2x}{dt^2}&=-2\Omega_0^2(2-q) x
\end{align}
hence, our effective gravitational potential which was initially unstable ($\partial_x\psi_{\mathrm{eff},\mathrm{Hill}}<0$)  is stabilised thanks to the conservation of angular momentum, provided that $q<2$ ($q=3/2$ for astrophysical discs).The oscillations described by this particle have a frequency
\index{Epicyclic frequency}
\begin{align*}
\omega^2&=2\Omega_0^2(2-q)\equiv \kappa^2.
\end{align*}
This characteristic frequency is named epicyclic frequency. In the particular case of a Keplerian disc ($q=3/2$) we find $\kappa^2=\Omega^2$ i.e. the epicyclic frequency coincides with the orbital frequency. As a result, orbits are closed, a well known property of the two body problem (e.g. Fig.~\ref{fig:epicyclic}).

\add{This shows that at the linear level, pure Keplerian flows are stable. This doesn't mean that nonlinear (or subcritical) instabilities cannot exist in these flows, given that the shear is a natural reservoir of free energy to trigger instabilities. Such nonlinear instabilities are well known to develop in non-rotating sheared flows and in pipe flows. This question of subcritical instabilities is at the origin of many experiments, numerical simulations and theoretical developments. Today, there seems to be a consensus on the fact that pure Keplerian flows appear to be stable for Reynolds number up to a few millions. However, thermal effects, such as heating, vertical stratification, etc. are known to affect this picture, leading to thermally-driven linear and nonlinear instabilities. This is a whole field of research, which will not be covered here. Interested reader may consult \cite{FL19} for a more detailed overview of this topic.}

\subsubsection{Linear Magnetohydrodynamic stability}

As we have seen, protoplanetary discs are hydrodynamicaly stable at the linear level. In MHD, however, things start to be a bit more interesting (see also \citealt{BH98} for a similar treatment). Let us embed our disc in an external and constant magnetic field $\bm{B}_0$ which we assume is vertical. Assuming we still consider infinitesimal displacements around the equilibrium position of the fluid particles, the velocities are infinitely small, and the induction equation for magnetic fluctuations $\delta \bm{b}$ reads
\begin{align*}
\frac{\partial \delta \bm{b}}{\partial t}&=B_0\frac{\partial \bm{v}}{\partial z}.
\end{align*}
Clearly, the stability will now depend on how we move the particles \emph{with respect to each other}. Let us consider a set of particles initially at $(x,y)=0$ and let us perturb these particles with a vertical harmonic perturbation
\begin{align*}
\bm{x}=\bm{x}_0\exp(ikz)    
\end{align*}
the resulting magnetic perturbation can be obtained by integrating the induction equation with respect to time:
\begin{align*}
\delta \bm{b}&=ik B_0\bm{x}.    
\end{align*}

In order to model how the field impacts the dynamics, we have to include the Lorentz force $\bm{F}_L$ in the equation of motion. In the horizontal direction, only the magnetic tension term $\bm{B\cdot\nabla B}$ appears, so we have

\begin{align*}
\frac{\bm{F}_L}{\rho}&=\frac{\bm{B}_0\bm{\cdot\nabla }\delta \bm{b}}{4\pi\rho}\\
&=-\frac{k^2B_0^2}{4\pi\rho}\bm{x}\\
&=-V_A^2k^2\bm{x},
\end{align*}
where $V_A$ is the Alfv\'en speed. The horizontal equation of motion are therefore reduced to
\begin{align*}
\frac{d^2x}{dt^2}&=2q\Omega_0^2 x+2\Omega_0\frac{dy}{dt}-V_A^2k^2x\\
\frac{d^2y}{dt^2}&=-2\Omega_0\frac{dx}{dt}-V_A^2k^2y
\end{align*}
where it is clear that the magnetic forces are acting as a \emph{restoring} force (hence the usual representation of a spring for the Lorentz force). Note also that angular momentum conservation is now broken by the azimuthal tension force. It is this effect which leads to an instability.

\begin{figure}
\begin{center}
    \includegraphics[width=0.70\linewidth]{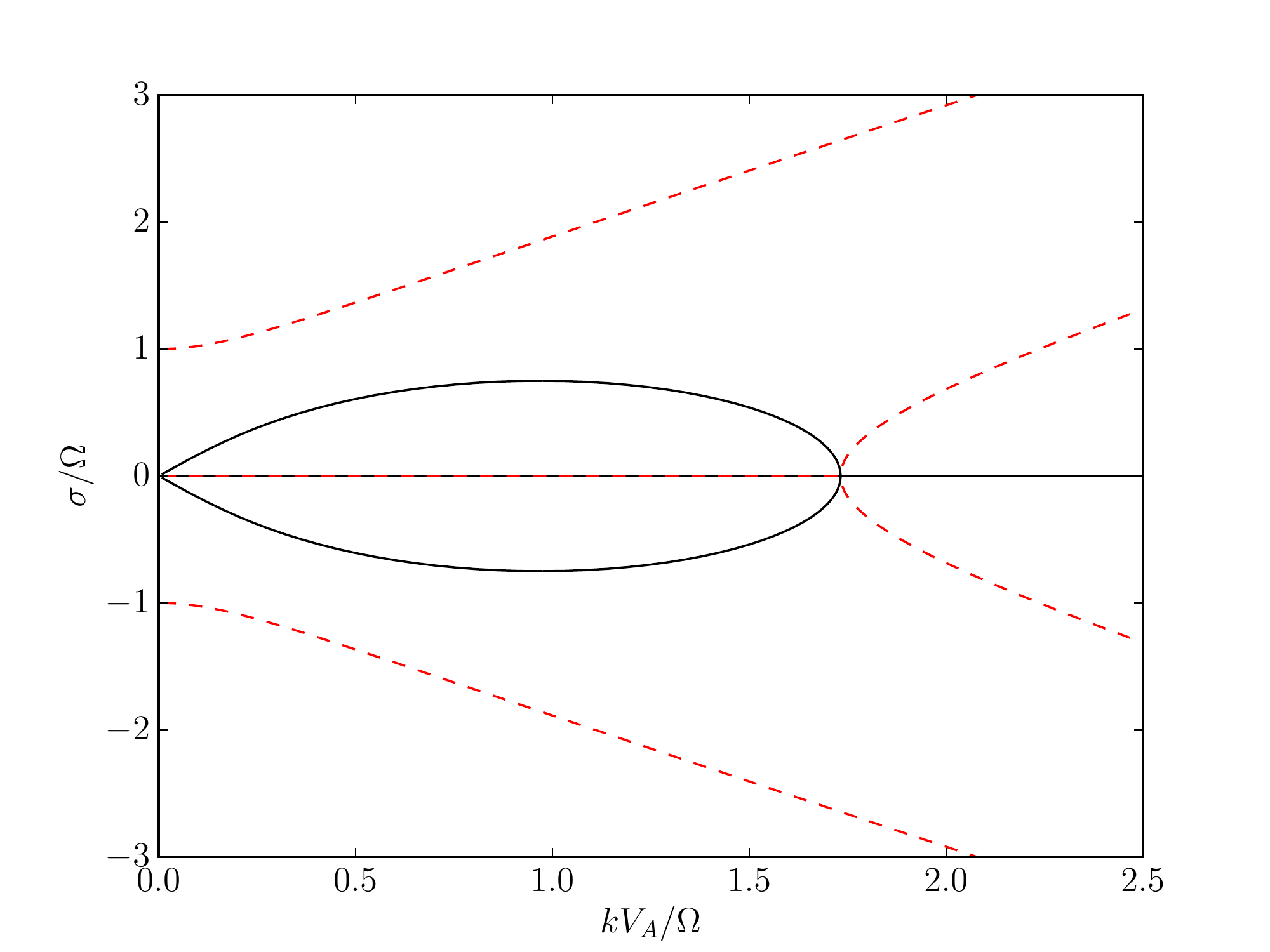}
    \caption{Real part (black) and imaginary part (red dashed line) of the solutions of (\ref{eq:MRI_disp}) with $q=3/2$. The MRI appears for weak enough fields $V_Ak<\sqrt{3}\Omega_0$. }
    \label{fig:mri_disp}
\end{center}
\end{figure}

To show this, let us assume $\bm{x}=\bm{x}\exp(\sigma t)$. The equations of motion lead to the following eigenvalue problem
\begin{align*}
(\sigma^2+V_A^2k^2)x&=2q\Omega_0^2x+2\Omega_0\sigma y,\\
(\sigma^2+V_A^2k^2)y&=-2\Omega_0\sigma x,
\end{align*}
which allows us to get the dispersion relation
\begin{align*}
(\sigma^2+V_A^2k^2)^2-2q\Omega_0^2(\sigma^2+V_A^2k^2)+4\Omega_0^2\sigma^2&=0,
\end{align*}
where we recover epicyclic oscillations when $V_A=0$ with $\sigma^2=-2\Omega_0^2(2-q)=-\kappa^2$ and pure Alfv\'enic oscillations when $\Omega_0=0$ with $\sigma^2=-V_A^2k^2$. Expanding this dispersion relation leads to
\begin{align}
\label{eq:MRI_disp}
\sigma^4+\sigma^2\Big(\kappa^2+2V_A^2k^2\Big)+V_A^2k^2\Big(V_A^2k^2-2q\Omega_0^2\Big)=0.
\end{align}
This dispersion relation describes a linear instability when $\sigma^2$ is real, i.e. when
\begin{align}
\label{eq:MRIcrit}
    V_A^2k^2-2q\Omega_0^2<0.
\end{align}
\index{MRI!Ideal MHD}
This instability is the magneto-rotational instability (or MRI). It appears when the magnetic tension force is not too strong, as suggested by (\ref{eq:MRIcrit}). It is possible to solve the full dispersion analytically to get the eigenvalues (see Fig.~\ref{fig:mri_disp}). When $V_Ak<\sqrt{2q}\Omega_0$, positive eigenvalues are found which are the signature of the MRI. The maximum growth rates are obtained for $V_Ak=\sqrt{2q}\Omega_0/2$ with $\sigma_\mathrm{max}=q\Omega/2$. Above the limit (\ref{eq:MRIcrit}), the unstable branch becomes a stable Alfv\'en wave, which shows that the MRI is mostly an Alfv\'enic perturbation. In addition to this pair of branches, we find a pair of epicyclic modes which are stable for all $kV_A$ (Fig.~\ref{fig:mri_disp}), and have a non-zero frequency for $V_A=0$. If compressibility is added, the degeneracy between slow magnetosonic and Alfv\'en waves is lifted. In this case, it can be shown that the MRI arises from the slow magnetosonic mode \citep{BH92}.

\begin{figure}
\begin{center}
    \includegraphics[width=0.70\linewidth]{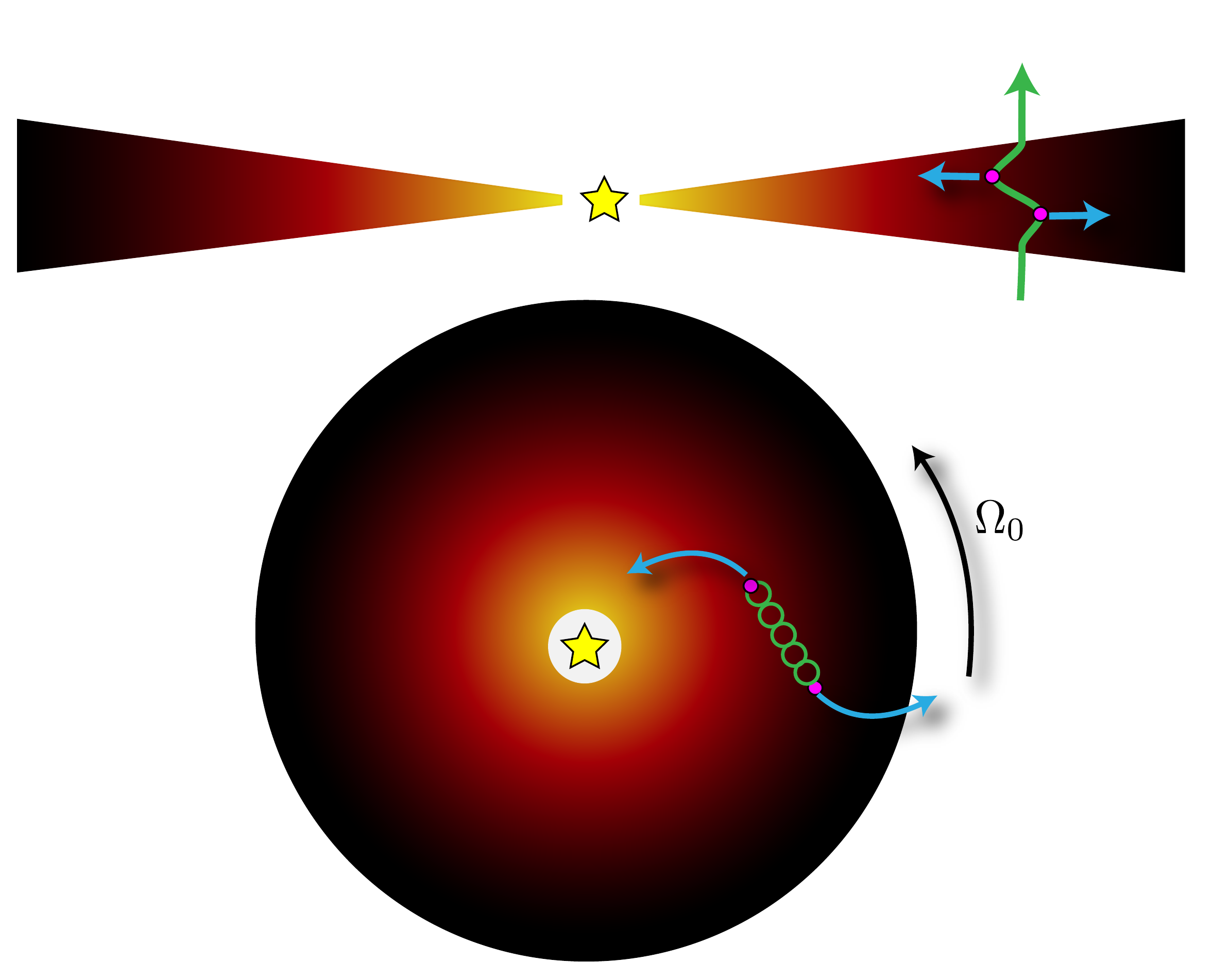}
    \caption{Physical representation of the MRI mechanism (see text).}
    \label{fig:mri_mech}
\end{center}
\end{figure}

The physical interpretation of the MRI is straightforward: consider 2 fluid particles attached to a vertical field line and assume we slightly move these particles radially. At first, they will start an epicyclic motion and drift azimuthally (Fig.\ref{fig:mri_mech}). As they drift away, the azimuthal magnetic tension will act as a spring bringing back the particles together, slowing down the inner particle and accelerating the outer particle. This results in a loss of angular momentum for the inner particle, which falls further down, and reversely for the outer particle. This mechanism can only work if the radial magnetic tension is sufficiently weak otherwise, the particles come back to their initial point resulting in an Alfv\'enic oscillation. It is this radial component of the Lorentz force which is the stabilising agent of the MRI.

\subsection{Application of the MRI to local models\label{sec:mri_application}}
\begin{figure}
\begin{center}
    \includegraphics[width=0.70\linewidth]{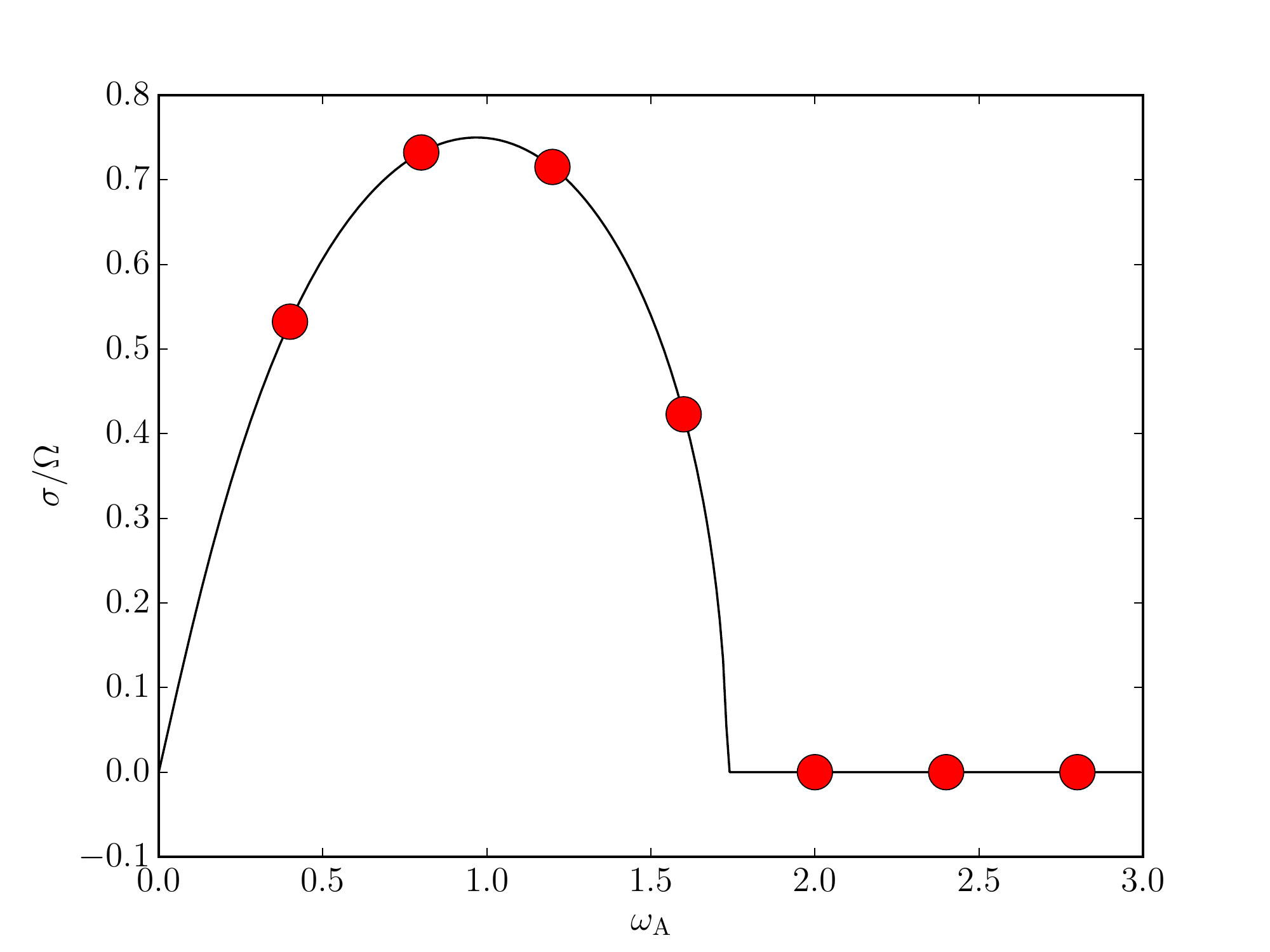}
    \caption{MRI growth rate as a function of the Alfv\'en frequency $\oA$ (black). Quantised modes accessible to a disc model with $\Vaz=0.2~\Omega H$ are shown in red with increasing $n$ from left to right. Only 4 modes are MRI-unstable in this example (see text).}
    \label{fig:MRI_quantised}
\end{center}
\end{figure}

In real disc models, the disc is characterised by a mean vertical and potentially a mean azimuthal magnetic field, as in the linear analysis. However, the vertical wavelength of the perturbation cannot be larger than the disc scale height\footnote{We will come back more quantitatively to this point in \S\ref{sec:linstrat}}. In other words, $k_{z,\mathrm{min}}\simeq 1/H$ which implies that any disc model threaded by a vertical field has a minimal Alfv\'en frequency $\oA\equiv k_z \Vaz$. In addition, the vertical wavenumbers accessible to a disc are quantised because of the limited vertical extension (see \S\ref{sec:linstrat}), hence $k_z=nk_{z,\mathrm{min}}$ and $\oA$ is also quantised. As an example, we show in Fig.~\ref{fig:MRI_quantised} the modes accessible to a disc threaded by a vertical field with $\Vaz=0.2~\Omega H$. In this particular example, only the four smallest $n$ are MRI-unstable, the most unstable mode being $n=2$.

This quantisation of MRI modes also indicates that the MRI can be stabilised for sufficiently strong fields. Indeed, as the field strength increases, quantised unstable modes drift to the right of Fig.~\ref{fig:MRI_quantised} since $k_{z,\mathrm{min}}$ increases. The MRI is entirely stabilised when the lowest $n$ gets into the stable regime, i.e. when $\Vaz>\sqrt{2q}\Omega/ k_{z,\mathrm{min}}$. With $k_{z,\mathrm{min}}\simeq 1/H$ this implies in Keplerian discs
\begin{align*}
    \Vaz\gtrsim \sqrt{3}\Omega H\quad\quad\rightarrow\quad\quad\textrm{Stability}.
\end{align*}
For this reason, the MRI is often seen as a ``weak field'' instability, even though technically, this is more a result of the geometrically thin disc approximation. Note also that this criterion is not an effect of compressibility, as it is sometimes thought. Indeed, since $\Omega H=c_s$, the criterion above implies that the disc stabilises for vertical field strength above equipartition. This is however just a coincidence resulting from the vertical equilibrium in a geometrically thin disc. The MRI also exists well above equipartition in a slightly modified form, provided that large enough wavelengths are allowed \citep{KO00}.

\subsection{Non-axisymmetric MRI}
We start from equations (\ref{eq:Hill_density})---(\ref{eq:Hill_induction}) in which we assume the disc is threaded by a mean field having a mean vertical and azimuthal components: $\bm{B}=B_{0,y}\bm{e}_y+B_{0,z}\bm{e}_z$. In this subsection, we relax the axisymmetry hypothesis used above, but we still assume the flow is incompressible, and neglect vertical gravity and stratification. We then consider the following set of equations:
\begin{align}
\partial_t\bm{w}+\bm{w\cdot \nabla w}-q\Omega x \partial_y \bm{w}&=-\bm{\nabla}\Pi+\frac{\bm{B\cdot \nabla B}}{4\pi\rho}+2\Omega w_y \bm{e_x}-(2-q)\Omega w_x\bm{e_y}\\
\partial_t \bm{B}+\bm{w\cdot\nabla B}-q\Omega x \partial_y \bm{B}&=\bm{B\cdot \nabla w}-q\Omega B_x \bm{e_y}\\
\nabla\cdot \bm{w}&=0
\end{align}
We are going to linearize this system, assuming the field can be decomposed as $\bm{B}=\bm{B}_0+\bm{b}$. The presence of the advection term $q\Omega x\partial_y$ however leads to some technical difficulties as it involves an explicit spatial dependency whenever modes are non-axisymmetric. We therefore follow \cite{K80} and \cite{CC86} using decomposition into time-dependent "waves"\footnote{Although these solutions are often called "waves", these are not normal modes of the physical system, but they are just a convenient way to decompose solutions in a sheared flow}, which, for any quantity $\tilde{X}$ assumes as spatial decomposition
\begin{align*}
	\tilde{\bm{X}}=\bm{X}(t)\exp(i\bm{k(t)\cdot x}).
\end{align*}
Using this decomposition, it is easy to show that
\begin{align*}
\partial_t \tilde{\bm{X}}-q\Omega x\partial_y \tilde{\bm{X}}=\dot{\bm{X}}+i\bm{X}\Big(\dot{\bm{k}}\bm{\cdot x}-q\Omega x k_y\Big). 	
\end{align*}
Since it is the only term which shows this explicit $x$ dependency in the equations of motion, and since these equations are assumed to be valid for all $x$, we are forced to conclude that the term in parenthesis cancels out:
\begin{align*}
	\dot{k}_x-q\Omega k_y=0\quad;\quad
	\dot{k}_y=0\quad;\quad 
	\dot{k}_z=0
\end{align*}
Without loss of generality, we can therefore assume a decomposition into "shearing waves"\index{Shearing waves}, defined as
\begin{align*}
\bm{k(t)}=\bm{k_0}+q\Omega t k_y\bm{e_x}
\end{align*}
which is solution to the set of equations above. We using this shearing wave decomposition we obtain
\begin{align}
\dot{\bm{w}}&=-i\bm{k}\Pi+i\frac{\bm{k \cdot B}_0}{4\pi\rho} \bm{b}+2\Omega w_y \bm{e_x}-(2-q)\Omega w_x\bm{e_y}\\
\dot{ \bm{b}}&=i(\bm{k \cdot B}_0)\bm{w}-q\Omega b_x \bm{e_y}\\
\label{solenoidal}\bm{k \cdot w}&=0
\end{align}
where we have dropped the explicit time dependency of $k$ and the $\tilde{\\}$ symbols for simplicity. We next take the time derivative of (\ref{solenoidal}):
\begin{align*}
q\Omega k_y w_x+\bm{k\cdot} \dot{ \bm{w}}=0.
\end{align*}
This allows us to express the generalised pressure
\begin{equation}
\Pi=-\frac{i}{k^2}\Big(2\Omega w_y k_x-2\Omega(1-q)w_xk_y\Big)
\end{equation}
So that in the end, the equations of motion read:
\begin{align*}
\dot{\bm{w}}&=i\frac{\bm{k\cdot B}_0}{4\pi\rho}\bm{b}+2\Omega w_y (1-g_{xx})\bm{e_x}+2(1-q)\Omega w_x g_{xy}\bm{e_x}\\
 &-q\Omega w_x g_{yy}\bm{e_y}-(2-q)\Omega w_x(1-g_{yy})\bm{e_y}-2\Omega w_y g_{xy}\bm{e_y}\\
& +2(1-q)\Omega w_x g_{yz}\bm{e_z} -2\Omega w_y g_{xz}\bm{e_z}
\end{align*}
where we have introduced $g_{ij}=k_ik_j/k^2$.

It is not possible to go any further without making any approximation. Indeed, although by construction $\bm{k\cdot B}_0$ does not have any time dependency, the pressure factors $g_{ij}$ do have one, so that a standard normal mode decomposition is prone to failure. It is possible to numerically integrate these equations as a function of time (see e.g. \citealt{BH92}). This always leads to transiently growing solutions, i.e. perturbations which only grow for a finite time. To understand why this is always the case, let us continue our analysis using a first order WKB approximation.

To get a dispersion relation, one needs to assume that $k$ is ``almost'' steady, i.e. $k_y\ll k_x$ so that $\mathrm{d} \log k/\mathrm{d}t\ll\Omega$. This limit is often described as a strongly leading or strongly trailing wave. This limit implies that we can neglect all the $g_{yj}$ terms in the above expansion. Assuming $X=\tilde{X}\exp[\sigma t+i \bm{k(t)\cdot x})]$, we then get
\begin{eqnarray}
\sigma \bm{w}&=&i\frac{\bm{k\cdot B}_0}{4\pi\rho}\bm{b}+2\Omega v_y (1-g_{xx})\bm{e_x}-(2-q)\Omega v_x\bm{e_y}-2\Omega v_y g_{xz}\bm{e_z}\\
\sigma \bm{b}&=&i(\bm{k \cdot B}_0)\bm{v}-q\Omega b_x \bm{e_y}
\end{eqnarray}
These equations clearly exhibit an Alfv\'en mode in the $z$ direction, with $\sigma=\pm i (\bm{k\cdot V}_{\mathrm{A},0})$ and $\bm{V}_{\mathrm{A},0}\equiv \bm{B}_0/\sqrt{4\pi\rho}$ as one would expect. We can then solve independently the horizontal problem to obtain the dispersion relation:
\begin{equation}
\sigma^4+\sigma^2(2\omega_A^2+2(2-q)\Omega^2g_{zz})+\omega_A^2(\omega_A^2-2q\Omega^2g_{zz})
\end{equation}
where we have defined the Alfven frequency $\omega_A=\bm{k\cdot V}_{\mathrm{A},0}$. This equation describes both the traditional MRI mode and the non axisymmetric MRI, as its close ressemblance with (\ref{eq:MRI_disp}) suggests. It should however be noted that in the $k_z/k\rightarrow 0$ limit, the instability is lost since $g_{zz}\rightarrow 0$ and the last term of  the dispersion relation, responsible for the MRI, vanishes. Physically, this happens because the pressure gradient balances the Coriolis force in the $x$ component of the equation of motion.

Now, it should be kept in mind that this dispersion relation is derived for a shearing wave whose $\bm{k}$ is slowly evolving with time. As $t\rightarrow \infty$, one then expects $|k_x|\rightarrow \infty$ and therefore $g_{zz}\rightarrow 0$. Hence, by nature, shearing waves automatically quench the growth of the MRI as they evolve. For this reason, non-axisymmetric structures, even without any dissipation, are necessarily transiently growing solution, and therefore never lead to a genuine linear instability. That being said, non-axisymmetry is \emph{required} when only a large-scale toroidal field is available in the system, since one needs $\bm{k\cdot B}_0\ne 0$. Therefore, there is no "toroidal field MRI" as it is sometimes seen in the literature. There is just a transient growth in the linear phase, which can be described as a temporary MRI in the WKB approximation, but only a non-linear feedback can re-excite new shearing waves to keep increasing the energy of the fluctuations.

\add{It should be stressed that the absence of any linear non-axisymmetric instability is observed only in the local limit (i.e. shearing box). If one considers a global disc, including curvature and radial boundaries, then a genuine linear instability can be recovered \citep[e.g.][]{CP96}, with properties similar to that found in the WKB approximation shown above.}

\subsection{MRI in non-ideal MHD}
\subsubsection{Historical background}

\index{Dead zone}
The linear MRI in the non-ideal MHD regime has been explored by many authors. After the discovery of the MRI in the disc context by \cite{BH91}, it was soon realised that protoplanetary discs, but also discs in cataclysmic variables could be in the non-ideal MHD regime, casting doubts on the applicability of this instability to these objects. \cite{BB94} were the first to consider this problem, by working out the ambipolar-dominated MRI in the two-fluid approximation. Ohmic diffusion was first considered by \cite{J96} in unstratified models and \cite{SM99} in stratified discs, which led to the dead zone model of protoplanetary discs \citep{G96}. Later, \cite{W99} considered the 3 non-ideal MHD effects in simple axial geometry while \cite{D04} considered also oblique modes and toroidal fields. \cite{BT01} isolated the physics of the Hall-MRI and \cite{K08} demonstrated that one of the Hall-MRI branches was actually a new instability: the HSI. Finally, ambipolar diffusion was revisited in the single-fluid approximation by \cite{KB04}, demonstrating the existence of oblique ambipolar modes and their origin. 

In the following, we revisit and discuss each of these effects with a unified notation and geometry. Note however that our approach and dispersion relation is formally identical to that of \cite{D04}. 

\subsubsection{Linearised equations}
As in the cases above, we start from equations (\ref{eq:Hill_density})---(\ref{eq:Hill_induction}) in which we assume the disc is threaded by a mean field having a mean vertical and azimuthal components: $\bm{B}=B_{0,y}\bm{e}_y+B_{0,z}\bm{e}_z$. In the following, we consider small axisymmetric\footnote{in the Hill's approximation, non-axisymmetric linear perturbations only lead to transiently growing solutions which ultimately decay \citep{BH92}. These transient ``modes'' are however important for the MRI dynamo once nonlinear feedback is taken into account\citep{LO08b}. } perturbations of the equilibrium. We seek solutions of the form $\bm{w}=\bm{u}\exp[\bm{k\cdot x}+\sigma t]$ and $\bm{B}=\bm{B}_0+\bm{b}\exp[\bm{k\cdot x}+\sigma t]$. Where $\bm{k}=k_x\bm{e}_x+k_z\bm{e}_z$ is the wave number, and $\sigma$ is the linear growth rate of the instability. We moreover neglect vertical stratification and vertical gravity and assume the flow is incompressible, which implies that we consider the small shearing box approximation. As we will show later, this is enough to capture most of the physics relevant to the problem. We will explore in \S\ref{sec:linstrat} the impact of the vertical stratification on linear modes.

Under these assumptions, the linearised equations read
\begin{align*}
\sigma\bm{u}&=-i\bm{k} \Pi+i\frac{\bm{k\cdot B}_0}{4\pi \rho} \bm{b}+2\Omega u_y \bm{e}_x-(2-q)\Omega u_x\bm{e}_y,\\
\nonumber \sigma \bm{b}&=i\Big(\bm{k\cdot}\bm{B}_0\Big) \bm{u} -q\Omega b_x \bm{e_y}-\eta_O k^2\bm{b}+\eta_H\Big(\bm{k\cdot}\widehat{\bm{B}}_0\Big) \bm{k\times b},\\
&\quad\quad  -\eta_A\Bigg[\Big(\bm{k\cdot}\widehat{\bm{B}}_0\Big)^2\bm{b}-\Big(\bm{b\cdot}\widehat{\bm{B}}_0\Big)\bigg(\Big[\bm{k\cdot}\widehat{\bm{B}}_0\Big]\bm{k}-k^2\widehat{\bm{B}}_0\bigg)\Bigg],\\
\bm{k\cdot u}&=0,    \\
\bm{k\cdot b}&=0.    
\end{align*}
The expression of the non-ideal terms can be interpreted in the following way. First, Ohmic diffusion acts as a pure linear damping operator, as expected. The Hall term is proportional to $\bm{k\times b}$, which means it \emph{rotates} the magnetic perturbation around the $\bm{k}$ direction, keeping its norm constant. Note that the direction of rotation is given by the sign of $\eta_H$, which shows that the handedness given by the Hall effect is directly connected to the microphysics of the plasma (see \S\ref{sec:twospecies}). Finally, ambipolar diffusion involves an anisotropic diffusion term which we will discuss in the following.

The solenoidal conditions can be used to eliminate $u_z$ and $b_z$ from the equations in favour of the horizontal components, leading to a $4^\mathrm{th}$ order problem
\begin{align*}
\sigma u_x&=i\frac{\bm{k\cdot B}_0}{4\pi \rho} b_x+2\Omega \frac{k_z^2}{k^2}u_y,\\
\sigma u_y&=i\frac{\bm{k\cdot B}_0}{4\pi \rho} b_y-(2-q)\Omega u_x,\\
\sigma b_x&=i\Big(\bm{k\cdot}\bm{B}_0\Big) u_x-\eta_O k^2b_x-\eta_H\Big(\bm{k\cdot}\widehat{\bm{B}}_0\Big)k_z b_y-\frac{\eta_A}{B_0^2}\Bigg(k^2B_{0,z}^2b_x-k_xB_{0,y}\Big(\bm{k\cdot B}_0\Big)b_y\Bigg),\\
\nonumber \sigma b_y&=i\Big(\bm{k\cdot}\bm{B}_0\Big) u_y-q\Omega b_x-\eta_O k^2b_y+\eta_H\Big(\bm{k\cdot}\widehat{\bm{B}}_0\Big)\frac{k^2}{k_z}b_x\\
&\quad\quad\quad\quad\quad-\frac{\eta_A}{B_0^2}\Bigg(\Big[(\bm{k\cdot B}_0)^2+k^2B_{0,y}^2\Big]b_y-\frac{k^2}{k_z^2}k_xB_{0,y}(\bm{k\cdot B}_0)b_x\Bigg).
\end{align*}
The role played by ambipolar diffusion is here a bit more self-explanatory. We observe that the diagonal terms in the 2\textsuperscript{nd} pair of equations are always negative definite, hence ambipolar diffusion is really acting as a diffusion term on the diagonal components with an amplitude controlled by the magnitude but also the orientation of $\bm{B}_0$. However, there are also off-diagonal terms proportional to $k_xB_{0,y}$. As we will see, these terms can lead to oblique unstable modes (see also \citealt{KB04}).

We next solve the above set of equations for $\sigma$ and look for unstable eigenvalues. We follow \cite{PW12} and first obtain an equation for the velocity fluctuations
\begin{align*}
\bm{u}=\frac{i\bm{k\cdot B}_0}{4\pi\rho(\sigma^2+\kappa^2)}
\left(\begin{array}{cc}
\sigma & 2\Omega \\
-(2-q)\Omega & \sigma
\end{array}\right)
\bm{b}
\end{align*}
So that the induction equation becomes in matrix form:
\begin{align}
\label{eq:linear_matrix_form}
\Bigg[
\left(\begin{array}{cc}
\sigma + \eta_O k^2+\tA k^2 V_{\mathrm{A}z}^2 & \lH \oA k_z-\tA k_xV_{\mathrm{A}y}\oA\\
q\Omega - \frac{k^2}{k_z^2}\Big(\lH\oA k_z+\tA k_xV_{\mathrm{A}y}\oA\Big)   & \sigma+\eta_O k^2+\tA(\oA^2+k^2V_{\mathrm{A}y}^2)
\end{array}\right)\quad\quad 
\\
+\frac{\oA^2}{\sigma^2+\frac{k_z^2}{k^2}\kappa^2}
\left(\begin{array}{cc}
\sigma & 2\Omega\frac{k_z^2}{k^2} \\
-(2-q)\Omega & \sigma
\end{array}\right)
\Bigg]\bm{b}=0
\end{align}
where we have introduced the Alv\'en speed $V_\mathrm{A}=B_0/(4\pi \rho)^{1/2}$, the Alfv\'en frequency $\oA\equiv \bm{k\cdot} \bm{V}_\mathrm{A}$, the Hall length $\lH\equiv \eta_H/V_\mathrm{A}$, the ambipolar time $\tA\equiv \eta_A/V_\mathrm{A}^2$ and the epicyclic frequency $\kappa^2\equiv 2\Omega^2(2-q)$.

After a long but straightforward calculation, one eventually gets the dispersion relation which can be written
\begin{align}
\label{eq:disp_gal}
\sigma^4+\mathcal{C}_3\sigma^3+\mathcal{C}_2\sigma^2+\mathcal{C}_1\sigma+\mathcal{C}_0=0    
\end{align}
with
\begin{align*}
\mathcal{C}_3&=2\eta_Ok^2+\tA(k^2\Va^2+\oA^2),\\
 \mathcal{C}_2&=\frac{k_z^2}{k^2}\kappa^2+2\oA^2+\eta_O^2k^4+\tA^2\oA^2k^2\Va^2+q\Omega\tA\oA  k_x\Vay+ \lH k_z\oA\Bigg(\frac{k^2}{k_z^2}\lH k_z\oA -q\Omega\Bigg),\\
\mathcal{C}_1&=\mathcal{C}_1\Bigg(\oA^2+\frac{k_z^2}{k^2}\kappa^2\Bigg),\\
\nonumber \mathcal{C}_0&=\oA^2\Bigg(\oA^2\underbrace{-2q\Omega^2\frac{k_z^2}{k^2}}_\textrm{MRI}\Bigg)+\kappa^2 k_z^2k^2\eta_O^2+\lH\oA k_z\Bigg(\underbrace{(4-q)\Omega\oA^2}_\textrm{ion-cyclotron instability}\underbrace{-q\Omega\frac{k_z^2}{k^2}\kappa^2}_\textrm{HSI}+\lH \oA k_z \kappa^2\Bigg),\\
&\quad\quad +\kappa^2\tA^2\oA^2k_z^2\Va^2+\underbrace{q\Omega \tA k_x \Vay\oA \Bigg(\frac{k_z^2}{k^2}\kappa^2+\oA^2\Bigg)}_\textrm{Oblique ambipolar modes}.
\end{align*}
The stability of this linear system can be analysed in the vicinity of $\sigma=0$. In this case, a necessary and sufficient condition for instability is $\mathcal{C}_0<0$. This allows us to identify three sources of instability: the usual MRI, the ion-cyclotron instability, the Hall-shear instability (HSI) and the term at the origin of Oblique ambipolar modes, which is not a genuine instability branch.

Before exploring the non-ideal regime, let us point out that in the ideal MHD limit, this dispersion relation shows that modes with $k_x\ne 0$ have a lower growth rate than $k_x=0$ modes. As a result, $k_x=0$ modes are always the most unstable eigenmodes of the system. These modes are often called ``channel modes'' as they don't have any horizontal spatial dependency. They are also exact non-linear solutions of the full MHD equations \citep{GX94}. For this reason, they are very robust and they often show up in the non-linear regime, as we will see later.

\subsubsection{Ohmic diffusion\label{sec:lin_ohmic}}
\begin{figure}
\begin{center}
    \includegraphics[width=0.70\linewidth]{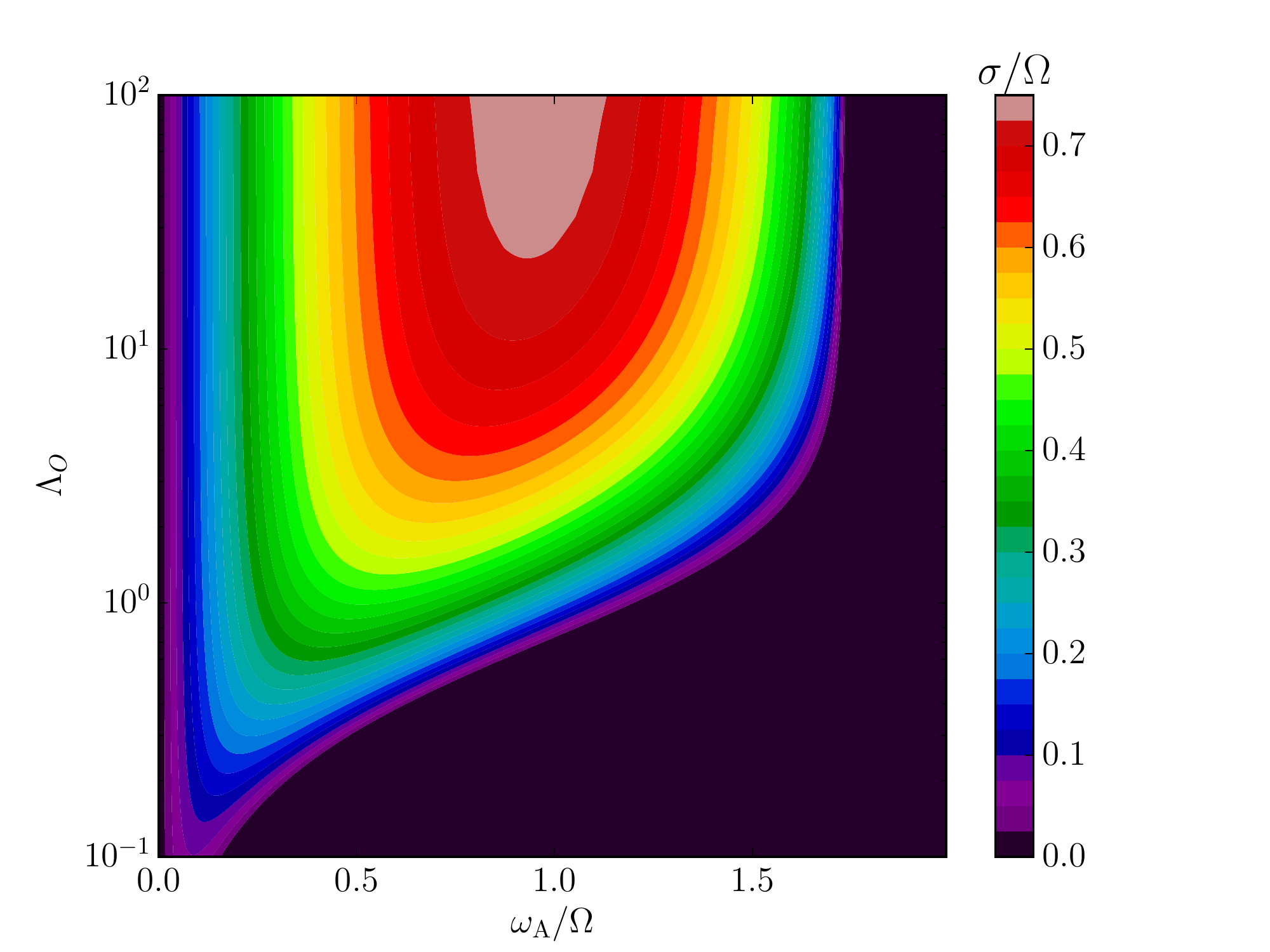}
    \caption{MRI growth rate as a function of the Ohmic Elsasser number $\Lambda_O$ for a Keplerian disc ($q=3/2$) with $k_x=0$. Note the damping of the most unstable mode for $\Lambda_O=1/\sqrt{3}$ and the survival of low growth rate modes in the limit $\oA\rightarrow 0,~ \Lambda_O\rightarrow 0$. }
    \label{fig:mri_ohm}
\end{center}
\end{figure}

\index{MRI!Ohmic}
The impact of Ohmic diffusion on the stability of the MRI is physically very intuitive: it stabilises MRI modes, starting from the largest $\oA$ of the system (see Fig.~\ref{fig:mri_ohm}). The stability condition in the presence of Ohmic diffusion is deduced from the condition $\mathcal{C}_0=0$ and reads

\begin{align}
    \frac{2q\Omega^2}{k^2\Vaz^2}-1<\kappa^2\frac{k^2}{k_z^2}\frac{\eta_O^2}{\Vaz^4}\quad\rightarrow\quad\textrm{Stability.}
\end{align}
Clearly, the modes with $k_x=0$ are the last ones to be stabilised when one increases Ohmic diffusion. Let us focus on this case and assume the flow is Keplerian so that $\kappa=\Omega$ and $q=3/2$. The stability of the most unstable ideal mode is often considered as a proxy for the stability of the flow. This mode has $\oA=\sqrt{3}\Omega/2$ so that the above inequality reduces to
\begin{align}
\label{eq:ohm_stability_optimimum}
\Lambda_O^2<\frac{1}{3}\quad \rightarrow\quad\textrm{Stability of the most unstable ideal MRI mode.}
\end{align}
Note however that his does not imply that \emph{all} of the modes available to the disc are stable, and that the disc is stable. A more constraining criterion results from this and requires that the mode with the largest length scale is MRI stable, i.e. that
\begin{align}
\label{eq:ohmic_stability_limi}    \frac{3\Omega^2}{k_{z,\mathrm{min}}^2\Vaz^2}-1<\Lambda_O^{-2}\quad\rightarrow\quad\textrm{General Disc Stability.}
\end{align}

\subsubsection{Hall effect\label{sec:lin_hall}}
\index{MRI!Hall}
\begin{description}
\item[Hall-driven linear waves: ]
\index{Whistler waves}
\index{Ion-cyclotron waves}
The Hall effect is known to be at the origin of new linear waves. These waves can be captured by letting $\Omega\rightarrow 0$ and neglecting Ohmic and ambipolar diffusion. In this case the dispersion relation (\ref{eq:disp_gal}) gives
\begin{align*}
0&=\sigma^4+\sigma^2\Big(2\oA^2+\lH^2\oA^2k^2\Big)+\oA^4\\
&=\big(\sigma^2+i\sigma\lH \oA k+\oA^2\big)    \big(\sigma^2-i\sigma\lH \oA k+\oA^2\big).    
\end{align*}
We recognise two waves with frequency $\omega\equiv i\sigma$ given by
\begin{align*}
\omega=\oA\Bigg[\pm\frac{\lH k}{2}+ \sqrt{\frac{\lH^2k^2}{4}+1}\Bigg]
\end{align*}
where "+" waves are known as whistlers or electron-cyclotron modes, while "-" waves are ion-cyclotron modes. The whistler frequency $\oH=\oA\lH k$ increases as $k^2$ in the limit $k\rightarrow\infty$ while the ion-cyclotron frequency tends to a constant $\omega_\mathrm{IC}=\oA/(\lH k)$. Hence these two waves behave very differently at small scale. For positive $\lH$, whistlers are right-handed polarised wave while ion-cyclotron are left-handed. Physically, whistlers are essentially an oscillation of the electrons fluid (or of the lightest charged particle), leaving all of the other components of the plasma unaffected. Whistler and ion-cyclotron waves become standard right and left handed circularly-polarised Alfv\'en wave in the limit $k\rightarrow 0$. 

\item[Hall-shear instability: ]
\index{Hall-shear instability (HSI)}
The Hall-shear instability\footnote{This instability is also named "diffusive instability" (DI) by \cite{PW12} to emphasise that rotation is unimportant for this instability, in contrast to the MRI.}  (HSI) is a new branch of instability \citep{K08} which is often confused with the traditional MRI, despite its different physical origin. It is essentially an instability of whistler waves under the action of shear.

To capture the HSI, one can let $\oA\rightarrow 0$ while keeping $\lH \oA>0$. This ``low magnetisation'' limit allows one to decouple the ions from the electrons, as is evident from (\ref{eq:linear_matrix_form}). Neglecting Ohmic and ambipolar diffusion, one gets the following dispersion relation
\begin{align*}
    \Bigg(\sigma^2+\frac{k_z^2}{k^2}\kappa^2\Bigg)\Bigg[\sigma^2+\lH\oA k_z\Bigg(\frac{k^2}{k_z^2}\lH k_z\oA-q\Omega\Bigg)\Bigg]=0,
\end{align*}
which exhibit a linear instability when
\begin{align}
\label{eq:HSI_limit}
 \lH \Vaz k_z^2\Big(k^2\lH \Vaz -q\Omega\Big)<0\quad \rightarrow\quad \textrm{HSI unstable}.
 \end{align}
Interestingly, the HSI shows up only when $q \lH \Vaz>0$, or in other words, when the vertical field points in the same direction as the rotation axis in Keplerian discs\footnote{Physically, it is not the rotation but the background shear of the flow $S\equiv \partial_x V_y$ which matters. The general criterion is therefore $S\Vaz<0$.}, assuming $\lH>0$. When the whistler frequency becomes too large ($k^2\lH \Vaz>q\Omega$), the instability disappears. For a given $k_z$, the most unstable mode has $k_x=0$ hence $k=k_z$. For this reason, the HSI often shows up as channel-like mode in simulations, in a way similar to the MRI.  Last, the maximum growth rate is identical to the MRI $\sigma_\mathrm{max}=q\Omega /2$ and is obtained for $k^2\lH \Vaz=q\Omega/2$

\begin{figure}
\begin{center}
    \includegraphics[width=1.0\linewidth]{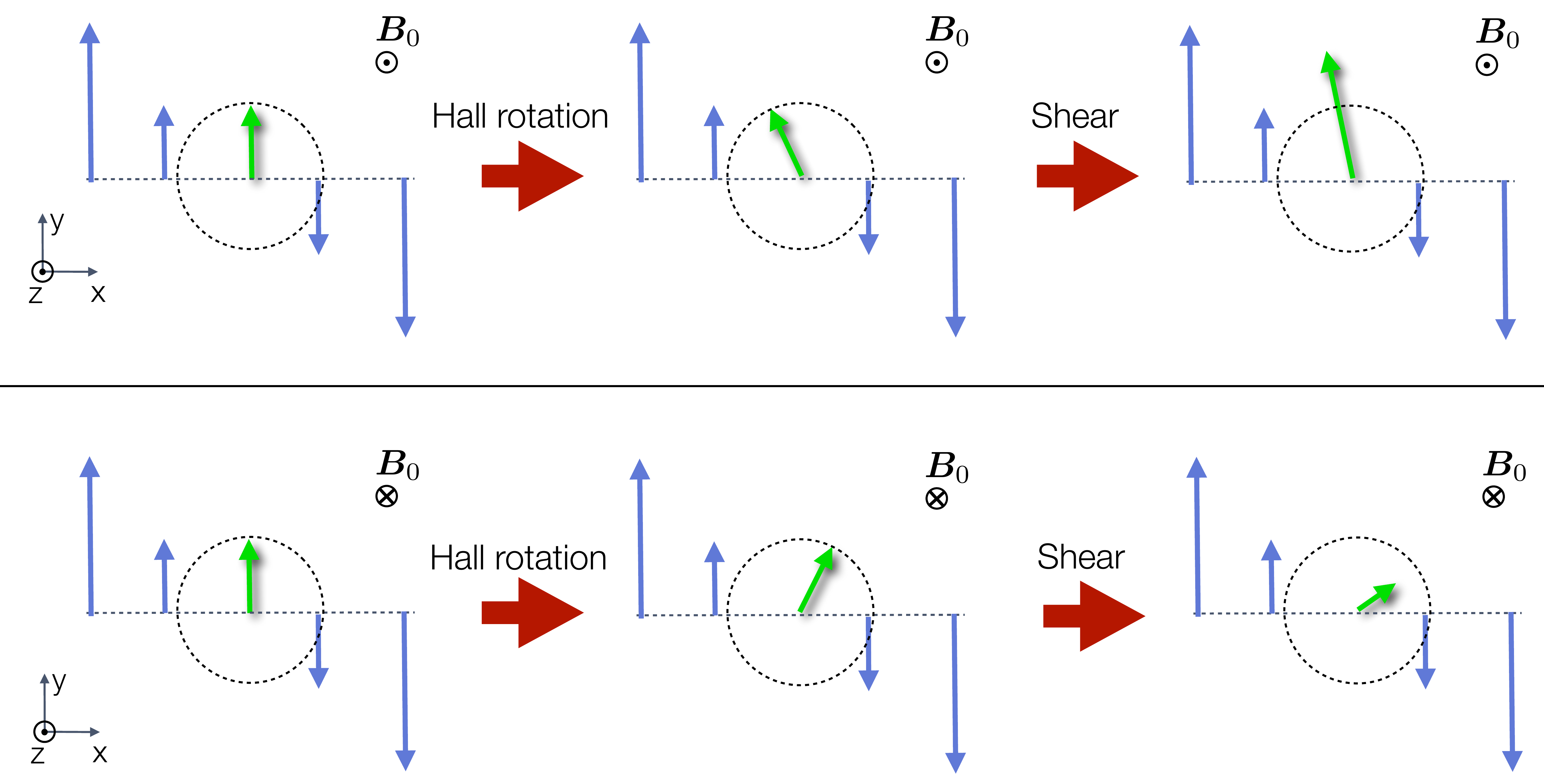}
    \caption{Physical principle of the Hall-shear instability (HSI). The magnetic perturbation (in green) is rotated clockwise or counter-clockwise by the Hall effect depending on the polarity of the mean field $B_0$ (top/bottom). When $B_0>0$, the rotated perturbation is amplified by the shear (in blue) while it is damped by the shear when $B_0<0$. } 
    \label{fig:HSI_principle}
\end{center}
\end{figure}

Physically, this instability is a result of sheared whistler waves. If we look at the disc from the top with the vertical field pointing towards us, the magnetic perturbation of a whistler wave will tend to rotate counter-clockwise (Fig.~\ref{fig:HSI_principle}). If the vertical field is positive (i.e. aligned with the rotation axis), the Keplerian shear is going to shear the perturbation in the opposite direction, amplifying the toroidal field from the radial field and feeding the instability. On the contrary, if the vertical field is anti-aligned with the rotation axis, the direction of rotation of whistler perturbations is that of the shear, resulting in a damped whistler wave.

\item[Ion-cyclotron instability: ]
\index{Ion-cyclotron instability}
The ion-cyclotron instability\footnote{This instability is also named "diffusive MRI" (DMRI) by \cite{PW12}} is more difficult to isolate compared to the HSI since ion inertia cannot be neglected in this case. However, it is still possible to filter out whistler waves by letting $k_z\rightarrow \infty$ and keeping constant $k_x\lesssim k_z$. In this case, the whistler wave frequency $\oH$ becomes infinite while the cyclotron frequency stays finite. Since we're looking for a finite growth rate, we will assume $\sigma$ is finite when $k\rightarrow \infty$. Neglecting ambipolar and Ohmic diffusion and keeping only $O(k^4)$ terms, the dispersion relation (\ref{eq:disp_gal}) becomes \citep[see also][]{SL15}:
\begin{align*}
    \sigma^2 \lH^2k^2\oA^2+\Big[\oA^2+(2-q)\Omega \lH \oA k_z\Big]\Big[\oA^2+2\Omega \lH \oA k_z\Big]=0.
\end{align*}
This relation clearly describes ion-cyclotron modes since in the limit $\Omega\rightarrow 0$, we recover the ion-cyclotron frequency $\sigma=\pm i\omega_{\mathrm{CI}}$. It describes unstable modes, resulting from the interaction between ion-cyclotron waves and epicyclic motions, provided that
\begin{align}
\label{eq:ICI_limit}
-\frac{\Vaz^2}{(2-q)\Omega}<\lH\Vaz<-\frac{\Vaz^2}{2\Omega}\quad \rightarrow\quad \textrm{ion-cyclotron unstable} 
\end{align}
This inequality brings up two remarks: 1- the ion cyclotron instability appears for \emph{anti-aligned} field configurations, i.e. field configuration opposite to the HSI. 2- this instability does not vanish in the limit $k_z\rightarrow\infty$. Therefore, there is no small scale quenching similar to the MRI and the HSI, but there is a field strength limit. 

\item[General case: ]

In the general case, one cannot compute the growth rate of the Hall-MRI easily. To illustrate the general growth rate of the Hall-MRI, we present in Fig.~\ref{fig:mri_Hall} the growth resulting from (\ref{eq:disp_gal}) in the Hall only case. To quantify the intensity of the Hall effect, we have defined a Lundquist number based on the vertical wavenumber 
\begin{align*}
\mathcal{L}_\mathcal{H}^*=\frac{1}{\lH k_z}=\frac{V_A}{\eta_H k_z}.
\end{align*}
We have chosen the most unstable modes $k_x=0$ and assumed $k_z>0$ so that $\oA<0$ corresponds to $B_{0z}$ anti-aligned with $\Omega$. As can be seen in Fig.~\ref{fig:mri_Hall}, we recover the MRI in the limit $\mathcal{L}_\mathcal{H}\rightarrow \infty$ which gives identical growth rates under the symmetry $\oA\rightarrow -\oA$. As $\mathcal{L}_\mathcal{H}^*$ decreases and the Hall effect increases, this symmetry is broken. 

\begin{figure}
\begin{center}
    \includegraphics[width=0.80\linewidth]{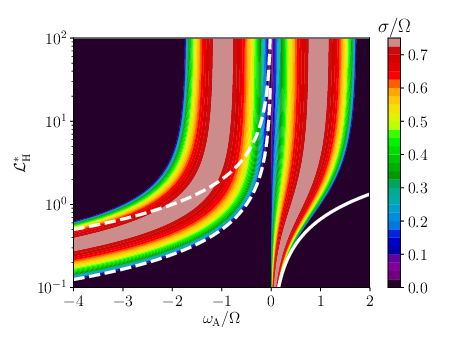}
    \caption{Hall-MRI growth rate as a function of the modified Hall Lundquist number $\mathcal{L}_\mathcal{H}^*=(\lH k_z)^{-1}$ for a Keplerian disc ($q=3/2$) with $k_x=0$. We have assumed $k_z>0$ so that $\oA<0$ corresponds to the anti-aligned case $\Vaz<0$. The white plain line corresponds to the HSI stability limit (\ref{eq:HSI_limit}) and the white dashed lines to the ion-cyclotron stability limits (\ref{eq:ICI_limit}). } 
    \label{fig:mri_Hall}
\end{center}
\end{figure}

For $\oA>0$ (aligned field configuration), the MRI becomes the HSI, and the optimum growth rate starts to move to lower $\oA$. This is expected from our analysis since the maximum growth rate of the HSI is found for $\oA/\Omega    \simeq \mathcal{L}_\mathrm{H}^* /2$. We show in white filled contours the stability limit of the HSI obtained from (\ref{eq:HSI_limit}). This limit matches the full dispersion relation for $\mathcal{L}_\mathrm{H}^*\lesssim 0.3$.

For $\oA<0$ (anti-aligned field configuration), the MRI becomes the ion-cyclotron instability and the optimum growth rate moves to higher $|\oA|$. The stability contours of the ion-cyclotron instability (\ref{eq:ICI_limit}) are shown in dashed white line. As for the HSI, they match the full dispersion relation for $\mathcal{L}_\mathrm{H}^*\lesssim 0.3$.

\item[Impact of Ohmic diffusion on the Hall-MRI\label{sec:lin_hall_revival}: ]
As shown above, Ohmic diffusion tends to damp MRI modes, starting from the largest $\oA$ and moving to lower $\oA$ as $\Lambda_O$ decreases. On the other hand, the Hall effect creates two distinct branches from the MRI, depending on the field alignment configuration. Since Ohmic diffusion suppresses first high $|\oA|$ modes, the ion-cyclotron instability will be first to be stabilised. On the other hand, since the HSI is living at lower $|\oA|$ compared to the MRI, it will be less affected by Ohmic diffusion than the MRI. This effect is illustrated in Fig.~\ref{fig:mri_Hall+Ohm}, demonstrating that HSI modes survive more easily to Ohmic diffusion, even when ideal MRI modes are suppressed.

\begin{figure}
\begin{center}
    \includegraphics[width=0.80\linewidth]{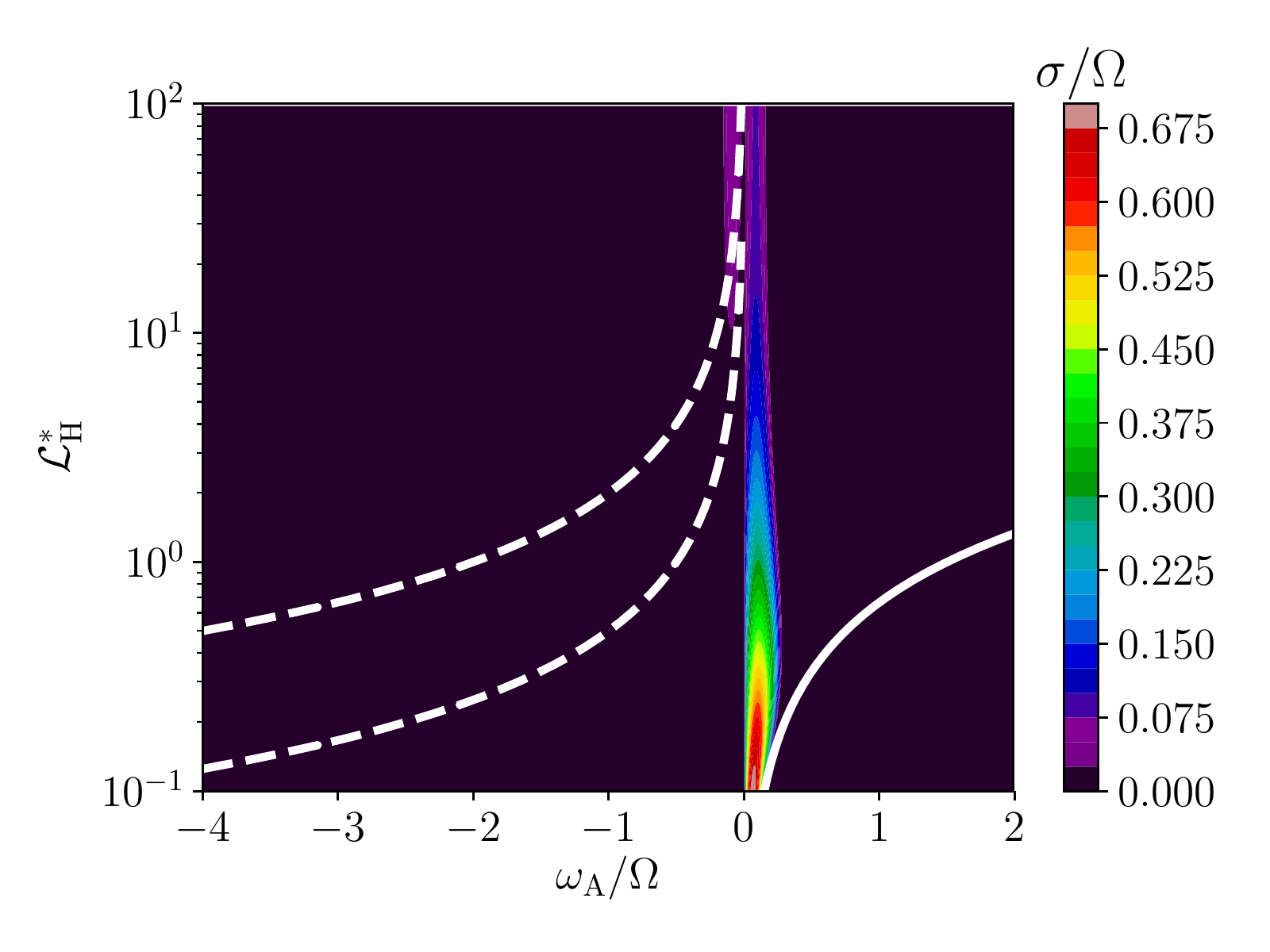}
    \caption{Same as Fig.~\ref{fig:mri_Hall} but including Ohmic diffusion with $\Lambda_O=0.1$. Note the complete stabilisation of the ion-cyclotron branch and the strong damping of ideal MRI modes. Only HSI modes subsist at low $\mathcal{L}_\mathrm{H}^*$ with growth rates comparable to the ideal MRI case.} 
    \label{fig:mri_Hall+Ohm}
\end{center}
\end{figure}

We can go further and estimate the stability limit of the HSI in the limit $\oA\rightarrow 0$ while keeping $\lH \oA>0$. In this limit, the dispersion relation of the HSI including Ohmic diffusion reads
\begin{align*}
    \Bigg(\sigma^2+\frac{k_z^2}{k^2}\kappa^2\Bigg)\Bigg[\Big(\sigma+\eta_Ok^2\Big)^2+\lH\oA k_z\Bigg(\frac{k^2}{k_z^2}\lH k_z\oA-q\Omega\Bigg)\Bigg]=0,
\end{align*}
from which we can deduce a general instability criterion. For the most unstable HSI mode, this criterion reads
\begin{align}
\label{eq:hall_ohm_stability_optimum}
\Lambda_O^2<\frac{1}{2}\Lambda_H^2    ,\quad \rightarrow\quad\textrm{Stability of the most unstable HSI mode.}
\end{align}
where $\Lambda_H$ is the Hall Elsasser number (see \S\ref{sec:diffu_application}). This expression can be compared directly to the Ohmic-only criterion (\ref{eq:ohm_stability_optimimum}), and shows that as expected, the HSI can make the system unstable even when $\Lambda_O\ll 1$, provided that $\Lambda_H\ll 1$ as well.

For this reason, some authors have proposed that the MRI could be "resuscitated" in regions having a strong Ohmic diffusion thanks to the presence of the Hall effect, which could be dominant in some parts of the disc \citep{WS12}. The physical interpretation of this effect is relatively simple. In the case of the MRI, the maximum growth rate is found for $\oA\sim \Omega$, in other words, the Alfvenic and rotation frequencies match. In the HSI case, it is the whistler and rotation frequencies $\oH\sim \Omega$ which have to match. In the limit of strong Hall effect, $\oH= \lH k_z \oA$. Therefore when $\lH kz\gg 1$,  the scale at which the whistler frequency matches the rotation frequency is much larger than the scale at which the Alfv\'en frequency matches the rotation frequency. In other words, the optimum HSI mode has a much larger wavelength than its MRI counterpart. For this reason, the HSI is less sensitive to diffusion than the MRI, since diffusion first damps small scales modes.

Although these statements are true in the linear regime, the non-linear saturation of the HSI is not guaranteed to be similar to that of the MRI. Effectively, the linear analysis is unable to predict the turbulent angular momentum transport one could obtain from the HSI. As we will see in \S \ref{sec:sb_unstrat_hall}, the HSI in the non-linear regime is indeed full of surprises\dots

\end{description}
\subsubsection{Ambipolar diffusion\label{sec:lin_ambipolar}}

\index{MRI!Ambipolar}
The linear ambipolar diffusion is a non-diagonal operator as shown in (\ref{eq:linear_matrix_form}). The non-diagonal terms are non-zero only when $k_x \Vay\ne 0$, i.e when \emph{both} non-axial wavevectors and guide fields are considered. Otherwise, ambipolar diffusion acts as a usual diffusion operator by damping magnetic perturbation. Note however that even in this case, it does not act as a scalar diffusion, unless $k_x=0$ and $\Vay=0$. 

\textbf{Diagonal case:} ($k_x \Vay= 0$) In this case, the stability criterion is very similar to that of Ohmic diffusion. By requiring $\mathcal{C}_0>0$, one gets
\begin{align*}
    \frac{2q\Omega^2}{k^2\Vaz^2}-1<\kappa^2\tA^2\Bigg(\frac{\Va}{\Vaz}\Bigg)^2\quad\rightarrow\quad\textrm{Stability.}
\end{align*}
This criterion shows that, as for Ohmic diffusion, the most unstable modes have $k_x=0$. Moreover, it shows that the stability condition is affected by the toroidal field strength, a stronger $\Vay$ leading to a more stable system. As for Ohmic resistivity, we can derive a criterion for the stability of the most unstable ideal MRI mode in Keplerian flows which reads

\begin{align}
\label{eq:ad_stability_optimimum}
\Lambda_A^2<\frac{1}{3}\Bigg(\frac{\Va}{\Vaz}\Bigg)^2\quad \rightarrow\quad\textrm{Stability of the most unstable ideal MRI mode,}
\end{align}
and a criterion for the stability of all the modes available smaller than the minimum vertical wavenumber $\mathrm{k_{z,\mathrm{min}}}$:
\begin{align}
\label{eq:ad_stability_limit}    \frac{3\Omega^2}{k_{z,\mathrm{min}}^2\Vaz^2}-1<\Lambda_A^{-2}\Bigg(\frac{\Va}{\Vaz}\Bigg)^2\quad\rightarrow\quad\textrm{General Disc Stability.}
\end{align}

This shows that the MRI can be stabilised even for $\Lambda_A>1$ provided that the toroidal field dominates over the poloidal one. The field topology is therefore a key point for the stability under the action of ambipolar diffusion. The physical reason for this is relatively simple. The MRI mechanism is only sensitive to the magnetic tension which is due to $B_z$ in the absence of non-axisymmetric perturbations. On the contrary, ambipolar diffusion increases with the total field strength. Hence, an increase in $\Vay$ results in an increase of the effective diffusion $\eta_A$, while the MRI feedback loop is left unperturbed. 

\begin{figure}
\begin{center}
    \includegraphics[width=0.80\linewidth]{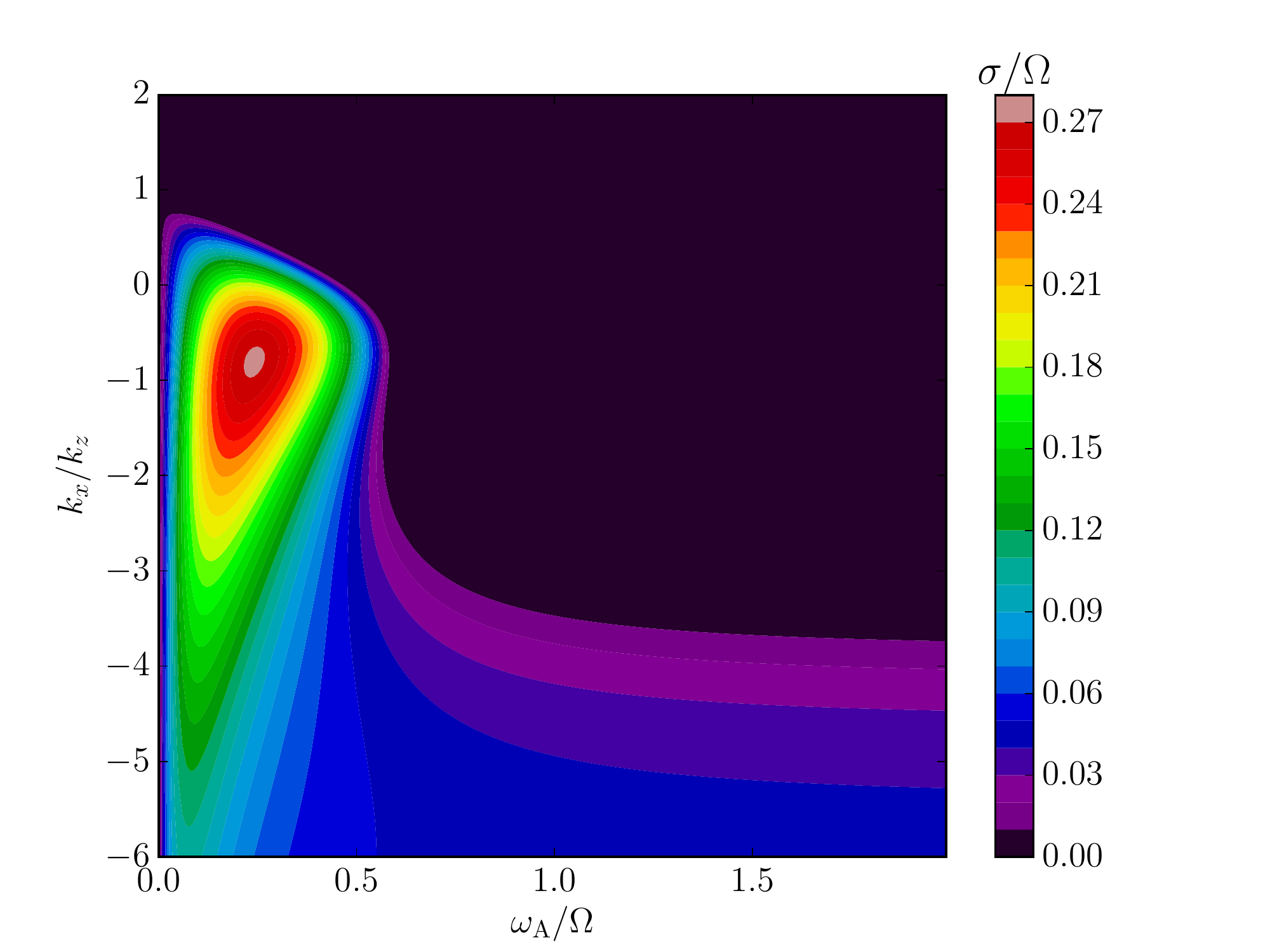}
    \caption{MRI growth rate with ambipolar diffusion and $\Lambda_A=0.4$. We have chosen $\Vay=\Vaz$ and $k_x\ne 0$ to illustrate the effect of non-diagonal ambipolar terms. The most unstable mode in this case is found for $k_x<0$ and a weakly unstable branch exists for $k_x/k_z\simeq 3.5$ in the limit $\oA\rightarrow\infty$. We name these modes "oblique ambipolar modes". Note however their low relative growth rate}
    \label{fig:mri_Ambi}
\end{center}
\end{figure}

\textbf{ Non-diagonal case:} ($k_x\Vay\ne 0$) In this case, the non-diagonal terms of ambipolar diffusion can act as a positive feedback loop in the induction equation. The criterion for this positive feedback is easily obtained from the dispersion relation (\ref{eq:disp_gal}): $q\Omega k_x k_z \Vay\Vaz<0$. An illustration of the impact of this term on the general stability property is shown in Fig.~\ref{fig:mri_Ambi} where we have assumed $\Vay=\Vaz$, $q=3/2$ and $\Lambda_A=0.4$. We observe that oblique modes ($k_x\ne 0$) tend to be more unstable than axial modes when $\Lambda_A<1$. It can be shown that oblique modes are always unstable, albeit with a vanishing growth rate, in the limit $\Lambda_A\rightarrow 0$ provided that $k_x/k_z$ is sufficiently large \citep{KB04}.

\subsubsection{Vertical stratification\label{sec:linstrat}}

All of the above was computed ignoring vertical stratification. It should however be pointed out that since discs are vertically stratified, unstratified results should be taken with care. Let us emphasize here a few important results regarding the effect of stratification on MRI modes. In this section, we will assume that the disc is vertically isothermal (see \S\ref{sec:equilibrium}) and is only threaded by a vertical field (no toroidal field). We quantify the intensity of the imposed field with the plasma beta parameter in the midplane:
\begin{align*}
\bmid &\equiv\frac{8\pi P_\mathrm{mid}}{B^2}=\frac{2c_s^2}{\Vaz^2 (z=0)}.
\end{align*}

\begin{description}
\item[Ideal MRI: ]

the ideal MRI case with an isothermal disc is presented in details by \cite{LFG10}. We will therefore just summarise here the main conclusions. First, stratified eigenvalues (growth rates) satisfy the same dispersion relation (\ref{eq:MRI_disp}) as non stratified modes with $\Va=\Va(z=0)$. Vertical wave numbers $k$ are quantised but eigenmodes are not simple harmonic functions in the $z$ direction. \cite{LFG10} have shown that the first eigenmodes have $k_nH=1.1584,~2.0796,~2.9829,~3.8798\dots$ for $n=1\dots 4$. Taking $k_1$ as the lowest wavenumber accessible to the system, the MRI is stable for all possible eigenmodes in Keplerian discs ($q=3/2$) if $\Vaz(z=0)>\sqrt{3}\Omega/k_1$ i.e. if
\index{MRI!Critical $\beta$}
\begin{align*}
\bmid<\frac{2(k_1H)^2}{3} \simeq 0.89,\quad\rightarrow\quad\textrm{Stability.}
\end{align*}
This value can vary slightly depending on the boundary conditions for the perturbation as $z\rightarrow \infty$.  Overall, it is safe to assume that the MRI exists provided that $\bmid\gtrsim 1$. 

\begin{figure}
\begin{center}
    \includegraphics[width=0.45\linewidth]{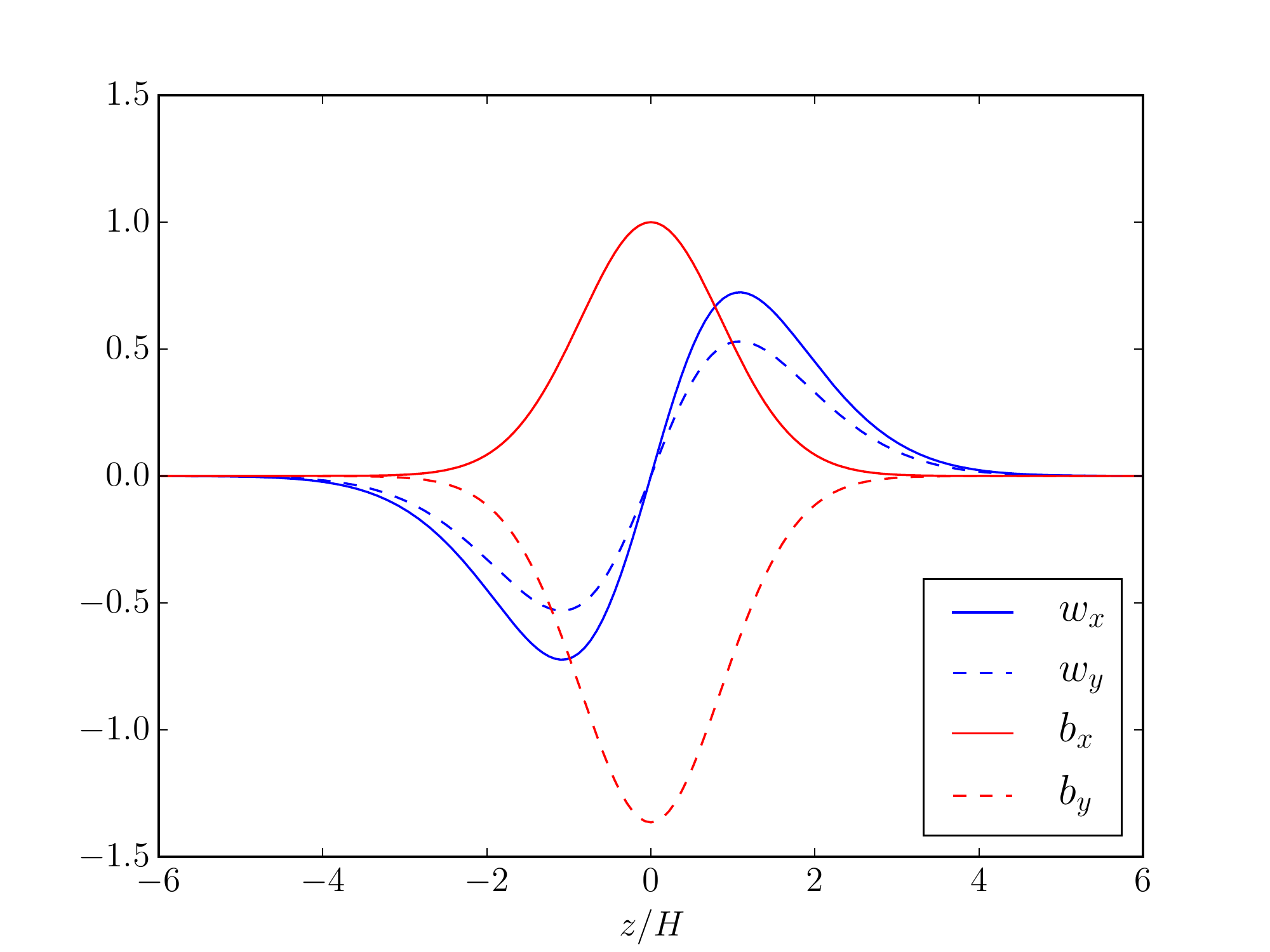}
    \includegraphics[width=0.45\linewidth]{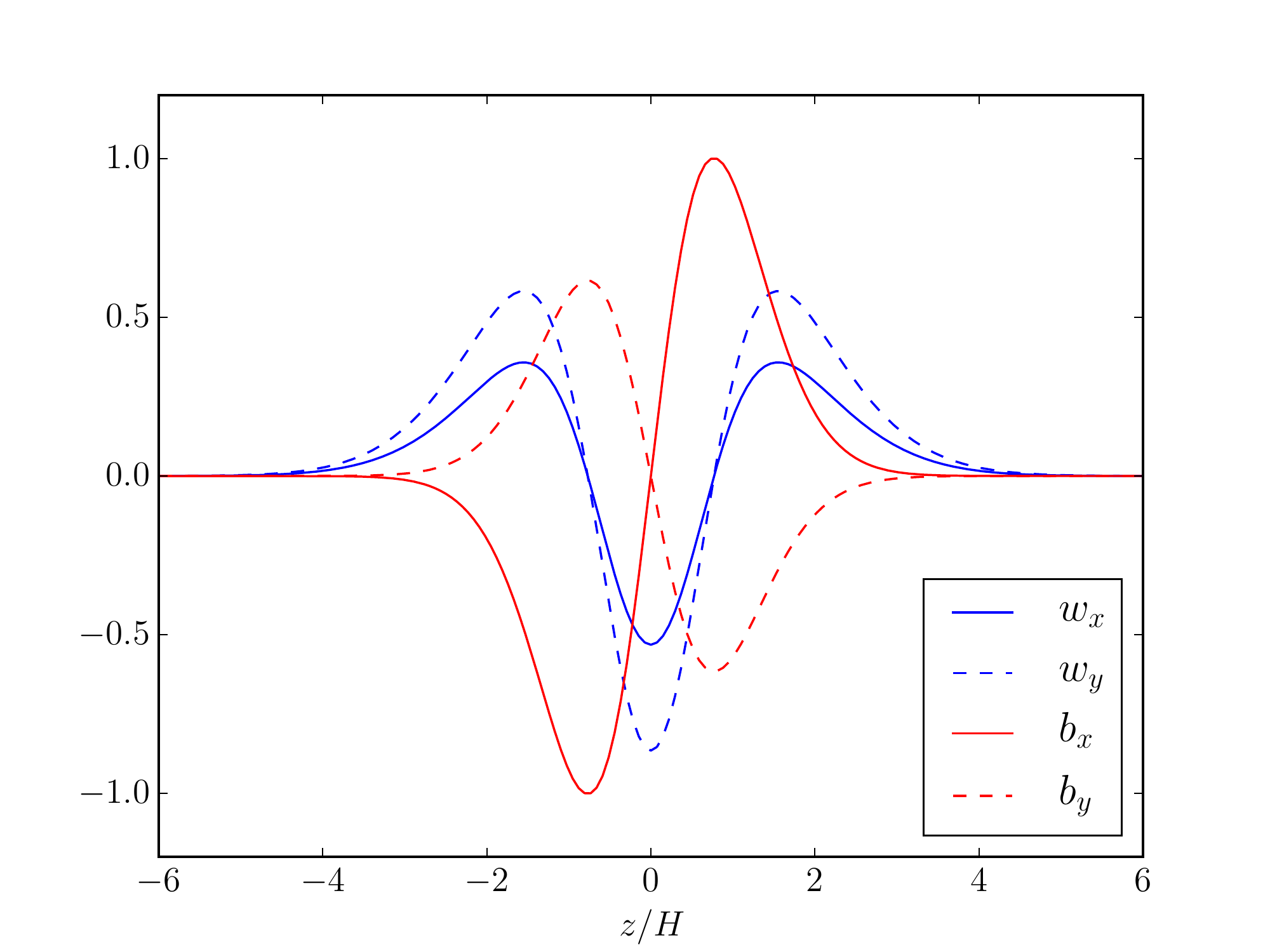}
    \includegraphics[width=0.80\linewidth]{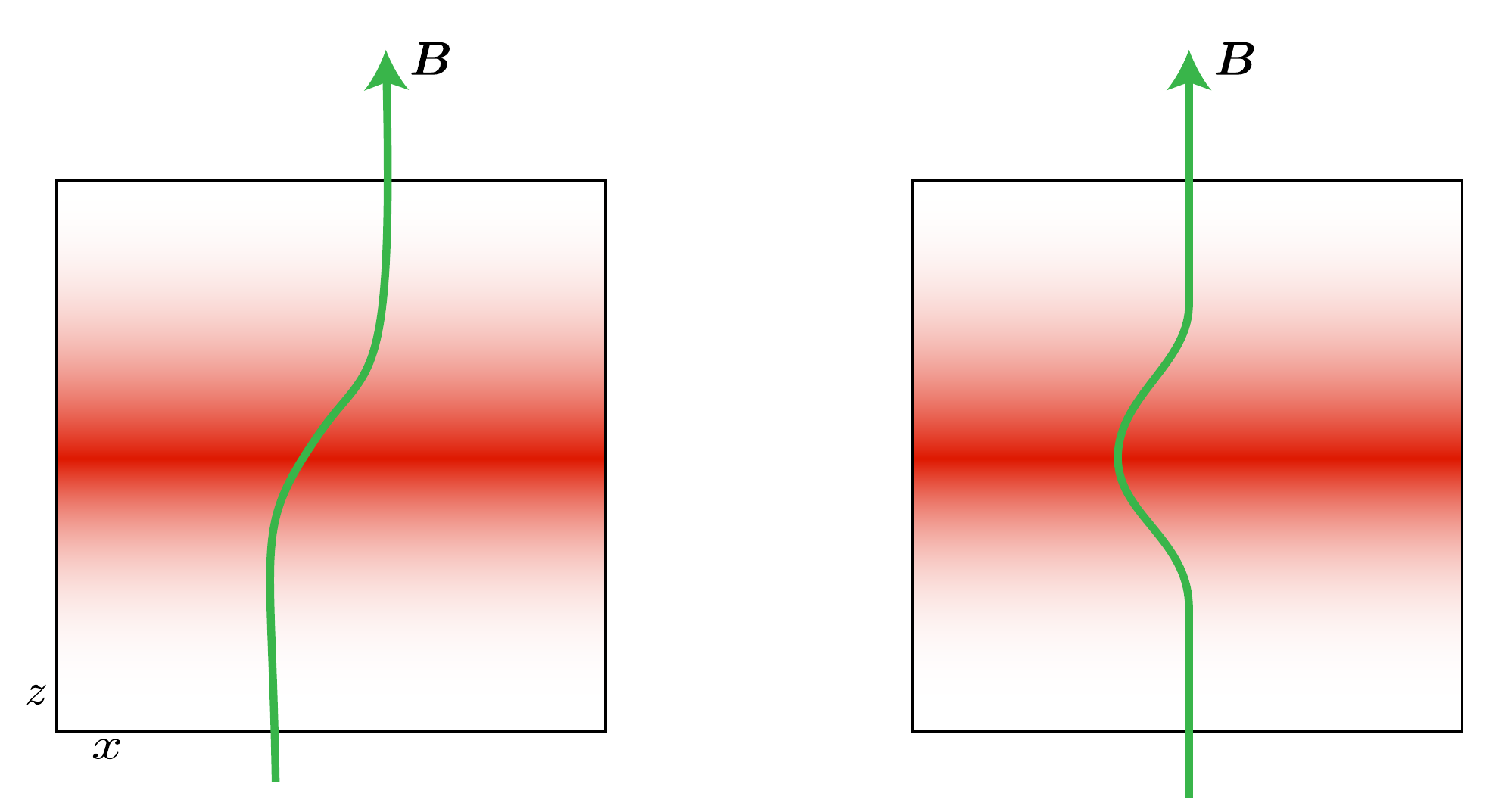}
    \caption{Top: MRI eigenmodes computed at $\bmid=5$ for $n=1$ (left, odd mode, $\sigma=0.72\Omega$) and $n=2$ (right, even mode $\sigma=0.67\Omega$). Bottom: schematic representation of the field perturbation due to odd (left) and even (right) modes.}
    \label{fig:mri_strat_ideal}
\end{center}
\end{figure}

Second, eigenmodes come in two symmetries: odd symmetry modes (which corresponds to odd $n$) have odd $v_{x,y}(z)$, even $B_{x,y}(z)$ and exhibit a maximal magnetic perturbation at the midplane and no velocity perturbation at this location. Even symmetry modes (even $n$) on the other hand have a velocity jet in the midplane and no magnetic perturbation at $z=0$ (see Fig.~\ref{fig:mri_strat_ideal}). The MRI does not choose specifically any symmetry in these local stratified models: both even and odd $n$ follow the same dispersion relation.

Third, the fastest growing modes tend to be more oscillatory in the disc midplane at higher $\bmid$. This is because qualitatively, one expects $k\Vaz\sim \Omega$ for the fastest growing mode, hence in the midplane, one gets $k_zH\sim \sqrt{\bmid}$. An example of such a mode is shown in Fig.~\ref{fig:mri_strat_high_beta} 

\begin{figure}
\begin{center}
    \includegraphics[width=0.55\linewidth]{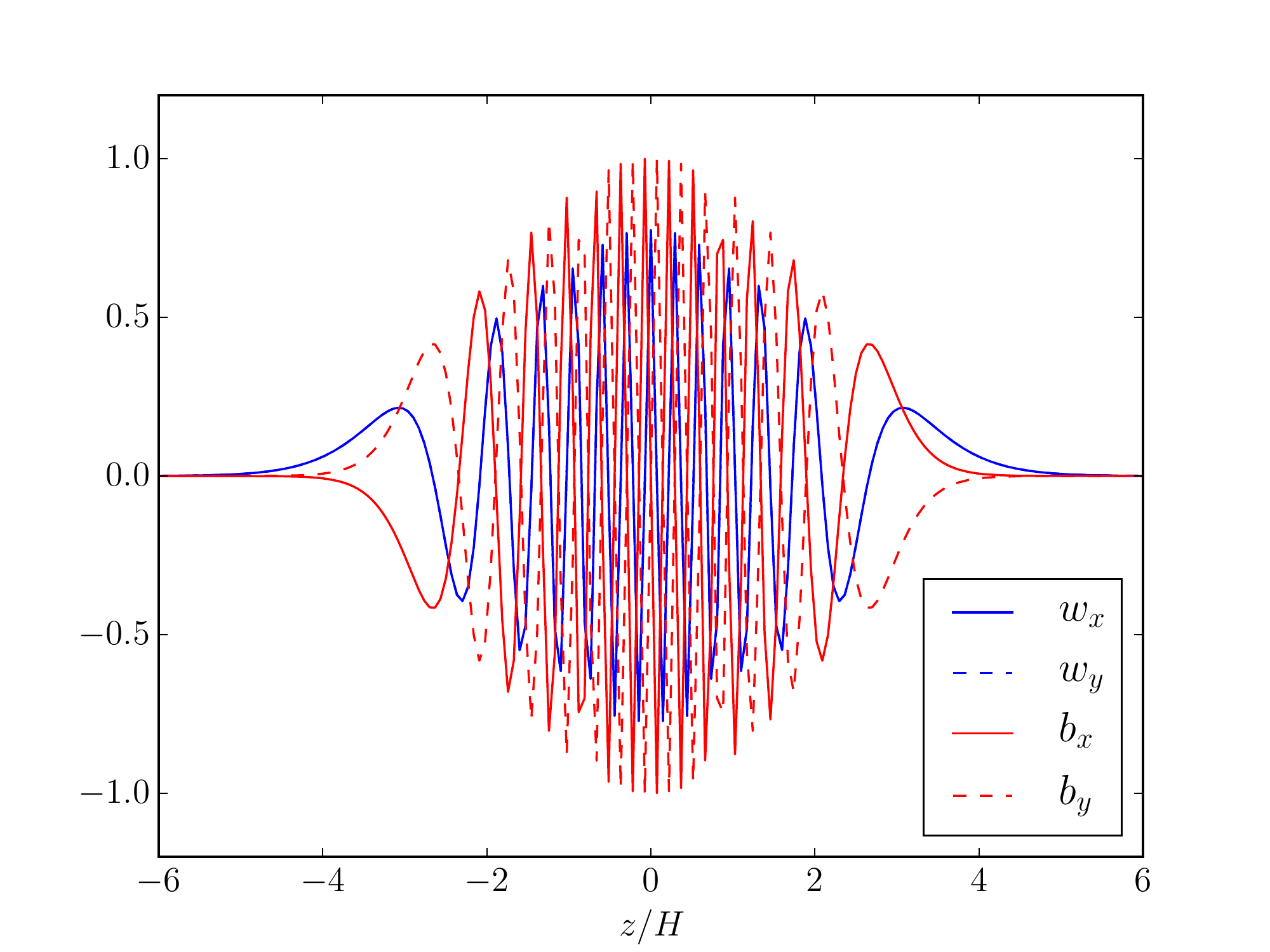}
    \caption{Fastest growing MRI eigenmode ($\sigma=0.75\Omega$) in a stratified model for $\bmid=10^3$. Note the strongly oscillating behaviour close to the midplane.}
    \label{fig:mri_strat_high_beta}
\end{center}
\end{figure}

\item[Ohmic and ambipolar diffusion: ]

as shown in the non-stratified section, Ohmic and ambipolar diffusion tend to suppress the MRI when the Elsasser numbers are less than $1$. Typically, this situation occurs in the densest regions of the disc (see Fig.~\ref{fig:partii:ni_effects}). We show in Fig.~\ref{fig:mri_strat_OA} the most unstable MRI eigenmode with $\beta=10^3$ resulting from our fiducial metal-free mode at $R=1~\mathrm{AU}$. As expected, the MRI is strongly suppressed where $\Lambda_O<1$. Ambipolar diffusion tends to reduce the overall growth rate but does not impact significantly the shape of the eigenmode. This is a perfect linear illustration of the historical "layered accretion" paradigm where the MRI still survives at the surface of the disc, leaving the disc midplane up to 1-2 scale-height essentially magnetically dead \citep{G96}. 

\add{Note that when the linear eigenmodes do not propagate in the disc, as it is the case here, there are no well-defined "odd" and "even" modes. Instead, there are two families of modes localised either at the top or the bottom side of the disc, with identical growth rates. It is in principle possible to combine these top and bottom modes to reconstruct even and odd modes, but here we have chosen here to stress the lack of propagation through the disc of the perturbation. }

\begin{figure}
\begin{center}
    \includegraphics[width=0.45\linewidth]{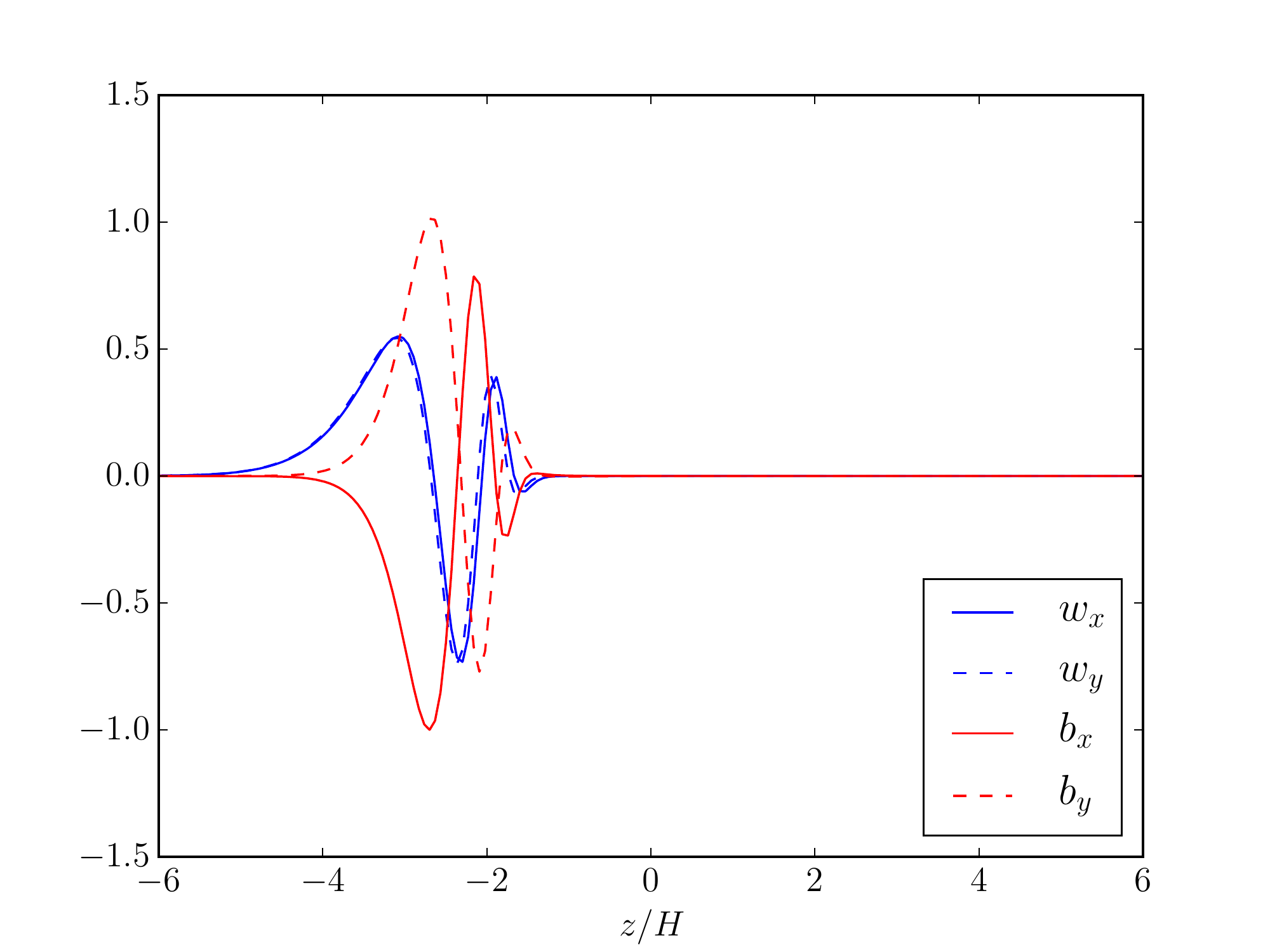}
    \includegraphics[width=0.45\linewidth]{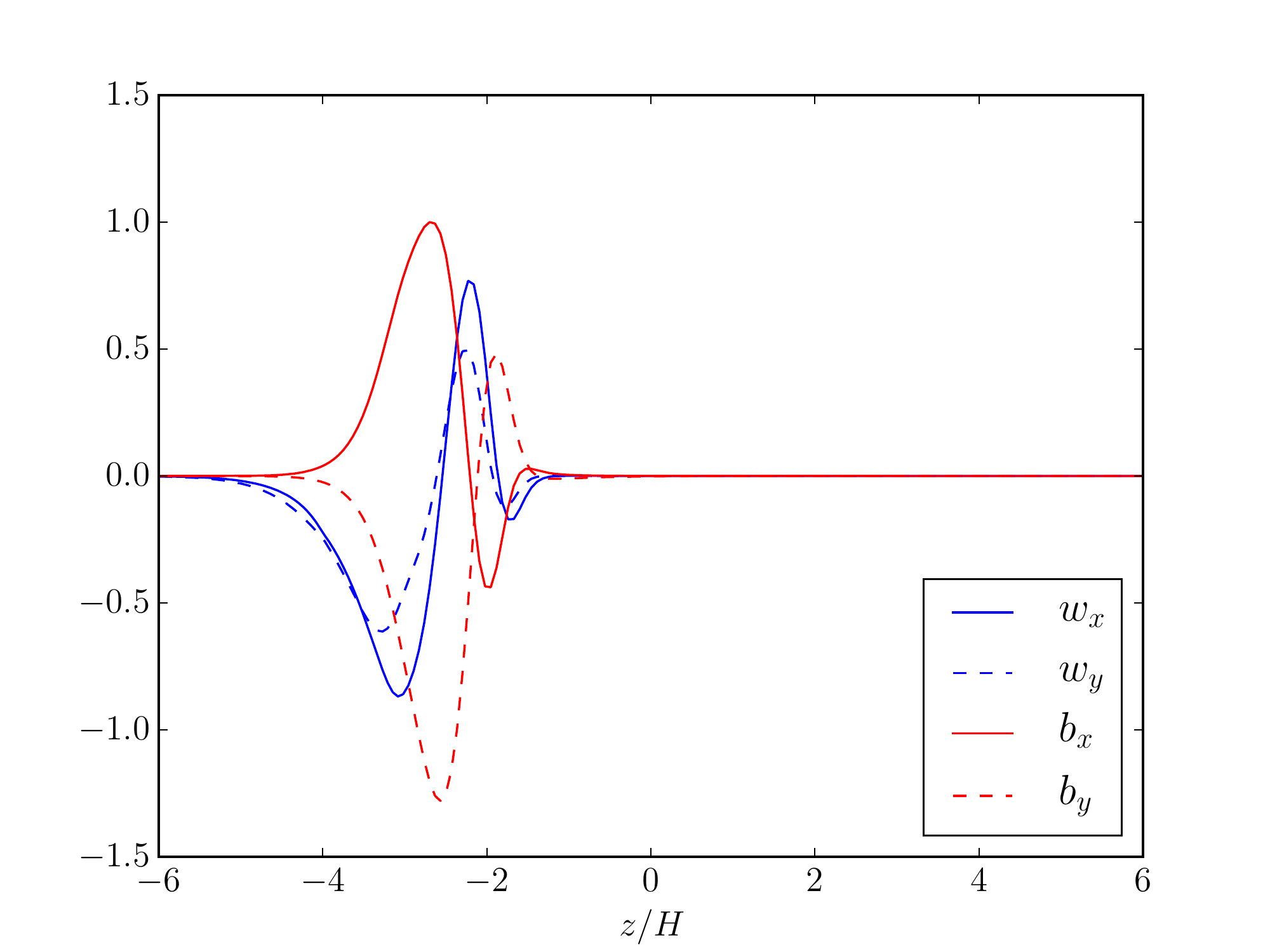}
    \caption{Fastest growing MRI eigenmode in a stratified model for $\bmid=10^3$ using the diffusivity profiles from Fig.~\ref{fig:partii:ni_effects} at $R=1~\mathrm{AU}$. Left: including Ohmic diffusion only ($\sigma=0.72\Omega$), right including Ohmic and ambipolar diffusion ($\sigma=0.56\Omega$). Note the lack of perturbations in the disc midplane due to Ohmic diffusion and partially to ambipolar diffusion. Equivalent eigenmodes with the same growth rates are found on the $z>0$ side of the disc.  }
    \label{fig:mri_strat_OA}
\end{center}
\end{figure}

\item[Hall effect\label{sec:Hall_lin_strat}: ]
as discussed above, the Hall effect can potentially revive dead-zones when the field is aligned with the rotation axis, thanks to the HSI branch of the Hall-MRI. This effect is illustrated in Fig.~\ref{fig:mri_strat_OAH} where the MRI eigenmode now propagates in the midplane resulting in a fully active disc column at $R=~1\mathrm{AU}$ when $\Vaz\Omega>0$. In this case, however, the growing perturbation is mostly a magnetic one and velocity perturbations are mostly absent in the midplane of the eigenmode. This is because the HSI is an unstable whistler mode, which is essentially an electronic perturbation leaving the ions (and the neutrals) unperturbed. When the field is anti-aligned, we recover results similar to the purely diffusive case, albeit with a slightly reduced growth rate.

\begin{figure}
\begin{center}
    \includegraphics[width=0.45\linewidth]{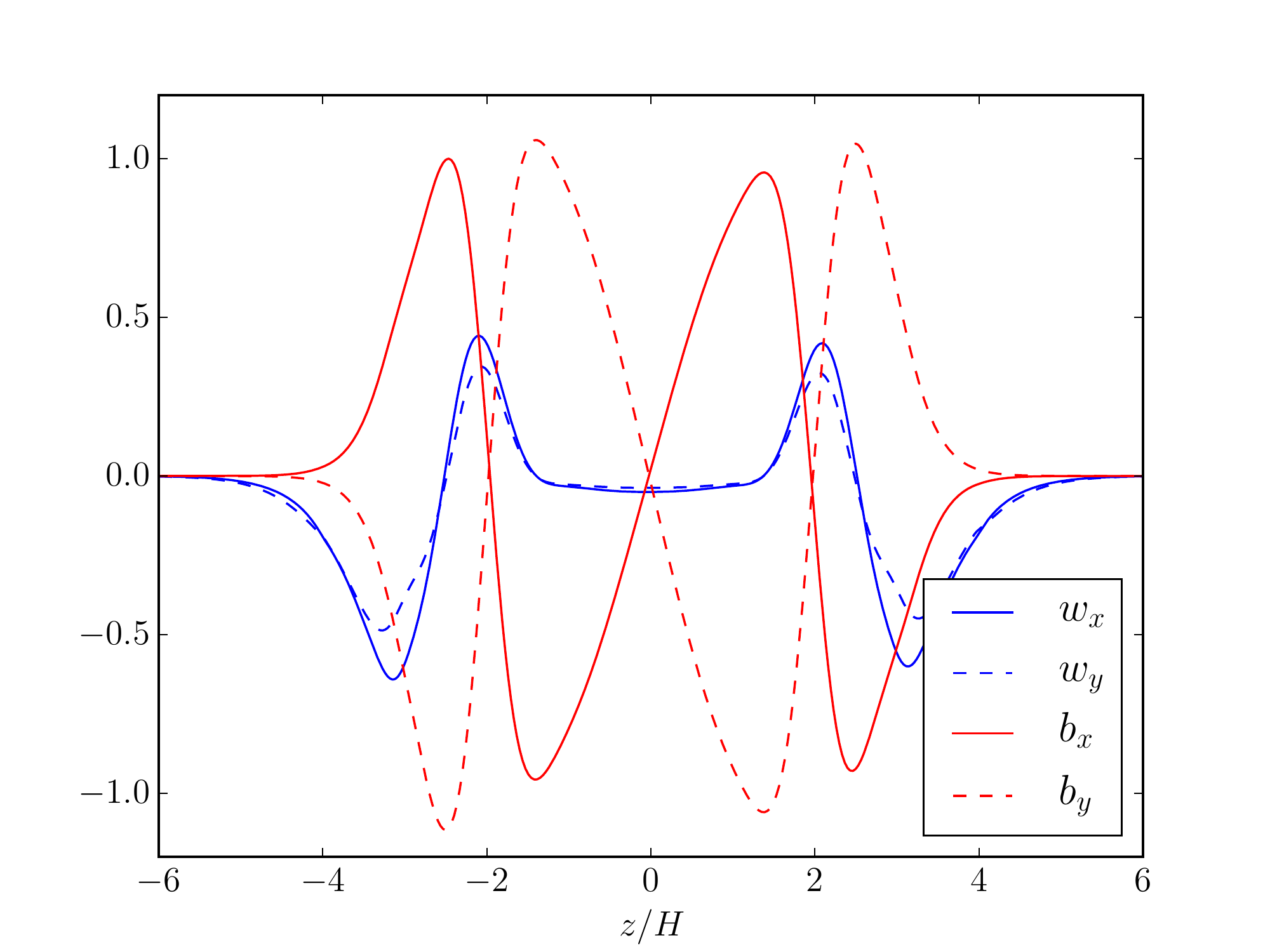}
    \includegraphics[width=0.45\linewidth]{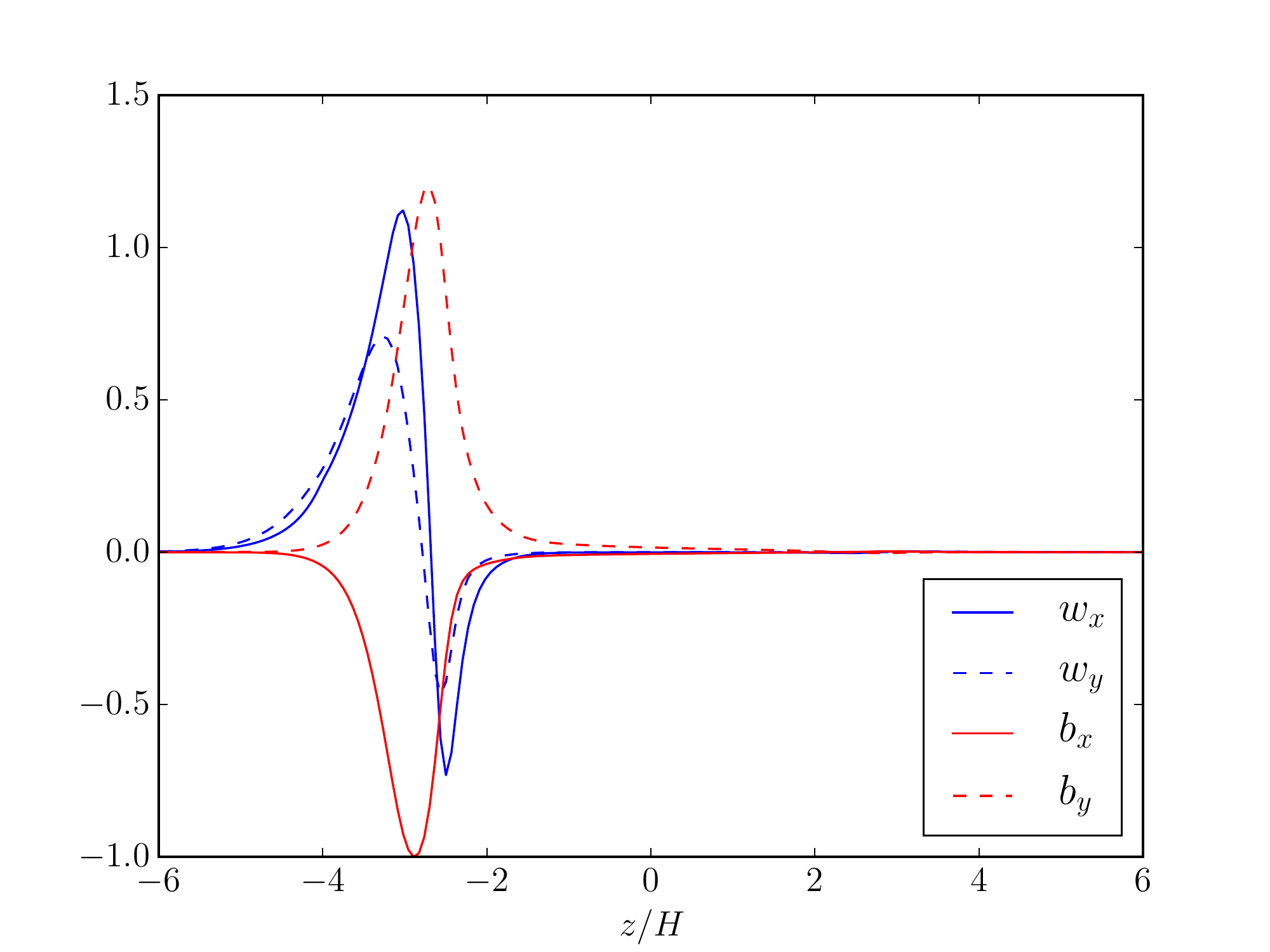}
    \caption{Fastest growing MRI eigenmode in a stratified model for $\bmid=10^3$ using the diffusivity profiles from Fig.~\ref{fig:partii:ni_effects} at $R=1~\mathrm{AU}$ including Ohmic, Hall and ambipolar diffusion. Left: assuming $\Vaz\Omega>0$ ($\sigma=0.65\Omega$), right: assuming $\Vaz\Omega<0$ ($\sigma=0.51\Omega$). The ``dead zone'' is now subject to long wavelength perturbations when $\Vaz\Omega>0$, as a result of the HSI. }
    \label{fig:mri_strat_OAH}
\end{center}
\end{figure}

\end{description}

\section{The Helicoidal MRI (HMRI)}
The Helicoidal MRI was first identified by \cite{HR05} using a spectral analysis of a rotating Taylor-Couette flow. This instability is known to work for arbitrarily low magnetic Reynolds numbers in rotating sheared flows, implying that it could work both in liquid sodium experiments and in weakly ionised astrophysical discs such as protoplanetary discs. For the sake of completness, let us show here the origin of this instability and discuss its application to Keplerian discs.

In contrast to the non-axisymmetric MRI (which is fully local), the HMRI is tightly linked to the presence of curvature in the physics of the system, hence it cannot be captured in the Hills approximation (even though a WKB analysis can capture it in the global geometry, see \citealt{KSF14} and below). Therefore, let us start back from the evolution equations in global geometry.

\subsection{Full set of equations in cylindrical geometry}

We consider the motion of a conductive fluid in cylindrical coordinates. We denote the velocity of the fluid $\bm{u}$ and the the magnetic field $\bm{B}$. We consider the inviscid, incompressible and ideal equations of magnetohydrodynamics, which reads, component by components:

\begin{align*}
\nonumber \partial_t u_R+&u_R\partial_R u_R+\frac{u_\phi}{R}\partial_\phi u_R-\frac{u_\phi^2}{R}+u_z\partial_z u_R\\
&=-\frac{1}{\rho_0}\partial_R\Big(P+\frac{B^2}{8\pi}\Big)+\frac{1}{4\pi\rho_0}B_R\partial_R B_R+\frac{1}{4\pi\rho_0}\frac{B_\phi}{R}\partial_\phi B_R-\frac{1}{4\pi\rho_0}\frac{B_\phi^2}{R}+\frac{1}{4\pi\rho_0}B_z\partial_z B_R\\
\nonumber \partial_t u_\phi+&u_R\partial_R u_\phi+\frac{u_\phi}{R}\partial_\phi u_\phi+\frac{u_Ru_\phi}{R}+u_z\partial_z u_\phi\\
&=-\frac{1}{\rho_0}\frac{1}{R}\partial_\phi\Big(P+\frac{B^2}{8\pi}\Big)+\frac{1}{4\pi\rho_0}B_R\partial_R B_\phi+\frac{1}{4\pi\rho_0}\frac{B_\phi}{R}\partial_\phi B_\phi+\frac{1}{4\pi\rho_0}\frac{B_RB_\phi}{R}+B_z\partial_z B_\phi\\
\nonumber \partial_t u_z+&u_R\partial_R u_z+\frac{u_\phi}{R}\partial_\phi u_z+u_z\partial_z u_z\\
&=-\frac{1}{\rho_0}\partial_z\Big(P+\frac{B^2}{8\pi}\Big)+\frac{1}{4\pi\rho_0}B_R\partial_R B_z+\frac{1}{4\pi\rho_0}\frac{B_\phi}{R}\partial_\phi B_z+\frac{1}{4\pi\rho_0}B_z\partial_z B_z
\end{align*}
for the equation of motion, which we combine to the induction equation:
\begin{align*}
 \partial_t B_R+&u_R\partial_R B_R+\frac{u_\phi}{R}\partial_\phi B_R+u_z\partial_z B_R=B_R\partial_R u_R+\frac{B_\phi}{R}\partial_\phi u_R+B_z\partial_z u_R\\ 
 \partial_t B_\phi +&u_R\partial_R B_\phi+\frac{u_\phi}{R}\partial_\phi B_\phi+u_z\partial_z B_\phi+\frac{u_\phi B_R}{R}=B_R\partial_R u_\phi+\frac{B_\phi}{R}\partial_\phi u_\phi+B_z\partial_z u_\phi+\frac{u_R B_\phi}{R}\\
  \partial_t B_z +&u_R\partial_R B_z+\frac{u_\phi}{R}\partial_\phi B_z+u_z\partial_z B_z=B_R\partial_R u_z+\frac{B_\phi}{R}\partial_\phi u_z+B_z\partial_z u_z
\end{align*}
and the continuity equation
\begin{align*}
\frac{1}{R}\partial_R R u_R+\frac{1}{R}\partial_\phi u_\phi+\partial_z u_z	=0
\end{align*}
\subsection{Linearisation}
We linearise the equations with respect to a background flow
\begin{align*}
\bm{u}_0\equiv R\Omega(R)\bm{e}_\phi	
\end{align*}
where $\Omega(R)$ is the angular velocity profile of the mean flow (arbitrary, for the moment). In addition, we consider the case with a mean magnetic field defined as
\begin{align*}
\bm{B}_0=B_{\phi,0}(R)\bm{e}_\phi+B_{z,0}\bm{e}_z	
\end{align*}
so the only spatial dependency is in the $R$ direction for the toroidal component of the field. In addition, we're going to assume that the flow is axisymmetric, so that we can cancel $\partial_\phi$ derivatives.

The velocity and magnetic fields are then expanded as
\begin{align*}
\bm{u}=\bm{u}_0+\bm{v}\\
\bm{B}=\bm{B}_0+\bm{b}	
\end{align*}
where deviations are assumed to be infinitely small compared to the means. The linearised equations of motion eventually read
\begin{align*}
\partial_t v_R&=-\partial_R \Pi -\frac{2B_{\phi,0}b_\phi}{4\pi\rho_0 R}+\frac{1}{4\pi\rho_0}B_{z,0}\partial_z b_R+2\Omega v_\phi\\
\partial_t v_\phi&=\frac{1}{4\pi\rho_0}b_R\frac{1}{R}\partial_RRB_{\phi,0}+\frac{1}{4\pi\rho_0}B_{z,0}\partial_z b_\phi-(2\Omega+R\partial_R\Omega) v_R\\
\partial_t v_z&=-\partial_z \Pi+\frac{1}{4\pi\rho_0}B_{z,0}\partial_z b_z
\end{align*}
while the induction equation reads
\begin{align*}
\partial_t b_R&=B_{z,0}\partial_z v_R\\
\partial_t b_\phi&=b_RR\partial_R\Omega+B_{z,0}\partial_z v_\phi-v_RR\partial_R\frac{B_{\phi,0}}{R}\\	
\partial_t b_z&=B_{0,z}\partial_z v_z
\end{align*}
In order to make the notations more concise and consistant with the usual MRI derivation, we define the following coefficients
\begin{align*}
q\equiv-\frac{d\log\Omega}{d\log R},\\
p\equiv -\frac{d\log B_{\phi,0}}{d\log R}.
\end{align*}
It should be noted that the particular case $p=1$ corresponds to a case where no axial current is present in the system. This is the reference case considered by \cite{HR05}

Using these notations the equations eventually reads
\begin{align*}
\partial_t v_R&=-\partial_R \Pi -\frac{2B_{\phi,0}b_\phi}{4\pi\rho_0 R}+\frac{1}{4\pi\rho_0}B_{z,0}\partial_z b_R+2\Omega v_\phi\\
\partial_t v_\phi&=-(p-1)\frac{B_{\phi,0}}{4\pi\rho_0 R}b_R+\frac{1}{4\pi\rho_0}B_{z,0}\partial_z b_\phi-\Omega(2-q) v_R\\
\partial_t v_z&=-\partial_z \Pi+\frac{1}{4\pi\rho_0}B_{z,0}\partial_z b_z
\end{align*}
while the induction equation reads
\begin{align*}
\partial_t b_R&=B_{z,0}\partial_z v_R\\
\partial_t b_\phi&=-q\Omega b_R+B_{z,0}\partial_z v_\phi+(p+1)\frac{B_{\phi,0}}{R}v_R\\	
\partial_t b_z&=B_{0,z}\partial_z v_z
\end{align*}
\subsection{WKB approximation and pressure}
We are going to look for a local solution, i.e. a solution with fast variation compared to $R$. We will therefore zoom on a tiny region around a fiducial radius $R_0$ defining $x\equiv R-R_0$, and expand the solution as
\begin{align*}
\bm{v,b}\propto\exp\Big[\sigma t+i(k_x x+k_z z)\Big]	
\end{align*}
we get
\begin{align*}
\sigma v_R&=-ik_R \Pi -2\omega_{A\phi}b_\phi+i\omega_{Az} b_R+2\Omega v_\phi\\
\sigma v_\phi&=-(p-1)\omega_{A\phi}b_R+i\omega_{Az} b_\phi-\Omega(2-q) v_R\\
\sigma v_z&=-ik_z \Pi+i\omega_{Az} b_z
\end{align*}
while the induction equation reads
\begin{align*}
\sigma b_R&=i\omega_{Az} v_R\\
\sigma b_\phi&=-q\Omega b_R+i\omega_{A\phi} v_\phi+(p+1)\omega_{A\phi}v_R\\	
\sigma b_z&=i\omega_{Az} v_z
\end{align*}
where we have defined the Alfv\'en frequency $\omega_{Az}=k_z B_{0,z}/\sqrt{4\pi\rho_0}$ and the toroidal Alfv\'en frequency $\omega_{A\phi}=B_{\phi,0}/(R_0\sqrt{4\pi\rho_0})$. It is obvious from here that the vertical equation of motion and induction are just fed by the horizontal problem, but they do not have any feedback in the horizontal plane. We will therefore consider only the horizontal equations without loss of generality. In order to solve for the pressure, we dot the equation of motion by $k$ to get an equation for $\Pi$. This gives
\begin{align*}
\Pi =	\frac{2ik_R\omega_{A\phi}b_\phi}{k^2}-2i\Omega v_\phi\frac{k_R}{k^2}.
\end{align*}
This eventually leads to the following linear problem

\begin{align*}
\begin{pmatrix}
-\sigma &   2\Omega g_{zz} & i\omega_{Az} & -2\omega_{A\phi}g_{zz} \\
-(2-q)\Omega & -\sigma & -(p-1)\omega_{A\phi} & i\omega_{Az}\\
i\omega_{Az} &     0      &  -\sigma & 0\\
(p+1)\omega_{A\phi} & i\omega_{Az} & -q\Omega & -\sigma
\end{pmatrix}
\begin{pmatrix}
v_r\\v_\phi \\ b_r\\ b_\phi	
\end{pmatrix}=0
\end{align*}
where $g_{zz}=k_z^2/k^2$.
This has non trivial roots provided that the matrix determinant cancels out. This condition leads to a dispersion relation on $\sigma$:
\begin{align*}
	\sigma^4+\sigma^2\Big[2\omega_{Az}^2+\kappa^2 g_{zz}+2(p+1)\omega_{A\phi}^2g_{zz}\Big]-8i\sigma\omega_{Az}\omega_{A\phi}\Omega g_{zz} \\
	+\omega_{Az}^2\Big[\omega_{Az}^2-2q\Omega^2g_{zz}+2(p-1)\omega_{A\phi}^2g_{zz}\Big]&=0
\end{align*}

We recognise the usual MRI-related form of the dispersion relation in the last term. The case $p=1$ which corresponds to the HMRI initially derived by \cite{HR05} therefore does not affect the original MRI criterion. As we will see later, the HMRI driving term is actually the linear term in $\sigma$.

\subsection{HMRI with resistivity}
Adding resistivity $\eta$ means that the linearized induction equations are modified as follows
\begin{align*}
\sigma b_R&=i\omega_{Az} v_R-\eta k^2 b_R\\
\sigma b_\phi&=-q\Omega b_R+i\omega_{A\phi} v_\phi+(p+1)\omega_{A\phi}v_R-\eta k^2 v_\phi\\	
\sigma b_z&=i\omega_{Az} v_z-\eta k^2b_z
\end{align*}
The resulting linear system is then transformed into
\begin{align*}
\begin{pmatrix}
-\sigma &   2\Omega g_{zz} & i\omega_{Az} & -2\omega_{A\phi}g_{zz} \\
-(2-q)\Omega & -\sigma & -(p-1)\omega_{A\phi} & i\omega_{Az}\\
i\omega_{Az} &     0      &  -\sigma-\eta k^2 & 0\\
(p+1)\omega_{A\phi} & i\omega_{Az} & -q\Omega & -\sigma-\eta k^2                                                                                                                                                                                                                                                                                                                                                                                                                                                                                                                                                                                                                                                                                                                                                                                                                                   
\end{pmatrix}
\begin{pmatrix}
v_r\\v_\phi \\ b_r\\ b_\phi	
\end{pmatrix}=0,
\end{align*}
which results into the following dispersion relation
 \begin{align*}
\nonumber \sigma^2(\sigma+\eta k^2)^2+\sigma(\sigma+\eta k^2)\Big[2\omega_{Az}^2+2(p+1)\omega_{A\phi}^2g_{zz}\Big]+(\sigma+\eta k^2)^2\kappa ^2g_{zz}\\-i\omega_{Az}\omega_{A\phi}\Omega g_{zz}\big[(8-2q)(\sigma+\eta k^2)+2q\sigma\big]+\omega_{Az}^2\Big[\omega_{Az}^2-2q\Omega^2g_{zz}+2(p-1)\omega_{A\phi}^2g_{zz}\Big]&=0.	
 \end{align*}
This can be combine into a standard $4^\mathrm{th}$ order polynomial in $\sigma$:
\begin{align*}
\nonumber \sigma ^4+2\sigma^3\eta k^2+\sigma^2\Big[2\omega_{Az}^2+\kappa^2g_{zz}+2(p+1)\omega_{A\phi}^2g_{zz}+\eta^2k^4\Big]
\\\nonumber -\sigma\Bigg[8i\omega_{Az}\omega_{A\phi}\Omega g_{zz}-2\eta k^2\Big(\omega_{Az}^2+\kappa^2g_{zz}\Big)\Bigg]
\\+\omega_{Az}^2\Big[\omega_{Az}^2-2q\Omega^2g_{zz}+2(p-1)\omega_{A\phi}^2g_{zz}\Big]-(8-2q)i\omega_{Az}\omega_{A\phi}\Omega g_{zz}\eta k^2+\eta^2k^4\kappa^2g_{zz}&=0	
\end{align*}
which is formally equivalent to \cite{KSF14}, Appendix B. Although this dispersion relation contains both the MRI and the HMRI in the resistive regime, it is difficult to disentangle the two instabilities. We therefore follow \cite{LGH06} and consider the limit where the resistivity is much larger than the other terms.

\subsection{Inductionless limit}
In the inductionless limit, we neglect magnetic induction by considering roots $\sigma\ll \eta k^2$. We therefore introduce a small parameter $\varepsilon=1/\eta k^2$, and expand the above dispersion relation at first order in $\varepsilon$. This suppresses the strongly damped magnetic modes which have $\sigma\sim -\eta k^2$ and which are primarily damped Alfv\'en waves, and on which the usual MRI lives. For this reason, the inductionless limit allows us to suppress (stabilised) MRI modes. Following this limit, we get a second order dispersion relation: 

\begin{align*}
\sigma^2+\sigma\varepsilon \Big[2\omega_{Az}^2+2(p+1)\omega_{A\phi}^2g_{zz}\Big]+\kappa ^2g_{zz}-2i\varepsilon \omega_{Az}\omega_{A\phi}\Omega g_{zz}(4-q)&=0
\end{align*}
The roots are simple to obtain and give at first order in $\varepsilon$:
\begin{align*}
\sigma=\pm i\kappa g_{zz}^{1/2}+\varepsilon\Big[\pm \frac{1}{\kappa}\omega_{Az}\omega_{A\phi}\Omega g_{zz}^{1/2}(4-q)-\omega_{Az}^2-(p+1)\omega_{A\phi}^2g_{zz}\Big].
\end{align*}
As it can be seen, this growth rate describes a small deviation from pure epicyclic oscillations. Therefore, in contrast to the MRI, which is an instability of the (slow) Alv\'en branch, the HMRI is an overstability of the epicyclic branch. This explains, in part, its survival in the limit of small magnetic Reynolds numbers.

The instability clearly arises for the $+$ sign of the roots, when
\begin{align}
\label{inscond}
	\frac{1}{\kappa}\omega_{Az}\omega_{A\phi}\Omega g_{zz}^{1/2}(4-q)-\omega_{Az}^2-(p+1)\omega_{A\phi}^2g_{zz}>0\quad\rightarrow\quad\textrm{INSTABILITY}
\end{align}
Since this condition is on a second order polynomial on $\omega_{A\phi}$, it is strictly equivalent to requiring that the discriminant of the polynomial is positive, i.e that
\begin{align*}
\frac{(4-q)^2}{2(2-q)}-4(p+1)>0	
\end{align*}
Assuming $\kappa^2>0$, this is again equivalent to asking that
\begin{align}
\label{qcond}
q^2+8qp-16p>0	
\end{align}
which is (again!) a second order polynomial in $q$, which we ought to resolve. The stability condition is then simply
\begin{align*}
	q>4(\sqrt{p(1+p)}-p)\quad\textrm{or}\quad q<-4(\sqrt{p(1+p)}+p)\quad\rightarrow\quad\textrm{INSTABILITY}
\end{align*}
In the current-free configuration ($p=1$), we recover the so called Liu limit \citep{LGH06} $q>4(\sqrt{2}-1)\simeq 1.657$, i.e. that a current-free toroidal field is stable in the Keplerian regime. It is the main reason why this instability has been mostly neglected in the astrophysical context.

However, if one allows for an axial current in the system ($p\ne 1$), it is possible to recover the instability for Keplerian rotation profiles. This can be deduced from  (\ref{qcond}):
\begin{align*}
	p<\frac{q^2}{8(2-q)}\quad\rightarrow\quad\textrm{INSTABILITY}
\end{align*}
In the Keplerian regime, we therefore require $p<9/16=0.5625$.

The maximum growth rates are obtained from the mean of the two roots of (\ref{inscond}) which is simply
\begin{align}
\label{eq:optimum_wap}
\omega_{A\phi,\mathrm{max}}=	\frac{1}{2\kappa}\omega_{Az}\Omega g_{zz}^{1/2}(4-q).
\end{align}
The maximum growth rate is then given by
\begin{align*}
\sigma_\mathrm{max}=i\kappa g_{zz}^{1/2}+\frac{V_{Az}^2}{\eta}\Big[ g_{zz}\frac{(4-q)^2}{2(2-q)(p+1)}-1\Big].
\end{align*}

which turns out to be spatially scale-free. From these results, we can deduce a few important results for Keplerian discs with $q=3/2$. First, one finds that the growth rate in units of the local orbital frequency scales like the Ohmic Elsasser number, i.e $\Im(\sigma_\mathrm{max})/\Omega\simeq \Lambda_O$. This implies relatively low growth rates in protoplanetary discs, unless they are hosting a dynamically strong vertical field (with $V_{Az}\sim \Omega H$ where $H$ is the disc thickness). In addition, these ``optimum'' growth rates can only be reached for $\omega_{A\phi}\sim \omega_{Az}$ (from Eq.~\ref{eq:optimum_wap}). Combining these constrains, we find that $V_{A\phi}\sim R\Omega Hk$. Given that by construction $Hk>1$ (because the vertical wavelength needs to fit in the disc), this shows that the azimuthal Alfv\'en velocity needs to be of the order of the Keplerian velocity (or larger). This is a \emph{very} strong azimuthal field, well above equipartition. Among other things, such a strong field implies that the disc is not Keplerian anymore as the radial tension and magnetic pressure forces are of the order of the central gravity.

For these reasons, the role played by the HMRI in the context of protoplanetary discs has been mostly ignored, as it lives in a parameter regime probably distant from that of real systems, which are known to be in Keplerian rotation, and hence for which $\omega_{A\phi}\ll \Omega$.

\newpage
\part{Nonlinear saturation of the MRI}
\index{MRI!Saturation}
The saturation of the MRI has been mostly studied in local shearing box simulations. Since the seminal paper by \cite{HGB95}, several studies have been dedicated to non-ideal effects and the role they play in the MRI saturation. 

Let us first distinguish unstratified and vertically stratified shearing box models. In the unstratified model, the box is periodic in the vertical direction, making the system much simpler to analyse. It corresponds to the small shearing box model of \cite{UR04} (see also \S\ref{sec:shearingbox}). When vertical stratification is included, the flow is \emph{usually} allowed to escape through the vertical boundary conditions, potentially leading to outflows. Here, we will first focus on unstratified models before moving to stratified ones.

It is also important to separate simulations with and without a mean field. Although the MRI (the linear instability) does require a mean field to exist, it's been shown that in the non-linear regime, MHD turbulence exists without any mean field. We call this case the ``MRI dynamo''. Although this was initially a mere curiosity, it turned out to be the fiducial configuration for many stratified and even global simulations due to the technical difficulties associated to simulations with a mean field.

In all of these models, one uses box averages, defined for a quantity $Q$ by
\begin{align*}
\langle Q\rangle=\iiint\,\mathrm{d}^3 \bm{x}\, Q\quad.
\end{align*}
One of the key element is then to quantify the $\alpha$ parameter from the turbulent stress, which is defined as
\begin{align*}
\alpha\equiv \frac{1}{\langle P\rangle}\Bigg\langle \rho w_xw_y-\frac{B_xB_y}{4\pi}\Bigg\rangle,    
\end{align*}
which is also sometimes averaged in time.

\section{Unstratified models}
Except in rare circumstances, unstratified models are periodic in the vertical direction. In this case, the energy conservation equation (\ref{eq:sb_mech_conservation}) reads
\begin{align*}
\partial_t \langle \mathcal{E}_\mathrm{Mech}\rangle &=\langle P\bm{\nabla \cdot w}\rangle +\langle\bm{E}_\mathrm{NI}\bm{\cdot J}\rangle +q\Omega_0 \Bigg\langle \rho w_xw_y-\frac{B_xB_y}{4\pi}\Bigg\rangle.
\end{align*}
In numerical experiments, since MRI turbulence is subsonic, the contribution of the $P\mathrm{d}V$ term is small while the $x-y$ stress term is definite positive (it is the term driving accretion!). Therefore, a quasi-steady turbulent state for which $\mathcal{E}_\mathrm{mech}\sim\mathrm{constant}$ necessarily implies that non-ideal effects (or viscous effect, which have been ignored here) are \emph{non negligible}. This statement can be understood in terms of Kolmogorov's turbulent cascade argument. In the energy equation, the stress is a source of mechanical energy. In the cascade argument, it corresponds to the energy injection rate. For the cascade to be in a steady state, some dissipative effect must enter the picture \emph{at some scale} to dissipate what has been injected at the beginning of the cascade. One concludes from this argument that there is no such thing as ideal MRI turbulence. 

Here, we will keep the terminology "ideal MRI" in cases where dissipation is sufficiently small to be negligible on large scales $\Lambda_{O,H,A}\gg 1$, or when no physical dissipation is explicitly considered in the numerical method (an approach referred to as ``Implicit Large Eddy simulations: ILES'')\index{Implicit large eddy simulation (ILES)}. Nevertheless, numerical dissipation is \emph{always actively playing a role} in these models, which is not without consequences, as we shall see\dots

Configurations with a mean field are often characterised by the plasma $\beta$ parameter of the mean field, as defined by
\index{Plasma!$\beta$}
\index{$\bmean$ plasma beta! unstratified shearing box}
\begin{align*}
    \bmean\equiv\frac{8\pi \langle P\rangle }{\langle B\rangle ^2}.
\end{align*}
This should be confused with the turbulent plasma $\beta$ parameter
\begin{align*}
    \langle \beta\rangle \equiv\Bigg\langle \frac{8\pi P }{ B ^2}\Bigg\rangle ,
\end{align*}
 which is \emph{not} a control parameter of the physical system since usually $B\gg\langle B\rangle$.
 
In principle, the vertical box extension $L_z$ does not necessarily match the pressure scale height $H=c_s/\Omega$ since vertical stratification is ignored in these models. However, in most of the simulations published today, it is the case, so that the two scales can be identified. This allows us to get an alternative expression for $\bmean$ which can be useful to interpret numerical simulations
\begin{align*}
\bmean\equiv \frac {2\Omega^2L_z^2}{\Vaz^2}.
\end{align*}
It is this expression for $\beta$ which is used in incompressible simulations.
\subsection{Ideal MHD\label{sec:ideal_unstrat}}
\subsubsection{With a mean field}
The first historical models of MRI turbulence \citep{HGB95} tested both mean vertical and toroidal fields and were computed in the ideal MHD framework. MRI unstable modes grow, breakup presumably thanks to secondary instabilities which are essentially Kelvin-Helmholtz type modes \citep{GX94,LL09} and end up in fully-developed MHD turbulence. 

In the mean vertical field case, \cite{HGB95} found that  and $\alpha\simeq 6.7 \bmean^{-1/2}$. However, it was later pointed out by \cite{BM08} that, in the presence of a mean vertical field, the box aspect ratio used by \cite{HGB95} was prone to recurrent channel mode solutions, which was absent in wider boxes (more elongated in $x$). This implies that $\alpha$ and $\langle \beta \rangle$ are overestimated by a factor of 2 compared to wider boxes \citep{BM08}. We therefore correct \cite{HGB95} for this and get the following scaling law in the range $400<\bmean<5\times 10^4$:
\begin{align}
\nonumber     \alpha\simeq 0.61\langle\beta\rangle^{-1},\\
\label{eq:alpha_ideal_estimate}    \alpha\simeq 3.3 \bmean^{-1/2},
\end{align}
a scaling which has also been verified in incompressible simulations \citep{LL10}. The extrapolation down to $\beta\rightarrow 1$ suggest that $\alpha\gtrsim 1$ could be reached. However, it is very difficult to explore this limit numerically since the MRI becomes in this case very strong and channel modes never break up into developed turbulence \citep{HGB95}. \cite{LL07} explored this limit in the incompressible regime and found that turbulence was in this case very intermittent with long ``quiet'' episodes of linear growth followed by strong and sudden bursts of turbulence. In this situation, a constant $\alpha$ is probably not a good model of the disc physics.

In the mean toroidal field case, the scaling deduced from \cite{HGB95} in the range $2<\bmean<1200$ gives lower $\alpha$ values:
\begin{align*}
    \alpha\simeq 0.51\langle\beta\rangle^{-1},\\
    \alpha\simeq 0.35  \bmean^{-1/2}.
\end{align*}

\index{MRI!dynamo}
\index{Subcritical instability}
\subsubsection{Zero mean field: the "MRI dynamo"} The existence of an MRI-driven dynamo was first demonstrated by \cite{HGB96}. In this configuration, no external field is imposed and turbulence regenerates a field on which the MRI can grow, feeding back turbulent motions. Because of this feedback loop, the system needs a finite amplitude perturbation to sustain turbulence \citep{RO08}, implying that the instability is in this case \emph{subcritical}. The dynamo feedback loop has been the subject of intense studies after the discovery of dynamo cycles \citep{LO08}, both based on a quasi-linear theory of the toroidal MRI \citep{LO08b} and on a dynamical system approach \citep{HRC11,RRC13}.

\index{MRI!Numerical convergence}
According to \cite{HGB96}, the MRI dynamo yields $\alpha\sim 0.01$. However, it was later realised that the value of $\alpha$ in ideal MHD simulations (where no physical dissipation is introduced) depends on the numerical resolution \citep{FP07}, with $\alpha\propto 1/N$ where $N$ is the number of grid points in one direction\footnote{The scaling of $\alpha$ on $N$ actually depends on the order of the spatial reconstruction scheme as shown by \cite{BCF11}.}. \add{This problem of \emph{numerical convergence} (because numerics do not seem to be converging as one increases the resolution) implies that the estimation for $\alpha$ from zero net flux simulation is intrinsically flawed.} Obviously, this is most probably due to numerical dissipation (the only dissipation channel of these simulations), which is not acting as a real physical dissipation operator. If one introduces explicit viscosity $\nu$ and resistivity $\eta_O$ in the system, $\alpha$ then converges to a finite value, which seems to depend only on $\mathrm{Pm}\equiv\nu/\eta_O$ \citep{F10} \emph{and not anymore on the resolution}. This, however, leads to another complication known as the ``Pm effect''. 

\subsubsection{The Pm effect}
\index{Pm effect}
\add{The Pm effect shows up in quasi-ideal simulations \emph{both} with \emph{and} without a mean vertical field.}
When one introduces only a small amount of viscosity $\nu$ and resistivity $\eta_O$, the large-scale linear MRI modes are largely unaffected. For this reason, this regime corresponds to a ``quasi ideal-MHD'' regime. However, it turns out that the saturation level depends on the magnetic Prandtl number. A lot of literature has been devoted to this effect \citep{LL07,FPLH07}. In simulations with a mean field \citep{LL07,SH09}, the Pm effect results in an increase of $\alpha$ when $\mathrm{Pm}$ increases. In the zero mean field case, the effect is even stronger since the MRI dynamo, and therefore MHD turbulence, simply disappears when $\mathrm{Pm}\lesssim 1$ \citep{FPLH07,WB17}, giving $\alpha=0$ in that regime.

This effect shows up even in situations where the Reynolds and Elsasser numbers are much larger than one, indicating that linear or quasi-linear theory cannot explain it \citep{LL10}. Instead, it's been proposed that non-local energy transfers in spectral space \citep{LL11} could be responsible for this effect. If true, since non-locality is necessarily bounded \citep{AE10}, the $\mathrm{Pm}$ effect should disappear as $(\eta_O,\nu)\rightarrow 0$. An alternative viewpoint is to separate the small ``turbulent'' scales from the large scales where the dynamo mechanism is presumably lying. By carefully computing the energy exchange between the scales, one finds that the small scales act as an effective viscosity when $\mathrm{Pm}<1$ and tend to damp the large-scale mechanism \citep{RRC15}. This, however, is not fully satisfactory as it does not prove that the MRI dynamo is inexistent in the limit $\mathrm{Pm}\rightarrow 0$ and $\mathrm{Rm}\rightarrow \infty$.

Despite numerous efforts, recent results indicate that the $\mathrm{Pm}$ effect is still very much alive in simulations, up to $\mathrm{Rm}=5\times 10^4$ and $\Lambda_O=O(100)$ \citep{PB17}. This effect is still a very open question regarding the MRI saturation level in the `` quasi-ideal'' regime.

Note finally that the Pm effect is relevant only for the innermost parts of protoplanetary discs where $\mathrm{Rm}\gg 1$ is expected. For the regions above $1~\mathrm{AU}$, large-scale physics is dominated by magnetic diffusion and the $Pm$ effect is not relevant.

\subsection{Ohmic diffusion}
\index{MRI!Ohmic}
In non-stratified shearing box simulations with a net vertical flux, the presence of at least one MRI unstable mode is given by (\ref{eq:ohmic_stability_limi}). Using the expression for $\bmean$, and assuming that $k_{z,\mathrm{min}}=2\pi/L_z$, the stability condition of a non-stratified shearing box reads
\begin{align}
\label{eq:MRI_stability_ohmic_nsbox}
\mathrm{Rm}>\frac{\bmean}{\sqrt{\frac{3}{2\pi^2}\bmean-4}}\quad\rightarrow\quad \textrm{instability}.
\end{align}
The impact of a strong Ohmic diffusion on the non-linear saturation of the MRI was first explored by \add{\cite{SIM98} in 2D} and \cite{FS00} in 3D. In the case with a mean field, it is found that when $\Lambda_O\lesssim 1$ (or equivalently $Rm\lesssim \bmean$) and (\ref{eq:MRI_stability_ohmic_nsbox}) is verified so that the system is MRI-unstable, MRI turbulence is affected: $\alpha$ becomes lower than in the ideal case, and turbulence becomes intermittent, with periods of linear growth followed by rapid decay due to reconnection events. As in the low-$\beta$ ideal case, it is unclear whether this regime can be modelled with a constant $\alpha$ coefficient.

In the zero mean-field case, \cite{FS00} found that the MRI was disappearing below a critical Reynolds number $10^3<\mathrm{Rm}_\mathrm{c}< 10^4$. Whether $\mathrm{Rm}_\mathrm{c}$ depends on resolution, as the saturation level of the MRI dynamo does, is an open question.

\begin{figure}
\begin{center}
    \includegraphics[width=0.70\linewidth]{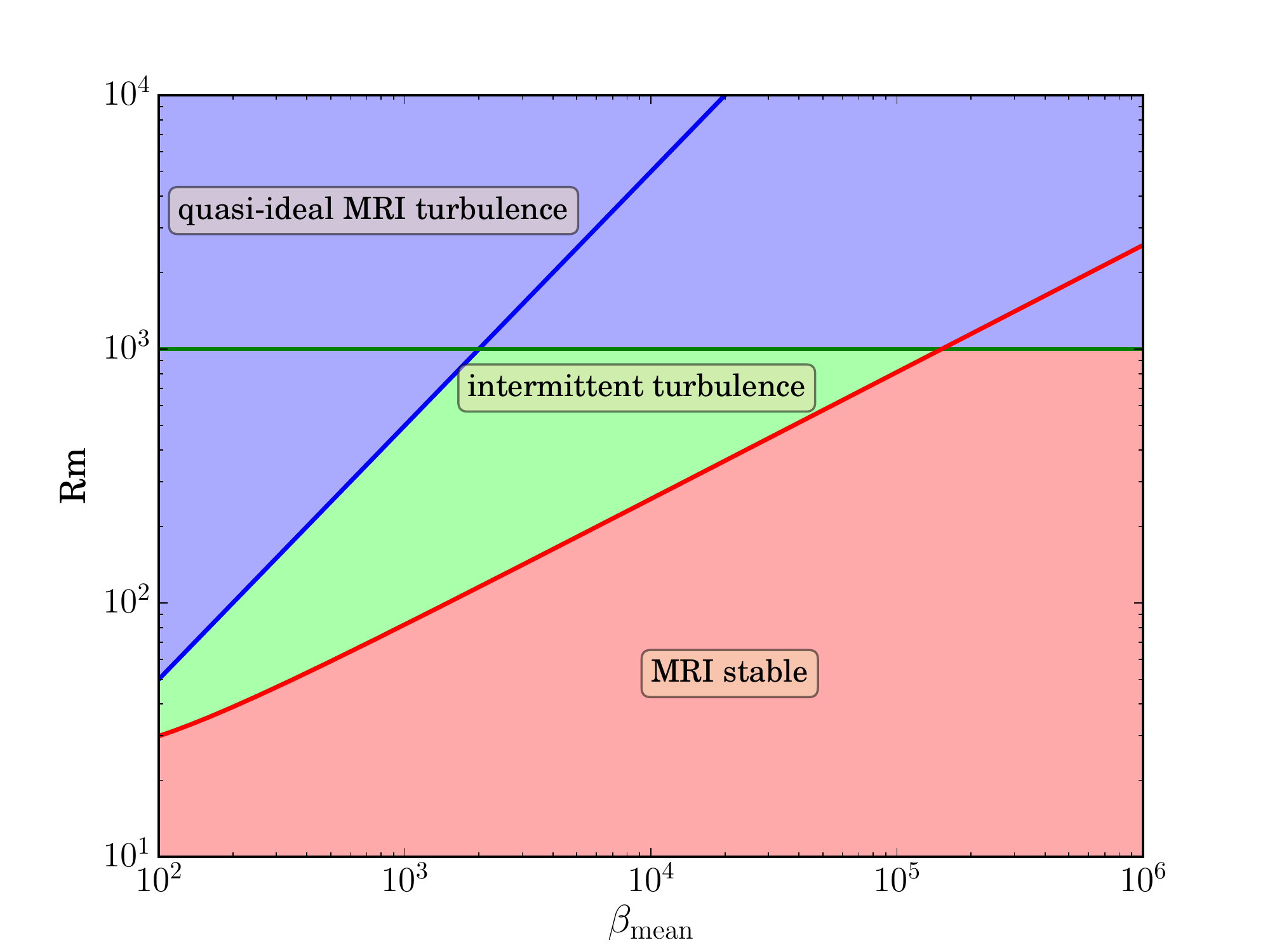}
    \caption{MRI turbulence regions as a function of Rm and $\bmean$ in the Ohmic diffusion case. The red line corresponds to the linear stability criterion (\ref{eq:MRI_stability_ohmic_nsbox}), the blue line to the limit $\Lambda_O=1$ and the green line to the zero net field limit $\mathrm{Rm}_\mathrm{c}=10^3$.}
    \label{fig:mri_sb_unstrat_ohm}
\end{center}
\end{figure}

One can combine these results into a global map showing how turbulence saturates in simulations with and without a net flux ( Fig.~\ref{fig:mri_sb_unstrat_ohm}). We have assumed that in the limit $\bmean \rightarrow \infty$, the zero net flux subcritical threshold was to be considered instead of the linear stability limit. As a result, the region $\bmean>10^5$ follows the zero mean field criterion, while $\bmean<10^5$ shows a transition region between fully developed turbulence with $\Lambda_O>1$ and a linearly stable flow, which we have named intermittent turbulence, in reference to \cite{FS00}. Since the data is sparse, we do not have any clear estimate of the values $\alpha$ in the intermittent turbulence region.

\subsection{Ambipolar diffusion}
\label{sec:ad_saturation}
\index{MRI!Ambipolar}
The role played by ambipolar diffusion on the saturation level of the MRI was first studied by \cite{HS98} in the two-fluid limit. They found that MRI turbulence was unaffected by ambipolar diffusion for $\Lambda_A\gtrsim 100$. Because protoplanetary discs are strongly collisional, the two-fluid approach is not very efficient since numerical time-steps are limited by the collision timescale. For this reason, this problem was revisited using the single fluid approach by \cite{BS11c}. They found that MRI is progressively suppressed by ambipolar diffusion as $\Lambda_A$ decreases.

Starting from \cite{BS11c} results, it is possible to create a simple phenomenology for MRI saturation under the impact of ambipolar diffusion. First, let us recall from (\ref{eq:ad_stability_limit}) that the linear stability criterion with $k_{z,\mathrm{min}}=2\pi/L_z$  is given by
\begin{align*}
\Lambda_A&>\Lambda_{A,\mathrm{crit}}\frac{\Va}{\Vaz}\quad\rightarrow\quad \mathrm{instability},\\
\mathrm{with}\quad \Lambda_{A,\mathrm{crit}}&\equiv \frac{1}{\sqrt{\frac{3}{8\pi^2}\bmean-1}}.
\end{align*}
The linear stability of a box with a pure vertical field is therefore very similar to the Ohmic case. However, as the instability grows, $\langle \Va\rangle$ increases which eventually leads to the violation of the stability criterion above. This self-suppression effect of MRI-turbulence allows us to deduce the saturation level by assuming that at saturation
\begin{align*}
\Lambda_A=    \Lambda_{A,\mathrm{crit}}\Bigg(\frac{\Va}{\Vaz}\Bigg)_{\mathrm{sat}}.
\end{align*}
The ratio of Alfv\'en speeds can be evaluated by $\langle \beta\rangle$:
\begin{align*}
\Bigg(\frac{\Va}{\Vaz}\Bigg)_{\mathrm{sat}}\simeq\Bigg(\frac{\bmean}{\langle \beta\rangle}+1\Bigg)^{\delta}.
\end{align*}
In principle, one would naively expect $\delta=0.5$. However, we find that this estimate does not fit the numerical results of \cite{BS11c}. This is because the large scale field $\Va$ is not necessarily proportional to the instantaneous $\langle \beta\rangle$ computed from the fluctuations at all scales. Instead, we therefore choose $\delta=1$ which presents a better correlation to the available data. Noting that, similarly to the ideal case of \cite{HGB95}, \cite{BS11c} found that $\langle \beta\rangle \simeq (2\alpha)^{-1}$, we get a simple estimate for $\alpha$ by combining the expression above:
\begin{align}
\label{eq:alpha_ambi_estimate}
\alpha\simeq\frac{1}{2\bmean}\Bigg(\frac{\Lambda_A}{    \Lambda_{A,\mathrm{crit}}}-1\Bigg)
\end{align}
\begin{figure}
\begin{center}
    \includegraphics[width=0.70\linewidth]{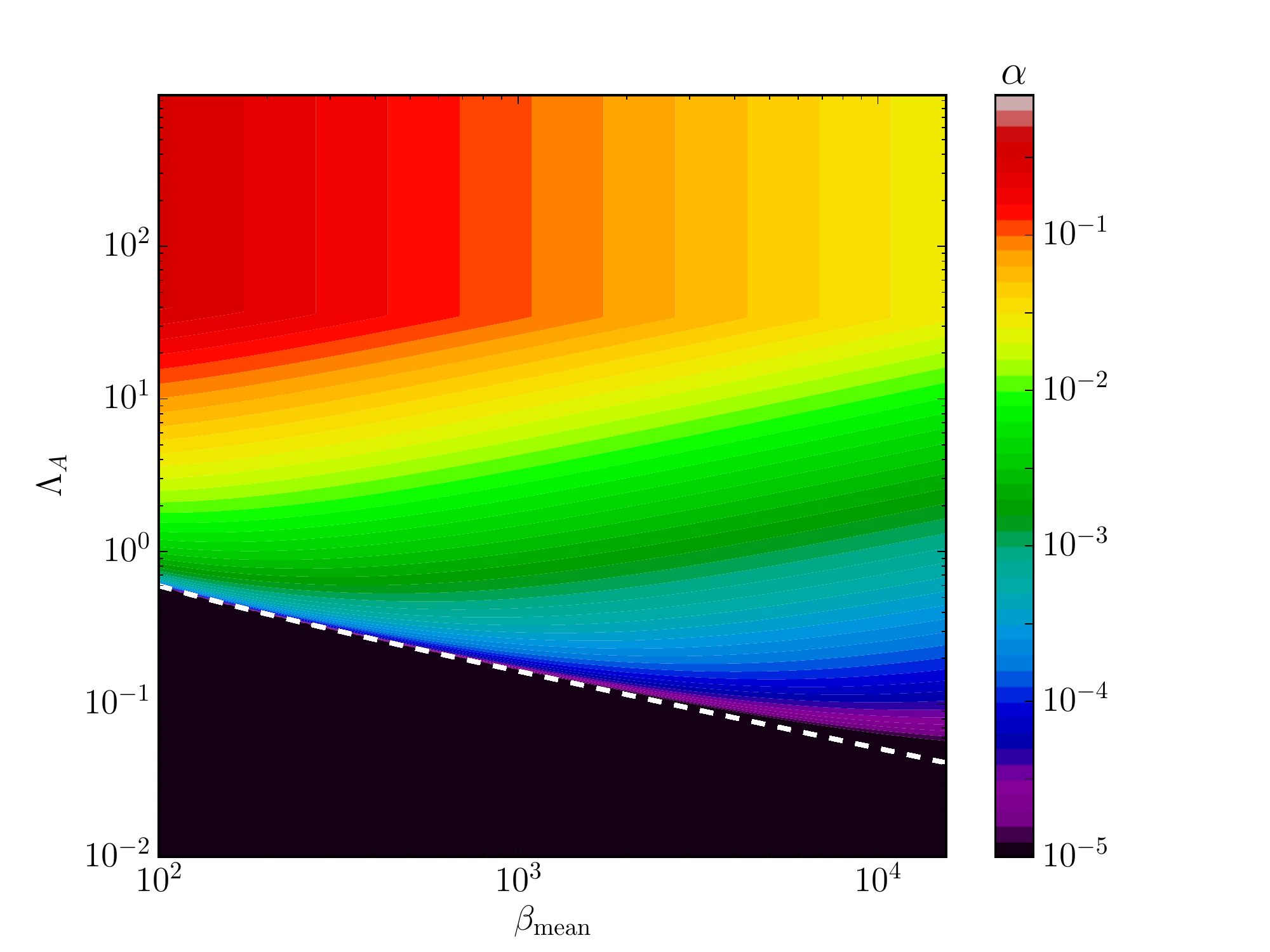}
    \caption{MRI turbulent transport deduced from (\ref{eq:alpha_ambi_estimate}) and (\ref{eq:alpha_ideal_estimate}) in the ambipolar-dominated regime with a mean vertical field. These estimates match the numerical values of \cite{BS11c} at $\pm 50\%$. The white dashed line corresponds to the marginal stability limit $\Lambda_A=\Lambda_{A,\mathrm{crit}}$. The region below this line has $\alpha=0$ in the net vertical field case, but can reach $\alpha\sim 10^{-4}$ when a mean toroidal field component is introduced, thanks to the presence of unstable oblique modes (see text).}
    \label{fig:mri_sb_unstrat_ambi}
\end{center}
\end{figure}

Of course, this estimates diverges as $\Lambda_A\rightarrow\infty$ since the saturation mechanism due to ambipolar diffusion becomes inexistent. In this case, the ideal-MHD estimate (\ref{eq:alpha_ideal_estimate}) should be used instead. A sample of predicted $\alpha$ from this saturation estimation is given in Fig.~\ref{fig:mri_sb_unstrat_ambi}. These estimates are within 50\% of the calculated value with a pure vertical field of \cite{BS11c} and can be used as a proxy to estimate the transport in ambipolar-dominated discs. This estimate also successfully recovers the ideal regime when $\Lambda_A\gtrsim 50$, as reported by numerical simulations.

In the case of a pure azimuthal field, \cite{BS11c} have shown that no turbulence is sustained below $\Lambda_A\lesssim 3$, and $\alpha$ progressively drops to 0 from the ideal MHD value at $\Lambda_A=100$.

Finally, in the case of a mixed mean vertical and azimuthal field, the subsistance of oblique modes (see \S\ref{sec:lin_ambipolar}) even at $\Lambda_A\lesssim 0.1$ creates a weak transport with $\alpha\sim 3\times 10^{-4}$ for $\Vay\sim \Vaz$ \citep{BS11c}. Above $\Lambda_A\sim 1$, the scaling (\ref{eq:alpha_ambi_estimate}) is approximately recovered as oblique modes become unimportant.

\subsection{Hall effect\label{sec:sb_unstrat_hall}}
\begin{figure}
\begin{center}
    \includegraphics[width=0.70\linewidth]{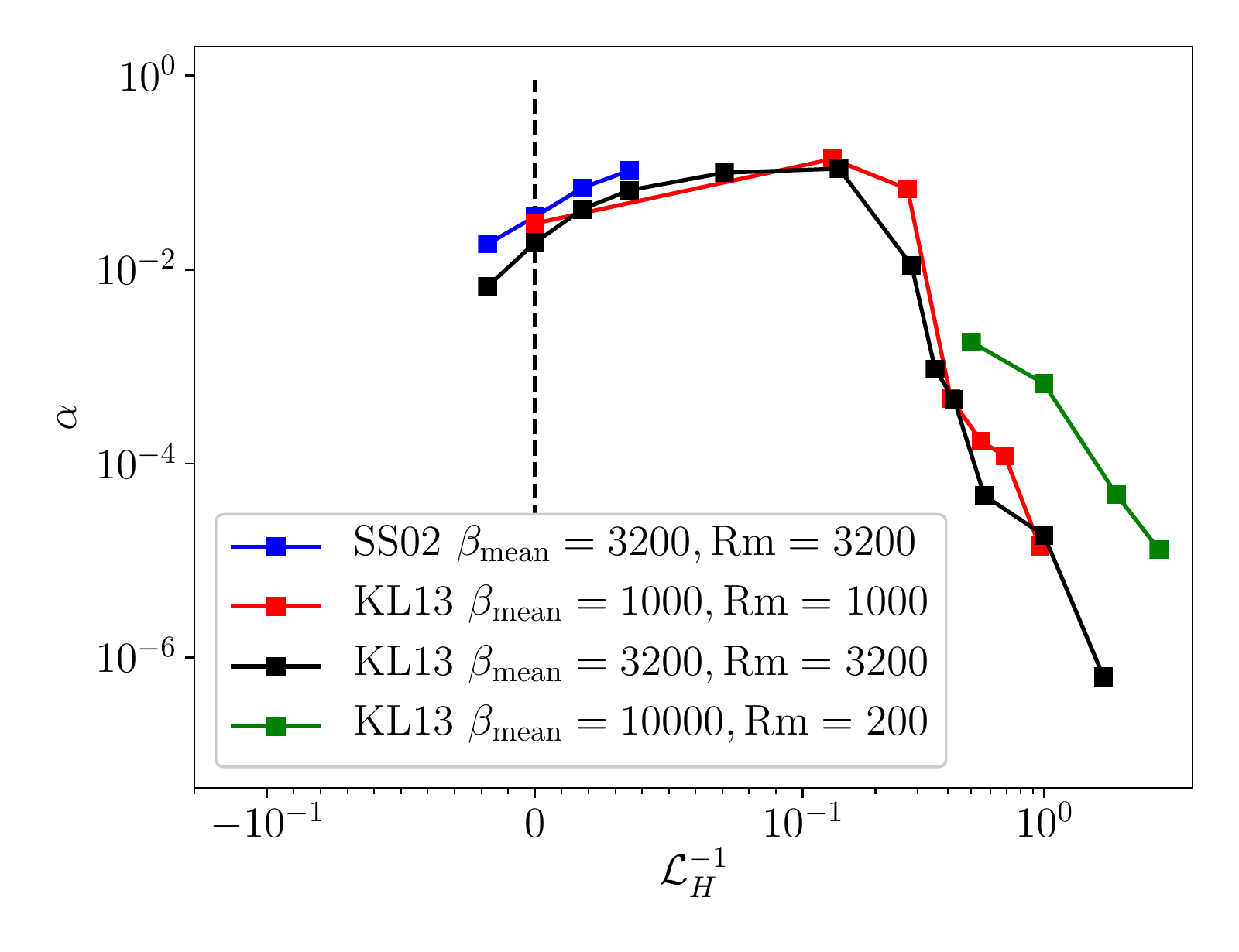}
    \caption{Evolution of the turbulent transport as a function of the intensity of the Hall effect in the mean vertical field case. Data from \cite{SS02} (SS02) and \cite{KL13} (KL13). $\mathcal{L}_H<0$ corresponds to anti-aligned field configuration. Note that the KL13 $\bmean=10^4$ case is linearly stable for $\mathcal{L}_H^{-1}=0$ because of Ohmic diffusion, and exemplify the reactivation of the linear MRI under the action of Hall (see \S\ref{sec:lin_hall_revival}). \add{Note that the $\alpha$ values from \cite{SS02} have been renormalised to match the definition of $\alpha$ in \cite{KL13}.}}
    \label{fig:mri_sb_unstrat_hall}
\end{center}
\end{figure}

\index{MRI!Hall}
The impact of the Hall effect on the saturation level of MRI turbulence was first explored by \cite{SS02} with a relatively weak Hall effect ($\mathcal{L}_H\gtrsim 20$). They found that with a mean vertical field, $\alpha$ increases with decreasing $\mathcal{L}_H$ in the aligned case, while $\alpha$  decreases in the anti-aligned case. However, protoplanetary discs are likely to have lower $\mathcal{L}_H$ than the ones studied by \cite{SS02} (typically $\mathcal{L}_H\sim 1$ as in Fig.~\ref{fig:partii:ni_effects}). For this reason, this problem was revisited by \cite{KL13} with simulations in the $\mathcal{L}_H\sim 1$ regime. In the case with a mean vertical field aligned with the rotation axis, it is found that despite being violently unstable from the linear point of view due to the HSI, the flow settles down into a quasi-laminar state for $\mathcal{L}_H\lesssim 5$, with negligible turbulent transport (Fig.~\ref{fig:mri_sb_unstrat_hall}). This ``low transport state'' is characterised by a self-organised flow where the vertical field is concentrated in a narrow region in the $x$ direction (which can be identified as a ring in global geometry). The mechanism behind self-organisation in Hall-MHD is detailed in \S\ref{sec:self-organisation}

\begin{figure}
\begin{center}
    \includegraphics[width=0.70\linewidth]{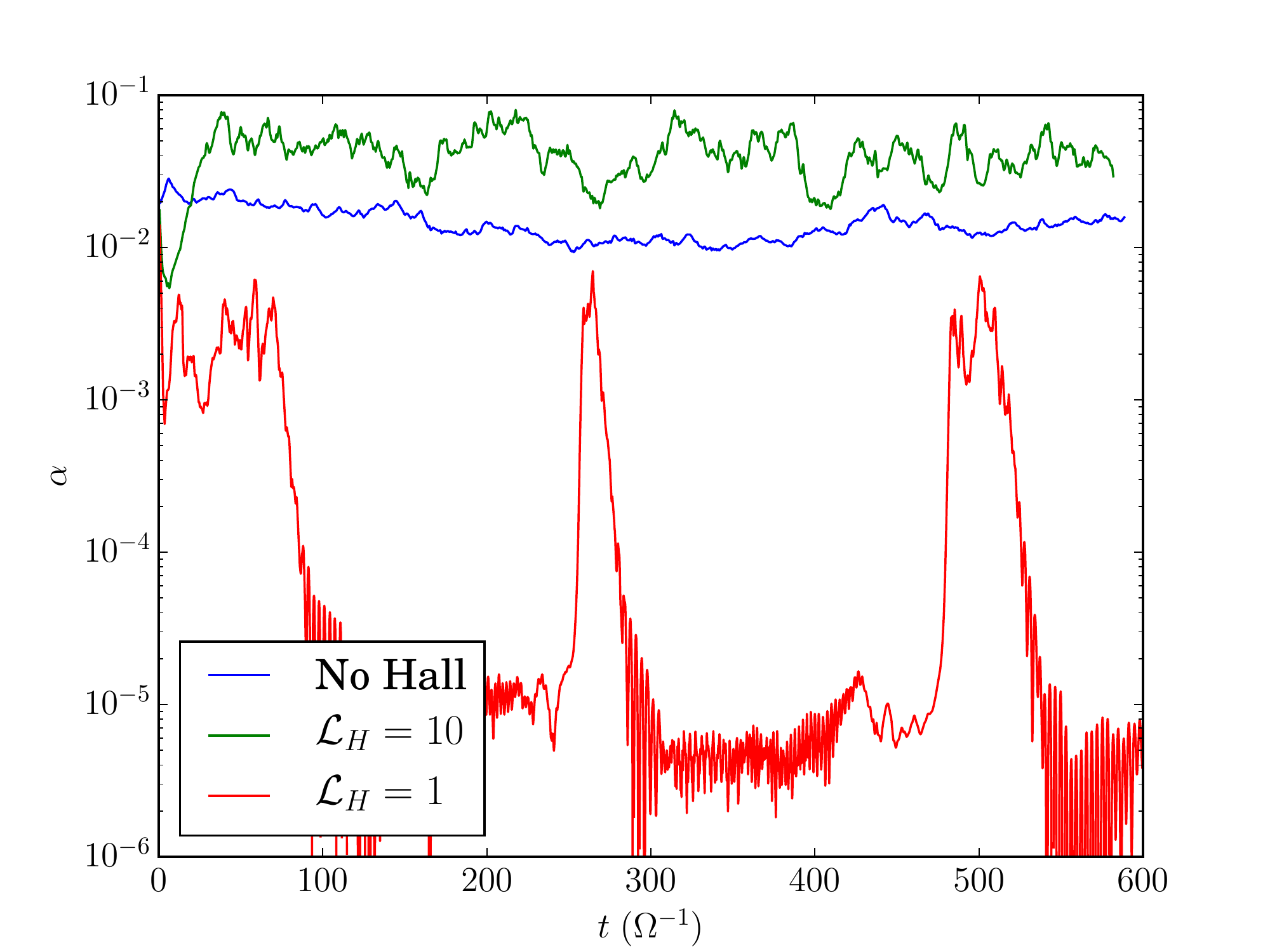}
    \caption{Turbulent transport $\alpha$ as a function of time in zero net flux simulations including the Hall effect. For weak Hall effect ($\mathcal{L}_H\gtrsim 10$), $\alpha$ is larger than in the case without Hall effect by a factor $\sim 3$. When Hall gets stronger, the system gets into the low transport state ($\mathcal{L}_H\lesssim 5$). In contrary to the net field case, it periodically switches back to a high transport state, resulting in bursts of $\alpha$. }
    \label{fig:mri_sb_unstrat_hall_ZNF}
\end{center}
\end{figure}

In the zero net-flux case, \cite{SS02} found evidence that $\mathrm{Rm}_c$ decrease from around $10^4$ in the ideal MHD regime possibly down to a few times $10^3$ for $\mathcal{L}_H\simeq 20$. At the same time, $\alpha$ increases by a factor of a few compared to the ideal case. For stronger Hall effects ($\mathcal{L}_H\lesssim 5$), a low transport state similar to the case with net flux is observed. However, in this case, the system switches back to a turbulent state periodically, resulting in short bursts of turbulence in the system (Fig.~\ref{fig:mri_sb_unstrat_hall_ZNF}). 

Overall, it is found that despite a powerful large-scale instability, the MRI in the Hall-dominated regime does not result into an efficient turbulent transport of angular momentum in unstratified boxes. \add{This surprising result, however, does not hold in stratified boxes, where the Hall effect effectively leads to an enhanced radial stress (see \S\ref{sec:nimhdstrat} ). Hence, the low transport state of \cite{KL13} is really a peculiarity of the unstratified setup. }

\subsection{Self-organisation\label{sec:self-organisation}}
\index{Self-organisation!Unstratified models}
Self-organisation is a process by which a disorganised (i.e. turbulent) flow creates large-scale and long-lived structures. There are several examples of self-organisation in nature, the most well-known one being probably the great red spot of Jupiter, resulting from small-scale turbulent motions which cascade to large scales forming a giant anticyclone. Self-organisation is a spontaneous symmetry breaking process: the system starts from a statistically homogenous state and ends up in a heterogeneous state with well-identified structures. As such, self-organisation should be distinguished from local instabilities such as Kelvin-Helmholtz or the Rossby wave instability (RWI, \citealt{LL99}) which result from a special location in the flow (e.g. vortensity extremum).

 Although self-organisation was clearly pointed out as a key phenomenon in Hall-MHD by \cite{KL13}, this phenomenon (or a weaker version of it) has been observed in ideal MHD simulations of MRI turbulence by several authors since early 2000. It is also a very promising mechanism to explain some of the structures observed in the sub-millimetric range (\S\ref{obs:struct}). Let us overview the different mechanisms which have been proposed to explain this phenomenon.
 
 \subsubsection{Ideal MHD}
 \cite{H01} and \cite{SP02} were among the first to notice the formation of ``ring-like'' structures in MRI simulations with net vertical flux. The simulations are in these cases semi-global: vertical stratification is neglected while the radial curvature is retained, leading to a cylindrical setup. It is found that the net vertical flux is trapped in a low-density region, forming a gap. In these gaps, $\alpha$ can reach values as high as 1, consistently with the fact that these gaps correspond to low $\bmean$ regions. \cite{H01} proposed that this could be the signature of a viscous instability: a local density minimum results in a local decrease of $\bmean$ (assuming the mean field is kept at its initial value). Since $\alpha\propto \bmean^{-1/2}$ (see \ref{eq:alpha_ideal_estimate}), $\alpha$ increases in this region, which removes mass from the region because of angular momentum conservation. Unfortunately, this proposition has never been investigated further, leaving the origin of self-organisation in these simulations unexplained.
 
\index{Turbulent resistivity}
\label{sec:self-org-ideal-uns}
 The same phenomenon was reported in local shearing box models with a mean vertical field by \cite{BS14}. They proposed that the non-diagonal components of the turbulent resistivity tensor \citep{LL09b} could be at the origin of the effect by acting as an ``effective negative diffusivity''. This explanation is however dubious since several authors have measured the turbulent resistivity tensor of MRI turbulence \citep{GG09,LL09b,FS09} and found that the effective resistivity was \emph{always positive}. Strikingly, \cite{BS14} simulations clearly show that despite having a globally concentrated field $B_z(x)$, the turbulent electromotive force $\bm{\mathcal{E}}=\langle \bm{w\times B}\rangle$ does not depend on $x$ \citep[][fig.~3]{BS14}. Hence, the turbulent resistivity prescription $\bm{\mathcal{E}}=\eta_\mathrm{turb}\bm{\nabla\times }\langle \bm{B}\rangle$ likely breaks down altogether and should be replaced with a more elaborated closure scheme.

In simulations without net flux, \cite{FN05} reported the spontaneous formation of giant anticyclones, though this could be a boundary condition artefact (Fromang, private communication). \cite{JYK09} also reported the formation of ``pressure bumps'' which are caused by large-scale fluctuations of $\alpha$. In contrast to the simulations with a mean field vertical field, the features observed in the zero net field case are transient and only survive for a limited time, which depends on the box size. \cite{JYK09} proposed a model based on a stochastic $\alpha$ which predicts long-lived axisymmetric structures in quasi-geostrophic equilibrium, as observed in their simulations.

\subsubsection{Hall-MHD}
\begin{figure}
\begin{center}
    \includegraphics[width=0.70\linewidth]{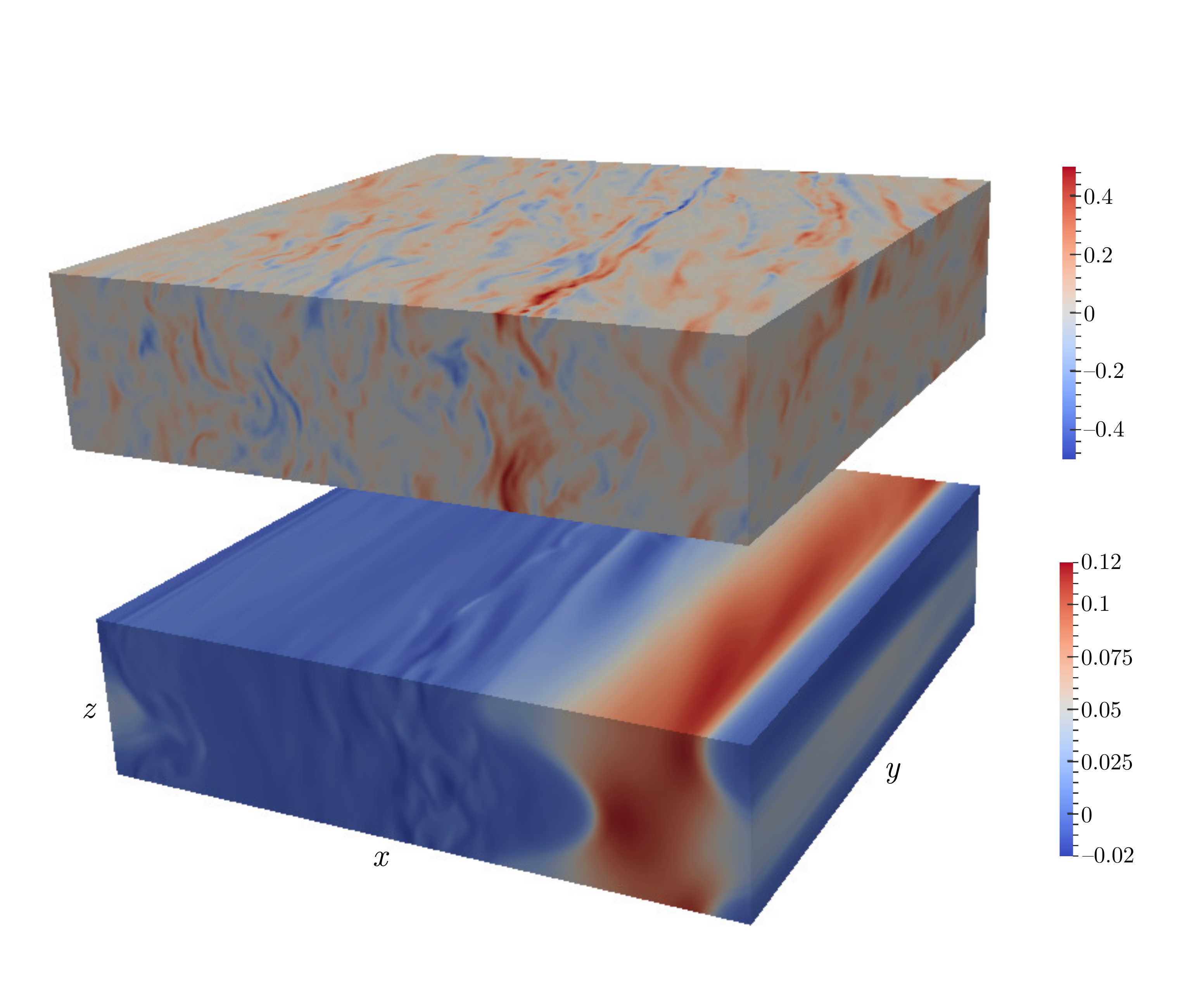}
    \caption{Vertical field component in snapshots of MRI turbulence. Top: ideal-MHD simulation with $\bmean=3200$, bottom: Hall-MRI simulation with $\bmean=3200$ and $\mathcal{L}_H=1.75$. Figure from \cite{KL13}. }
    \label{fig:mri_sb_unstrat_hall_organised}
\end{center}
\end{figure}

 \index{Self-organisation!Hall MHD}
The first mention of self-organisation in Hall-MHD appears in \cite{KL13}, where self-organisation has a dramatic impact on the saturation level of Hall-dominated MRI (see \S\ref{sec:sb_unstrat_hall}). Self-organisation appears in an obvious manner by looking at the vertical field component of the flow (Fig.~\ref{fig:mri_sb_unstrat_hall_organised}). While in the ideal-MHD case, self-organisation appears as a ``second order'' effect on top of MRI turbulence, in the Hall-MHD case, self-organisation is the main saturation mechanism of the MRI. In other words, turbulence is mostly suppressed by self-organisation. Self-organisation shows up when $\mathcal{L}_H \lesssim 5$, and $\alpha$ essentially vanishes as a result of the lack of turbulence (Fig.~\ref{fig:mri_sb_unstrat_hall}). This surprising result also holds in the case with zero-net flux or in the mixed case having both a mean azimuthal and vertical magnetic flux.

\begin{figure}
\begin{center}
    \includegraphics[width=0.85\linewidth]{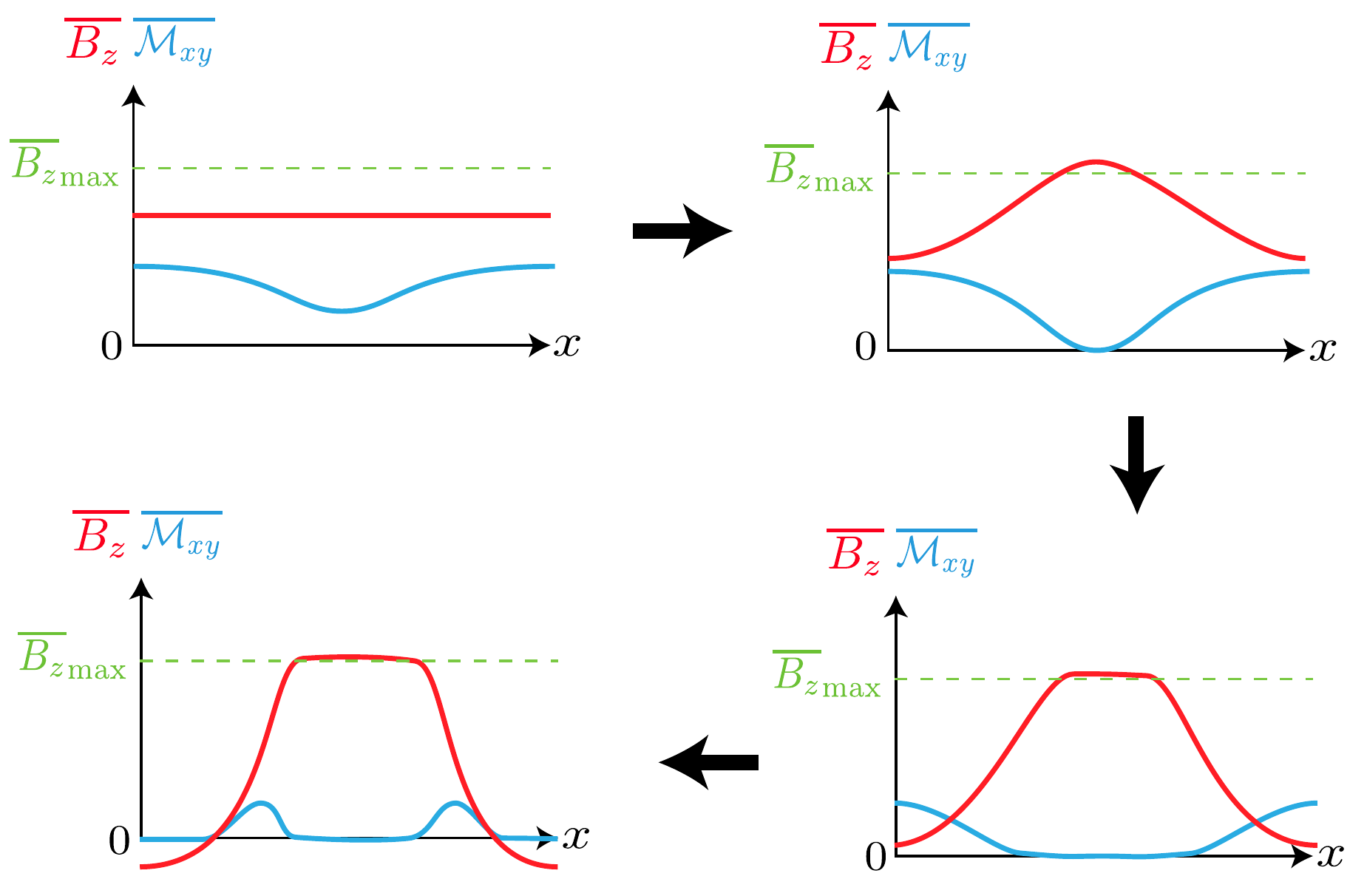}
    \caption{Hall self-organisation phenomenology. We start from a local fluctuation of the stress $\mathcal{M}_{xy}$ (top left). This minimum creates a local maximum of $\overline{B_z}$ which overshoot the maximum $\overline{B_z}$ allowed by the HSI. Because the flow becomes locally stable, the stress vanishes (top right). On the boundaries of this stable region, the Maxwell stress still transport magnetic flux towards the stable region, making it larger and emptying the rest of the domain (bottom right). At some point, the total flux in the rest of the domain becomes negative, and it becomes HSI-stable. The stress therefore vanishes in this region as well, leaving only the interface with a minimal stress (bottom left). Figure inspired by \cite{KL13}. }
    \label{fig:Hall-SO}
\end{center}
\end{figure}

The origin of Hall-driven self-organisation can be tracked down to the induction equation in the presence of Hall effect. Indeed, When the Hall length is constant, the induction equation reads
\begin{align}
\partial_t\bm{B}=\bm{\nabla\times}(\bm{w\times B})+\lH\bm{\nabla\times}\Big(\bm{\nabla \cdot}\frac{-\bm{BB}}{4\pi}\Big)
\end{align}
which highlights the role of the Maxwell stress in the induction equation. Guided by numerical simulations which show the appearance of a vertical magnetic field with variations in the $x$ direction, let us define an average
\begin{align*}
\overline{Q}=\iint \,\mathrm{d}y\,\mathrm{d}z\, Q    
\end{align*}
so that the induction equation for the ``mean'' vertical field reads
\begin{align}
\partial_t\overline{B_z}=\partial_x \Big(\overline{w_z B_x-w_x B_z}\Big )+\lH    \partial_x^2 \frac{-\overline{B_xB_y}}{4\pi}
\end{align}
where we recognised the radial Maxwell stress term $\mathcal{M}_{xy}=-B_xB_y/4\pi$, also present in the angular momentum conservation equation (\ref{eq:angular_final}). This demonstrates that in Hall-MHD, the transport of magnetic flux is tightly linked to the the transport of mass. 
Due to energetic constraints, $\mathcal{M}_{xy} >0$ (see \S\ref{sec:conservation}) in shear-driven instabilities/turbulence. Therefore, a concentration of magnetic field due to the Hall effect is possible at local stress \emph{minimum}. In a turbulent flow, short-lived stress minima occur randomly in the flow, and these minima tend to accumulate vertical magnetic flux according to the equation above. When the Hall effect is strong enough, a local minimum can accumulate enough flux to become stable for the HSI. In this case, the flow becomes stable and the local turbulent stress vanishes, becoming a permanent minimum. This minimum continues to accumulate magnetic flux thanks to the remaining stress present on both sides until the flux outside of the minimum of stress becomes negative. At this point, the stress also vanishes in the regions $\overline{B_z}<0$ and the systems settles down into a quasi-stationary state with very low-stress level (see also Fig.~\ref{fig:Hall-SO}).

Although this process was initially identified in unstratified shearing box simulations, it was later unambiguously identified in cylindrical unstratified simulations \citep{BLF16}. However, vertically stratified simulations do not seem to exhibit this process, for reasons not yet identified to date \citep{LKF14}.

\subsubsection{Ambipolar diffusion}

\begin{figure}
\begin{center}
    \includegraphics[width=0.85\linewidth]{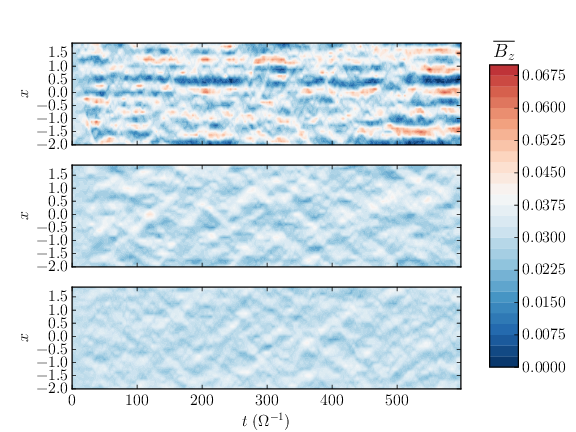}
    \caption{$y-z$ average $B_z$ as a function of time for non stratified MRI simulations with $\Lambda_A=3$ and $\bmean=1000$. Top: $L_x=L_y=4H$, middle: $L_x=4H,\ L_y=8H$, bottom $L_x=4H, L_y=16H$. Note the disparition of zonal flows when $L_y>L_x$. All three simulations have been computed with a fixed resolution per scale height in the 3 spatial directions $n_{x,y,z}=64\ \mathrm{pts}/H$ using the Snoopy code \citep{LL07}. }
    \label{fig:MRI_sb_unstrat_Zonal_ambi}
\end{center}
\end{figure}

Self-organisation due to ambipolar diffusion was first mentioned by \cite{BS14} in unstratified shearing-boxes. It was also observed in cylindrical simulations including ambipolar diffusion by \cite{BLF16}, albeit at a very low level. However, whether ambipolar diffusion plays an active role in the self-organisation mechanism is an open question. Clearly, the ''strength'' of self-organisation (quantified by the ratio $B_{z\mathrm{max}}/B_{z\mathrm{min}}$) in the ideal-MHD case is larger than the ambipolar diffusion case by a factor $10$ (Fig.~2 in \citealt{BS14}). In addition, the mechanism proposed for self-organisation only involves ideal-MHD terms \citep{BS14}. Finally, the existence of zonal flows in this configuration (non-stratified, ambipolar diffusion dominated) largely depends on the box aspect ratio. \cite{BS14} largely explored the situation with $L_x=L_y=4 H$. However, a choice of box with $L_y>L_x$ tends to break zonal flows (Fig.~\ref{fig:MRI_sb_unstrat_Zonal_ambi}). Overall, it is very possible that self-organisation in ambipolar-dominated unstratified shearing boxes is a mere numerical artefact.

\section{Stratified shearing boxes}
Stratified models have been mostly explored in the zero mean field configuration. When a mean vertical field is included, simulations lead to ``high magnetic pressures that disrupt the vertical structure of the disk before the flow makes the transition to MHD turbulence" \citep[quoting][]{SHGB96}. Although the situation is less dramatic when more adapted boundary conditions are used (see below), this statement explains the lack of simulations with a net vertical field until recently. The situation with a mean toroidal field is less interesting since the toroidal component can usually escape through the vertical boundaries, eventually leading to a situation identical to that without a mean field.

As in the non-stratified case, we define the plasma $\beta$ parameter of the mean vertical field threading the disc
\index{$\bmean$ plasma beta!stratified systems}
\begin{align*}
    \bmean\equiv\frac{8\pi \langle P\rangle_{z=0} }{\langle B_z\rangle_{z=0} ^2}.
\end{align*}
where averages are taken in the disc midplane. 
In the following, we use a definition of $\alpha$ using the box averaged pressure, i.e.
 \begin{align*}
\alpha\equiv \frac{1}{\langle P\rangle}\Bigg\langle \rho w_xw_y-\frac{B_xB_y}{4\pi}\Bigg\rangle,    
\end{align*}
which does not depend on the vertical extension of the simulation domain (provided that all of the significant stress is contained in the box). However, some authors such as Stone and coworkers usually defines $\alpha$ from the vertically-averaged stress and the midplane pressure $P_0\equiv P(z=0)$, which leads to predicted $\alpha$ values 2-3 times smaller than the ones obtained with the definition above \citep{DSP10}. Note also that Stone's definition yields $\alpha$ which decreases as the box size increases. Indeed, the stress being concentrated in the region $z\lesssim 2H$, the vertically averaged stress decreases as the box size increases above $2H$. These differences in the definition of $\alpha$ should be kept in mind when comparing results from different groups.
 
\subsection{Zero mean field}
\subsubsection{Ideal MHD}
\begin{figure}
\begin{center}
    \includegraphics[width=1.15\linewidth]{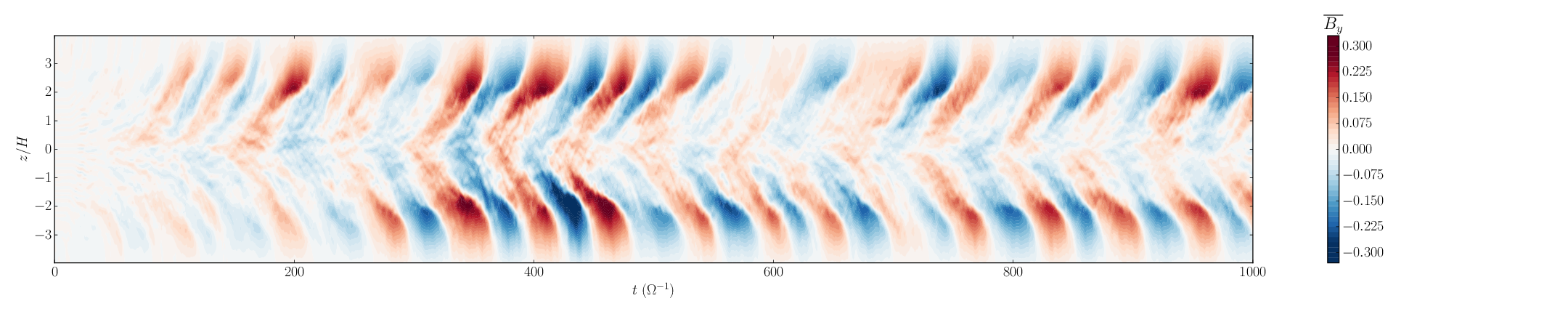}
    \caption{$x-y$ average of $B_y$ ($\overline{B_y}$) as a function of time and $z$ for zero vertical net flux stratified MRI simulations. Note the presence of a butterfly pattern indicating a periodic reversal of the the toroidal field followed by its rise away from the disc midplane.}
    \label{fig:MRI_sb_strat_Butterfly}
\end{center}
\end{figure}

\index{Butterfly diagram}
The zero mean field stratified case was first explored by \cite{BNS95} and \cite{SHGB96}. They first noted the spontaneous appearance of a ``butterfly'' diagram when looking at the space-time behaviour of the $x-y$ average toroidal magnetic field (Fig.~\ref{fig:MRI_sb_strat_Butterfly}, see also \citealt{DSP10} and \citealt{SBA12}). This diagram shows quasi-periodic flip of the toroidal field, with a periodicity close to 10 local orbital periods. Whether this butterfly diagram leads to observational counterpart by modulating the turbulent transport is still debated (e.g. \citealt{HR16} for an example). On a theoretical side, it is well reproduced by mean field models \citep{G10,GP15}, but we're still lacking a first principle theory for this dynamo. 

In principle, once vertical stratification is included, one needs to specify both the vertical density and temperature profile as well as an equation of state. However, most of the literature has focused on isothermal simulations, in which the temperature is constant. In this case, it is found that 
\begin{align}
    \alpha\simeq 0.02\pm 0.01\quad\quad\textrm{(stratified, isothermal, zero vertical flux)}
\end{align}
\citep[e.g.][]{SHGB96,DSP10}.  In the outer region of protoplanetary discs, radiative transfer calculations tell us that the vertical temperature profile is approximately constant, so the isothermal approximation is probably a good model in this case.

Nevertheless, some authors have explored non-isothermal models, such as Hirose and collaborators \citep{HKS06,HBK09,HBK14}, Flaig and collaborators \citep{FKK10,FRK12} and Bodo and collaborators \citep{BCM13,BCM15}. It is found that when the vertical profile becomes unstable for convection (i.e. it violates the Schwarzschild criterion), the turbulent transport of angular momentum can increase by up to an order of magnitude \citep{HBK14}, thanks to a mechanism which is yet to be fully elucidated.

\index{MRI!Numerical convergence}
Following the numerical convergence issue pointed out by \cite{FP07} in unstratified simulations (see \S\ref{sec:ideal_unstrat}), numerical convergence in stratified setups has also been tested. Initial explorations were limited by numerical resources to 128 points per scale height \citep{DSP10} and showed the convergence of $\alpha$ as a function of the number of grid points, which led to the conclusion that ``stratification saves the day''. However, this issue was tackled again with higher resolution simulations: 200 points per scale height \citep{BCMR14} and 256 points per scale height \citep{RGFK17}. These recent results show a weak dependence on the resolution with $\alpha\propto N^{-1/3}$ at the resolution explored, albeit with a different vertical boundary condition (\citealt{DSP10} used periodic boundary conditions while \citealt{RGFK17} used outflow boundary conditions). Overall, these results indicate that \emph{numerical convergence is also an issue in stratified models}.

\subsubsection{Non-ideal MHD}

Realistic ohmic diffusion profiles were introduced in zero net flux stratified simulations for the first time by \cite{FS03}. These simulations exemplified \cite{G96} layered accretion model with a mid-plane dead-zone and an active surface layer. The debate then crystallised on the thickness of the active layer, which, depending on the model, could lead (or not) to accretion rates compatible with observations.

This layered accretion paradigm, however, omits ambipolar diffusion and the Hall effect. Ambipolar diffusion is particularly important at low densities, right in the active layers of \cite{G96}. It was quickly realised that ambipolar diffusion could dramatically reduce the strength of MRI turbulence in this layer, and even suppress it, since $\Lambda_A\lesssim 1$ in this region \citep{PC11b,PC11a,DTHK13}. This is confirmed by direct numerical simulations including Ohmic and ambipolar diffusion but no Hall effect \citep{BS13,SBS13}. These simulations suggest dramatically low values for $\alpha$, with $\alpha\simeq 3\times 10^{-6}$ at 1 AU \citep{BS13} to $\alpha\simeq 10^{-3}$ at 100 AU \citep{SBS13}.

These results imply accretion rates $\dot{M}\lesssim 10^{-10}\ M_\odot/\mathrm{yr}$ in the region $1-30$ au, which are too low by at least 2 orders of magnitude compared to observational constraints. However, the zero mean field configuration is rather artificial. Indeed, protoplanetary discs are expected to form from a collapsing cloud which has dragged some fraction of its initial magnetic flux. Therefore, a non-negligible poloidal flux is likely to be present in these objects, which motivated the need for simulations including a mean poloidal field.

\subsection{Mean field \& outflows\label{sec:meanfield_outflows}}
\subsubsection{Ideal MHD}
The first viable shearing-box simulation of a vertically stratified disc with a mean vertical field was performed by \cite{SI09} and \cite{SI10}, thanks to more robust numerical techniques compared to the initial attempts of \cite{SHGB96} and carefully designed vertical boundary conditions. Their simulations have a relatively weak field $\bmean\gtrsim 10^4$ but show significant deviations from the zero mean field scenario, and in particular, the presence of a strong outflow, quantified by the mass flux leaving the disc. In a box with $L_z=8H$, they find a vertical mass flux 100 times larger with $\bmean=10^4$ compared to the zero net field case \citep{SI09}.

\begin{figure}
\begin{center}
    \includegraphics[width=0.45\linewidth]{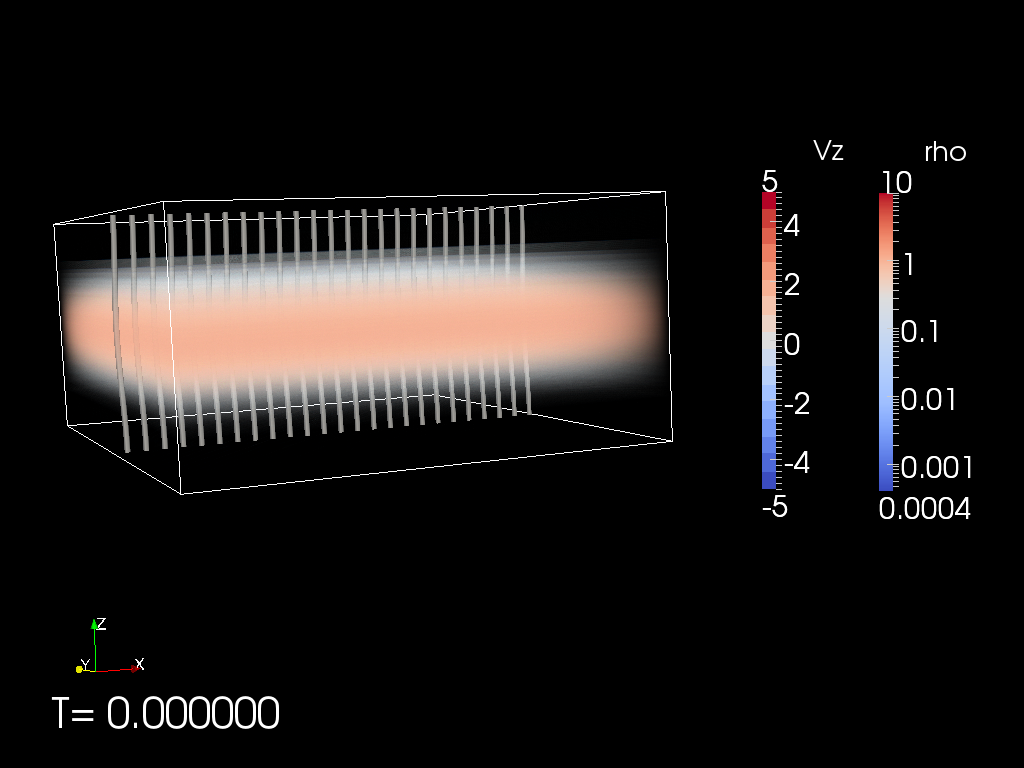}
    \includegraphics[width=0.45\linewidth]{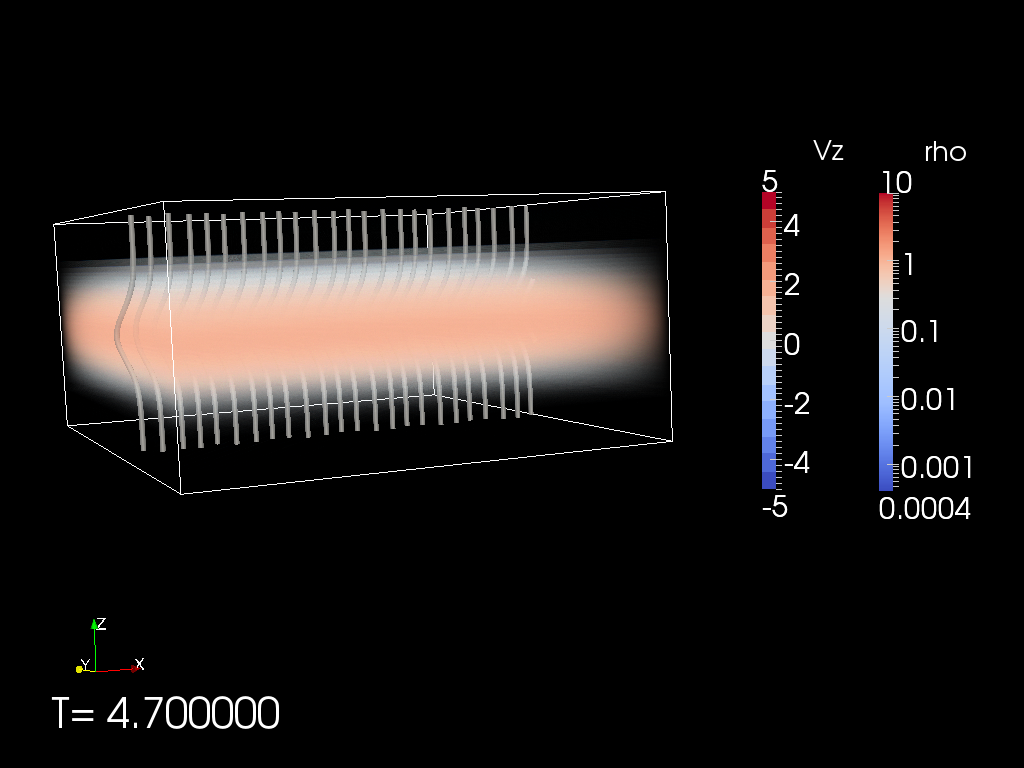}\\
    \includegraphics[width=0.45\linewidth]{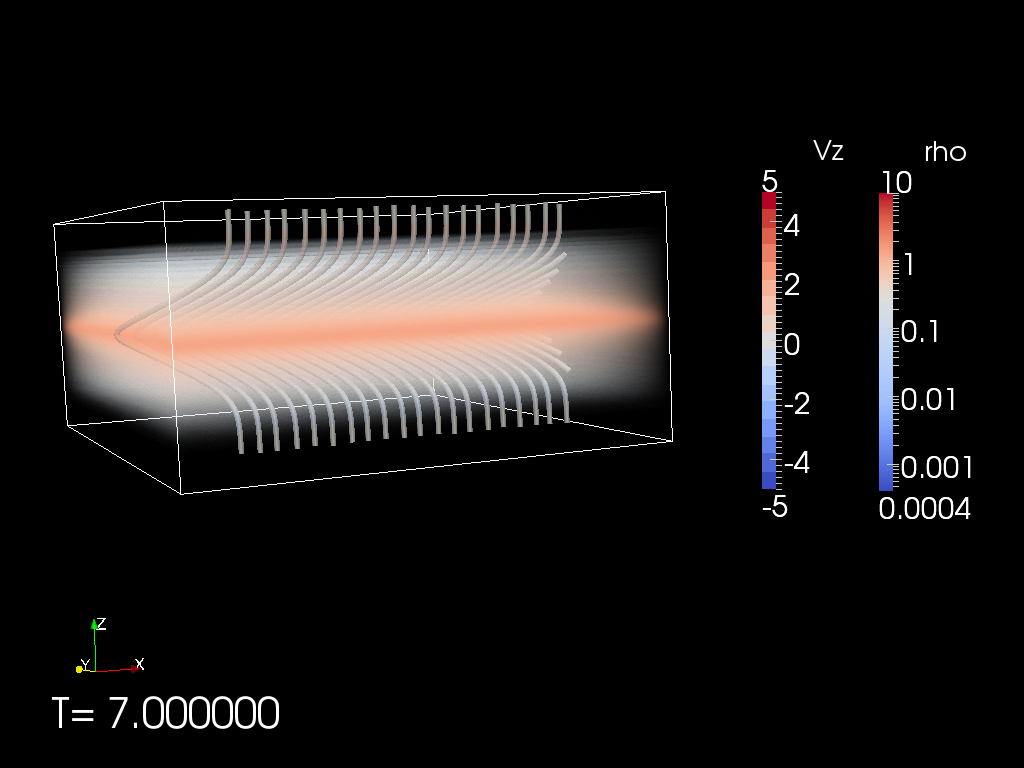}
    \includegraphics[width=0.45\linewidth]{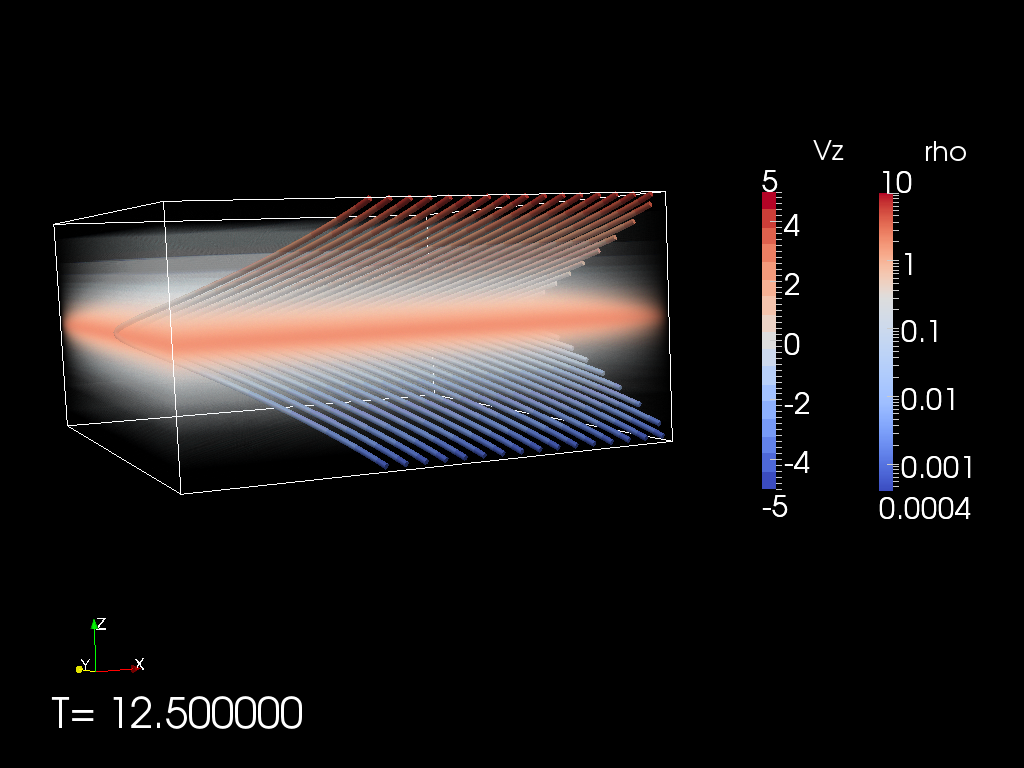}
    
    \caption{Growth and saturation of the MRI in the presence of a strong vertical field ($\bmean=10$) in ideal MHD. Tubes are magnetic field lines while gas density is represented with volume rendering. Note the initial growth of the linear mode in the disc midplane which eventually saturates into a quasi-laminar outflow configuration similar to \cite{BP82} paradigm. From \cite{LFO13}.  }
    \label{fig:mri_sb_outflow}
\end{center}
\end{figure}

\index{Magneto-centrifugal outflow}
This problem was revisited by a large number of authors, both in the strong field limit $\bmean\simeq 1$ \citep[Fig.~\ref{fig:mri_sb_outflow}]{O12,M12,LFO13}, and in the intermediate regime $\bmean=10^3-10^4$ \citep{FLLO13,BS13b}. It was quickly demonstrated that these outflows were a variation of \cite{BP82} magneto-centrifugal paradigm \citep{LFO13,FLLO13} with a strong time-dependency. Despite this connection to well-known launching mechanisms, some properties of shearing box outflows are intrinsically flawed. 

First, the mass loss rate depends on the vertical box extension \citep{FLLO13}, taller boxes leading to lower mass loss rates. This surprising result is actually expected from the shape of the gravitational potential in a shearing box (\S\ref{sec:Hills}) which is unbounded when $z\rightarrow \infty$. Some authors have proposed to fix this issue by adding higher order terms to the vertical gravity force, making possible to escape at $z=\pm\infty$ with a finite amount of mechanical energy \citep{SI10,MP15}. This approach, however, violates the conservative nature of gravitation, since the resulting force does not derive from a potential anymore. A more rational approach would be to include all of the third order terms in the Hill's approximation. This however also introduces radial curvature terms, making the shear-periodic boundary conditions unadapted. 

\begin{figure}
\begin{center}
    \includegraphics[width=0.80\linewidth]{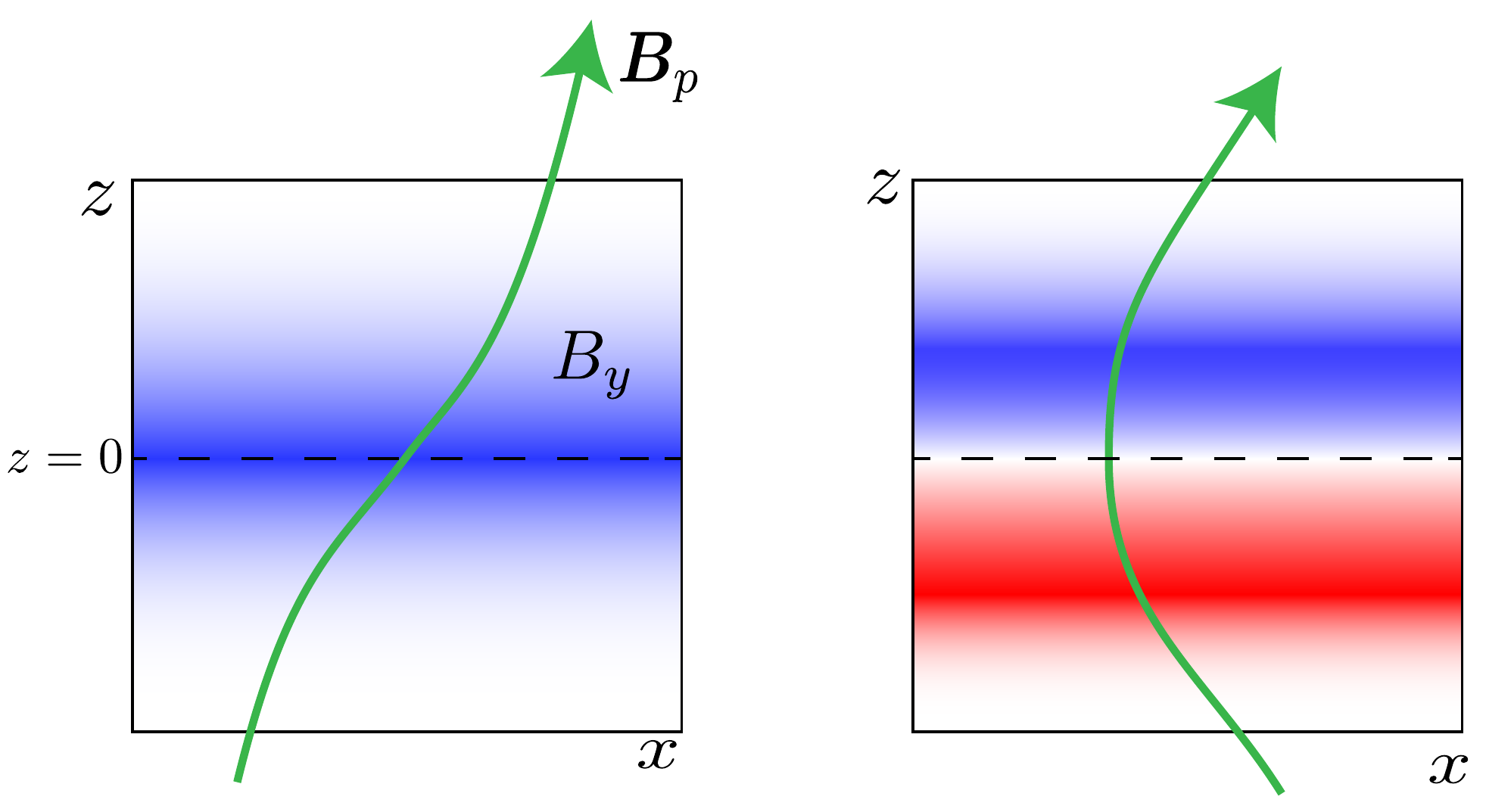}
    
    \caption{Symmetries of the outflow configuration allowed in a shearing box. Poloidal field lines are represented in green while the toroidal field component $B_y<0$ is shown in blue and $B_y>0$ in red. Left: odd symmetry configuration, right: even symmetry configuration. The usual \cite{BP82} picture corresponds to the even symmetry in a shearing box model.}
    \label{fig:Field_symmetry_outflow}
\end{center}
\end{figure}

Second, the geometry of the outflow is problematic. In principle, shearing boxes allow both vertically even and odd magnetic configuration (Fig.~\ref{fig:Field_symmetry_outflow}). At the linear level (see \S\ref{sec:linstrat}), the two symmetries show similar growth rates and properties. However, in the nonlinear regime, the odd symmetry is problematic for the physical interpretation of the outflow. By simply looking at the poloidal field topology (Fig.~\ref{fig:Field_symmetry_outflow}), it is clear that the odd symmetry connects the bottom side of the disc to the central star, while the top side is connected to $R\rightarrow +\infty$. Physically, it means that no angular momentum is actually extracted from the disc: angular momentum comes from an unknown source at $z\rightarrow -\infty$ and flows \emph{through} the disc up to $z\rightarrow +\infty$. The angular momentum conservation equation (\ref{eq:angular_final}) clearly describe this situation: in the odd symmetry case, one has $B_y B_z(z)=B_yB_z(-z)$, so the surface stresses on both sides of the disc cancel out and the disc is not accreting any material!

Naturally, this situation is rather unrealistic and the only physical configuration is the even symmetry one (or at least, a configuration where $B_y B_z$ have a different sign on both sides of the disc). However, it turns out that shearing boxes tend to settle into the odd configuration when $\bmean\lesssim 10^3$ \citep{SSAB16}. The reason for this unexpected result is still debated. It is certainly related to the fact that the shearing box has ``too many symmetries'', and does not differentiate $x\rightarrow \pm \infty$. Some authors have enforced the even symmetry manually \citep{LFO13} which leads to physical outflow configurations. However, this is not satisfactory as the flow symmetry should be enforced by the global geometry of the disc and its surrounding, which is not captured in shearing box models.

Despite these difficulties with outflows, it is possible to measure the turbulent stress in these simulation. A systematic exploration of shearing box models with $\bmean\in [10,\infty]$ \citep{SSAB16} shows that
\begin{align}
\label{eq:ideal_strat_alpha}
\alpha=10.1\, \bmean^{-0.53} \quad\textrm{for}\quad \bmean<10^5    
\end{align}
$\alpha$ recovering the zero net flux value when $\bmean>10^5$. This scaling leads to $\alpha$ values about 3 times larger than the unstratified estimate (\S\ref{sec:ideal_unstrat}). As in the unstratified case, the level of turbulent stress depends on the magnetic Prandtl number, $\alpha$ increasing with increasing Pm \citep{FLLO13}, so caution is still needed when using these scalings in phenomenological models.

The characterisation of winds in a shearing box is always a bit difficult because of the symmetry issues noted above. In addition, the usual MHD outflow invariants (\S\ref{sec:outflow_definitions}) are only defined in global geometry, making them inaccessible in local models (but see \citealt{LFO13} for local equivalents of the global invariants). Despite these difficulties, one usually defines an outflow rate $\zeta$ and a torque parameter $\upsilon$ which respectively measure the mass and angular momentum evacuated by the magnetised wind.

It is customary to define the outflow rate as
\begin{align}
\label{eq:zeta_definition}
\zeta\equiv \frac{\overline{\rho w_z}|_\mathrm{top}-\overline{\rho w_z}|_\mathrm{bottom}}{\rho_\mathrm{mid}c_s}
\end{align}
\index{$\zeta$ wind mass loss rate}
where "top" and "bottom" subscripts stand for the top and bottom of the disc (which in principle can be chosen freely) and $\overline{\quad}$ denotes a horizontal averaging procedure. In a real (global) disc, the local domain is emptied by the wind, but it is also replenished by the divergence of the accretion flow, so that a steady state is locally achieve. Since there is technically no accretion flow in shearing boxes, this replenishment is absent, leading to boxes which slowly loose mass. The typical timescale over which a box looses a significant fraction of it mass is given by $\tau_\mathrm{loss}\sim (\zeta\Omega)^{-1}$. This implies that the shearing box model is really valid in the limit $\zeta\ll 1$.

The outflow rate is known to depend on several key parameters. First, it should be noted that the outflow rate depends on the box extension, both horizontal and vertical. In the horizontal plane, it seems that convergence is reached for $L_x,L_y\gtrsim 4H$ \citep{FLLO13}. In the vertical direction however, as the box gets higher, the outflow rate decreases. \cite{FLLO13} and \cite{LFO13} show that doubling the vertical extension of the box divides $\zeta$ by a factor 3 for $\bmean=10^4$ and by a factor 1.6 for $\bmean=10$. In any case, this indicates that the outflow rate \emph{does not converge in shearing boxes}. This is because it is not possible to escape to infinity in the gravitational potential of the Hill's approximation (see discussion in \S\ref{sec:Hills}). Hence, the "border" of the gravitational potential is set artificially by the location of the physical boundary in the $z$ direction of the shearing box. 

Assuming $L_z\simeq 10H$, the combination of the data published by \cite{SI09}, \cite{BS13b} and \cite{FLLO13} leads to:
\begin{align}
\label{eq:ideal_strat_zeta}
\zeta\simeq 5\times 10^{-5}+\frac{10}{\bmean}\quad\mathrm{for}\quad	\beta\gtrsim 10^2
\end{align}

This relation gets shallower for $\bmean<10^2$ \citep{BS13b,LFO13}, and eventually leads to a sharp decrease of $\zeta$ when $\beta\lesssim 1$ (see Fig.9 in \citealt{LFO13}). 
It is also expected that the prefactor is a decreasing function of $L_z$ and could trace the geometrical aspect ratio $H/R$ of the disc, which is not captured in shearing boxes \citep{BS13b}. However, this dependence is complex since it likely depends on the magnetisation. It should also be pointed out that this relation is close to the one found in RMHD simulation of the MRI in the context of dwarf novae \citep{SLDF18}. Hence, the disc thermodynamics do not seem to greatly affect the outflow rate.
Nevertheless, this scaling should be taken with caution, and is probably only an upper bound to the real outflow rates one would obtain solving for the global problem.

The second parameter characterising the outflow is the angular momentum extracted by the wind, $\upsilon$, defined as:
\begin{align}
\label{eq:upsilon_definition}
\upsilon \equiv \frac{\overline{T_{yz}}|_\mathrm{top}	-\overline{T_{yz}}|_\mathrm{bottom}	}{P_\mathrm{mid}}
\end{align}
\index{$\upsilon$ wind torque parameter}
 where we have defined the stress tensor $T_{yz}=\rho w_y w_z-B_yB_z/4\pi$. This quantity directly enters the angular momentum conservation law (see \S\ref{sec:alpha-upsilon-disc}) so that, once it is known, one can automatically compute the accretion rate associated with the wind. Evaluating $\upsilon$ is however notoriously difficult in shearing boxes because its sign is not well-defined (see the symmetry discussion above). A naive temporal averaging therefore leads to a negligible value of the angular momentum extracted by the wind. In order to circumvent this difficulty, several strategies have been used: \cite{FLLO13} computed $\upsilon$ on a short time period, during which the polarity remains fixed, while \cite{BS13b} and \cite{SLDF18} computed the time-average absolute value of $\upsilon$. Another difficulty lies in the altitude at which this quantity is evaluated. \cite{BS13b} evaluated it at the box boundary, while \cite{SLDF18} computed its maximum as a function of $z$. In all these cases, the results behave like the scaling inspired from \cite{SLDF18}:
 \begin{align}
 \label{eq:ideal_strat_upsilon}
 \upsilon \simeq \frac{(4\pm 3)\times 10}{\bmean}\Bigg[\Big(\frac{\bmean}{4.7\times 10^4}\Big)^2+1\Bigg]^{0.3}	
 \end{align}
where the uncertainty is due to the different method of measurements found in the literature. As for $\zeta$, this scaling probably gets shallower for $\bmean<10^2$. However, this has never been properly evaluated in the literature.

It  should be noted that for $\bmean>1$, $\upsilon<\alpha$. Hence, the vertical stress is always weaker than the radial stress, by a factor $O(\bmean^{1/2})$. However, the respective contribution of these two terms to the mass accretion rate in the disc is proportional to $(R/H) (\upsilon/\alpha)$ (see \S\ref{sec:alpha-upsilon-disc}), implying that the vertical stress is the dominant mass accretion mechanism whenever $\bmean\lesssim(R/H)^2$ \citep{FLLO13,BS13b}.

\subsubsection{Non-ideal MHD}
\label{sec:nimhdstrat}

\begin{figure}
\begin{center}
    \includegraphics[width=1.0\linewidth]{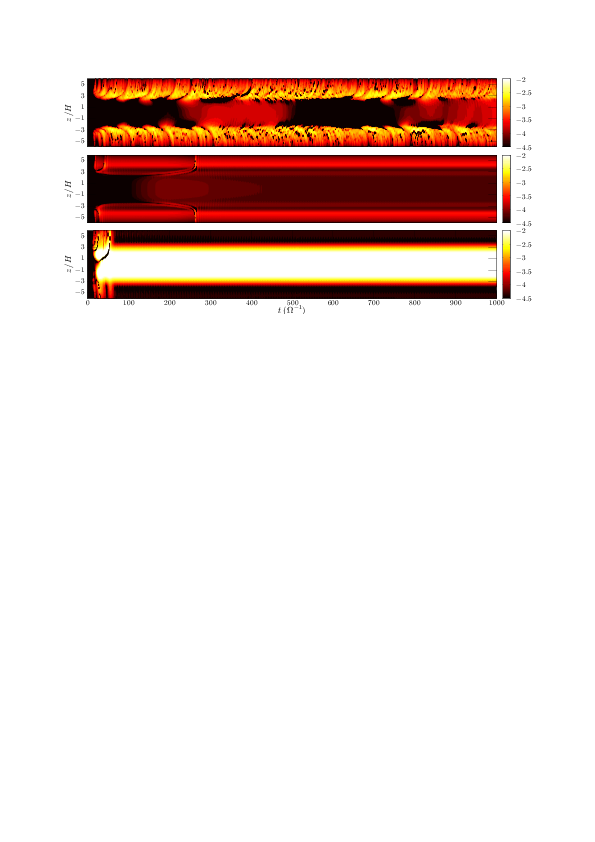}
    
    \caption{Maxwell stress $M_{xy}$ averaged horizontally as a function of $t$ and $z$ in simulations with Ohmic diffusion only (top), Ohmic and ambipolar diffusion (middle) and Ohmic, ambipolar and Hall effect (bottom). Simulations computed at 1 AU assuming a minimum mass solar nebula, with a mean vertical field $\bmean=10^5$. From \cite{LKF14}. }
    \label{fig:MRI_strat_ni}
\end{center}
\end{figure}

Simulations with a mean vertical field first focused on the impact of ohmic and ambipolar diffusion \citep{BS13,SBA13}, neglecting the Hall effect for technical reasons. It was found that the presence of a mean field allows the formation of a magnetised outflow at the disc surface ionised by far UV radiation, leading to accretion because of the angular momentum extracted by the outflow. Complete models including also the Hall effect \citep{LKF14,B14,SL15,B15} showed similar outflows, with in addition the presence of a midplane laminar stress due to the Hall effect \citep{LKF14}.

The difference between Ohmic only, Ohmic+Ambipolar, and Ohmic+ambipolar+Hall effect is demonstrated in Fig.~\ref{fig:MRI_strat_ni} for an MMSN disc at 1 AU. In the Ohmic only case, one recovers the turbulent surface layer and a dead midplane. Adding ambipolar diffusion leads to a different picture where the disc surface becomes mostly laminar, with a weakly magnetised outflow. This might be surprising at first sight since the linear analysis shows little difference between the Ohmic and Ohmic+Ambipolar cases regarding the localisation and growth rates of the eigenmodes (see Fig.~\ref{fig:mri_strat_OA}). The difference between Ohmic and Ambipolar cases is due to the fact that $\eta_A\propto B^2$ while $\eta_O$ does not depend on $B$. As eigenmodes grow, the horizontal field grows as well and $\eta_A$ increases. This rapidly leads to the saturation of the eigenmode by diffusion, in a way similar to \S\ref{sec:ad_saturation}. In the end, the perturbation never grows to the large enough amplitudes required to break up in developed turbulence, as in the Ohmic case.

When the Hall effect is eventually added, the disc midplane is subject to a \emph{laminar stress} in the region 1-10 AU, in addition to the weak surface outflow which subsists. The midplane stress is tightly linked to the field polarity: a negative mean field ($\bm{\Omega\cdot B}<0$) leads to its disappearance. This laminar stress is due to the HSI \citep{LKF14} which only shows up for positive polarities (see discussion in \S\ref{sec:lin_hall} and \S\ref{sec:Hall_lin_strat}). Hence, the Hall effect is indeed able to revive the MRI in dead zones provided that the mean field has a positive polarity, as already deduced from the linear theory. However, it saturates as a laminar stress. This effect is recovered up to $R\sim 30\,\mathrm{AU}$ but disappears at 100 AU in MMSN models \citep{SL15} since the Hall effect is much weaker in the outermost regions of the disc. The presence of this laminar stress suggests that the correlation lengths of the midplane structures are larger than the horizontal box size, which contradicts the spirit of the shearing box approximation and calls for global simulations to properly characterise this transport process.

The top/down symmetry in non-ideal MHD models is also problematic in simulations including non-ideal MHD effects. When $\bmean<10^5$, these simulations exhibit most of the time an odd symmetry with respect to the disc midplane, which implies that the outflow is not extracting any net angular momentum from the disc. There is today no clear explanation for this trend. It could be that the midplane current layer required by the odd symmetry configuration is expelled by the strong ohmic or ambipolar diffusion \citep{BS13}, or that the HSI spontaneously saturates in an even configuration \citep{LKF14}, or a combination of these effects.

\begin{figure}
\begin{center}
    \includegraphics[width=0.49\linewidth]{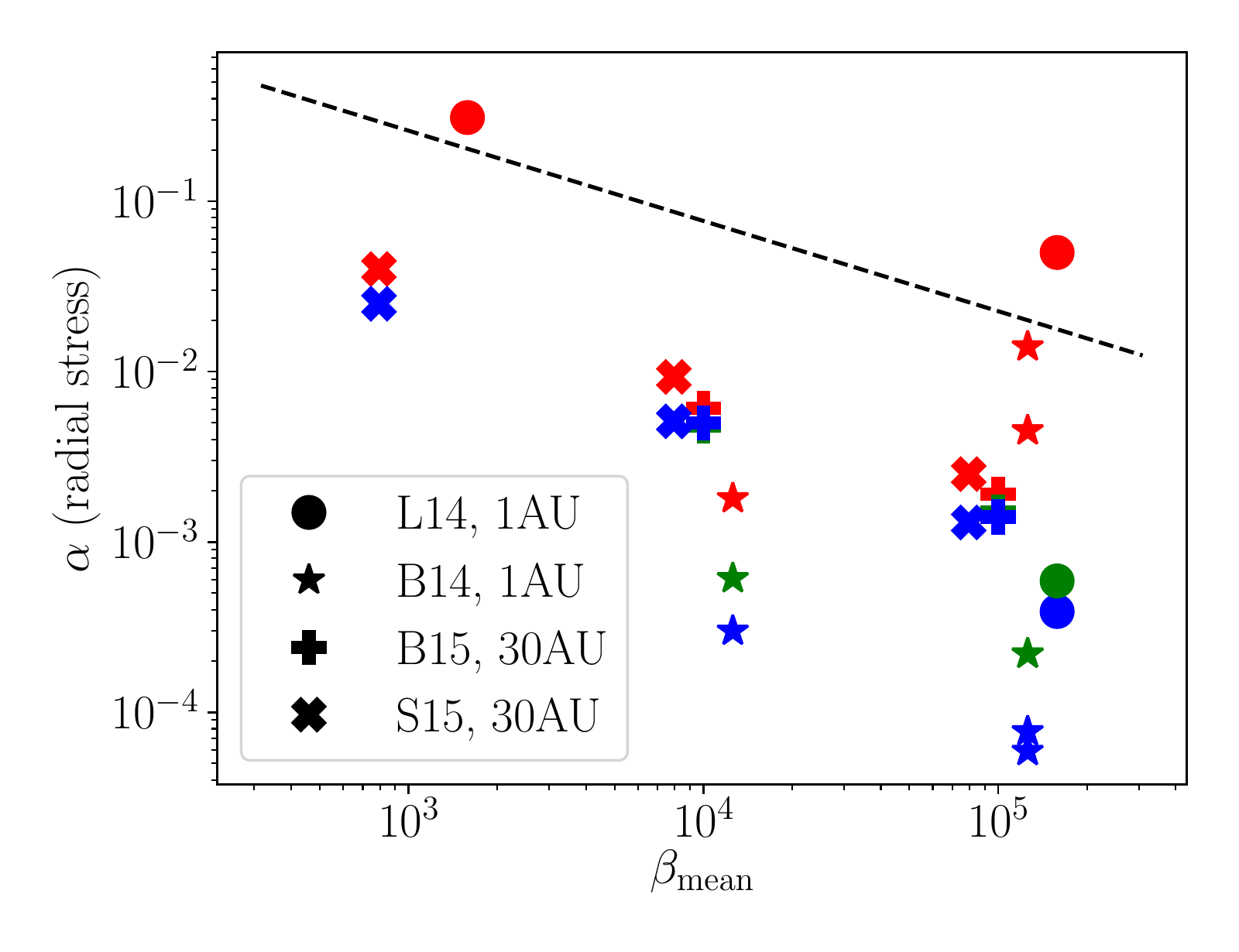}
    \includegraphics[width=0.49\linewidth]{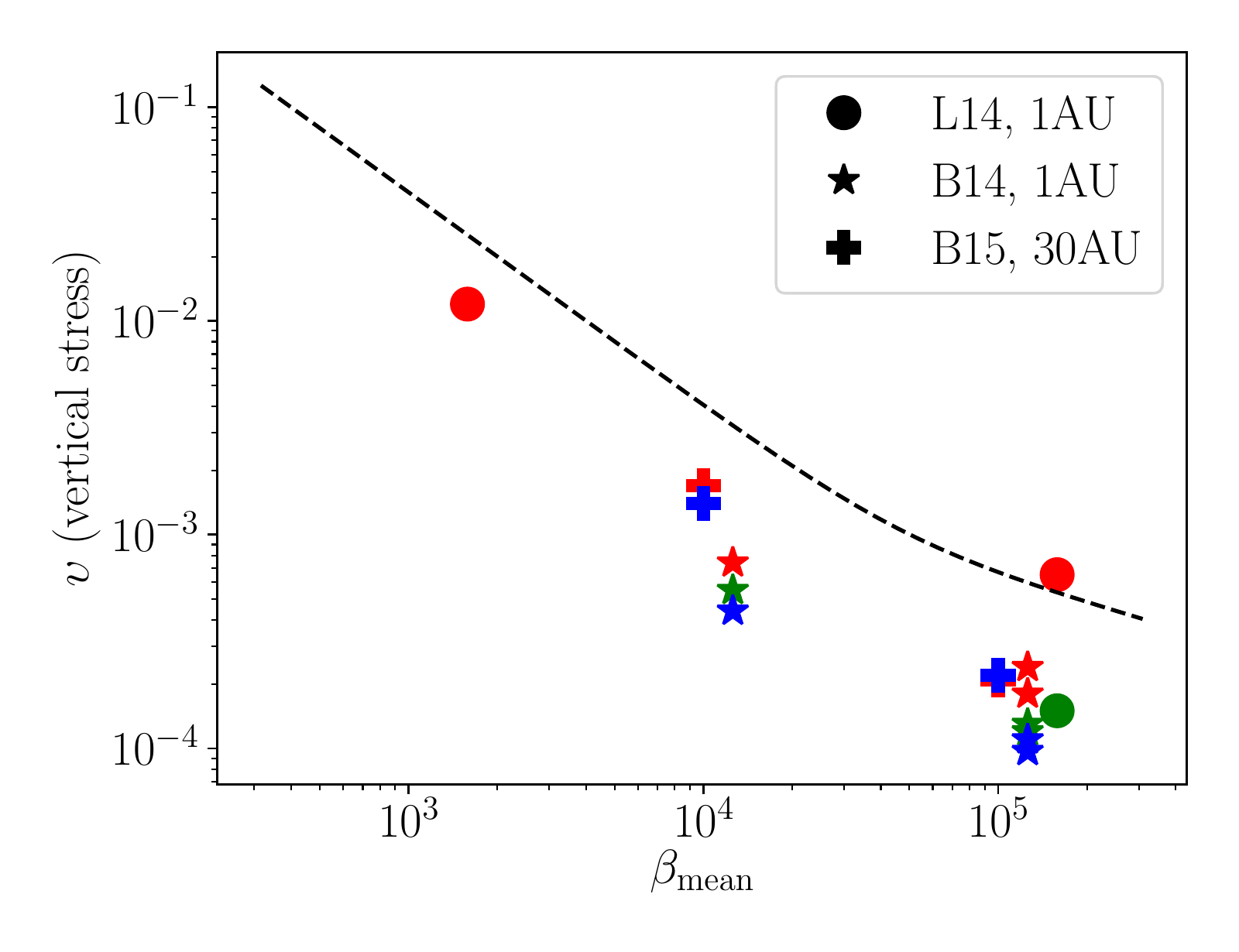}
    \includegraphics[width=0.49\linewidth]{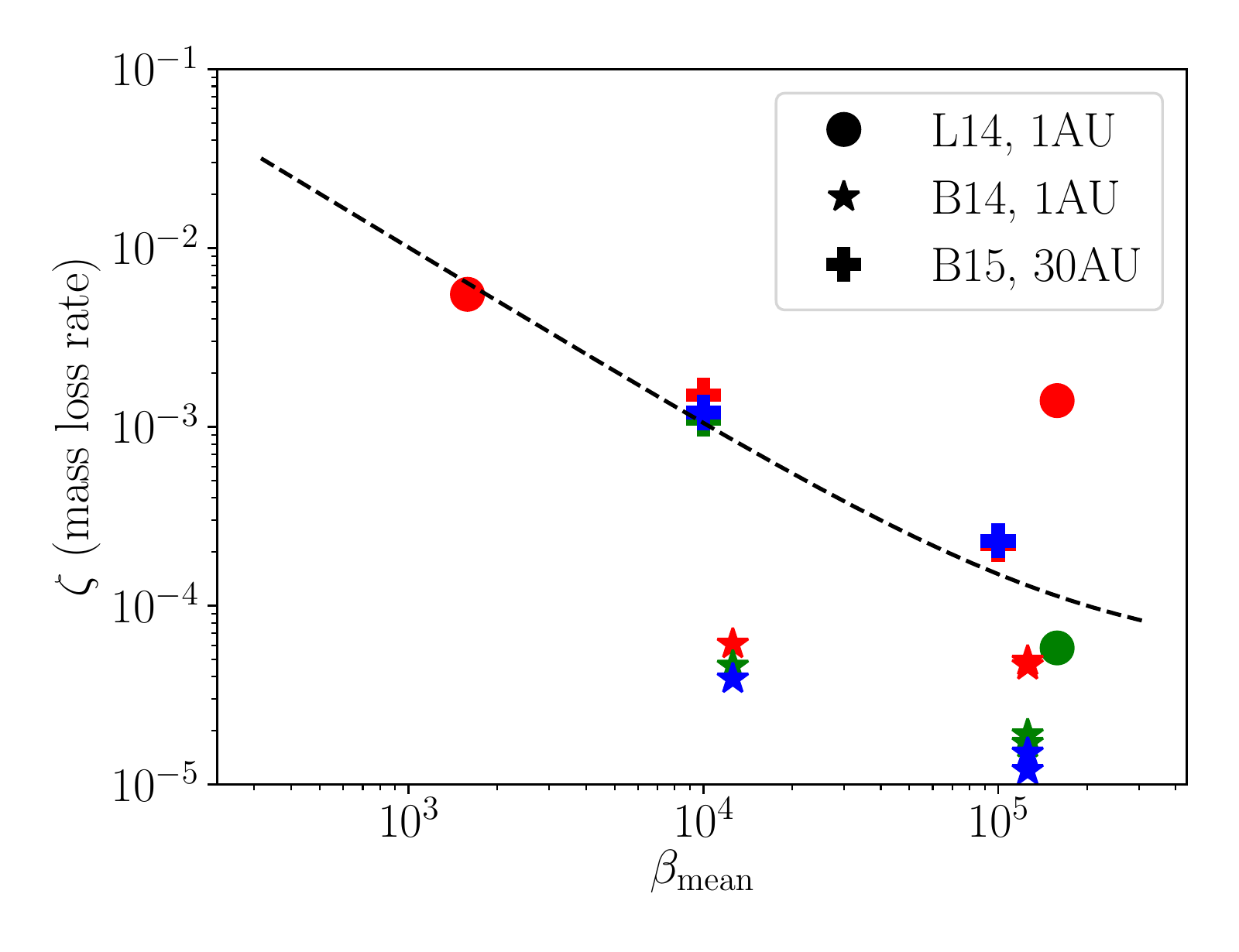}
    
    \caption{Measurements of transport coefficients in the literature, assuming ionisation structures at 1 AU and at 30 AU. The ideal MHD relations (\ref{eq:ideal_strat_alpha}), (\ref{eq:ideal_strat_zeta}) and (\ref{eq:ideal_strat_upsilon}) are shown in black dashed lines. We have used data from \cite{LKF14}=L14, \cite{B14}=B14, \cite{B15}=B15 and \cite{SL15}=S15. Simulation with the vertical field aligned with the rotation axis are in red, simulations with vertical field anti-aligned are in blue, and simulations without Hall effect are in green. Note that \cite{B14} has two chemical models at 1 AU, with and without grains, hence the presence of two sets of points. Note that points on the same $\bmean$ have been slightly shifted horizontally to improve readability.}
    \label{fig:NI_shearinbox_stats}
\end{center}
\end{figure}

Using similar technics to the one used in ideal-MHD shearing box, several groups have managed to measure the transport coefficient in non-ideal MHD simulations, computing the vertical stress on one side, or computing its absolute value. A representative summary of the results is given in Fig.~\ref{fig:NI_shearinbox_stats}

Several trends can be seen in this figure. First and foremost, all of the transport coefficients are reduced by the non-ideal MHD effects (with the notable exception of one run from \citealt{LKF14} with a very strong Hall effect at 1 AU). The most reduced coefficient is the radial angular momentum transport $\alpha$, while the vertical (wind) transport $\upsilon$ appears to be the less affected. This difference in behaviour is the main reason why winds are today favoured in protoplanetary discs: \emph{ the efficiency of wind-driven transport is less affected by non-ideal MHD effects than radial angular momentum transport.} In addition to these remarks, one note that the sensitivity to the field polarity is more pronounced at 1 AU than at 30 AU, as expected from the profile of dimensionless Hall number (Fig.~\ref{fig:partii:ni_effects}). As already guessed in the discussion above, aligned field tend to have a larger $\alpha$, $\zeta$ and $\upsilon$. It is interesting to also notice that, for simulations at 30 AU, the scaling with $\bmean$ is similar to the one found in ideal MHD, apart from a constant offset.

\subsubsection{Outflow induced self-organisation}
\index{Self-organisation!Stratified shearing box}

There are several evidence of self-organisation in stratified shearing boxes. The first piece of evidence can be seen in the space-time diagrams of \cite{SA14}. \cite{B15} then explicitly reported the spontaneous formation of self-organised flows, which is particularly strong for $\bmean\lesssim 10^4$. The same process was later identified by \cite{SB18} and \cite{RL19}. Self-organisation is clearly dissociated from the Hall effect in stratified flows, as it appears also in simulations where the Hall effect is absent (\citealt{SA14,SB18,RL19}).

\begin{figure}
\begin{center}
    \includegraphics[width=0.49\linewidth]{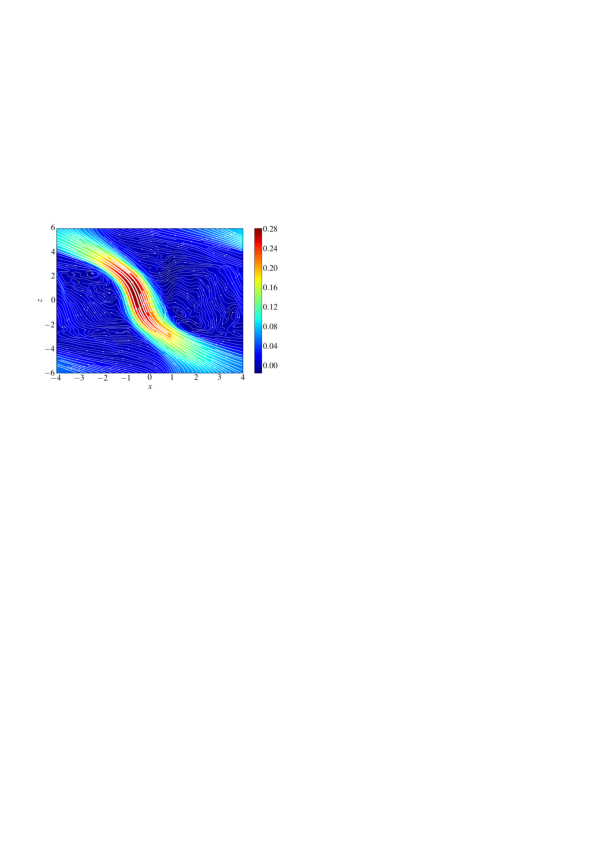}
    \includegraphics[width=0.49\linewidth]{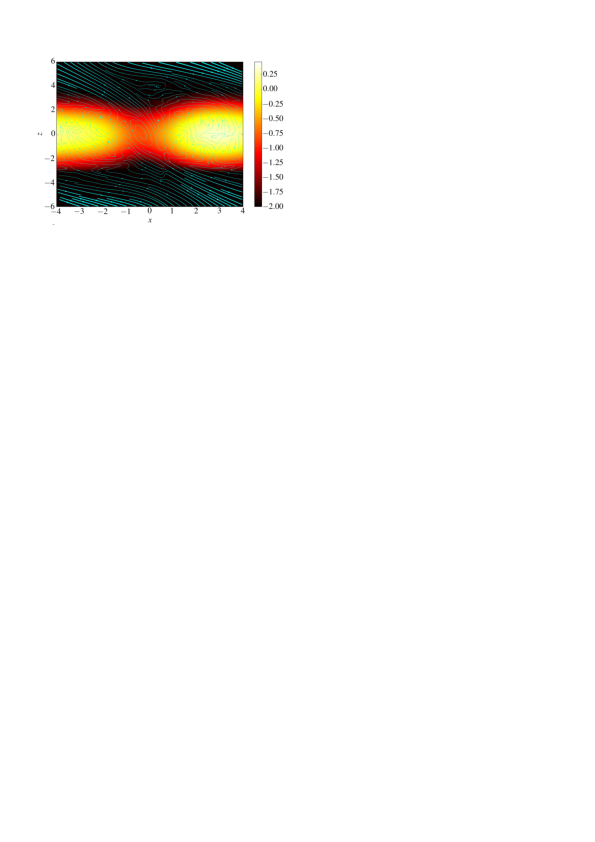}
    
    \caption{Shearing box simulation exhibiting self-organisation with ambipolar diffusion only. Left: Magnetic configuration, with poloidal field lines and colours showing the toroidal field component amplitude. Right: density map (colour) and poloidal velocity streamlines. Notice the poloidal field concentration in the region $(x,z)\simeq 0$, associated to a minimum of density (="gap"). Figure from \cite{RL19}.}
    \label{fig:MRI_strat_selforg}
\end{center}
\end{figure}

This self-organisation shows again a very tight intrication between the field and the gas: magnetic field lines tends to concentrate into "gaps", leaving regions with a lot of mass but no poloidal field (Fig.~\ref{fig:MRI_strat_selforg}). Several scenarios have been proposed to explain this effect. \cite{BS14} proposed that this is a result of the anisotropy of the turbulent diffusivity tensor, following the unstratified box argument (\S \ref{sec:self-org-ideal-uns}).

\begin{figure}
\begin{center}
    \includegraphics[width=0.99\linewidth]{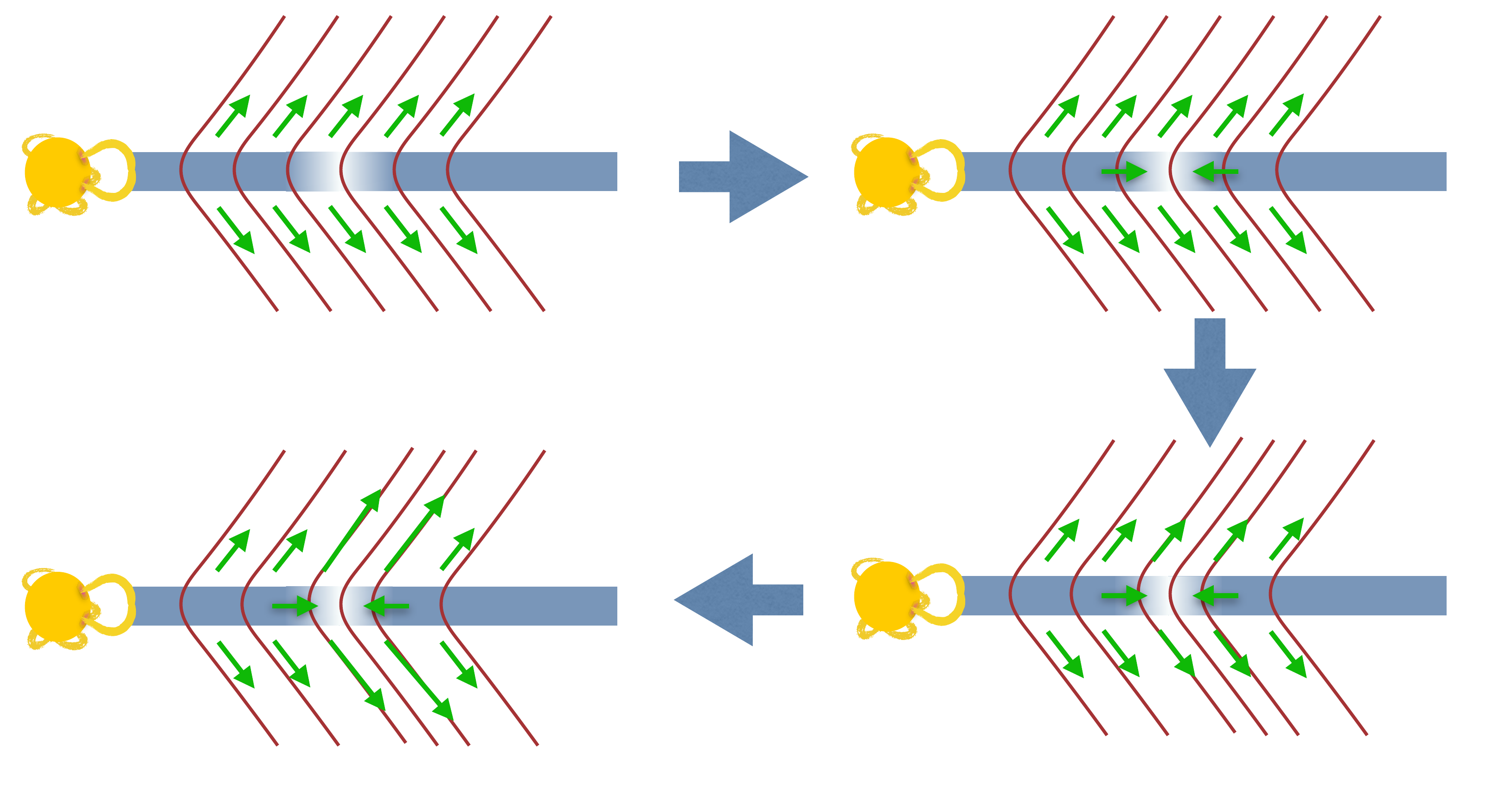}
    \caption{Self-organisation feedback loop proposed by \cite{RL19}. Consider a small density deficit (top left), this deficit induces a radially converging flow (top right), which drags the poloidal field line inwards (bottom right). The stronger field, leads to a more efficient local ejection index which empties even more the region (bottom left).}
    \label{fig:self_orgRL19}
\end{center}
\end{figure}

An alternative viewpoint was proposed by \cite{RL19} who noted that, in the presence of vertical stratification, the \emph{average} radial flow was converging \emph{towards} the region of flux accumulation, and hence in the gap. This counter-intuitive finding implies that gaps are necessarily emptied by the outflow, and the radial flow is simply trying to replenish the gaps. This radial flow drags the poloidal field towards the gap (cf Fig.~\ref{fig:self_orgRL19}), and is therefore a pure advection process, not a turbulent anti-diffusion effect as proposed by \cite{BS14}. The stronger field in the gap leads to a more efficient ejection, as observed in the ejection "plume" (cf fig.~\ref{fig:MRI_strat_selforg}), leading to a quasi-steady state from the disc viewpoint . This feedback loop turns out to be a linear instability of the wind-emitting disc, with predictable growth rate and optimum disturbance. Let us note that the linear instability criterion is simply 
\begin{align*}
	-\frac{\mathrm{d}\log \zeta}{\mathrm{d}\log \bmean}>-\frac{\mathrm{d}\log \alpha}{\mathrm{d}\log \bmean}\qquad \rightarrow \qquad \textrm{instability}
\end{align*}
which essentially compares the time needed to empty the disc by the outflow to the time needed to refill the disc by "viscous" diffusion \citep{RL19}. As shown above, this criterion is generally satisfied in stratified shearing box models, so this "wind instability" mechanism could potentially explain the self-organisation observed in \emph{stratified} models.

\section{Conclusion}
In ideal MHD, the MRI is found to be a robust angular momentum transport mechanism, but in the absence of an external poloidal field, it fails to account for observed accretion rates if one includes the relevant non-ideal MHD effects. 

When a mean field is added, magnetised outflows are found as a result of the saturation of the MRI.  \add{In ideal MHD, outflows are expected to be the dominant angular momentum transport mechanism when $\bmean\lesssim (R/H)^2$. When non-ideal effects are introduced however, the radial transport of angular momentum is significantly reduced, while outflows are much less affected. As a result, when non-ideal MHD effects are strong, magnetised outflows are expected to become the dominant mechanism of angular momentum transport, even for relatively weak fields.}

We note however that the shearing box approach cannot model these outflows properly because of several inconsistencies (top/down symmetry of the outflow, mass loss rate depends on boundary conditions). In addition, the presence of a laminar midplane stress due to the HSI and the existence of a self-organisation mechanism associated to the outflow make the shearing box ill-suited to study the dynamics of these objects. The only way to connect our local understanding of the physics to the global dynamics of these objects is therefore to perform global simulations, in which outflows are properly modelled.  

\part{Global models of protoplanetary discs}

In this section, we revisit the physical concepts underlying ejection in accretion discs. We then use these concepts to interpret the most recent numerical models of protoplanetary discs which exhibits accretion and outflows.

\section{Some outflow definitions and properties\label{sec:outflow_definitions}}
Before exploring the connection between a weakly ionised disc and an outflow, let us underline a few important definitions and properties on outflows. To this end, let us work first in the ideal MHD approximation and assume for the moment that the outflow is stationary and axisymmetric. In a cylindrical frame $(R,\phi,z)$, the conservation of mass and momentum read
\begin{align}
\label{glob:mass}
\bm{\nabla \cdot}\rho\bm{u}_p&=0\\
\label{glob:momr}
    \rho \bm{u}_p\bm{\cdot \nabla} u_R&=\rho\Omega^2 r-\partial_R P+\frac{J_\phi B_z}{c}-\frac{J_z B_\phi}{c}-\rho\partial_R\psi\\
\label{glob:momphi}
    \frac{1}{R}\rho \bm{u}_p\bm{\cdot \nabla} \Omega R^2&=\frac{1}{R}\bm{\nabla\cdot }\Bigg(R\frac{\bm{B}_pB_\phi}{4\pi}\Bigg)\\
    \label{glob:momz}
    \rho \bm{u}_p\bm{\cdot \nabla} u_z&=-\partial_z P-\partial_z\Bigg(\frac{B_\phi^2+B_R^2}{8\pi}\Bigg)+\frac{B_R\partial_R B_z}{4\pi}-\rho\partial_z\psi
\end{align}
 where the index $p$ denotes the poloidal components $(R,z)$ of the vector fields, $\Omega\equiv u_\phi/R$ ($\Omega$ is not necessarily Keplerian) and $\psi=GM/(R^2+z^2)^{1/2}$ is the gravitational potential, assumed to be due solely to the central star. (\ref{glob:momphi}) can be recast in conservative form to obtain an angular momentum conservation equation
\begin{align}
\label{glob:ang}
\bm{\nabla\cdot }\Bigg(\rho \bm{u}_p\Omega R^2-R\frac{\bm{B}_p B_\phi}{4\pi}\Bigg)&=0
\end{align}
which shows that the magnetic field can carry angular momentum in the Maxwell stress $\bm{B}_pB_\phi$. 

In addition to the equation of motion, one needs to solve the induction equation in the ideal regime in the poloidal and azimuthal directions
\begin{align}
\label{glob:indp}
\bm{\nabla \times }\Big(\bm{u}_p\bm{\times B}_p\Big)&=0,\\
\label{glob:indt}
\bm{\nabla \cdot}\frac{1}{R}\Big(\Omega R \bm{B}_p-B_\phi\bm{u}_p\Big)&=0.
\end{align}
Finally, the magnetic field has to satisfy the solenoidal condition $\bm{\nabla\cdot B}=0$. In an axisymmetric configuration this can be used to express the poloidal field components as a function of a scalar streamfunction $a$:
\begin{align}
\bm{B}_p&=    \frac{1}{R}\bm{\nabla}a\bm{\times e}_\phi.
\end{align}
By construction, the value of $a$ is constant along a poloidal field line. Hence, we can label each poloidal field line with the value of its streamfunction. From these equations, it is possible to derive several physical constant of motion which are useful to interpret outflow solutions.

\subsection{Frozen in condition}
\index{Invariant!Frozen in}
The frozen in condition is derived from mass conservation (\ref{glob:mass}) and the poloidal induction equation (\ref{glob:indp}). Let us first start with the vertical component of the induction equation, which reads
\begin{align*}
\frac{1}{R}\frac{\partial}{\partial R} R\Bigg(u_zB_R-u_RB_z\Bigg)&=0,
\end{align*}
from which one deduces that
\begin{align*}
    u_zB_R-u_RB_z&=\frac{\beta(z)}{R},
\end{align*}
where $\beta$ is an unknown scalar function (physically, $\beta/R$ is the $\phi$ component of the electromotive force $E_\phi$). Using the radial induction equation, we may see that $\partial_z\beta=0$ so that $\beta$ is a constant. To avoid any singularity for $E_\phi$ at $R=0$, we are then forced to have $\beta=0$ and hence $\bm{B}_p$ and $\bm{u}_p$ are parallel, i.e.
\begin{align}
\label{eq:glob_frozen_mu}
\bm{u}_p&=    \mu(R,z)\bm{B}_p
\end{align}
We can inject this relation in the mass conservation equation to obtain
\begin{align*}
\bm{B}_p\bm{\cdot\nabla} (\rho\mu) &=0,
\end{align*}
which indicates that $\eta\equiv \rho\mu$ is a constant along magnetic field lines, i.e. $\eta=\eta(a)$. We can then recast (\ref{eq:glob_frozen_mu}) into
\begin{align}
\label{eq:glob_frozen}
\bm{u}_p&=\eta(a)\frac{\bm{B}_p}{4\pi\rho}    
\end{align}
\index{Alfv\'en surface}
which constitutes the frozen in condition where $\eta$ describes the amount of mass loaded along a poloidal field line. $\eta$ is a direct measure of the gas density $\rho_\mathrm{A}$ at the Alfv\'en surface, which is the region defined by $\bm{u}_p=\bm{V}_{\mathrm{A},p}=\bm{B}_p/\sqrt{4\pi\rho}$:
\index{$\eta$, frozen in invariant}
\begin{align}
\label{eq:eta_def}
    \eta(a)=\sqrt{4\pi \rho_\mathrm{A}(a)}.
\end{align}

Note regarding the applicability of the frozen-in condition in shearing boxes: when deriving the frozen-in condition, we have used a regularity condition at $R=0$ for the electromotive force. Such a condition does not exist in the shearing box approximation so that $E_\phi$ can be non-zero in principle. This implies that in a stationary shearing box solution, the poloidal field and velocity are not necessarily parallel \citep{LFO13}. Physically, it means that field lines can be indefinitely dragged radially inward or outward, without affecting the stationarity condition, thanks to the assumed radial periodicity. This is physically impossible from a global point of view, but it illustrates once again the numerous drawbacks of the shearing box model when dealing with outflows.

\subsection{Magnetic surface rotation}
\index{Invariant!Magnetic surface rotation}
The rotation of magnetic surfaces is obtained by substituting the frozen-in condition (\ref{eq:glob_frozen}) into the azimuthal component of the induction equation (\ref{glob:indt}) :
\begin{align*}
\bm{B}_p\bm{\cdot \nabla}\Bigg(\Omega-\frac{B_\phi\eta(a)}{4\pi \rho R}\Bigg)&=0,
\end{align*}
which allows us to define a new invariant along field lines
\index{$\Omega^*$, magnetic surface rotation invariant}
\begin{align}
\label{eq:mag_surface_inc}\Omega^*(a)&\equiv \Omega-\frac{\eta(a)}{4\pi \rho R}B_\phi=\Omega_\mathrm{A}-\frac{V_{\mathrm{A}\phi}(R_\mathrm{A})}{R_\mathrm{A}}
\end{align}
\index{Alfv\'en radius}
where the A indices denotes quantities evaluated at the Alfv\'en surface and $R_\mathrm{A}$ is the Alfv\'en radius, where the outflow becomes super-Alfv\'enic . It can easily be checked that in a frame rotating at $\Omega^*(a)$, the \emph{total} field is parallel to the total velocity. In other words, there is no induced electromotive force in the frame rotating at $\Omega^*(a)$. $\Omega^*$ can therefore be interpreted as the rotation speed of magnetic surfaces. By combining (\ref{eq:glob_frozen}) and (\ref{eq:mag_surface_inc}), we can express the \emph{total} velocity as a function of the magnetic field,
\begin{align}
\label{eq:u_tot_outflow}
\bm{u}&=\frac{\eta(a)}{4\pi \rho}\bm{B}+R\Omega^*(a)\bm{e}_\phi    .
\end{align}
\index{Ferraro's iso-rotation}
This illustrates the fact that \emph{only the poloidal components} of the field and the velocity are parallel in general. In the particular case where $\eta=0$ (no motion along the poloidal field lines), we recover Ferraro's iso-rotation law $\Omega=\Omega^*(a)$.

\subsection{Angular momentum}
\index{Invariant!Angular momentum}
The angular momentum invariant is easily deduced from the angular momentum conservation equation (\ref{glob:ang}):
\begin{align*}
\bm{B}_p\bm{\cdot \nabla}\Bigg(\Omega R^2-\frac{RB_\phi}{\eta(a)}\Bigg)&=0    
\end{align*}
from which we deduce the angular momentum invariant $\mathcal{L}(a)$:
\begin{align}
\label{eq:inv_angular}
\mathcal{L}(a)&\equiv     \Omega R^2-\frac{RB_\phi}{\eta(a)}\\
&=\Omega^*(a)R_A^2
\end{align}
where the last equality has been obtained at the Alfv\'en surface using (\ref{eq:eta_def}). Hence, the amount of angular momentum transported by the outflow is an invariant made of two parts: the classical kinetic contribution and a magnetic part stored in $B_\phi$.

\subsection{Bernoulli invariant}
The Bernoulli invariant is obtained from the scalar product of the poloidal equations of motion (\ref{glob:momr}---\ref{glob:momz}) with $\bm{u}$. One gets
\begin{align*}
\bm{u}_p\bm{\cdot \nabla}\Bigg[{\frac{u^2}{2}+\psi_G}\Bigg]&=-\bm{u}_p\bm{\cdot}\frac{\bm{\nabla}P}{\rho}+\bm{u \cdot}\frac{\bm{J\times B}}{\rho c},
\end{align*}
where $\psi_G=GM/(R^2+z^2)^{1/2}$ is the gravitational potential. We then use (\ref{eq:u_tot_outflow}) to obtain the work of the Lorentz force:
\begin{align*}
    \bm{u \cdot}\frac{\bm{J\times B}}{c}&=R\Omega^*(a)\frac{\bm{J}_p\bm{\times B}_p}{\rho c}\\
    &=\Omega^*(a)\frac{\bm{B}_p\bm{\cdot \nabla} \Big(RB_\phi\Big)}{4\pi\rho }\\
    &=\bm{u}_p\bm{\cdot\nabla}\Bigg(\frac{R\Omega^*(a)B_\phi}{\eta(a)}\Bigg)
\end{align*}
Additionally, we may express the work of pressure forces using the enthalpy per unit mass $\mathcal{H}$
\begin{align*}
\bm{u}_p\bm{\cdot}\frac{\bm{\nabla}P}{\rho}&=\bm{u}_p\bm{\cdot\nabla}\mathcal{H}+\delta \mathcal{Q}
\end{align*}
in which we have also considered the effect of an additional heating term due to radiative heating/cooling $\delta \mathcal{Q}$. Putting all the terms together and integrating along one particular streamline $a$, we find that the quantity
\index{Invariant!Bernoulli}
\begin{align}
\mathcal{B}&\equiv    \frac{u^2}{2}+\psi_G+\mathcal{H}-\frac{R\Omega^*(a)B_\phi}{\eta(a)}+\int_{\mathcal{S}(a)} \delta\mathcal{Q}\,\mathrm{d}s
\end{align}
is conserved along poloidal field lines and streamlines. Note that since the heating term is not a proper differential, it depends on the integral of the heating term along the chosen streamline $\mathcal{S}(a)$. Of course, for an outflow to escape up to $z\rightarrow \infty$, on needs $\mathcal{B}>0$ on the streamlines (assuming $\mathcal{H}>0$). In the midplane of a Keplerian disc, one has $u^2/2+\psi_G=-v_K^2/2$ where $v_K$ is the Keplerian velocity, so that additional ingredients are required to produce an outflow. Two extreme situations can be identified:
\index{Thermal wind}
\index{Cold MHD wind}
\begin{description}
\item[Thermal winds : ] where magnetic effects are neglected and ejection is possible because the disc is hot (large initial enthalpy) or because a lot of heating is applied along the streamlines (named photo-evaporation in the protoplanetary discs community).
\item[Cold MHD winds :] where thermal effects are neglected and the toroidal field acts as an energy reservoir to launch the outflow.
\end{description}

\subsection{Dimensionless numbers characterising an outflow}

Based on the MHD invariants, it is possible to define a series of dimensionless numbers which characterise an outflow streamline unambiguously. These numbers use the physical properties at the base of the outflow. Let us therefore write $\Omega_0$ the rotation rate at the base of the outflow and $R_0$ its cylindrical radius (one typically has $\Omega_0=\Omega_K(R_0)$), and $B_0$ the poloidal field strength threading the disc at the location where the outflow is launched. We then define
\begin{align*}
\kappa&\equiv \eta \frac{\Omega_0 R_0}{B_0}\\
\omega^*&\equiv \frac{\Omega^*}{\Omega_0}\\
\lambda&\equiv \frac{\mathcal{L}}{\Omega_0 R_0^2}=\omega \Bigg(\frac{R_A}{R_0}\Bigg)^2\\
e&\equiv\frac{\mathcal{B}}{\Omega_0^2R_0^2/2} 
\end{align*}
The Bernoulli invariant can be easily expressed in terms of the other invariants, using the dimensionless rotation rate $\omega\equiv\Omega/\Omega_0$:
\begin{align}
e=\frac{u_p^2}{\Omega_0^2R_0^2}+\omega^2\frac{R^2}{R_0^2}-\frac{2R_0}{\sqrt{R^2+z^2}}+2\omega^*\Bigg(\lambda-\omega\frac{R^2}{R_0^2}\Bigg)+\theta
\end{align}
where $\theta=2\Big[\mathcal{H}+\int_\mathrm{a=cste} \delta\mathcal{Q}\,\mathrm{d}s\Big]/\Omega_0^2R_0^2$ is the dimensionless thermal energy content of the flow.
\index{$\lambda$, magnetic level-arm}
This expression clearly demonstrates the contribution of $\lambda$ to the energy content of the flow. This parameter, often called the \emph{magnetic level-arm}, is of key importance, as can be seen if one computes its value at the outflow base, assuming $u_p\ll \Omega_0R_0$, $\omega=1$ and $R=R_0$:
\begin{align}
e\simeq	2\omega^*(\lambda-1)+\theta -1.
\end{align}
Obviously, for the outflow to be able to propagate up to infinity, one needs $e>0$. We recover here the two extreme limits discussed above: purely thermal winds, which have $\lambda=1$ and require $\theta>1$, or cold MHD winds with $\theta=0$ which need
\begin{align*}
	2\omega^*(\lambda-1)>1
\end{align*}
For all practical applications of MHD outflows, one has $\omega^*\simeq 1$ to a very good approximation. This implies that an outflow can exist only if
\begin{align}
\label{glob:lambda_energy}
\lambda>\frac{3}{2}    
\end{align}


\subsection{Ejection efficiency}
The existence of an outflow is tightly linked to the process of accretion happening inside the disc since the energy of the wind is obtained from the accretion power of the disc. To understand this connection, let us first define the accretion rate of the disc
\begin{align*}
\dot{M}_\mathrm{acc}(R)\equiv-\int_{z^-}^{z^+}\mathrm{d}z \int_0^{2\pi} R\mathrm{d}\phi\, \rho u_r
\end{align*}
where the integration is performed on the box vertical extension, defined by $z^-$ and $z^+$. It is also useful to define the outflow rate of the wind between the inner radius of the disc $R_\mathrm{in}$ and the radius under consideration
\begin{align*}
\dot{M}_\mathrm{wind}(R)\equiv \int_{R_\mathrm{in}}^R  \mathrm{d}R \int_0^{2\pi} R\mathrm{d}\phi\, [\rho u_z]_{z^-}^{z^+}.
\end{align*}
These two quantities are connected by the continuity equation (\ref{eq:mass}):
\begin{align*}
2\pi\frac{\partial \Sigma}{\partial t}+\frac{1}{R}\frac{\partial}{\partial R}\Big(\dot{M}_\mathrm{wind}-	\dot{M}_\mathrm{acc}\Big)=0
\end{align*}
At this stage, it is useful to introduce  the ejection efficiency index
\index{$\xi$ ejection efficiency index}
\begin{align}
\label{eq:xi_def}
\xi&\equiv \frac{\mathrm{d}\log \dot{M}_\mathrm{acc}}{\mathrm{d}\log R}\\
&=\frac{1}{\dot{M}_\mathrm{acc}}\frac{\mathrm{d} \dot{M}_\mathrm{wind}}{\mathrm{d}\log R},
\end{align}
which quantifies what fraction of the mass is being lost in the wind, the second line being obtained from the continuity equation, assuming stationarity. As expected, $\xi=0$ corresponds to a situation without any wind.

One can also relate the accretion rate to the ejection rate using the angular momentum conservation equation (\ref{eq:angular_final}) as
\begin{align*}
\dot{M}_\mathrm{acc} &=\frac{4\pi}{\Omega_K}\Bigg[\underbrace{\frac{1}{R}\frac{\partial}{\partial R} R^2\overline{W_{R\phi}}}_{\tau_{R}} + \underbrace{R\langle W_{z\phi}\rangle_{z^-}^{z^+}}_{\tau_{z}}\Bigg],
\end{align*}
where we have define the stress tensor $W_{i\phi}\equiv \rho v_iv_\phi-B_iB_\phi/4\pi$. This allows us to define the radial and vertical contribution to the accretion of the disc $\tau_{R\phi}$ and $\tau_{z\phi}$. It is then useful to introduce the ratio of these two quantities
\begin{align*}
\Lambda\equiv \frac{\tau_{z}}{\tau_{r}}	
\end{align*}
so that the accretion rate is simply
\begin{align*}
	\dot{M}_\mathrm{acc} &=\frac{4\pi}{\Omega_K}\tau_{z}\frac{1+\Lambda}{\Lambda}
\end{align*}
It is then possible to relate the vertical torque $\tau_{z\phi}$ to the mass ejection rate by noting that the torque is evaluated high above the disc midplane, so that the kinetic contribution to the stress is negligible, which implies $\tau_{z\phi}\simeq - [RB_\phi B_z/4\pi]_{z^-}^{z^+}\simeq -RB_0 B_\phi(z^+)/2\pi$ where the second equality assumes the outflow is top/down symmetric and that the poloidal field strength doesn't vary much between the midplane and the disc surface. It is then simple to show that the vertical torque is directly connected to the MHD invariants
\begin{align*}
	\tau_{z}\simeq 2 (\Omega^*R_A^2-\Omega_0 R_0^2)[\rho u_z](z_+).
\end{align*}
The vertical mass flux being directly to the radial derivative of the outflow rate, we can express the accretion rate as
\begin{align*}
	\dot{M}_\mathrm{acc} &\simeq\frac{2}{R_0^2\Omega_0}\frac{1+\Lambda}{\Lambda}(\Omega^*R_A^2-\Omega_0 R_0^2)\frac{\mathrm{d}\dot{M}_\mathrm{wind}}{\mathrm{d}\log R}
\end{align*}
From which we get an expression for the mass ejection index
\begin{align}
\label{eq:xi_constraint}
\xi\simeq \frac{\Lambda}{2(\Lambda+1)}\frac{1}{\lambda-1}	
\end{align}
This relation reveals several key features of MHD outflows. First, it shows that large level-arms $\lambda$ are associated to small ejection indices. Interestingly, this result does not depend on the radial contribution to the angular momentum budget $\Lambda$. Second, the energy constraint (\ref{glob:lambda_energy}) imposes an upper bound on $\xi$ in cold MHD winds: $\xi\lesssim \Lambda/(\Lambda+1)\lesssim 1$. Note that this relation allows for outflows approximately as massive as the mass accretion rate, but it does not allow for outflows vastly more massive than this. Outflows with $\xi\gg 1$ therefore necessarily require some thermal energy driving to escape the potential well, as one would expect.

\subsection{Connection to shearing box simulations}

In the previous part, we had to introduce several \emph{local} quantities to characterise outflows in local shearing box models. As pointed out, however, shearing boxes lack global constrains, which implies that some of the solutions are likely unphysical. As a first step, it is therefore useful to relate these local quantities to global MHD invariants to test the domain of validity of shearing box solutions. In this subsection, we will make the assumption that the global outflow is top/down symmetric, so that the invariant are the same on both sides of the disc.

We first start with the outflow rate $\zeta$ (defined in \ref{eq:zeta_definition}), which can be easily connected to the mass loading parameter $\kappa$
\begin{align*}
\kappa=\frac{1}{4}\frac{R}{H}\bmean \zeta	.
\end{align*}
The magnetic level-arm can also be related to $\upsilon$ (defined in \ref{eq:upsilon_definition}), provided that we neglect the kinetic contribution to the vertical stress, which is valid if we choose the disc upper boundary to lie high enough above the midplane
\begin{align*}
\lambda=1+\frac{H}{R}\frac{\upsilon}{\zeta}.
\end{align*}
The energetic constrain (\ref{glob:lambda_energy}) therefore leads to a new constrain in shearing boxes: $2\zeta/\upsilon< H/R$. This constrain cannot be satisfied by the shearing box scalings (\ref{eq:ideal_strat_zeta}) and (\ref{eq:ideal_strat_upsilon}) when $H/R\lesssim 0.2$. Hence, shearing box wind solutions always eject too much mass with too little energy to escape the global potential well of discs with realistic aspect ratios. In other words, if one puts a shearing box wind solution in a global disc configuration, the ejected material should fall back onto the disc.

Finally, we can relate the stress rate $\Lambda$ to $\alpha$ and $\upsilon$. For this, let us assume that the disc surface density follows a power law: $\Sigma=\Sigma_0(R/R_0)^{-p}$, that $\alpha$ and $H/R$ are constant with radius, and that the disc is vertically isothermal. Under these assumptions, the contributions to the mass accretion are:
\begin{align*}
\tau_R&=\alpha \Omega_0^2R_0^2\Big(\frac{H}{R}\Big)^2(1-p)\Sigma\\
\tau_z&=\frac{1}{\sqrt{2\pi}}\upsilon\Omega_0^2R_0^2\frac{H}{R}\Sigma
\end{align*}
so that the ratio simply reads
\begin{align*}
\Lambda=\frac{1}{\sqrt{2\pi}(1-p)}\frac{\upsilon}{\alpha}\frac{R}{H}.
\end{align*}
The scalings (\ref{eq:ideal_strat_zeta}) and (\ref{eq:ideal_strat_upsilon}) then suggest that $\Lambda\rightarrow 0$ when $\bmean\rightarrow \infty$. Combining these results with \ref{eq:xi_constraint}, this implies that the ejection index $\xi$ tend to decrease as $\bmean\rightarrow \infty$, as one would expect.

\section{Outflow phenomenology}
\begin{figure}
\begin{center}
    \includegraphics[width=0.90\linewidth]{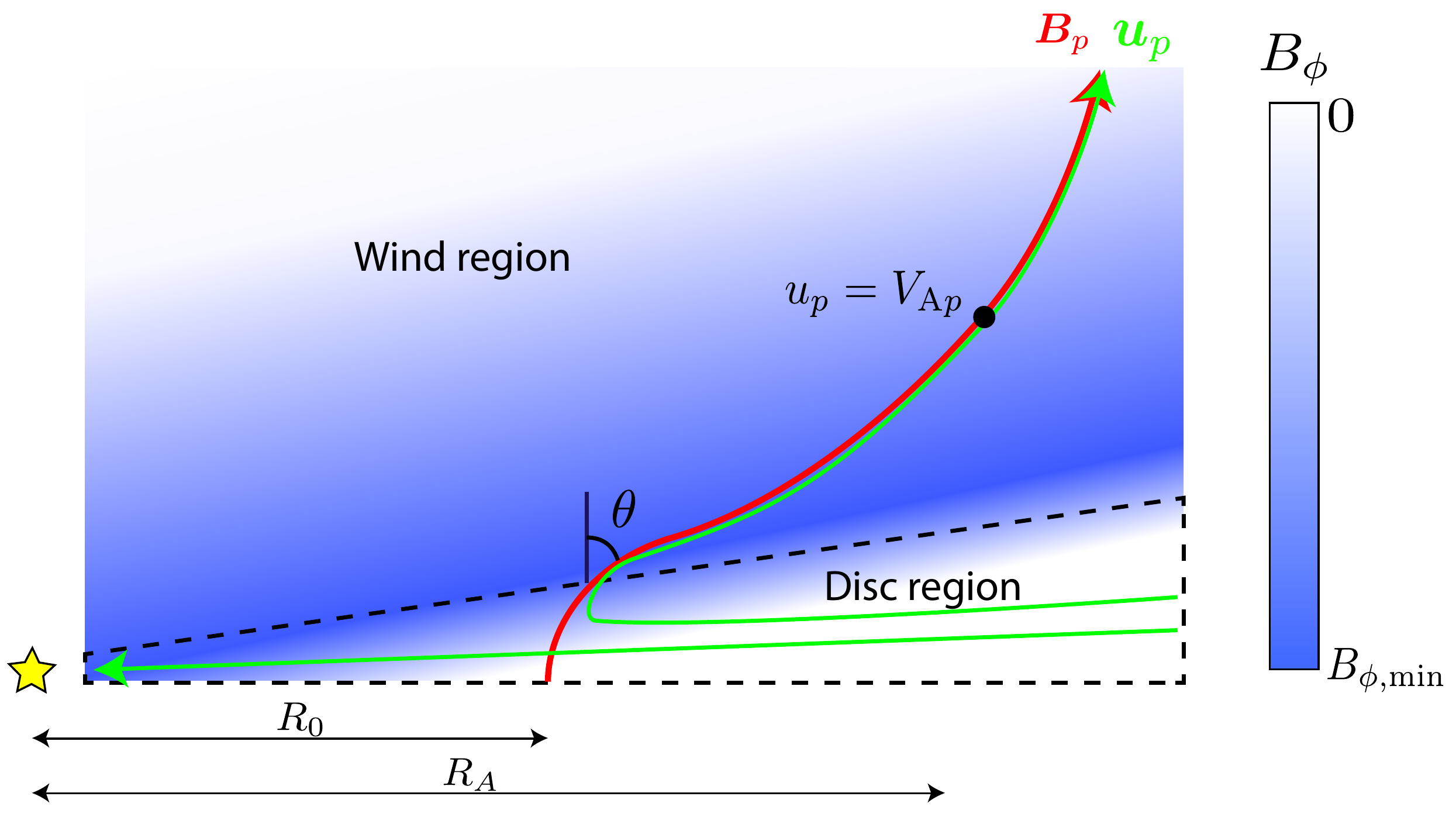}
    
    \caption{Global disc-wind interaction scheme. We distinguish a disc region in dashed lines from the outflow region. The poloidal field line is represented in red and the poloidal streamlines in green. In addition, the toroidal field component is shown in blue in a frame corotating at $\Omega_K(R_0)$  }
    \label{fig:global_outflow_phenomenlogy}
\end{center}
\end{figure}

The launching of an outflow is tightly linked to physics of the accreted material since the outflow energy eventually comes from the accretion power of the disc. Let us divide the overall structure into a ``disc region'' and an ``outflow'' region (Fig.~\ref{fig:global_outflow_phenomenlogy}) and describe the physical processes in each region.

\subsection{Disc region}
In the disc region, the flow is mostly accreted, while the field lines are stationary. The fact that poloidal field and streamlines are not parallel implies that non-ideal MHD effects are \emph{necessarily} present in the disc. Historically, these non-ideal effects have been treated in two ways: 1- assuming that the disc ionisation fraction is extremely low as in protoplanetary discs, so that ambipolar and ohmic diffusion are very large \citep{K89,WK93}, or 2- assuming that the disc was turbulent, the turbulence leading to a ``turbulent resistivity'' modelled as a non-isotropic diffusivity tensor \citep{FP93}.

The toroidal component of the magnetic field $B_\phi$ is the key ingredient of the interaction between the disc and the outflow. In the disc, the toroidal field is produced by the shearing of the radial field by the differential rotation of the disc $\partial_tB_\phi\simeq B_RR\partial_R \Omega$. This effect is actually the main energy source for the outflow, which converts shear energy into magnetic energy stored in $B_\phi$ at the disc surface. It is often assumed that outflows are top-down symmetric to satisfy the global symmetries of the system, so that $B_R(z=0)=0$ and $B_\phi(z=0)=0$ follows. The bending of poloidal field lines in the disc implies $\partial_z B_R>0$ for $z>0$ and hence $\partial_zB_\phi<0$ up to the disc surface. The actual value of $B_\phi$ depends on the competition between the shearing of $B_R$ and the magnetic diffusion which damps the shear amplification. Overall, one can estimate $B_\phi^+$ at the disc surface by balancing shear and diffusion
\begin{align*}    
B_\phi^+\sim B_z\tan(\theta)\frac{\Omega h^2}{\eta}
\end{align*}
where $h$ is the disc scale height, $\eta$ is the magnetic diffusivity and $\theta$ is the inclination of the poloidal field at the disc surface (see Fig.~\ref{fig:global_outflow_phenomenlogy}). Accretion of the disc material naturally follows from the profile of $B_\phi(z)$. The toroidal field is responsible for an azimuthal magnetic tension force $B_z\partial_z B_\phi/4\pi$ which slows down the rotating material and lead to accretion. We can deduce the accretion rate from (\ref{eq:angular_final}):
\begin{align}
\label{eq:outflow_accrate}
    \rho u_R \sim \frac{B_\phi^+ B_z}{4\pi \Omega h} 
             \sim \frac{B_z^2\tan(\theta) h}{4\pi \eta}
\end{align}
where $\rho$ is measured in the middle of the accretion flow (which in general is the disc midplane, but can also be off-midplane for dissymmetric outflows). This accretion is physically due to a transfer of angular momentum from the accreted material to the toroidal field. In the end, the angular momentum is stored in $B_\phi^+$ and is eventually used to launch the outflow.

The outflow base has to be replenished by the disc. Hence, a net positive vertical acceleration is required in the disc region to push material upward, even if at modest velocities. A careful examination of (\ref{glob:momz}) allows us to isolate the role played by each term in the vertical acceleration: the magnetic pressure is necessarily directed downward since $B_R^2$ and $B_\phi^2$ are both increasing functions of $z$. Gravity is also directed downward. The magnetic tension term is usually small (it involves a radial derivative while the other terms involve vertical derivatives, which are larger by a factor $R_0/h$), and typically directed downward if we assume the most natural situation with $B_z$ decreasing with radius. Hence, the \emph{only term leading to an upward acceleration is the vertical thermal pressure gradient}. The role played by the thermal pressure at the base of the outflow has often been missed. However, it has some important consequences. For instance, $B_\phi^+$ cannot reach arbitrarily large values which would otherwise prevent thermal pressure from pushing materials to the wind base and vertically squeeze the disc. Ejection therefore requires $(B_\phi^+)^2/8\pi\lesssim P^+$ \citep{F97}.

To summarise the physics of the disc region:
\begin{itemize}
\item The region is non-ideal as the gas has to be allowed to stream through poloidal field lines which are stationary.
\item the toroidal field at the disc surface stores the energy and angular momentum taken from the accreted material
\item the upward motion needed to replenish the outflow base is due to the vertical thermal pressure gradient. This sets a limit to the amount of toroidal field that can be stored at the disc surface since magnetic pressure prevents this upward motion.    
\end{itemize}

\subsection{Outflow region}
In order to understand the dynamics of the outflow from its base, it is convenient to introduce the Alfv\'enic Mach number $\xi$ defined as
\begin{align*}
\xi\equiv \frac{u_p}{V_{\mathrm{A},p}}=\eta \sqrt{\frac{4\pi }{\rho}}
\end{align*}
where the second equality is deduced from (\ref{eq:glob_frozen}). In the outflow, we can expect a continuous acceleration of the flow, so that $\xi$ is an increasing function as one moves along a poloidal streamline, with $\xi\sim 0$ at the outflow base. We can eliminate $B_\phi$ in favour of the angular velocity by combining the magnetic surface rotation invariant (\ref{eq:mag_surface_inc}) and the angular momentum invariant (\ref{eq:inv_angular}) to obtain
\begin{align*}
\Omega=\Omega^*\Bigg(\frac{1-\xi^2(R_A/R)^2}{1-\xi^2}\Bigg).
\end{align*}
This expression looks singular for $\xi=1$ which corresponds to the Alfv\'en point. However, at this particular point, $R=R_A$ so that $\Omega$ is actually smooth across this point. Second, in the limit of low Mach numbers $\xi\ll 1$ which corresponds to the base of the outflow, we can expand this expression to get
\begin{align*}
\Omega\simeq \Omega^*\Bigg[1-\xi^2\Big(\frac{R_A^2}{R^2}-1\Big)+\mathcal{O}(\xi^4)\Bigg]    
\end{align*}
Hence, the outflow is rotating at a \emph{constant angular velocity} up to the point where $\xi^2\simeq 1/(R_A^2/R^2-1)$. Physically, the angular momentum stored in $B_\phi^+$ by the disc is progressively used to accelerate the flow via the azimuthal magnetic tension force, leading to the apparent solid rotation profile. This works until most of the toroidal field has been used and the angular momentum is all in kinetic form. On the opposite limit $\xi\rightarrow\infty$, we find $\Omega\simeq \Omega^*R_A/R^2$, i.e. a constant angular momentum rotation profile, as expected.

The vertical acceleration of the outflow results from angular momentum conservation. As demonstrated above, angular momentum stored in $B_\phi$ is converted into kinetic angular momentum. Hence $B_\phi$ decreases along the streamline. This leads to a magnetic pressure force $\partial_z B_\phi^2$ directed upward in (\ref{glob:momz}) and hence a vertical acceleration of the outflow. As the outflow bends toward the vertical axis, $B_R$ decreases as well, leading to an additional magnetic pressure term $\partial_z B_R^2$ accelerating the outflow vertically. Overall, the vertical acceleration is a \emph{magnetic pressure effect} due to the decrease of $B_\phi$ and $B_R$ along the streamlines.

One can use the Bernoulli invariant to characterise the topology of the outflow close to the wind base. For simplicity and following \cite{BP82}, let us assume that the wind base is located at $(R_0,z=0)$. We consider a fluid particle, initially following a Keplerian rotation orbit at $(R_0,z=0)$ and we assume this particle follows the streamline of the outflow so that at a later time, the particle is located at $(R_0+\delta R,\delta z)$. Since we focus on the launching region of the outflow, we assume that $\Omega\simeq \Omega^*$. During this displacement, the Bernoulli invariant should be conserved, so we should have
\begin{align*}
-\frac{1}{2}(\Omega^*R_0)^2+\psi_G(R_0,0)>    -\frac{1}{2}[\Omega^*(R_0+\delta R)]^2+\psi_G(R_0+\delta R,\delta z)\\
\rightarrow \quad \frac{1}{2}(\Omega^*R_0)^2\Bigg(-3\Big(\frac{\delta R}{R_0}\Big)^2+\Big(\frac{\delta z}{R_0}\Big)^2\Bigg)<0,
\end{align*}
\index{Magneto-centrifugal outflow}
where the inequality comes from the assumption that $u_p(R_0+\delta R, \delta z)>u_p(R_0,0)$ (i.e. the flow accelerates). It can be converted into a criterion on $\theta$: $\tan\theta>1/\sqrt{3}$ or $\theta>30^\circ$. This well-known criterion, initially derived by \cite{BP82} is valid provided that thermal effects (enthalpy, heating) are negligible, i.e. in cold winds. This critical angle is usually interpreted in the framework of the ``magneto-centrifugal acceleration'' of \cite{BP82}: magnetic field lines are assumed to behave as rigid poloidal wires on which fluid particles are drifting (``bead on a wire'' analogy). If the field lines are sufficiently inclined, the particles can overcome the vertical gravity and are accelerated by the centrifugal force. 

\index{Magnetically-driven outflow}
This simple physical interpretation based on the Bernoulli invariant close to the wind base is very useful as a first approach to outflow physics, but it should not be taken too strictly as it leads to several misconceptions about the very nature of outflows. Let us underline the main physical differences between the magneto-centrifugal acceleration picture and the processes at work in an outflow: 
\begin{enumerate}
\item     Magnetic field lines \emph{do not behave as rigid poloidal wires}. As shown above, the magnetic configuration has a strong toroidal field at the wind base $B_\phi^+$ which indicates that the field is wound up in this region.
\item The constant angular velocity approximation is only valid close to the wind base (and far from the Alfv\'en radius). It is a result of the toroidal field tension which accelerates the flow azimuthally, hence the necessity of a wound field configuration. Winds having $R_A/R_0\sim 1$ will therefore experience a very small region of solid rotation (if any).
\item As discussed above, \emph{vertical acceleration is always a magnetic pressure effect} which does not rely on gravity, centrifugal acceleration or magnetic tension.
\end{enumerate}
As proposed by \cite{CL94} and \cite{F97}, it is therefore preferable to qualify outflows as ``magnetically driven'' instead of ``centrifugally driven''.

Let us summarise here the physics of a cold outflow close to the launching region up to the Alfv\'en point
\begin{itemize}
\item The flow is accelerated azimuthally by the magnetic tension due to $B_\phi$. This results in a transfer of angular momentum from $B_\phi$ to the gas as it accelerates.
\item Since $B_\phi^2$ (and possibly $B_R^2$) decrease with $z$, the flow is accelerated vertically by magnetic pressure.
\item Close to the launching point and far from the Alfv\'en radius, the outflow is approximately in solid rotation.
\item In a cold wind, energy conservation at the wind base implies $\theta>30^\circ$. 
\end{itemize}

\section{Global numerical models}

Many numerical models of protoplanetary discs have been published in the literature. Here, we will focus on models tackling the impact of non-ideal MHD effects and winds on the dynamics of a disc, since we focus on the outer parts of protoplanetary discs. We therefore exclude ``ideal MHD'' models and simulations without a large scale magnetic field.

This first model to investigate this regime was published by \cite{GTN15}. However, the limited vertical extension of this work (typically 4 scale heights) makes the interpretation of outflow properties difficult. Therefore, we focus on models with a larger vertical extension such as the ones published by \cite{BLF17} and \cite{B17}.

\subsection{Global topology}

\begin{wrapfigure}{r}{0.5\linewidth}
\begin{center}
    \includegraphics[width=0.9\linewidth]{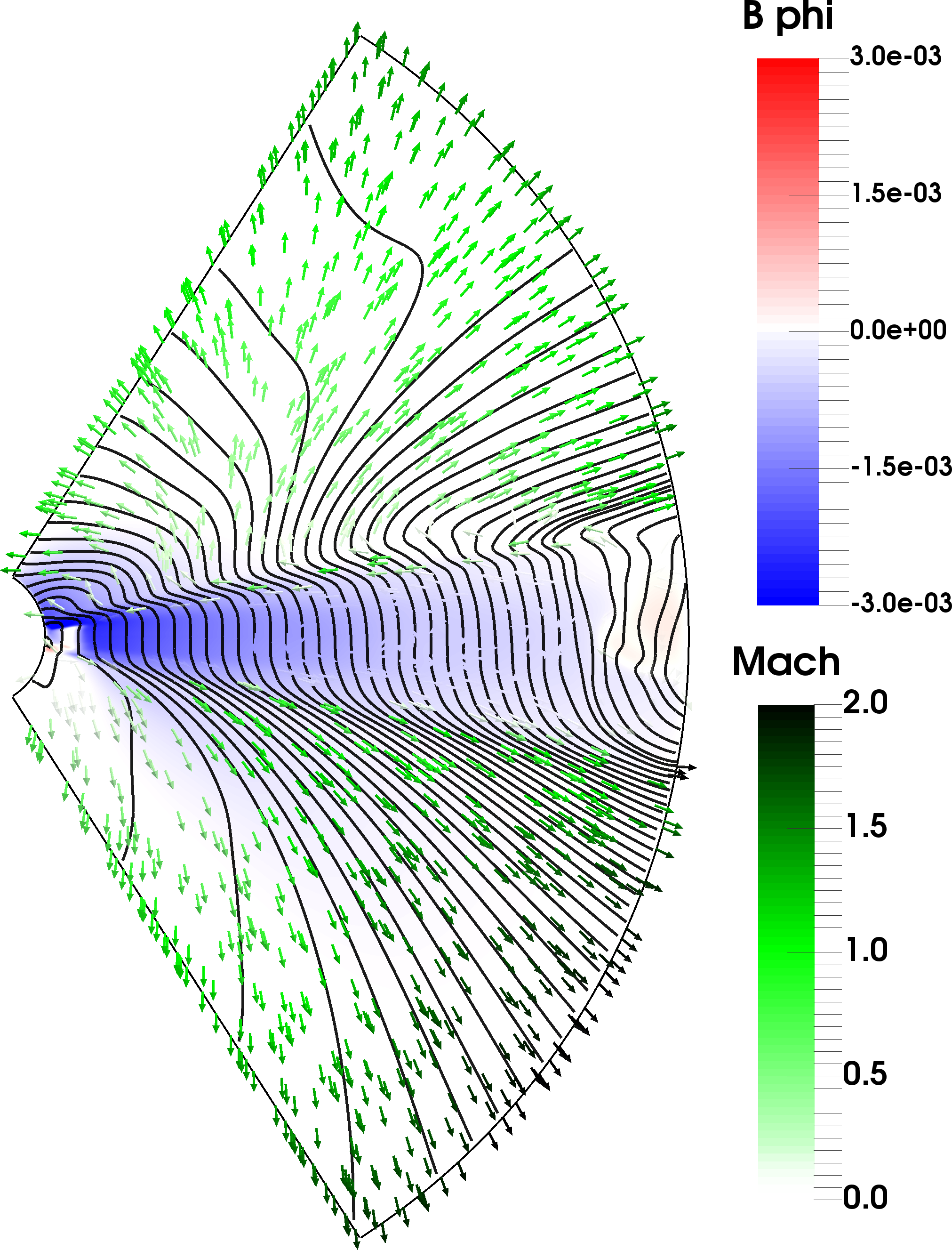}
    
    \caption{Global simulation for a wind in a protoplanetary disc which exhibits a dissymmetric outflow. Black lines are poloidal magnetic field lines, green arrows represent the poloidal velocity and the background colour traces the azimuthal field $B_\phi$. Close to the midplane, the configuration has an odd symmetry, as found in shearing box models and a current sheet is found at $z\sim 3H$. Figure from \cite{BLF17}.}
    \label{fig:glob_symmetry}
\end{center}
\end{wrapfigure}

One of the main problems with shearing-box models is the presence of an odd symmetry for the solution, leading to difficulties to interpret the role played by the outflow. The use of global models avoids this problem since in this case, the gravitational potential is not symmetrical with respect to $r$. Many models seem to converge towards \emph{dissymmetric} outflows, i.e. outflows which are neither even or odd, but are essentially odd for $-3H<z<3H$ and exhibit a strong current sheet on one side of the disc. In this configuration, the outflow is dissymmetric, and more mass ejected from one side of the disc than the other (with ratios of $\dot{M}_w$ reaching a factor of a few). This kind of solution was found both with only Ohmic and ambipolar diffusion \citep{GTN15} and with all the three non-ideal effects \citep{BLF17,B17}.

The presence of an odd symmetry in the midplane region is reminiscent of the shearing box solutions described in \S\ref{sec:meanfield_outflows}. However, this symmetry is not verified far away from the midplane as global effects enter the scene. \cite{GTN15} have proposed that the global field structure was playing a significant role in shaping the outflow. Indeed, by choosing initially a field configuration with $\partial_RB_z>0$, it is possible to produce outflows directed inwards (towards the star). It's been therefore proposed that the global radial magnetic pressure gradient was responsible for the symmetry breaking observed in these models. Also, the vertical rotation profile (and therefore the vertical temperature structure) might be playing a role by shearing the poloidal field \citep{GTN15}. The question of the origin of the global outflow configuration therefore remains open.

Outflows are not always dissymmetric. Indeed, both \cite{BLF17} and \cite{B17} have reported symmetric (even) outflow configurations. These symmetric configurations seem to be found mostly when $B_z$ is weak enough ($\beta_\mathrm{mid}\gtrsim 10^4$) \citep{BLF17,B17} and for anti-aligned cases ($B_z\Omega<0$). The sensitivity of the outflow configuration on the field polarity suggests that the HSI is partly responsible for the outflow configuration. However, there is no one-to-one correspondence between the field alignment/strength and the outflow configuration, and the choice of initial conditions also seems to be playing a non-negligible role. If true, it might be desirable to start from a magnetic configuration as close as possible to the configuration expected from core collapse calculations \citep{B17}.

\subsection{Accretion}

\begin{figure}
\begin{center}
    \includegraphics[width=0.60\linewidth]{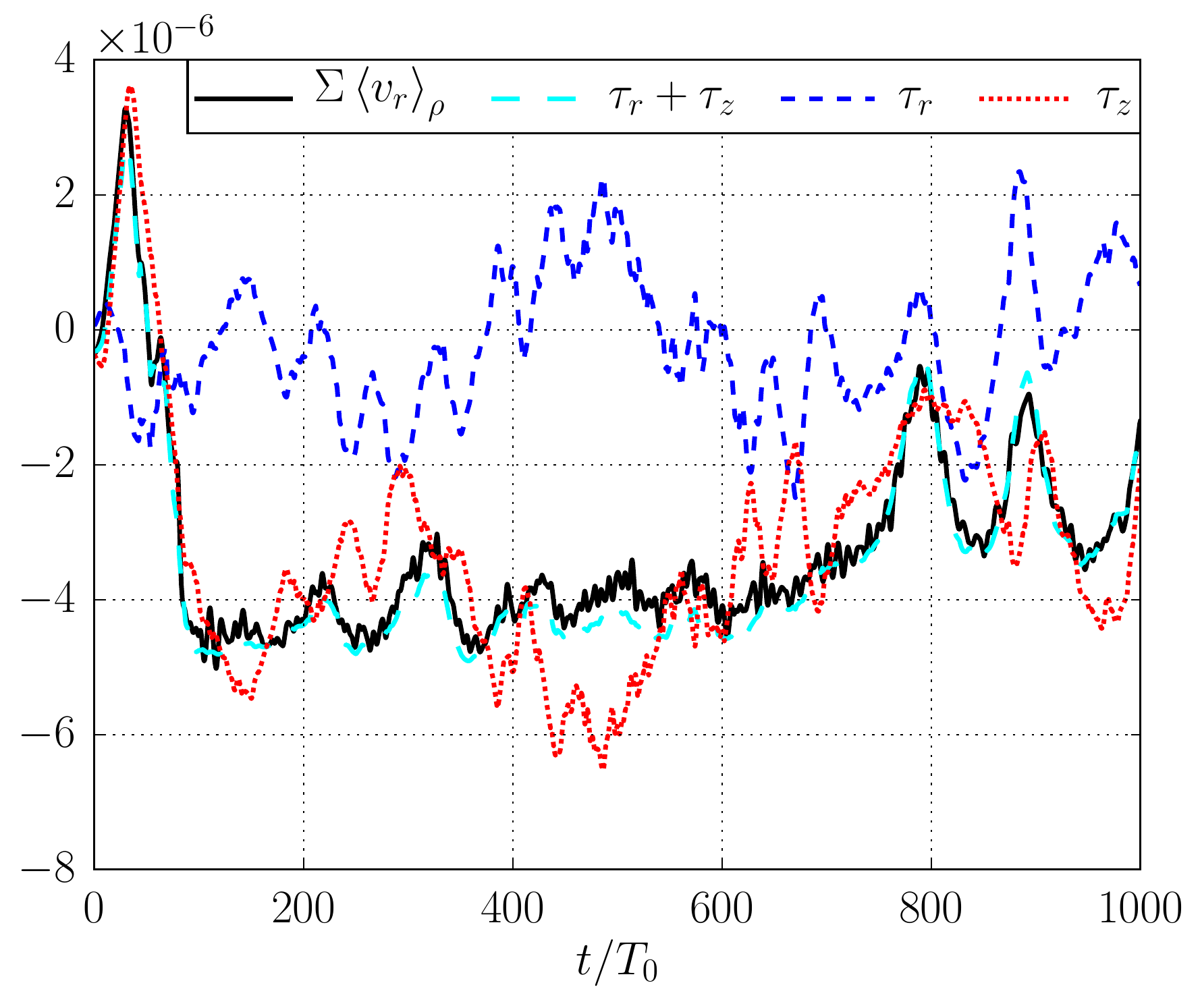}
    \caption{Measured accretion rate as a function of time in a non-ideal model. The accretion rate and torque contributions have been average radially. Most of the accretion is due to the wind ($\tau_z$ term) while the radial stress does not seem to contribute to the angular momentum budget. Figure from \citep{BLF17}.}
    \label{fig:glob_accretion}
\end{center}
\end{figure}

The question of the engine driving accretion can be directly addressed in global simulations. Indeed, one can measure individually each term in the angular momentum conservation equation
\begin{align}
\label{eq:torques}
\dot{M}_\mathrm{acc}=\frac{4\pi}{R\Omega_K}\Bigg\{\underbrace{\frac{\partial}{\partial R} R^2\Bigg[\overline{\rho v_\phi v_r}-\frac{\overline{B_\phi B_r}}{4\pi}\Bigg]}_{\tau_r}+\underbrace{\Bigg[R^2\rho v_\phi v_z-R^2\frac{B_\phi B_z}{4\pi}\Bigg]_{z=-h}^{+h}}_{\tau_z}\Bigg\},
\end{align}
where we have defined the mass accretion rate $\dot{M}_\mathrm{acc}=-2\pi R\overline{\rho v_r}$ and the radial and vertical torques $\tau_{r,z}$. An example of such a measure is given in Fig.~\ref{fig:glob_accretion}. We find that accretion is mostly due to the wind, the surface torque being the main contribution to angular momentum extraction in the disc.

Note however that the fact that the radial \emph{torque} is negligible does not imply that the radial \emph{stress} is also negligible. As can be seen from (\ref{eq:torques}), one can cancel the radial torque if the radial stress is proportional to $1/R^2$. In \cite{BLF17}, a strong laminar radial stress is indeed present in the disc, with effective $\alpha$ values reaching a few times $10^{-2}$. However, because of the surface density profile, the net torque exerted on the disc mostly cancels out. Similar behaviours have been obtained by \cite{B17}, with disc regions exhibiting positive and negative torques.

Overall, it is clear that in these models, the wind torque is playing a significant if not dominant role in the mass accretion rate. The radial torque, on the other hand, is less straightforward as it depends on the initial conditions chosen for the model. Accretion, decretion or both can be obtained from the radial torque, despite the presence of a relatively strong positive laminar radial stress when the field is aligned with the vertical rotation axis, as in shearing box models (see \S\ref{sec:nimhdstrat}). 

Despite the uncertainties, these models predict mass accretion rates $\dot{M}_\mathrm{acc}\sim 10^{-8}-10^{-7}\,M_\odot/\mathrm{yr}$. However, even these values should be interpreted with care, as they are usually measured in the middle of the simulation domain, typically 5-20 AU from the central star. If an outflow is indeed present and carrying mass away, the mass accretion rate \emph{onto the star} can be significantly smaller than the one derived in the bulk of the disc. From the definition of the ejection efficiency (\ref{eq:xi_def}), we have
\begin{align*}
\dot{M}_\mathrm{acc}(R)=\dot{M}_\mathrm{acc}(R_0)\Bigg(\frac{R}{R_0}\Bigg)^{\xi}    .
\end{align*}
Large ejection efficiencies ($\xi=O(1)$) as the one found in recent global models therefore lead to dramatically reduced mass accretion rates at the inner radius of the disc.

Unless one assumes an ionisation rate much higher than the one expected in these objects (see \S\ref{sec:nonideal}), the flow is mostly laminar, with a very low time-dependency. Hence, the radial stress measured in these models is not the usual turbulent stress found in ideal simulations, but really a purely magnetic term with no velocity counterpart. This implies that dust grains present in the disc will be less subject to turbulent fluctuations and will therefore settle towards the disc midplane more rapidly.

\subsection{Ejection and mass loss rate}

The outflow is not only responsible for carrying angular momentum away from the disc, but it also contributes significantly to the mass loss of the disc. All of the simulations published up to now find that the mass loss rate in the outflow is, broadly speaking, comparable to, or even larger than the mass accretion rate in the disc ($\dot{M}_\mathrm{w}\gtrsim 10^{-8}-10^{-7}\,M_\odot/\mathrm{yr}$), which implies $\xi\sim 1$. 

The mass outflow rate is tightly connected to the amount of flux threading the disc, with $\dot{M}_\mathrm{w}\propto \beta_\mathrm{mid}^{-1/2}$ \citep{BLF17}, indicating that the mass flux is proportional to the magnetic flux threading the disc. Interestingly, similar scalings are obtained in non-ideal shearing box models \citep{BS13,LKF14} while steeper dependencies are found in ideal shearing box models \citep{SI09,BS13b}. 

The engine driving ejection can be isolated first by looking at the magnetic level arm of these outflows. Most of the models published up to now find level arms $\lambda < 2$. This is coherent with the very high mass loss rates found in these simulations (high $\xi$, see Eq.~\ref{eq:xi_constraint}). 
\add{While it is in principle possible to obtain cold outflows with low $\lambda$ and high $\xi$ in discs threaded by a weak field ($\bmean\gg 1$, see for instance \citealt{JFL19}), some of the outflows published to date in protoplanetary discs have $\lambda< 3/2$ \citep[e.g.][]{BLF17}, which violate the cold MHD wind constraint (Eq.~\ref{glob:lambda_energy}). Hence these outflows are \emph{not purely magnetically-driven}.}

\begin{figure}
\begin{center}
    \includegraphics[width=0.60\linewidth]{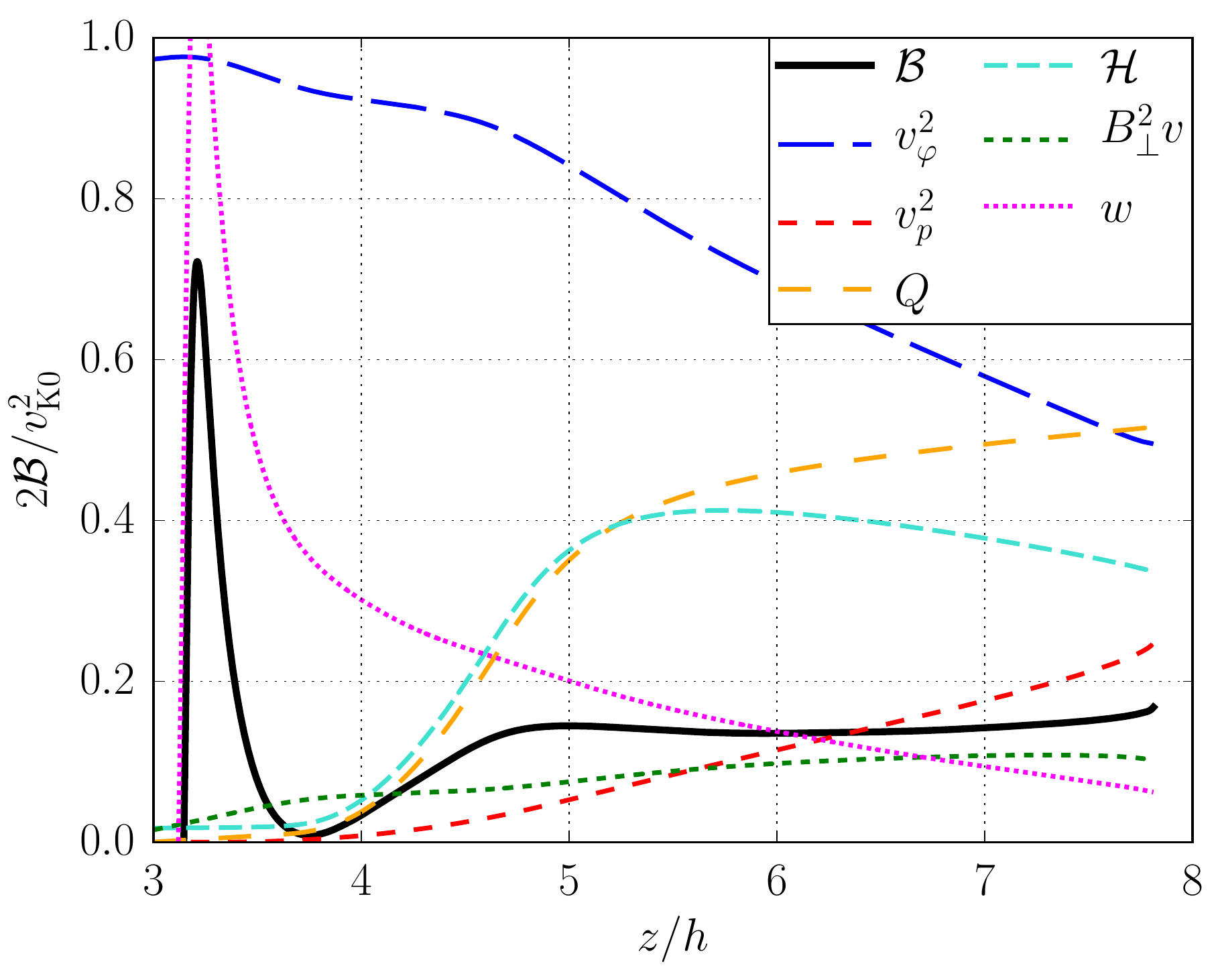}
    \caption{Bernoulli invariant $\mathcal{B}$ measured in an outflow driven from a non-ideal protoplanetary disc.  Magnetic contribution are $B_\perp^2v$ and $w$, thermal contributions comes from the enthalpy $\mathcal{H}$ and external heating $\mathcal{Q}$ while kinetic energy terms are represented by $v_\phi^2$
     and $v_p^2$. Magnetic terms contribute significantly close to the launching point, while thermal energy becomes important higher up in the outflow. The ideal MHD region starts for $z\gtrsim 5h$. Figure from \cite{BLF17}.}
    \label{fig:bernie_will}
\end{center}
\end{figure}

\begin{figure}
\begin{center}
\begin{tabular}{cc}
    \includegraphics[width=0.45\linewidth]{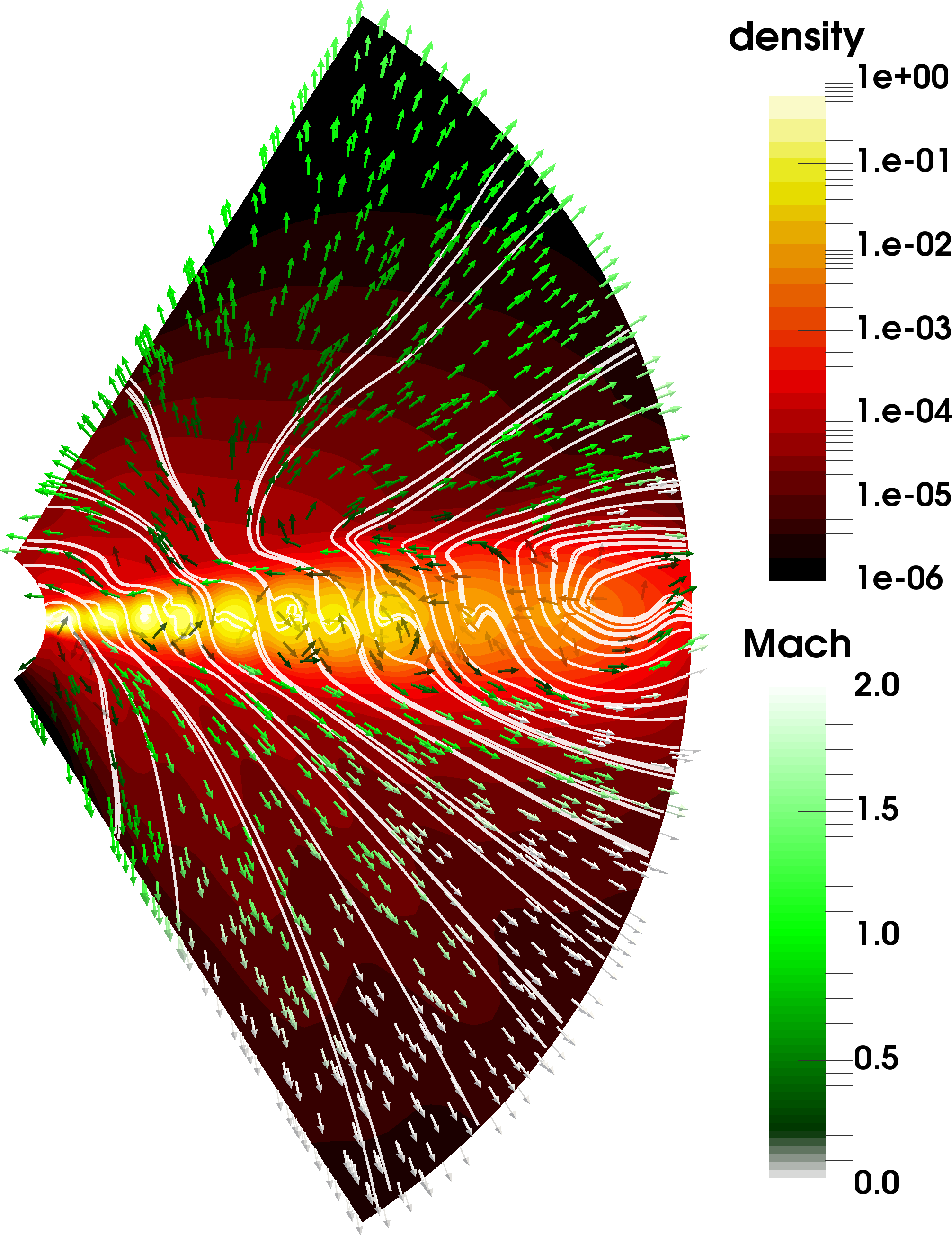}& 
    \includegraphics[width=0.45\linewidth]{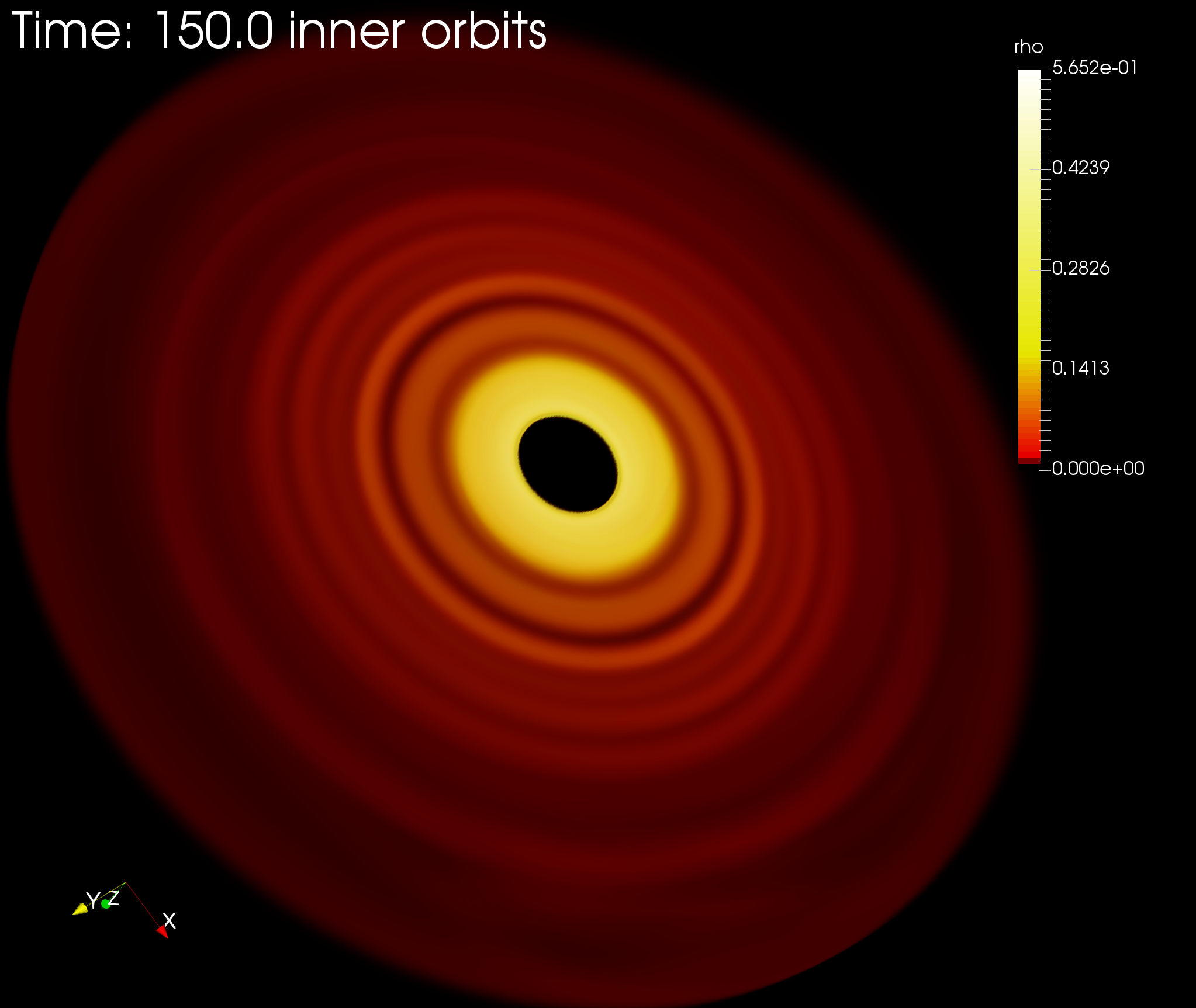}
    \vspace{0.5cm}
\end{tabular}
    \caption{Left: self organisation in a simulation with $\bmid=10^2$ computed in 2.5D. The density is represented in colormap while magnetic field lines are in white lines and velocity field is shown in green arrows. Notice that field lines are accumulated in regions of reduced density in the midplane. Figure from \citep{BLF17}. Right: volume rendering of a similar model, this time computed in full 3D. Note that the flow remains axisymmetric.}
    \label{fig:selforg_glob}
\end{center}
\end{figure}

This conclusion can also be reached by analysing directly the Bernoulli function of the outflow (e.g. fig.~\ref{fig:bernie_will}). Such an analysis shows that thermal effects (enthalpy and heating terms) both contribute significantly to the energetic of the outflow \citep{BLF17}. Still, magnetic effects are clearly not negligible at the base of the outflow, where magnetic pressure helps pushing the flow upward. 

\index{Magneto-thermal outflow}
For these reasons, these outflows have been labelled ``magneto-thermal''. This kind of outflow has already been identified in self-similar solutions by \cite{CF00}. Compared to historical cold wind solutions, they are (obviously) warmer, denser and slower. They reach high $\xi$ values (typically $\xi>0.1$) and have moderate $\lambda$. Of course, the fact that they extract angular momentum from the disc and that the initial acceleration is due to magnetic effects implies that they are not purely thermal.

\subsection{Self-organisation}

\index{Self-organisation!global models}
Self-organisation was unambiguously identified in global simulations by \cite{BLF17} for simulations with $\bmid\lesssim 10^3$, but it is absent from the models of \cite{GTN15} and \cite{B17}, who only considered $\beta\gtrsim 10^4$. It is most of the time found in simulations exhibiting dissymmetric wind configurations but does not seem to prefer one given field polarity, as would be expected from Hall-driven self-organisation. A careful examination of the flow shows that the poloidal field lines are concentrated in low-density regions. Hence, magnetic effects are playing a very important role in the mechanism.

It is possible to identify which process is responsible for self-organisation by looking closely at the non-ideal induction equation. It is then found that ambipolar diffusion is the only term responsible for the accumulation of magnetic flux in narrow regions, while Ohm and Hall effects are both diffusing the field away \citep{BLF17}. Hence, despite the presence of a rather strong Hall effect in these simulations, it is not the Hall-driven self-organisation which is at work in these models but ambipolar driven self-organisation. In essence, the mechanism seems to be similar to the one driving self-organisation in stratified shearing box models subject to ambipolar diffusion only \citep{B15}. The local configuration found in the global simulations is indeed identical to the configuration found in shearing boxes (see Fig.~\ref{fig:MRI_strat_selforg}), making the shearing box model a valuable tool to understand self-organisation in this regime.

Unfortunately, there is today no general theory predicting in which situation self-organisation is occurring nor what are the general properties of the structures which are formed.

\section{Conclusions}

In conclusions, the global modelling of the weakly ionised part of protoplanetary discs ($R\gtrsim 1\,\mathrm{AU}$) is still in its infancy. The inclusion of non-ideal MHD effects confirms that the disc midplane is mostly laminar, while the disc is still accreting thanks to magnetised outflows. Because of the low magnetisations used in these models ($\bmid\gtrsim 10^2-10^3$), the outflow is not purely magnetically-driven as external heating and thermal pressure are found to contribute to the global energy budget. This implies that it is difficult to draw systematic conclusions regarding the accretion rate and mass loss rate, as these can depend both on the ionisation structure but also on the thermodynamics of the disc wind, both of which are plagued by huge uncertainties.

It would be tempting to consider stronger field models with $\beta=O(1)$ which are known to lead to historical ``cold wind'' models \citep{F97}. However, such models would also have larger average accretion velocities \add{which are typically sonic}. If one wants to keep an accretion rate approximately compatible with observation, such a model would imply a disc surface density reduced by at least one order of magnitude (most likely two), which would be incompatible with the disc masses inferred from observations. Hence, the fact that the disc is massive, with an average accretion velocity much smaller than the speed of sound implies that if outflows exist in these regions, they must be due to a weak field: $\bmean \gg 1$.

Self-organisation is also a very promising mechanism to explain some of the observables in protoplanetary discs. However, the theoretical background for these features remains limited. The fact that they are seen both in global and local simulations is encouraging, but clearly, a detailed theoretical work is needed before any satisfactory prediction can be made and tested against observational data.

\newpage
\part{Summary and future directions}
\section{Summary}

These are exciting time for protoplanetary disc modelling and planet formation theory in general. Indeed, we now start to have direct observational constraints for these astrophysical objects, ranging from large \emph{resolved} structures, such as rings and non-axisymmetric bumps, to turbulent velocity dispersion measurements. Even magnetic field strength and topology are now beginning to be probed at the disc scale, giving more constraints to models.

The plasma in these discs is however relatively cold and therefore weekly ionised, reaching ionisation fraction $\xi\sim 10^{-13}$ in the disc midplane around one astronomical unit. This implies that non-ideal MHD effects (Ohmic, Ambipolar and Hall effect) are essential to get a proper description of the plasma. The amplitude of these effects, however, is poorly constraint. Because ionisation is mostly non-thermal, the amplitude of these non-ideal effects depends on the details of the ionisation sources, the disc structure, and the plasma composition (especially the abundance of tiny dust grains). Overall, there is an uncertainty of several orders of magnitude on these effects, implying that very detailed models including complex reaction networks are likely unnecessary at this stage since the input parameters (disc composition \& environment) are largely unknown.

In order to explain the observed accretion rates in these discs, and since angular momentum is a conserved quantity, one needs to find a way to remove the disc angular momentum, either by transporting it radially outwards in the disc bulk, or by transporting it vertically away in a magnetised wind. While radial transport has been historically favoured thanks to its simplicity and elegance (the well-known $\alpha$ disc model), vertical transport is now believed to be key in several astrophysical objects because of its high efficiency. 

The most favoured mechanism to explain angular momentum transport in discs is the magnetorotational instability (MRI), a linear, ideal MHD instability found in rotating sheared flows. This instability has been the subject of intense studies since the early 90s and it is known that it is strongly affected by the non-ideal effects present in protoplanetary discs. Most notably, it is suppressed in the regions where Ohmic and ambipolar diffusion are strong, and it gives a new branch, known as the Hall shear instability (HSI), when the Hall effect is dominant, in the case where the poloidal field is aligned with the rotation axis.

In the non-linear regime, the MRI behaviour strongly depends on the presence and strength of a mean vertical field threading the disc. Historically, most of the simulations published until early 2010 were in a regime without a mean field, commonly known as the "MRI dynamo". In ideal MHD, this regime is known to produced vigorous 3D turbulence and radial transport of angular momentum, but in non-ideal MHD, turbulence is suppressed, leading to a laminar flow, no angular momentum transport and no accretion. This "dead-zone" problem is circumvented by considering a mean vertical field threading the disc. Doing so, the MRI can indeed be revived in the upper layers, as expected from the linear analysis, but it then saturates into magnetised outflows, and the flow remains mostly laminar. In this situation, angular momentum is transported, mostly in the vertical direction, so accretion is saved, but its physical description then becomes fairly different from that of an $\alpha$ disc.

This connection between the MRI and magnetised outflow in discs threaded by a mean field was only realised during the past 10 years. In the presence of outflows, local models are insufficient since the dynamics of outflow is dictated by the global geometry of the system. In the case of protoplanetary discs, it is found that accretion driven by magnetised winds are compatible with observed accretion rates for relatively weak mean fields, $\bmean=O(10^4)$, which are compatible with the upper bounds on the field strength from observations. Because the outflow is in a weakly magnetised regime, it is usually found that the mass loss rate can be of the order of, or even larger than the mass accretion rate measured at the inner radius of the disc (this result being perfectly consistant with mass, angular momentum, and energy conservation). In addition the outflow is highly sensitive to thermal effects, which contribute significantly to the flow energetics, as is found in many models. Hence, these outflows have been called "magneto-thermal". 

\section{Perspectives}

Research on this topic is now following several paths. First, the fact that thermal effects can play a significant role implies that they should be modelled accurately. Several groups are now working actively on this problem \citep[e.g.][]{WBG19,G20}. It should be realised that thermal driving is a very complicated problem, as it involves heating by X-rays and UV photons, in addition to cooling, mostly by molecular and atomic lines. The computation of thermal processes therefore rely on complex chemical networks, coupled to radiative transfer codes, which are all computationally very intensive. Eventually, one hope is to find a way to simplify this physics using prescribed heating and cooling functions tested on complete models. This would allow a more systematic exploration of the long term impact of thermodynamics on these systems, and make a connection to winds observed in the sub-millimetric range.

A second question is the dynamical evolution of the mean magnetic field threading the disc. As shown above, this mean field is key  for magnetised outflows. It is strongly suspected that such a field should be present, as a direct result of the disc formation process, which relies on the collapse of a magnetised molecular cloud. During the collapse, a fraction of the magnetic field is trapped in the forming disc, and then plays the role of the mean field for outflows. However, once the disc is formed, it would be desirable to describe how this mean field evolves with time. It could be advected inwards by the accretion flow, leading to a strongly magnetised inner region, or it could inversely diffuse outwards because of non-ideal MHD effects. At the time of writing, this question is not settled, even qualitatively. Numerical models suggest that the field is diffusing outwards \citep{BS17,G20} while analytical models suggest inwards transport \citep{LO19}. If magnetised outflows are the dominant mechanism of accretion, then it is essential to address quantitatively this question in order to be able to model the long term dynamics of these discs, since mass and magnetic flux are tightly linked. This flux transport can be at the origin of complex dynamics in the disc, such as time variability or even eruptions, which are also observed in these systems.

A third axis of research is the impact of this dynamics on planet formation, from the dynamics and growth of dust grains to the migration of giant planets. Most of the literature published to date rely on the $\alpha$ disc paradigm, which itself assumes that the disc is turbulent. However, if accretion is driven by magnetised winds in a mostly laminar flow, this framework has to be revised. Indeed, the lack of turbulence affects the dynamics of large grains ($\gtrsim 10\mu\mathrm{m}$): vertical and radial settling, coagulation and disruption efficiency, etc, are all strongly modified. In addition, the fact that accretion is driven by surface stress, and not by turbulence is also going to reshape planet migration. It is not clear yet how type I and type II migration processes react to this shift in accretion paradigm, but the first attempts at including the wind stress in 2D planet migration numerical models already show a very significant impact \citep{KDK20}. Even the long term evolution of the disc is quite different from that of a viscous disc, as viscous spreading is absent for a wind-driven disc. These are only a few example, but it shows that many things which were thought to be well established are now standing on wobbly foundations.

\section*{Acknowledgements}
I thank the two anonymous referees who took the time to carefully read this work and whose remarks and questions greatly improved the initial version of the manuscript. I also thank Antoine Riols, Jonatan Jacquemin-Ide and Etienne Martel for their contribution in proof-reading several sections of the manuscript. This work has received funding from the European Research
Council (ERC) under the European Union’s Horizon 2020 research and innovation programme (Grant agreement No. 815559
(MHDiscs)).

\part*{Appendix}



\addcontentsline{toc}{part}{References}
\bibliographystyle{jpp}
\bibliography{biblio}
\printindex

\end{document}